\documentclass[epj-spec]{svjour}
\input psfig
\usepackage{epsfig}
\begin{document}
\title{A perturbative nonequilibrium renormalization group method for dissipative quantum mechanics}
\subtitle{Real-time RG in frequency space (RTRG-FS)}
\author{Herbert Schoeller\inst{1,2}\thanks{\email{schoeller@physik.rwth-aachen.de}}}
%
\institute{Institut f\"ur Theoretische Physik A, RWTH Aachen, 52056 Aachen, Germany
\and JARA-Fundamentals of Future Information Technology}
%
%
\abstract{
We study a generic problem of dissipative quantum mechanics,
a small local quantum system with discrete states coupled in an 
arbitrary way (i.e. not necessarily linear) to several infinitely 
large particle or heat reservoirs. For both bosonic or fermionic reservoirs 
we develop a quantum field-theoretical
diagrammatic formulation in Liouville space by expanding systematically
in the reservoir-system coupling and integrating out the
reservoir degrees of freedom. As a result we obtain a
kinetic equation for the reduced density matrix of the
quantum system. Based on this formalism, we present a formally exact 
perturbative renormalization group (RG) method from which the kernel
of this kinetic equation can be calculated. It is demonstrated how the 
nonequilibrium stationary state (induced by several reservoirs kept at different
chemical potentials or temperatures), arbitrary observables such as the
transport current, and the time evolution into the stationary state can be calculated.
Most importantly, we show how RG equations for the relaxation and dephasing rates 
can be derived and how they cut off generically the RG flow of the vertices. 
The method is based on a previously derived real-time RG technique 
\cite{hs_koenig_PRL00,keil_hs_PRB01,hs_lecture_notes_00,korb_reininghaus_hs_koenig_PRB07}
but formulated here in Laplace space and generalized to 
arbitrary reservoir-system couplings. Furthermore, for fermionic reservoirs with
flat density of states, we make use of
a recently introduced cutoff scheme on the imaginary frequency 
axis \cite{jakobs_meden_hs_PRL07}
which has several technical advantages. Besides the formal
set-up of the RG equations for generic problems of dissipative quantum
mechanics, we demonstrate the method by
applying it to the nonequilibrium isotropic Kondo model. We present a
systematic way to solve the RG equations analytically in the weak-coupling 
limit and provide an outlook of the applicability to the strong-coupling case.
} 
\maketitle
\section{Introduction}
\label{sec:1}
\begin{figure}
  \centerline{\psfig{figure=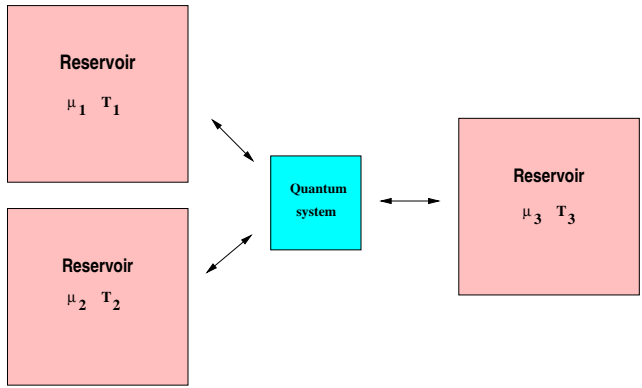,scale=0.5}}
  \caption{A small quantum system coupled to several infinitely large
    reservoirs via energy and/or particle exchange. The reservoirs are
    characterized by temperatures $T_i$ and chemical potentials $\mu_i$.}
\label{fig:sketch}
\end{figure}

{\bf General remarks.} Dissipative quantum mechanics is a fundamental field in 
theoretical physics combining the concepts of quantum mechanics
and nonequilibrium statistical mechanics \cite{diss_QM}. The aim is to develop
a microscopic description of how a small quantum-mechanical system in
contact with large reservoirs (see Fig.\ref{fig:sketch} for a sketch of the system)
evolves into a stationary state and 
what this stationary state looks like. In equilibrium statistical
mechanics for large quantum systems in contact with a single reservoir,
the strength and nature of the system-reservoir interaction is not important,
since it is a surface effect and negligible compared to the bulk of the
quantum system. Therefore, in this case, depending on which conserved
quantities are exchanged, the stationary state is always a canonical
or grandcanonical ensemble.
For quantum systems in contact with several reservoirs which are kept at
different temperatures or chemical potentials (inhomogeneous boundary
conditions), the stationary state is a nonequilibrium state and possibly
current-carrying. However, if the quantum system is large, again only the
interactions in the bulk are important which can be treated perturbatively via standard
quantum Boltzmann equations \cite{QBE}. After a short crossover time local 
equilibrium establishes and the further time evolution can be described
by hydrodynamic equations \cite{Hydro}. These standard tools of nonequilibrium
statistical mechanics break down for small quantum systems for several reasons.
First the system-reservoir coupling is no longer a negligible surface
effect but can change the states on the system considerably via quantum
fluctuations. Even for weak coupling renormalization of the coupling parameters
and the level positions of the quantum system can occur at low temperatures. 
Strinkingly new effects can occur for strong coupling such as a localization 
transition in the spin-boson model \cite{diss_QM} or resonant transmission
for a local spin coupled via exchange to reservoir spins (the
so-called Kondo effect) \cite{kondo_exp,kondo_theo}.
Secondly, it becomes energetically difficult to put several electrons
on the quantum system due to the large capacitive interaction $E_C\sim e^2/L$ ($L$ being
the length of the quantum system), the so-called
charging energy. Typical experimental values for semiconductor quantum 
dots, metallic islands or carbon nanotubes are $E_C\sim 1-10\mbox{K}$ and much larger than typical
temperatures 
$T\sim 10-100 \mbox{mK}$ (for single molecules coupled to leads the charging energy 
can be even larger reaching typical atomic values). A simple perturbative expansion
in the interaction is no longer possible and usual quantum Boltzmann equations
can not be used. Furthermore, for low-dimensional systems, the Coulomb interaction 
can lead to completely new physical phenomena such as Luttinger liquid behaviour
in 1-dimensional quantum wires \cite{Luttinger}. Concepts like local equilibrium 
are not applicable. Phase coherence is maintained over the whole system size and
the quantum system acts like a scattering region for electrons entering and leaving 
the system rather than a region where particles can relax, dephase and equilibrate.

For these reasons, new theoretical tools have been developed to understand
relaxation, dephasing, and nonequilibrium quantum transport through small quantum
systems coupled to external reservoirs. For noninteracting systems, the standard
tool is the Landauer-B\"uttiker formalism \cite{landauer_buettiker} where the particle
current is expressed by the scattering matrix together with the occupation of the scattering
waves determined by the chemical potentials of the reservoirs. In this case, it is possible
to consider arbitrary coupling between reservoirs and quantum system and the coherent
properties of the quantum system are fully taken into account. For interacting systems
(or systems with spin degrees of freedom), the situation is much more complicated
and no unique analytical or numerical formalism is available which can cover all regimes
of interest. Numerical methods for nonequilibrium are currently been developed, such as
time-dependent density matrix renormalization group (TD-DMRG) \cite{TD-DMRG}, time-dependent
numerical renormalization group \cite{TD-NRG}, numerical renormalization group at finite
bias voltage using scattering waves \cite{nonequilibrium-NRG}, quantum Monte Carlo with
complex chemical potentials \cite{QMC}, and iterative path-integral approaches \cite{thorwart_egger}.
Exact solutions using scattering Bethe-ansatz are available for the resonant level model
\cite{bethe}. Concerning analytical methods two perturbative approaches are commonly used,
depending on whether one expands in the interaction parameter inside the quantum system
or in the reservoir-system coupling. Expanding in the interaction has the advantage that
the unperturbed part of the Hamiltonian is quadratic in the field operators and standard
Keldysh-Green's function techniques can be applied \cite{caroli,jauho}. Furthermore, rather
large quantum systems can be treated since the effort scales with the number of single-particle
levels rather than with the number of many-particle states. However, this 
method has its limitations since for typical quantum dots the Coulomb interaction is the 
largest energy scale of the problem. In contrast, expanding in the reservoir-system coupling has the
advantage that arbitrary interaction strength  on the quantum sytem can be treated. Furthermore,
the reservoir-system coupling is often tunable in experiments and in most cases the lowest energy scale.
Therefore it seems reasonable to expand around the point where reservoirs and quantum system
are decoupled. In this case the unperturbed part of the Hamiltonian contains the full 
interacting quantum system and standard Green's function techniques can not be applied (a generic
problem for all strongly correlated systems). One way out of this problem is the use of
slave particle techniques where the interacting system is expressed in a quadratic form
using creation and annihilation operators of many-particle states \cite{slave_particles_general,barnes}.  
Standard Keldysh-Green's function methods can then be used by expanding in the reservoir-system
coupling \cite{slave_particles_wingreen}. However, technical complications arise due to an 
additional constraint for the
slave particle number and, most importantly, diagrammatic approximations are often doubtful
due to the unphysical nature of the slave particles. So even the noninteracting case is quite
nontrivial \cite{barnes} and vertex corrections are essential to obtain the 
relaxation and dephasing rates for the physical particles 
\cite{paaske_rosch_kroha_woelfle_PRB04}. In contrast, 
the most standard method of dissipative quantum
mechanics is to integrate out only the noninteracting reservoirs and describe the dynamics
of the reduced density matrix of the quantum system via a kinetic equation. This can be
achieved by projection operator techniques in Liouville space \cite{zwanzig} or via 
path-integral methods \cite{diss_QM,grifoni}. Most recently, a quantum field-theoretical version
of this strategy has been developed in 
Refs.~\cite{hs_schoen_PRB94,koenig_hs_schoen_EPL95,koenig_schmid_hs_schoen_PRB96},
see Ref.~\cite{hs_habil} for a review.
The advantage is that Wick's theorem is used to integrate out the reservoirs and an exact
diagrammatic representation of the kernel determining the kinetic equation is obtained. This
allows for a direct calculation of this kernel in terms of irreducible diagrams whereas with
projection operator techniques the calculation is complicated by cancellations of reducible expressions.
Furthermore, the usage of representations similiar to Feynman diagrams simplifies the implementation
of renormalization group ideas.

\begin{figure}
  \centerline{\psfig{figure=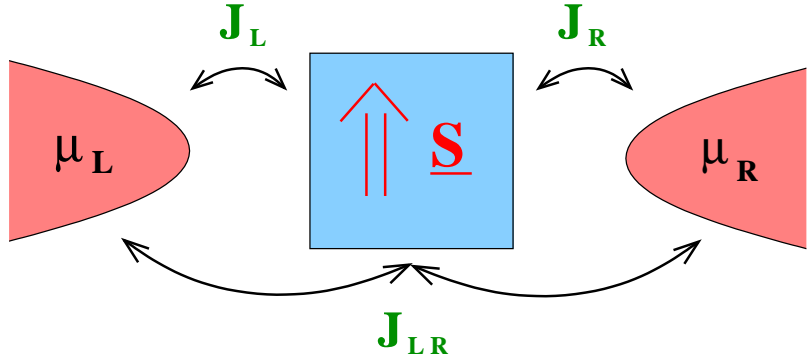,scale=1}}
  \caption{A spin-${1\over 2}$ quantum system coupled via exchange
to two reservoirs. $J_L = J_{LL}$ and $J_R =J_{RR}$ involve exchange between
the spins of the left/right reservoir with the local spin, and 
$J_{LR}=J_{RL}$ transfers a particle from one reservoir to the
other during the exchange process.} 
\label{fig:kondo}
\end{figure}

{\bf RG for the Kondo model.} The analytical methods expanding either in the interaction or the reservoir-system coupling 
can of course be used to calculate all physical quantities of interest in perturbation theory.
In this way many time-dependent and nonequilibrium phenomena have been described in various 
fields, such as spin boson models, quantum optics, mesoscopic systems, quantum information
theory, etc. However, perturbation theory is often plagued by various diverging terms in higher
orders if some low energy scale like temperature, voltage, magnetic field, etc. becomes too small.
In this limit perturbation theory becomes ill-defined and the natural question arises whether 
renormalization group methods can be generalized to the nonequilibrium situation to resum the
original perturbation theory in an appropriate way so that it becomes well-defined again. To be
specific let us introduce a simple but nontrivial example, the nonequilibrium Kondo model.
This model is currently one of the basic unsolved problems of nonequilibrium condensed matter physics 
and serves as an important benchmark model for theoreticians to test the applicability of their
nonequilibrium techniques. The model consists of the most simplest fermionic quantum system one can imagine,
namely a spin-${1\over 2}$ system, which interacts with two fermionic reservoirs (being kept at temperature
$T$ and two different chemical potentials $\mu_L$ and $\mu_R$) via exchange processes, 
see Fig.\ref{fig:kondo}.
The Hamiltonian for the reservoir-system coupling reads
\begin{equation}
\label{kondo}
V \,=\, {1\over 2}\sum_{\alpha\alpha'=L,R}\,\,\,\sum_{\sigma\sigma'=\uparrow ,\downarrow}\,\,
\sum_{kk'}\, J_{\alpha\alpha'}\,\underline{S}\cdot\underline{\sigma}_{\sigma\sigma'}\,\,
a^\dagger_{\alpha\sigma k} a_{\alpha'\sigma' k'}\quad .
\end{equation}
Here, $\underline{S}$ denotes the spin operator of the quantum system, $\underline{\sigma}$
are the Pauli matrices, and $a,a^\dagger$ are the fermionic annihilation and creation operators
for the particles in the reservoirs characterized by reservoir index $\alpha=L,R$, spin 
$\sigma=\uparrow,\downarrow$ and state index $k$. The exchange couplings are denoted by
$J_{\alpha\alpha'}$. For a symmetric system, there are only two different exchange couplings
\begin{equation}
\label{exchange_couplings_symmetric}
J_d = J_{LL} = J_{RR} \quad , \quad 
J_{nd} = J_{LR} = J_{RL} \quad.
\end{equation}
The diagonal exchange coupling $J_d$ characterizes spin exchange which involves only the
spin of one reservoir, whereas the nondiagonal coupling $J_{nd}$ describes a process where
one particle is transferred between the reservoirs (see e.g. 
Ref.~\cite{korb_reininghaus_hs_koenig_PRB07} for a derivation of this form of the 
interaction via a standard Schrieffer-Wolff transformation
from a conventional tunneling Hamiltonian). The latter process leads to a particle 
current at finite bias. For simplicity we consider only the isotropic case without magnetic field. 
The reservoirs are characterized by a noninteracting Hamiltonian
\begin{equation}
\label{reservoir_discret}
H_{res} \,=\, \sum_{\alpha=L,R}\,\,H_\alpha \,=\,
\sum_{\alpha=L,R}\,\,\epsilon_{\alpha \sigma k}\,\,
a^\dagger_{\alpha\sigma k} a_{\alpha \sigma k}\quad ,
\end{equation}
and the statistics is given by an equilibrium grandcanonical distribution
\begin{equation}
\label{grandcanonical}
\rho_{res} \,=\, \prod_{\alpha=L,R} \,\, \rho_\alpha \quad , \quad
\rho_\alpha \,=\, {1\over Z_\alpha}\,e^{-{1\over T_\alpha}(H_\alpha - \mu_\alpha N_\alpha)}\quad,
\end{equation}
where $T_L=T_R=T$ denotes the temperature and $\mu_L = -\mu_R = V/2$ the chemical potentials
of the reservoirs ($V$ is the bias voltage and we use units $e=\hbar=1$). 

Even in equilibrium $V=0$ the Kondo model is a highly nontrivial model and
numerous many-body techniques have been used to study its properties, see e.g.
\cite{hewson} for a review. We summarize here shortly its basic properties and
consider the simplest case $J=J_d=J_{nd}$. Higher-order perturbation theory in $J$
for transition rates (or the linear conductance) leads generically to logarithmic 
divergencies $\sim (J \,\mbox{ln}(D/\Lambda_c))^n$, where 
$D$ denotes the bandwidth of the reservoirs and $\Lambda_c\sim T$ is the 
low-energy scale. For the linear conductance, the
logarithmic terms start in $O(J^3)$  
\begin{equation}
\label{G_pert}
G\,=\,G_0\,{3\pi^2 \over 4}\,J^2\, (1+4J\,\mbox{ln}(D/\Lambda_c))\quad,
\end{equation}
with $G_0=2e^2/h$. To resum the most divergent terms in each order of 
perturbation theory (the so-called leading-order analysis), poor man scaling
methods have been developed \cite{poor_man_scaling} where the high energy scales
of the reservoirs are successively integrated out. If $\Lambda$ denotes the
effective bandwidth of the reservoirs (with initial value $D$), an infinitesimal
reduction $\Lambda \rightarrow \Lambda - d\Lambda$ is compensated by a 
renormalization of the exchange coupling $J$ while keeping the scattering
t-matrix invariant (in leading order). As a result one finds the so-called
poor man scaling equation (valid for $J\ll 1$)
\begin{equation}
\label{rg_kondo}
{d J \over d\Lambda}\,=\, - {2 J^2 \over \Lambda}\quad,
\end{equation}
with the solution
\begin{equation}
\label{J_poor_man_scaling}
J(\Lambda)\,=\, {1\over 2 \,\mbox{ln}(\Lambda / T_K)} \quad , \quad 
T_K\,=\, \Lambda \, e^{-{1\over 2J}}\quad,
\end{equation}
where $T_K$ denotes the Kondo temperature which is an invariant of the RG equation
(\ref{rg_kondo}). In the antiferromagnetic case $J>0$, the effective coupling
diverges at $\Lambda = T_K$ indicating a complete screening of the local spin
by the reservoir spins. This leads to a spin-singlet ground state and it can
be shown that the remaining potential scattering terms lead to unitary conductance
(Kondo effect). The Kondo effect has been measured for semiconductor quantum dots, 
carbon nanotubes, and molecules \cite{kondo_exp} (for theoretical works
see Refs.~\cite{kondo_theo,slave_particles_wingreen,koenig_schmid_hs_schoen_PRB96} or
Ref.~\cite{glazman_pustilnik_05} for a review). This is the so-called strong-coupling regime 
where the perturbative RG equation (\ref{rg_kondo}) is no longer valid. However, the poor man 
scaling equation has to be cut off by the low-energy scale $\Lambda_c\sim T$ and the strong coupling regime 
can not be reached for $\Lambda_c \gg T_K$. The weak coupling regime is defined by
$J_c=J(\Lambda_c)\ll 1$, where 
perturbation theory has to be carried out in the renormalized coupling $J_c$ 
with an effective bandwidth given by $\Lambda_c$. Replacing $J\rightarrow J_c$ and
$D\rightarrow \Lambda_c$ in (\ref{G_pert}), one obtains
\begin{equation}
\label{G_poor_man_scaling}
G\,=\,G_0\,{3\pi^2 \over 4}\,J_c^2 \,=\,G_0\,{3\pi^2 \over 16}\,{1\over \mbox{ln}^2(T/T_K)}\quad. 
\end{equation}
Since $J_c$ is logarithmically increased compared to the bare coupling $J_0 = J(D)$,
the onset of the Kondo effect is indicated by a logarithmic enhancement of the
conductance as function of temperature. Note that simple perturbation theory
in the original coupling $J_0$ can already break down in this regime since
the two conditions 
\begin{equation}
\label{break_down}
J_0 \,\mbox{ln}(D/\Lambda_c) \,=\, {1\over 2} - J_0 \,\mbox{ln}(\Lambda_c/T_K) \,\sim\, O(1)
\quad \mbox{and} \quad J_c \,=\, {1\over 2\mbox{ln}(\Lambda_c/T_K)}\,\ll\, 1  
\end{equation}
can easily be fulfilled, provided that $1\ll \mbox{ln}(\Lambda_c/T_K)\ll 1/J_0$
(although the typical experimental situation is rarely strictly in this regime, it provides a
well-defined regime where perturbative renormalization group methods can be tested).

In nonequilibrium $V\ne 0$ and $V>T$, the situation is not so clear. The standard
poor man scaling procedure suggests that the low-energy scale $\Lambda_c$ is given by
the maximum of $T$ and $V$, i.e. the voltage serves as a cutoff parameter in the same
way as temperature. Alternatively it was conjectured in Ref.~\cite{coleman} that $J_d$
is cut off by $T$ (since it involves only the spins of one reservoir) and $J_{nd}$ is
cut off by $\mbox{max}\{T,V\}$, opening up the possibility of a strong-coupling fixed
point for $V \gg T_K \gg T$. Finally, in Refs.~\cite{kaminski_etal,rosch_kroha_woelfle_PRL01} 
it was proposed that also spin relaxation and dephasing rates can cut off the RG flow. For
the isotropic Kondo model in the absence of a magnetic field the relaxation and 
dephasing rates are the same and are given by the Korringa rate
\begin{equation}
\label{Korringa}
\Gamma \,=\, \pi\,J^2\,\mbox{max}\{T,V\}\quad.
\end{equation}
This has the consequence that $\Gamma \gg T_K$ if $V\gg T_K$ (note that $T_K$ is
exponentially small for $J\ll 1$ whereas $\Gamma$ scales quadratic with $J$) 
and a strong-coupling fixed point can not be reached for $V \gg T_K \gg T$.
In this paper we will provide a microscopic nonequilibrium RG approach clarifying
all these issues. Indeed, it turns out that the Korringa rate cuts off
the RG flow of $J_d$ and $J_{nd}$. Roughly the poor man scaling equation
(\ref{rg_kondo}) has to be replaced by the RG equations 
(see Eqs.~(\ref{rg_J_two_reservoirs_diagonal_strong}) and (\ref{rg_J_two_reservoirs_nondiagonal_strong})
of Sec.~\ref{sec:5.3})
\begin{eqnarray}
\label{rg_Jd_appr}
{d J_d \over d\Lambda} &=& -\theta_T\,
\left({J_d^2 \over \Lambda + \Gamma}
+ J_{nd}^2\,\,\mbox{Re}{1\over \Lambda+\Gamma+iV}\right)\quad,\\
\label{rg_Jnd_appr}
{d J_{nd} \over d\Lambda} &=& -\theta_T\,
J_d J_{nd}\,\,\mbox{Re}\left\{
{1\over \Lambda+\Gamma}+{1\over \Lambda+\Gamma+iV}\right\}\quad,
\end{eqnarray}
with $\theta_T = \theta(\Lambda - T)$, where we have omitted additional frequency 
dependencies of the vertices (see Sec.~\ref{sec:5.3} for more details where also
the RG equation for $\Gamma$ is shown). As one
can see, temperature and Korringa rate cut off all terms on the r.h.s of the
RG equation whereas the voltage is only present in certain terms. This shows
that nonequilibrium induces new features into the RG equations. Neither $J_d$
nor $J_{nd}$ are completely cut off by the voltage and from these RG equations
alone there is no justification to cut off the RG flow by the voltage if $V>T$.
Instead, the RG flow should be cut off by $\Gamma$ leading to additional 
$J_c^2 \,\mbox{ln}(J_c)$
contributions for the exchange couplings. However, as will be discussed in detail in
Sec.~\ref{sec:5.3}, there is an additional RG equation for the current rate
which is cut off by the voltage and not by $\Gamma$. As a consequence the cutoff
parameter which enters into the conductance is indeed the voltage and we obtain
precisely the result (\ref{G_poor_man_scaling}) with temperature replaced by
bias voltage. However, it is important to recognize that the fact that
$\Gamma$ does not enter the final result for the conductance does not mean
that it can be neglected. For $T=0$ and $V\gg T_K$, the RG equations (\ref{rg_Jd_appr}) and
(\ref{rg_Jnd_appr}) would diverge for $\Lambda\rightarrow 0$ if the cutoff
parameter $\Gamma$ were not present, leading also to a divergence of the conductance
since the RG equation for the current rate is cut off smoothly by the
voltage, see Sec.~\ref{sec:5.3}.
Therefore, it is an important issue for nonequilibrium RG methods to incorporate
the physics of relaxation and dephasing rates into the RG equations.

The situation becomes even more interesting if an additional energy scale $\Delta$ is
present, such as magnetic field, external driving frequency,
level spacing, charge excitation energy etc. In this case, it turns out that additional
logarithmic terms can occur which are generically of the form 
$\mbox{ln}(D/|n{V\over 2}-\Delta|)$ with $n$ being an integer number, see 
Sec.~\ref{sec:4.3} and Refs.~\cite{reininghaus_hs_preprint,oppen_1}. At $V=2\Delta/n$
these logarithmic terms diverge corresponding to certain resonance positions. 
Relaxation and dephasing rates will also cut off these divergencies. Thus,
their effect is two-fold: relaxation and dephasing rates cut off the RG flow for the
vertices and also the RG flow for all physical observables, such as the current rate,
susceptibilities, occupation probabilities, etc. 
Furthermore, an interesting question is what happens in the strong-coupling
regime, where the RG flow of the relaxation and dephasing rates is cut off by
themselves. It is expected that they saturate to the Kondo temperature and
prevent the RG flow from diverging. This issue is discussed in Sec.~\ref{sec:5.3}.

{\bf RG methods.} Motivated by these considerations and due to the progress
in experimental techniques to study quantum transport through small
nanodevices, there is an increased interest in the development of
renormalization group methods for nonequilibrium systems (which are either
driven out of equilibrium with an external bias or are
prepared in an out of equilibrium initial state). The first proposal
for a nonequilibrium RG technique, commonly called real-time RG (RTRG),
was provided in Ref.~\cite{hs_koenig_PRL00} for problems of dissipative quantum
mechanics, see Ref.~\cite{hs_lecture_notes_00} for
a tutorial introduction. The method is perturbative in
the reservoir-system coupling and is based on the diagrammatic kinetic 
equation approach reviewed in Ref.~\cite{hs_habil}. Instead of the bandwidth,
the relative time of the reservoir Green's function was used as a cutoff
parameter and the kernel of the kinetic equation was kept invariant during
the RG flow by setting up a formally exact hierarchy of RG equations. The 
method has been applied to charge fluctuations in the
single electron box \cite{koenig_hs_PRL98}, transport through metallic quantum dots 
\cite{pohjola_koenig_hs_schoen_PRB99,hs_koenig_kuczera_schoen_JLTP00}, 
charge fluctuations in semiconductor quantum dots 
\cite{hs_koenig_PRL00,boese_hofstetter_hs_PRB01},
the polaron problem \cite{keil_hs_PRB00}, the influence of
acoustic phonons on transport through coupled quantum dots
\cite{keil_hs_PRB02}, and to the dynamics of the
spin boson problem \cite{keil_hs_PRB01}. 
Within all these applications, a linear coupling
between the quantum system and the particle or heat reservoir was
considered. In this case, it was sufficient to use the RG method
with a cutoff defined in time space. For models like the nonequilbrium
Kondo model, where spin fluctuations are important, a quadratic
coupling involving one creation and one annihilation operator of
the reservoirs occurs, see (\ref{kondo}). To treat this case in a convenient
way, a version of the RTRG method
with a cutoff defined in real frequency space has been developed recently
\cite{korb_reininghaus_hs_koenig_PRB07}. An essential ingredient of 
these techniques is the generation of non-Hamiltonian dynamics during
the RG flow in terms of an effective Liouville operator of the quantum
system with nonzero imaginary parts of its eigenvalues representing the
relaxation and dephasing rates. This effective Liouville operator appears
also in the RG equations of the vertices cutting them off at these
energy scales. However, a problem of this approach is the fact that a
certain time-ordering procedure has to be defined for the renormalized
vertices which leads to the generation of terms which are not present
in the original perturbation theory and have to be corrected by 
counter terms in higher orders. This leads to the problem
that the cutoff of the RG flow by relaxation and dephasing rates
can only be seen in leading order and a proof that this effect
happens in all orders is not possible (leaving the question open
whether a strong-coupling fixed point is in principle possible).
In addition, it is not unambigiously clear what the precise scale
of the rates cutting off the RG flow is. Finally, concerning the numerical
stability for solving the RG equations, it turns out that cutoff
functions which are defined in real time or real frequency space are not the
best choice. In this paper, we will show
how these problems can be circumvented by some new technical tricks which
have been used recently in Ref.~\cite{reininghaus_hs_preprint} to calculate 
analytically the nonlinear conductance, the magnetic susceptibility, the
spin relaxation and dephasing rates, and the renormalized g-factor up to 2-loop 
order for the nonequilibrium anisotropic 
Kondo model at finite magnetic field in the weak-coupling regime.
The first technical step is to show that the RG approach can be set up 
purely in frequency space \cite{korb_private_communication} using
the diagrammatic rules of Ref.~\cite{hs_habil}. As a
consequence the occurence of correction terms can be avoided and
the rates determining the cutoff of the RG flow obtain the right scale.
Secondly, it turns out that it is important to integrate out the symmetric
part of the Fermi or Bose distribution function of the reservoirs before
starting the RG flow. With this choice it is possible to show generically for all models 
of dissipative quantum mechanics that relaxation and dephasing rates cut off the 
RG flow in all orders of perturbation theory and within all truncation schemes 
of the RG equations. Another technical advantage is that the dependence on the
Keldysh indices can be avoided, minimizing the effort to solve the 
RG equations. Finally, it is more convenient to define a cutoff not in
real but in imaginary frequency space by introducing the cutoff into
the Matsubara poles of the reservoir distribution function. 
This idea was first proposed in Ref.~\cite{jakobs_meden_hs_PRL07} in 
the context of nonequilibrium functional renormalization group within 
the Keldysh formalism. This choice improves the numerical stability 
considerably. Furthermore, for fermionic reservoirs with a flat density of
states, it is 
possible to reformulate the RG equations in pure Matsubara space (all
integrals over the real frequencies can be performed analytically) with
the difference to equilibrium that a whole set of Matsubara axis occurs 
shifted by multiples of the chemical potentials of the reservoirs and
by the real part of the Laplace variable determining the time evolution
of the system. With this technique it is possible to provide a generic
procedure how to solve the RG equations analytically in a well-controlled 
way in the weak-coupling regime, see Ref.~\cite{reininghaus_hs_preprint}. 
Whether it is also applicable to the
strong-coupling regime is an open question discussed in Sec.~\ref{sec:5.3}
for the special case of the nonequilibrium isotropic Kondo model.
For later reference, we will call this version of nonequilibrium RG
real-time RG in frequency space (RTRG-FS).

Especially for the Kondo model, the development of nonequilibrium RG has
been pioneered in 
Refs.~\cite{rosch_kroha_woelfle_PRL01,rosch_paaske_kroha_woelfle_PRL03}
(for perturbation theory see
Refs.~\cite{paaske_rosch_woelfle_PRB04,paaske_rosch_kroha_woelfle_PRB04};
a poor man scaling approach is described in Ref.~\cite{glazman_pustilnik_05}).
In these works the slave particle approach was used in connection with Keldysh
formalism and quantum Boltzmann equations. A real-frequency cutoff was
used and the RG was formulated purely on one part of the Keldysh-contour
disregarding diagrams connecting the upper with the lower branch. This procedure
turns out to be sufficient for the Kondo model to calculate logarithmic terms 
in leading order but the cutoff by relaxation and dephasing rates can not
be obtained in this way. In these works it was investigated for the first time
how the voltage and the magnetic field cuts off the RG flow and 
how the frequency dependence of the vertices influences various
logarithmic contributions 
$\sim \mbox{ln}({\mbox{max}(V,h)\over |V-h|})$ or $\sim\mbox{ln}({V\over h})$ 
for the susceptibility and the nonlinear conductance. 

An alternative approach to RTRG-FS for combining relaxation and
dephasing rates microscopically with nonequilibrium RG was proposed in
Ref.~\cite{kehrein_PRL05} using flow-equation methods \cite{flow_eq}.
Specifically for the isotropic Kondo model without magnetic field it 
was shown that the cutoff of the RG flow by the decay rate $\Gamma$ 
occurs due to a competition of 1-loop and
2-loop terms on the r.h.s. of the RG equation for the vertex. This is
a completely different picture compared to RTRG-FS, where the cutoff
parameter $\Gamma$ occurs already in the 1-loop terms as an additional
term in the denominator of the resolvents, see the RG-equations
(\ref{rg_Jd_appr}) and (\ref{rg_Jnd_appr}). Thus, RTRG-FS is closer
to conventional poor man scaling equations and the physics of relaxation
and dephasing rates occurs naturally as a resummation of a geometric
series similiar to self-energy insertions in Green's function techniques.
Nevertheless, the flow equation method is well-definied and controlled,
and represents a technical alternative to RTRG-FS.

Nonequilibrium RG methods which expand perturbatively in the Coulomb 
interaction of the quantum system have also been developed recently
\cite{jakobs_meden_hs_PRL07,gezzi_pruschke_meden_PRB07,jakobs_diplom,jakobs_pletyukhov_hs_preprint}.
In these works, the Keldysh formalism has been combined with 
functional RG techniques known from quantum field theory \cite{wetterich,salmhofer}.
It turns out that a real-frequency cutoff is not
useful since it violates causality and leads to various 
numerical instabilites \cite{jakobs_diplom,gezzi_pruschke_meden_PRB07}.
For fermionic models in 1d a more useful cutoff scheme on the imaginary 
frequency axis has been proposed \cite{jakobs_meden_hs_PRL07}, which 
is also very useful for RTRG-FS (see the discussion above). 
An interesting result was obtained that Luttinger liquid
exponents can depend on the nonequilibrium distribution function of
the quantum system \cite{jakobs_meden_hs_PRL07}. 
For zero-dimensional systems (quantum dots) with a single 
spin-degenerate level coupled by tunneling to two reservoirs (the
so-called single-impurity Anderson model), the situation is
more complicated because there is no controlled truncation scheme
for a perturbative RG in the Coulomb interaction $U$, especially in the
interesting regime where the Coulomb interaction becomes larger than
the energy scale $\Gamma$ of the reservoir-system coupling. Furthermore, the
Green's function in the absence of the Coulomb interaction has a
pole in the complex plane with imaginary part $\Gamma$ (corresponding to energy 
broadening of the dot level due to coupling to the reservoirs). This pole 
is close to the real-axis compared to the cutoff-parameter
of the RG flow. As a consequence the cutoff
procedure defined on the Matsubara axis is not sufficient for this problem
and $\Gamma$ itself was proposed as the flowing cutoff parameter 
\cite{jakobs_pletyukhov_hs_preprint}. 
With this choice it is possible to study the nonequilibrium Anderson impurity-model
up to values of $U\sim 2-4\Gamma$ even in the Kondo regime $T,V \ll T_K$
\cite{jakobs_pletyukhov_hs_preprint} (the truncation scheme in this work
neglects the 3-particle vertex but takes the important parts of the
frequency-dependence of the 2-particle vertex into account).
For $U\gg \Gamma$ charge fluctuations are suppressed and the model is
equivalent to the Kondo model with $J\sim \Gamma/U$. However, it is not yet
possible to approach the limit $J\ll 1$, where the Kondo temperature
shows the typical exponential behaviour.
It remains an interesting and open question whether this approach can
be improved and a full solution of the nonequilibrium Anderson-impurity
model including the nonequilibrium Kondo model can be obtained.

Finally, we mention that in this introduction we have only
summarized the nonequilibrium RG approaches relevant for problems of
dissipative quantum mechanics. There is also a rapidly increasing interest
in the development of nonequilibrium RG methods for bulk quantum systems
motivated by the interesting possibilities to measure the time evolution 
of strongly correlated quantum systems in cold atom gases. For completeness
we mention some of the most interesting recent developments, e.g. the
study of quantum phase transitions in nonequilibrium \cite{millis,berges},
the far-from-equilibrium quantum field dynamics of ultracold atom systems
\cite{gasenzer_pawlowski}, and the time-evolution of
fermionic system after initial interaction quenches \cite{moeckel_kehrein_PRL08}.

{\bf Outline.} 
The present paper concentrates on the scheme of RTRG-FS which is especially useful for
a generic study of problems in dissipative quantum mechanics, where the 
expansion parameter is the reservoir-system coupling. We will describe the
formal technique for a generic model but will always accompany the formalism by an
illustration for fermionic models where charge, spin or orbital fluctuations
dominate. Especially we will apply the formalism in all details to the
nonequilibrium isotropic Kondo model. The paper is written
for students with some basic knowledge of many-body theory. Besides
second quantization and Wick's theorem nothing special is required. The paper
is organized as follows. In Sec.~\ref{sec:2} we will introduce the generic
model and some specific examples. In Sec.~\ref{sec:3} we present a nonequilibrium 
quantum field-theoretical diagrammatic description in Liouville space by introducing
creation and annihilation superoperators acting in Liouville space (i.e. with an
additional Keldysh-index). This is especially useful for obtaining 
compact diagrammatic rules. Based on this diagrammatic expansion we present
a derivation of a formally exact kinetic equation for the reduced density
matrix of the quantum system and provide the rules to calculate observables. 
In Sec.~\ref{sec:4} we develop the nonequilibrium RG approach. We discuss the general 
idea of invariance, various ways to define the cutoff function, and set up simple rules to
obtain the RG equations in arbitrary order in the coupling. We discuss their
general properties and prove the central theorem that the RG flow is always 
cut off by relaxation and dephasing rates. For fermionic models with spin or
orbital fluctuations, following Ref.~\cite{reininghaus_hs_preprint}, we will 
describe the generic scheme how to solve the RG equations analytically 
in the weak-coupling regime.
Finally, in Sec.~\ref{sec:5} we apply the technique to the nonequilibrium
isotropic Kondo model, show the results of Ref.~\cite{reininghaus_hs_preprint}
in 1-loop order, and present preliminary results for the strong-coupling case. 
We close with a summary and an outlook in Sec.~\ref{sec:6}.

\section{The model}
\label{sec:2}
\subsection{Generic case}
\label{sec:2.1}
Generically any Hamiltonian of a problem in dissipative quantum mechanics can
be decomposed into a part for the reservoirs, the local quantum system, and the
coupling between reservoirs and quantum system
\begin{equation}
\label{H_total}
H\,=\,H_{res}\,+\,H_S\,+\,V\,=\,H_0\,+\,V \quad.
\end{equation}
The unperturbed part is denoted by 
\begin{equation}
\label{unperturbed}
H_0\,=\,H_{res}\,+\,H_S\quad.
\end{equation}

For the local quantum system $H_S$ we make no assumption and represent it via
its eigenstates $|s\rangle$ and eigenenergies $E_s$
\begin{equation}
\label{H_S}
H_S\,=\,\sum_s\,E_s\,|s\rangle\langle s|\quad.
\end{equation}
In practice the eigenstates $|s\rangle$ can easily be found for a quantum system
with a low number of single particle states including arbitrary interactions. 
In reality each quantum system has an infinite number of states. However, if the 
system is very small (as we assume) the single particle level spacing is very 
large and high-lying states can be neglected (for sufficiently small temperature
and bias voltage). Therefore, it is often sufficient to take only a few single-particle
levels into account and $H_S$ can be easily diagonalized. Since
we want to include the possibility that particles can be exchanged between reservoirs
and quantum system, the states have to be classified according to their particle number. We
denote by $N_s$ the particle number for state $|s\rangle$. We include both possibilities
that the particles can be bosons or fermions. E.g. bosonic particles can be realized
in cold atom gases whereas fermionic particles (electrons) occur typically in
nanoelectronic systems.

The reservoirs are assumed to be noninteracting and infinitely large with a continuum density of states.
Therefore, we use a continuum notation for the creation and annihilation operators of
the particles in the reservoirs and the reservoir Hamiltonian reads in second quantization
\begin{equation}
\label{H_res}
H_{res}\,=\,\sum_\alpha H_\alpha\,=\,\sum_{\nu\equiv \alpha\sigma\dots}
\int d\omega\,(\omega+\mu_\alpha)\,a^\dagger_\nu(\omega)a_\nu(\omega)\quad,
\end{equation}
with the continuum commutation relations for the field operators
\begin{equation}
\label{commutation}
[a_\nu(\omega),a^\dagger_{\nu'}(\omega')]_\mp\,=\,\delta_{\nu\nu'}\delta(\omega-\omega')\quad,
\end{equation}
where the upper (lower) sign correponds to bosons (fermions) and $[A,B]_\mp = AB\mp BA$ is
the (anti-)commutator. $\nu$ is a discrete index which labels all remaining quantum numbers 
of the reservoir particles
\begin{equation}
\label{nu_index}
\nu\,=\, \alpha\sigma \dots \quad,
\end{equation}
where, by convention, $\alpha$ is the index numerating the reservoirs (e.g. $\alpha=L,R\equiv\pm$
for two reservoirs) and $\sigma$ is the spin index (e.g. $\sigma=\uparrow,\downarrow\equiv \pm$ 
for a spin-${1\over 2}$ system). Further possibilities are 
orbital indices (if orbital symmetries are present), channel indices
(for transverse modes), etc. The chemical potential of reservoir $\alpha$ is denoted by $\mu_\alpha$,
and $\omega$ is the energy of the reservoir state measured relative to $\mu_\alpha$ (for 
phonon or photon baths, the chemical potential is absent).
The reservoirs are assumed to be described by a grandcanonical distribution
function
\begin{equation}
\label{res_distr}
\rho_{res}\,=\,\prod_\alpha\,\rho_\alpha \quad,\quad
\rho_\alpha \,=\, {1\over Z_\alpha}\,e^{-{1\over T_\alpha}(H_\alpha - \mu_\alpha N_\alpha)}\quad,
\end{equation}
where $T_\alpha$ is the temperature of reservoir $\alpha$.

We note that the crossover from a discrete to a continuum notation for the reservoir field
operators can be formally achieved by the definition
\begin{equation}
\label{continuum}
a_\nu(\omega)\,=\,{1\over\sqrt{\rho_\nu(\omega)}}\sum_k\,\delta(\omega-\epsilon_{\nu k}+\mu_\alpha)\,
a_{\nu k}\quad,
\end{equation}
where $k$ is a discrete index labelling the states of the reservoirs, $\epsilon_{\nu k}$ is
their corresponding energy, and $\rho_\nu(\omega)$ is the density of states in reservoir $\alpha$
(which can depend on energy and possibly (for ferromagnetic leads) on the spin index). It is easy 
to show that with this definition the reservoir Hamiltonian (\ref{H_res}) in the continuum form is
equivalent to the discrete version
\begin{equation}
\label{H_res_discret}
H_{res}\,=\,\sum_{\nu k}\,\epsilon_{\nu k}\,a_{\nu k}^\dagger a_{\nu k}\quad.
\end{equation}
Thereby, we have assumed that the relation between $k$ and $\epsilon_{\nu k}$ is unique, otherwise
different branches of the dispersion relation have to be distinguished and labelled by additional
indices.

For later convenience we introduce the following more compact notation for the various indices 
characterizing the reservoir field operators
\begin{equation}
\label{formal_notation}
1\,\equiv\,\eta\nu\omega \quad,\quad 
a_1\,\equiv\,a_{\eta\nu}(\omega)\,=\,
\left\{
\begin{array}{cl}
a_{\nu}^\dagger(\omega) &\mbox{for }\eta=+ \\
a_{\nu}(\omega)&\mbox{for }\eta=-
\end{array}
\right.
\,\quad.
\end{equation}
Here, $\eta=\pm$ indicates either a creation or annihilation operator. If no
ambiguities occur we use $1\equiv \eta\nu\omega$ and $1'\equiv \eta'\nu'\omega'$, whereas for
several indices we take $1\equiv \eta_1\nu_1\omega_1$, $2\equiv \eta_2\nu_2\omega_2$, etc.
Furthermore we define by
\begin{equation}
\label{transpose}
\bar{1}\equiv -\eta,\nu\omega \quad,\quad a_{\bar{1}}\,=\,(a_1)^\dagger \quad,
\end{equation}
the index corresponding to the hermitian conjugate. With these notations the commutation relation
(\ref{commutation}) can be written in the compact form
\begin{equation}
\label{compact_commutation}
[a_1,a_{1'}]_\mp\,=\,\delta_{1\bar{1}'} \quad,
\end{equation}
where $\delta_{11'}$ stands for 
\begin{equation}
\label{delta}
\delta_{11'}\,\equiv\,\delta_{\eta\eta'}\delta_{\nu\nu'}\delta(\omega-\omega')\quad.
\end{equation}
In all expressions we sum (integrate) implicitly over the indices, i.e. we sum over
the discrete part $\eta$ and $\nu$, and integrate over the continuum variable $\omega$.
The reservoir Hamiltonian can be written in the compact form
\begin{equation}
\label{H_res_compact}
H_{res}\,=\,(\omega+\mu_\alpha)\,\delta_{1\bar{1}'}\,\delta_{\eta +}\,a_1 a_{1'}\quad,
\end{equation}
and the reservoir correlation function (also called reservoir contraction) reads
\begin{equation}
\label{contraction}
{a_1\,a_{1'}
  \begin{picture}(-20,11) 
    \put(-22,8){\line(0,1){3}} 
    \put(-22,11){\line(1,0){12}} 
    \put(-10,8){\line(0,1){3}}
  \end{picture}
  \begin{picture}(20,11) 
  \end{picture}
}
\,\equiv\,
\langle a_1 a_{1'}\rangle_{\rho_{res}}\,=\,\delta_{1\bar{1}'}\,f^\eta_\alpha(\omega)
\,=\,\delta_{1\bar{1}'}\,
\left\{
\begin{array}{cl}
\eta \\ 1
\end{array}
\right\}\,
f_\alpha(\eta\omega)\quad,
\end{equation}
where $f_\alpha^+(\omega)\equiv f_\alpha(\omega)$, 
$f_\alpha^-(\omega)\equiv 1\pm f_\alpha(\omega)$, and
\begin{equation}
\label{fb_function}
f_\alpha(\omega)\,=\,{1\over e^{\omega/T_\alpha}\mp 1}\,=\,\mp \,f_\alpha^-(-\omega)
\end{equation}
is the Bose (Fermi) function corresponding to temperature $T_\alpha$ (note that
the chemical potential $\mu_\alpha$ does not occur in this formula because $\omega$
measures the energy relative to $\mu_\alpha$). The upper (lower) case in (\ref{contraction})
refers to bosons (fermions), a convention which we will use throughout this paper.

For the coupling between reservoirs and quantum system we take the generic form
\begin{equation}
\label{coupling}
V\,=\,{1\over n!}\,\,g_{12\dots n}:a_1\,a_2\, \dots \,a_n: \quad,
\end{equation}
where $g_{12\dots n}$ is an arbitrary operator acting on the quantum system, and 
we sum implicitly over $n=1,2,\dots$ and all variables $\eta_i,\nu_i,\omega_i$,
$i=1,2,\dots,n$, which are contained in the formal notation $i\equiv \eta_i,\nu_i,\omega_i$.
We note that we explicitly exclude the case $n=0$ because this corresponds to
an operator of the local quantum system which can be incorporated in $H_S$.
The symbol $:\dots:$ denotes normal-ordering of the reservoir field operators with respect to the
reservoir correlation function, i.e. in any Wick-decomposition of an average over a sequence of 
reservoir field operators with respect to $\rho_{res}$, no contraction is allowed which
connects field operators occuring in the same normal-ordered block. As a consequence the average
over a normal-ordered block is defined as zero $\langle :a_1 a_2\dots a_n:\rangle_{\rho_{res}}=0$.
The field operators within the normal-ordering can be arranged in an
arbitrary order (up to a sign for fermions). The prefactor ${1\over n!}$ accounts for all
permutations of reservoir field operators which do not change the value of the diagrams in
perturbation theory (see later) since  the coupling vertex can always be chosen such that it is 
(anti)symmetric under exchange of two indices
\begin{equation}
\label{g_symmetry}
g_{1\dots i\dots j\dots n}\,=\,\pm\,g_{1\dots j\dots i\dots n}\quad.
\end{equation}
For simplicity, we assume here that either fermionic or bosonic
field operators occur for the reservoirs. If both are present (e.g. for combinations of electronic
particle reservoirs and phonon heat baths), one simply writes the coupling as a sum of several terms,
each one being of the form (\ref{coupling}). In principle also mixed couplings can be treated
containing fermionic and bosonic field operators in the same term but we will not consider this
case here because it just complicates the notation (in this case there is a factor 
${1\over n_B!\,n_F!}$ in front of Eq.~(\ref{coupling}), where $n_B$ ($n_F$) is the number of
bosonic (fermionic) field operators).

The form (\ref{coupling}) of the coupling includes all cases of charge, spin, and energy
transfer between reservoirs and quantum system and, for $n>1$, also includes nonlinear coupling.
The coupling vertex $g_{12\dots n}$ describes the change of the state of the quantum system
and is an operator. We have included the possibility that it depends in an arbitrary way
on the frequencies of the reservoir field operators. Although such a general form (especially
for $n>2$) is hardly necessary for realistic models, this general form with all possible values
for $n$ has to be considered since the RG procedure described in Sec.~\ref{sec:4} will generate
an effective coupling which includes all these terms. 

The total operator $V$ is of bosonic nature since the parity of fermion number
must be conserved. However, for fermions and $n$ odd, 
$g_{12\dots n}$ in (\ref{coupling}) is of fermionic nature and the sequence
of $g_{12\dots n}$ and $:a_1 a_2 \dots a_n:$ is important (the coupling $V$ in the form 
(\ref{coupling}) is even not hermitian in this case). To be precise, in the general case
$V$ should be written as
\begin{equation}
\label{coupling_fermions}
V\,=\, {1\over n!}\,
\left\{
\begin{array}{cl}
1 \\ \eta_1\dots\eta_n
\end{array}
\right\}\,
:a_n a_{n-1} \dots a_1:\,\,g_{12\dots n} \quad.
\end{equation}
With this choice it can easily be shown that $V$ is hermitian, 
provided the coupling vertex fulfils the condition
\begin{equation}
\label{g_hermiticity}
(g_{12\dots n})^\dagger \,=\, g_{\bar{n} \dots \bar{1}} \quad.
\end{equation}
Furthermore, it is shown in Appendix A that the form (\ref{coupling_fermions}) is only
important for fixing the value of the coupling vertex $g_{1\dots n}$ for a concrete model.
After having defined $g_{1\dots n}$ in this way, one can proceed with the much simpler form 
(\ref{coupling}) and disregard
all sign factors emerging from interchanging fermionic dot and reservoir operators. The reason
is that by calculating the reduced density matrix of the quantum system or averages of observables,
only expressions have to be evaluated where an average over the
reservoir distribution $\rho_{res}$ is taken, for details see Appendix A. Thus, in the 
following we work with the simpler form (\ref{coupling}) and commute dot and reservoir operators.
This simplifies the problem of sign factors considerably.

Finally, we mention that it is sometimes more convenient to include the density of states
of the reservoirs into the reservoir contraction (\ref{contraction}). In this case we
use the definition
\begin{equation}
\label{alternative_field_operators}
b_1\,=\,\sqrt{\rho_\nu(\omega)}\,\,a_1 \,\quad
\end{equation}
for the reservoir field operator and obtain for the contraction
\begin{equation}
\label{alternative_contraction}
{b_1\,b_{1'}
  \begin{picture}(-20,11) 
    \put(-20,9){\line(0,1){3}} 
    \put(-20,12){\line(1,0){10}} 
    \put(-10,9){\line(0,1){3}}
  \end{picture}
  \begin{picture}(20,11) 
  \end{picture}
}
\,\equiv\,
\langle b_1 b_{1'}\rangle_{\rho_{res}}\,=\,\delta_{1\bar{1}'}
\,\rho_\nu(\omega)\,f^\eta_\alpha(\omega) \quad.
\end{equation}
Correspondingly, the coupling $V$ is written in the form
\begin{equation}
\label{alternative_coupling}
V\,=\,{1\over n!}\,\,g_{12\dots n}:b_1\,b_2\, \dots \,b_n: \quad,
\end{equation}
where dot and reservoir operators commute, and 
\begin{equation}
\label{alternative_coupling_fermions}
V\,=\, {1\over n!}\,
\left\{
\begin{array}{cl}
1 \\ \eta_1\dots\eta_n
\end{array}
\right\}\,
:b_n b_{n-1} \dots b_1:\,\,g_{12\dots n} 
\end{equation}
for the determination of $g_{12\dots n}$.

\subsection{Specific examples}
\label{sec:2.2}
Here, we present some specific and experimentally relevant examples for charge, 
spin, orbital, or energy fluctuations induced by the coupling  between reservoirs 
and quantum system.

{\bf Charge fluctuations}. 
Electronic quantum transport through nanodevices or quantum dots is characterized by charge fluctuations
and is conveniently described by a tunneling Hamiltonian, i.e. the coupling is of the form
\begin{equation}
\label{tunneling}
V\,=\,\sum_{\alpha\sigma}\sum_k\sum_l\,t^{\alpha\sigma}_{kl}\,
a^\dagger_{\alpha\sigma k}\,c_{\sigma l} \,+\,(h.c.)\quad.
\end{equation}
Here, $\alpha$ and $\sigma$ are the reservoir and spin indices, respectively,
and $l$ is an index for an arbitrary single-particle state basis of the
quantum system. $t^{\alpha\sigma}_{kl}$ are the tunneling matrix elements and
$c_{\sigma l}$ annihilates a particle on the quantum system in state $l$ with
spin $\sigma$. In the continuum notation and with $\nu\equiv \alpha\sigma$, we obtain
\begin{equation} 
\label{tunneling_continuum}
V\,=\,\sum_{\eta\nu}\int d\omega\,\eta\,a_{\eta\nu}(\omega)\, g_{\eta\nu}(\omega) 
\,\equiv\, \eta_1\,a_1 \,g_1\quad,
\end{equation}
with $a_{\eta\nu}(\omega)$ defined by (\ref{formal_notation}) and (\ref{continuum}), 
and the coupling vertex is given by
\begin{equation}
\label{coupling_tunneling}
g_{\eta\nu}(\omega)\,=\,\sqrt{\rho_{\nu}(\omega)}\left\{
\begin{array}{cl}
\sum_l \,t^\nu_l(\omega)\, c_{\sigma l}\, &\mbox{for }\eta=+ \\
\sum_l \,t^\nu_l(\omega)^* \,c^\dagger_{\sigma l}\,&\mbox{for }\eta=-
\end{array}
\right.\quad,
\end{equation}
where $t^\nu_l(\omega)\equiv t^\nu_{kl}$ with $\omega=\epsilon_{\nu k}-\mu_\alpha$. 

To obtain dimensionless coupling vertices,
one can define for each $\nu$ and $\omega$ some reference tunneling matrix element
$t^\nu(\omega)\equiv t^\nu_k$ (independent of the state index $l$ of the quantum system), with
a corresponding level broadening parameter $\Gamma_\nu(\omega)$ defined by
\begin{equation}
\label{gamma}
\Gamma_\nu(\omega)\,=\,2\pi\,\rho_\nu(\omega)|t^\nu(\omega)|^2
\,=\,2\pi\,\sum_k\,|t^\nu_k|^2\,\delta(\omega-\epsilon_{\nu k}+\mu_\alpha)\quad.
\end{equation}
With this reference energy, we define the dimensionless reservoir field operators
\begin{equation}
\label{reservoir_field_operators_dimensionless}
b_{\eta\nu}(\omega)\,=\,\sqrt{\Gamma_\nu(\omega)/(2\pi)}\,a_{\eta\nu}(\omega)
\,=\,\left\{
\begin{array}{cl}
\sum_k \,|t^\nu_k|\,\delta(\omega-\epsilon_{\nu k}+\mu_\alpha)\, 
a^\dagger_{\nu k}\, &\mbox{for }\eta=+ \\
\sum_k \,|t^\nu_k|\,\delta(\omega-\epsilon_{\nu k}+\mu_\alpha)\, 
a_{\nu k}\, &\mbox{for }\eta=- 
\end{array}
\right.\quad,
\end{equation}
and the corresponding dimensionless coupling vertex
\begin{equation}
\label{coupling_dimensionless_charge}
g_{\eta\nu}(\omega)\,=\,\left\{
\begin{array}{cl}
\sum_l \,t^\nu_l(\omega)/|t^\nu(\omega)|\, c_{\sigma l}\, &\mbox{for }\eta=+ \\
\sum_l \,t^\nu_l(\omega)^*/|t^\nu(\omega)| \,c^\dagger_{\sigma l}\,&\mbox{for }\eta=-
\end{array}
\right.\quad,
\end{equation}
such that we obtain again the form (\ref{tunneling_continuum}) with $a\rightarrow b$. 
The reservoir contraction with respect to the $b$-operators reads
\begin{equation}
\label{Gamma_contraction}
{b_1\,b_{1'}
  \begin{picture}(-20,11) 
    \put(-20,9){\line(0,1){3}} 
    \put(-20,12){\line(1,0){10}} 
    \put(-10,9){\line(0,1){3}}
  \end{picture}
  \begin{picture}(20,11) 
  \end{picture}
}
\,\equiv\,
\langle b_1 b_{1'}\rangle_{\rho_{res}}\,=\,\delta_{1\bar{1}'}
\,{1\over 2\pi}\Gamma_\nu(\omega)\,f^\eta_\alpha(\omega) \quad.
\end{equation}

{\bf Spin and orbital fluctuations}.
If the charging energy of a quantum dot is very large (compared to temperature and
bias voltage), the gate voltage determining the position of the charge
excitation energies of the quantum dot relative to the electrochemical potentials
of the reservoirs can be adjusted such that charge fluctuations are suppressed. In 
this case, spin and orbital fluctuations dominate transport. The elementary processes
are given by an electron tunneling on (off) the quantum system, occupying a virtual
intermediate state, and in a second step tunneling off (on) the quantum system. As 
shown in detail in Ref.~\cite{korb_reininghaus_hs_koenig_PRB07}, the virtual 
intermediate state (which has a very high energy due to the large charging energy)
can be integrated out via a standard Schrieffer-Wolff transformation on a Hamiltonian 
level, and the result is a coupling of the form
\begin{equation}
\label{spin_orbital}
V\,=\,{1\over 2}\,g_{11'}\,:a_1 a_{1'}:\quad.
\end{equation}
In most cases (except for systems with negative charging energies \cite{oppen_2}), 
one creation and annihilation operator is present, i.e. $\eta=-\eta'$ (note that the forms
(\ref{coupling}) and (\ref{coupling_fermions}) are equivalent in this case).

Whereas (\ref{spin_orbital}) describes a generic model for spin and/or orbital
fluctuations on a quantum dot, a special case is the Kondo model (\ref{kondo}) 
discussed in the introduction. In this case, only fluctuations of a 
spin-${1\over 2}$ of the quantum system are considered and, comparing
(\ref{spin_orbital}) with (\ref{kondo}), we obtain for the dimensionless 
coupling vertex in the continuum form
\begin{equation}
\label{kondo_vertex}
g_{11'}\,=\,{1\over 2}\,\left\{
\begin{array}{cl}
J_{\alpha\alpha'}\,\underline{S}\cdot\underline{\sigma}_{\sigma\sigma'}\, 
&\mbox{for }\eta=-\eta'=+ \\
-J_{\alpha'\alpha}\,\underline{S}\cdot\underline{\sigma}_{\sigma'\sigma}\, 
&\mbox{for }\eta=-\eta'=- 
\end{array}
\right.\quad,
\end{equation}
such that the antisymmetry and hermiticity conditions 
\begin{equation}
\label{kondo_symmetries}
g_{11'}\,=\,-g_{1'1}\quad,\quad g_{11'}^\dagger\,=\,g_{\bar{1}'1}
\end{equation}
are fulfilled, compare (\ref{g_symmetry}) and (\ref{g_hermiticity}).

Since, due to the Schrieffer-Wolff transformation, high-lying charge
excitations have already been integrated out in the model (\ref{spin_orbital}),
one has to consider the reservoirs with a finite bandwidth $D$ (of the order of
the charging energy). Thus, all frequencies $\omega$ have to be smaller than
$D$, which can be achieved by various cutoff functions. In this paper, we will
use a Lorentzian cutoff defined by the function
\begin{equation}
\label{cutoff}
\rho(\omega)\,=\,{D^2\over \omega^2+D^2}\,=\,{D\over 2i}
({1\over \omega-iD}-{1\over \omega+iD})\quad.
\end{equation}
Most elegantly, we introduce this function by a modification of the 
reservoir contraction (\ref{contraction})
\begin{equation}
\label{contraction_cutoff}
{a_1\,a_{1'}
  \begin{picture}(-20,11) 
    \put(-22,8){\line(0,1){3}} 
    \put(-22,11){\line(1,0){12}} 
    \put(-10,8){\line(0,1){3}}
  \end{picture}
  \begin{picture}(20,11) 
  \end{picture}
}
\,\rightarrow \,\delta_{1\bar{1}'}\,\rho(\omega)\,f^\eta_\alpha(\omega)
\end{equation}
Although this is not done here by a formal redefinition of the field operators,
the effect of this modification of the reservoir correlation function is that
all frequencies are suppressed if they lie above the band cutoff $D$. This will 
become clear within the diagrammatic expansion described in Sec.~\ref{sec:3}, where 
the reservoir degrees of freedom are integrated out.

{\bf Energy fluctuations}.
Within the traditional field of dissipative quantum mechanics, a heat
bath or energy-exchange with the environment is considered, mostly within 
the so-called spin-boson model defined by the Hamiltonian
\begin{eqnarray}
\label{sb_res}
H_{res} &=& \sum_q\,\omega_q \, a^\dagger_q a_q \quad,\\
\label{sb_system}
H_S &=& -{\Delta\over 2}\sigma_x \,+\, {\epsilon \over 2}\sigma_z \quad,  \\
\label{sb_coupling}
V &=& {1\over 2}\sigma_z\,\sum_q\,\gamma_q\,(a^\dagger_q + a_q) \quad.
\end{eqnarray}
Here, the quantum system consists of a two-level system with tunneling 
coupling $\Delta$ between the two levels and bias $\epsilon$ determining
the detuning. The bath consists of a set of harmonic oscillators (phonons)
with bosonic creation and annihilation operators $a^\dagger_q,a_q$. The 
phonon energy is $\omega_q>0$ and the bath couples linearly (via the spatial
coordinate $\sim a^\dagger_q + a_q$) to the two-level system. The coupling
parameters $\gamma_q$ and the density of states of the phonon modes are 
conventionally parametrized by the spectral function
\begin{equation}
\label{sb_spectral_function}
J(\omega) \,=\, \pi\,\sum_q\,\gamma_q^2\,\delta(\omega-\omega_q)\,=\,
2\pi\,\alpha\,\omega^{n+1}\,e^{-\omega/D}\quad.
\end{equation}
$n=0$ describes the important ohmic case, whereas $n>0$ and $n<0$ 
are called the super-ohmic and sub-ohmic case. $D$ is the bandwidth of the bath,
and $\alpha$ describes the coupling parameter (which is dimensionless for the
ohmic case).

For our continuum notation, we define in analogy to 
(\ref{reservoir_field_operators_dimensionless}) 
the dimensionless bosonic field operators
\begin{equation}
\label{bosonic_field_operators_dimensionless}
b_\eta(\omega)\,=\,\left\{
\begin{array}{cl}
\sum_q \,g_q\,\delta(\omega-\omega_q)\, a^\dagger_q\, &\mbox{for }\eta=+ \\
\sum_q \,g_q\,\delta(\omega-\omega_q)\, a_q\, &\mbox{for }\eta=-
\end{array}
\right.\quad,
\end{equation}
and obtain with the dimensionless coupling vertex
\begin{equation}
\label{sb_coupling_dimensionless}
g_\pm\,=\,{\sigma_z \over 2}
\end{equation}
the form
\begin{equation}
\label{sb_continuum_coupling}
V\,=\,g_1\,b_1 
\end{equation}
for the coupling, and
\begin{equation}
\label{sb_contraction}
{b_1\,b_{1'}
  \begin{picture}(-20,11) 
    \put(-20,9){\line(0,1){3}} 
    \put(-20,12){\line(1,0){10}} 
    \put(-10,9){\line(0,1){3}}
  \end{picture}
  \begin{picture}(20,11) 
  \end{picture}
}
\,=\,\delta_{1\bar{1}'}
\,{1\over \pi}J(\omega)\,f^\eta(\omega) \quad
\end{equation}
for the bath contraction, where $f^+(\omega)=f(\omega)$, $f^-(\omega)=1+f(\omega)$ and 
$f(\omega)=1/(e^{\beta\omega}-1)$ is the Bose function.
Note that only positive frequencies $\omega$ are allowed since
the phonon energies are positive.

Formally, one can rewrite the expressions such that both positive and negative
frequencies are allowed by using the definiton
\begin{equation}
\label{absorption_emission}
d_\eta(\omega)\,=\,\left\{
\begin{array}{cl}
b_\eta(\omega)\, &\mbox{for }\omega>0 \\
b_{-\eta}(-\omega)\, &\mbox{for }\omega<0
\end{array}
\right.\quad.
\end{equation}
Since $d_\eta(\omega)=d_{-\eta}(-\omega)$, we get
\begin{equation}
\label{full_integral}
\sum_\eta\,\int_0^\infty d\omega\,b_\eta(\omega) \,=\,
\sum_\eta\,\int_0^\infty d\omega\,d_\eta(\omega) \,=\,
{1\over 2}\,\sum_\eta\,\int_{-\infty}^\infty d\omega\,d_\eta(\omega) \quad,
\end{equation}
and $V$ can be written as (note that $g_1$ does not depend on $\eta$)
\begin{equation}
\label{V_full_frequency}
V\,=\,\bar{g}_1\,d_1 \quad,\quad \bar{g}_1\,=\,{\sigma^z\over 4}\quad.
\end{equation}
Using $f^-(\omega)=1+f(\omega)=-f(-\omega)$, we get for the contraction
in the new representation
\begin{equation}
\label{sb_contraction_antisym}
{d_1\,d_{1'}
  \begin{picture}(-20,11) 
    \put(-20,9){\line(0,1){3}} 
    \put(-20,12){\line(1,0){10}} 
    \put(-10,9){\line(0,1){3}}
  \end{picture}
  \begin{picture}(20,11) 
  \end{picture}
}
\,=\,\delta_{11'}
\,{1\over \pi}\,\mbox{sign}(\omega)\,J(|\omega|)\,f^\eta(\omega) \quad,
\end{equation}
where all frequencies are allowed now.

\section{Quantum field theory in Liouville space}
\label{sec:3}
In this section we will develop a quantum field-theoretical
diagrammatic expansion in Liouville space. Conventional quantum
field theory in nonequilibrium is based on the Keldysh-formalism,
where all field operators are integrated out via Wick's theorem
or via Gaussian integrals within path integral formalism. This
is not possible here because the unperturbed part of the 
Hamiltonian, consisting of the reservoirs $H_{res}$ and the
quantum system $H_S$, contains arbitrary interaction terms in $H_S$.
Therefore, this part is not quadratic and Wick's theorem does not
apply. For this reason, only the degrees of freedom of the reservoirs
can be integrated out. However, the coupling vertex $g_{1\dots n}$
remains an operator and the dynamics between the vertices is described
by $H_S$. Therefore, a diagrammatic representation is obtained where
the vertices are operators and one has to keep track of their
time-ordering on the Keldysh-contour. In contrast to previous versions
of this procedure \cite{hs_lecture_notes_00}, we introduce here a
slightly different way of integrating out the reservoirs. Instead
of expanding the time evolution on the Keldysh contour and applying
Wick's theorem for the reservoir field operators, we will introduce
quantum field superoperators acting in Liouville space for the reservoirs. This
has the advantage that the two branches of the Keldysh contour can be taken
together from the very beginning in a compact way. Subsequently,
we will apply Wick's theorem for the quantum field superoperators. This
procedure combines in an efficient way the advantages of 
Liouville superoperators (the traditional formalism of dissipative quantum
mechanics) and quantum field theory on the Keldysh contour (the common
way to treat nonequilibrium systems). The traditional way of 
deriving kinetic equations by using projection operators in Liouville space
\cite{zwanzig} is a very formal procedure and much less useful. Although one obtains
a very compact and analytic notation for the kernel of the kinetic
equation without any need of a diagrammatic representation, the
reservoirs are not integrated out and everything is hidden in formal
projectors which finally have to be evaluated by decomposing all 
expressions into many artificial reducible terms where numerous cancellations
occur. Integrating out the reservoirs from the very beginning has the
advantage that the kernel can be defined via irreducible diagrams only 
and the calculations simplify considerably. Furthermore,
it is shown in Sec.~\ref{sec:4} that a systematic renormalization group
method generates an additional frequency dependence of the Liouvillian
and the vertices which can only be described within a diagrammatic
representation.

The main result of this section are the diagrammatic rules how
to evaluate the kernel of the kinetic equation and averages of observables,
which is the basis for the renormalization group
method developed in Sec.~\ref{sec:4}. 

\subsection{Dynamics of the reduced density matrix}
\label{sec:3.1}
{\bf General considerations.} The dynamics of the total density matrix is governed by the von
Neumann equation
\begin{equation}
\label{von_Neumann}
\dot{\rho}(t)\,=\,-i[H,\rho(t)]_-\,=\,-i\,L\rho(t)\quad,
\end{equation}
where $L$ is the so-called Liouville operator acting on usual
operators $A$ via the commutator
\begin{equation}
\label{Liouville_operator}
L\,A\,\equiv\,[H,A]_-\quad.
\end{equation}
Thus, in matrix notation, the Liouvillian is a super-matrix
with four indices
\begin{equation}
\label{matrix_1}
(LA)_{nm}\,=\,\sum_{n'm'}\,L_{nm,n'm'}\,A_{n'm'}\quad,
\end{equation}
with
\begin{equation}
\label{matrix_2}
L_{nm,n'm'}\,=\,\langle n|\,L\left(|n'\rangle\langle m'|\right)\,|m\rangle
\,=\,H_{nn'}\delta_{mm'}-\delta_{nn'}H_{m'm}\quad.
\end{equation}
Therefore the operators acting in Liouville space are also called superoperators.

The initial state of the density matrix at time $t_0$ is assumed to be an
independent product of an arbitrary part $\rho_S(t_0)$ for the quantum system
and an equilibrium part $\rho_{res}$ for the reservoirs
\begin{equation}
\label{initial_state}
\rho(t_0)\,=\,\rho_S(t_0)\,\rho_{res}\quad,
\end{equation}
where $\rho_{res}$ is the grandcanonical distribution defined in 
(\ref{res_distr}). This means that we assume the reservoirs
and the quantum system to be initially decoupled, and the coupling $V$ is switched
on suddenly at time $t_0$.

The von Neumann equation can be formally solved by
\begin{equation}
\label{solution_von_Neumann}
\rho(t)\,=\,e^{-iH(t-t_0)}\,\rho(t_0)\,e^{iH(t-t_0)}\,=\,
e^{-iL(t-t_0)}\,\rho(t_0)\quad.
\end{equation}
As a result we obtain the following formal expression for the reduced density
matrix of the quantum system
\begin{equation}
\label{reduced_dm}
\rho_S(t)\,=\,\mbox{Tr}_{res}\,\rho(t)\,=\,
\mbox{Tr}_{res}\,e^{-iH(t-t_0)}\,\rho(t_0)\,e^{iH(t-t_0)}\,=\,
\mbox{Tr}_{res}\,e^{-iL(t-t_0)}\,\rho_S(t_0)\,\rho_{res}\quad,
\end{equation}
where $\mbox{Tr}_{res}$ denotes the trace over the reservoir degrees of freedom.

The average of an arbitrary observable $R$ can be written in two equivalent ways,
either in Heisenberg picture or with Liouville operators as
\begin{equation}
\label{observable}
\langle R \rangle(t)\,=\,\mbox{Tr}\,R\,\rho(t)\,=\,
\mbox{Tr}\,e^{iH(t-t_0)}\,R\,e^{-iH(t-t_0)}\,\rho(t_0)\,=\,
\mbox{Tr}\,(-iL_R)e^{-iL(t-t_0)}\,\rho_S(t_0)\,\rho_{res}\quad,
\end{equation}
where the Liouvillian corresponding to the observable $R$ is defined by
\begin{equation}
\label{Liouville_observable}
L_R\,=\,{i\over 2}\,[R,\cdot]_+
\end{equation}
via the anticommutator. If $R=|s\rangle\langle s'|$ denotes a Hubbard operator
for the eigenstates $|s\rangle$ of the quantum system, (\ref{observable}) is
an expression for the matrix element $\langle s|\rho_S(t)|s'\rangle$ of the 
reduced density matrix of the quantum system.

{\bf Diagrammatic expansion.} 
Analogous to the Hamiltonian (\ref{H_total}), the Liouvillian can be decomposed
into a part for the reservoirs, the quantum system, and the coupling
\begin{equation}
\label{L_total}
L\,=\,L_0\,+\,L_V \quad,\quad L_0\,=\,L_{res}\,+\,L_S
\end{equation}
with
\begin{equation}
\label{L_decomposed}
L_{res}\,=\,[H_{res},\cdot]_-\quad,\quad
L_S\,=\,[H_S,\cdot]_-\quad,\quad
L_0\,=\,[H_0,\cdot]_-\quad,\quad
L_V\,=\,[V,\cdot]_-\quad.
\end{equation}
Note that we take a different definition for $L_V$ (containing the commutator) compared
to $L_R$ (containing the anticommutator and a different prefactor, see (\ref{Liouville_observable})).
From the context it should always be clear whether the coupling $V$ or an observable $R$ is
considered.

\begin{figure}
  \centerline{\psfig{figure=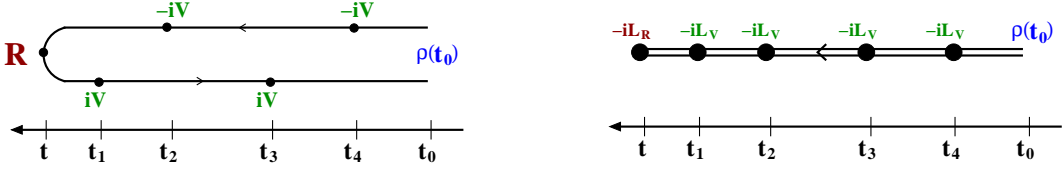,scale=0.43}}
  \caption{Expansion of the time evolution of an arbitrary observable $R$ in
the coupling on the Keldysh contour (left figure) or in Liouville space
(right figure), using the two different ways to write an average over an
observable $R$ either in Heisenberg picture or with Liouville operators, see
Eq.~(\ref{observable}). For the Keldysh contour, operators
are ordered along the contour in the direction indicated. The propagators 
between the vertices contain the unperturbed Hamiltonian $H_0$. They are given
by $e^{- iH_0(t_i-t_j)}$ for the upper branch of the Keldysh contour (forward 
propagation) and $e^{+iH_0(t_i-t_j)}$ for the lower branch (backward propagation), 
where $t_i>t_j$ in both cases. In Liouville space all
superoperators are time-ordered along one axis, the two parts of the Keldysh contour
have been taken together. Correspondingly, the intermediate state in Liouville
space has to be characterized by two states of the quantum system, one for the
upper and one for the lower part of the Keldysh contour. For convenience, we have
chosen the time direction to the left since with this convention time-ordering 
coincides with the sequence of Liouville operators (operators with larger time
stand to the left of operators with smaller time argument).}
\label{fig:Keldysh}
\end{figure}

Since we are aiming at a diagrammatic representation which is perturbative in 
the coupling, we insert (\ref{L_total}) in (\ref{reduced_dm}) or
(\ref{observable}) and expand in $L_V$ using time-dependent perturbation theory. 
This leads to a compact notation in Liouville 
space. Within the usual Keldysh formalism, one would insert the Hamiltonian (\ref{H_total})
in (\ref{reduced_dm}) or (\ref{observable}) and expand the forward and backward propagators
$e^{\mp iH(t-t_0)}$ in the coupling $V$. As a consequence, an additional
label (the so-called Keldysh-index) is needed to distinguish between the
forward and the backward propagation and the time evolution can be visualized
on the Keldysh-contour, see Fig.~\ref{fig:Keldysh}. The notation in 
Liouville space is much more compact, the Keldysh-indices are hidden in the larger
dimension of Liouville space (which is the square of the dimension of the quantum system)
represented by the super-matrix notation
(\ref{matrix_2}). Alternatively, one can say that the two parts of the
Keldysh-contour have been taken together in Liouville space, see 
Fig.~\ref{fig:Keldysh}.
However, instead of performing the perturbative expansion in time space, it is much
easier to do it in Laplace space (at least for explicitly time-independent Hamiltonians
as we are considering here). First, we define the reduced density matrix in Laplace
space by
\begin{equation}
\label{laplace}
\tilde{\rho}_S(E)\,=\,\int_{t_0}^\infty dt\,e^{iE(t-t_0)}\,\rho_S(t)\quad,
\end{equation}
where $E$ is the Laplace variable, having a positive imaginary part to
guarantee convergence. Using (\ref{reduced_dm}), we obtain
\begin{equation}
\label{reduced_dm_laplace}
\tilde{\rho}_S(E)\,=\,\mbox{Tr}_{res}\,{i\over E-L}\,\rho(t_0)\,=\,
\mbox{Tr}_{res}\,{i\over E-L_{res}-L_S-L_V}\,\rho_S(t_0)\rho_{res}\quad.
\end{equation}
This expression can easily be expanded in $L_V$ by a geometric series leading
to terms of the form
\begin{equation}
\label{perturbative_series}
i\,\mbox{Tr}_{res}\,{1\over E-L_{res}-L_S}\,L_V\,{1\over E-L_{res}-L_S}\,L_V\,
\dots \,L_V\,{1\over E-L_{res}-L_S}\,\rho_S(t_0)\rho_{res}\quad.
\end{equation}

The next step is to integrate out the reservoirs, i.e. the trace
$\mbox{Tr}_{res}$ over the reservoir degrees of freedom has to be performed in 
(\ref{perturbative_series}). 
This is achieved by decomposing (\ref{perturbative_series}) into a product
of a reservoir and system part and applying 
Wick's theorem to evaluate the average over the reservoir distribution. 
We first exhibit explicitly all parts of the reservoir operators in (\ref{perturbative_series})
by finding a representation of the coupling $L_V$ in Liouville space similiar to
the form of the coupling $V$ in Hilbert space, given by (\ref{coupling}). We use
the form
\begin{equation}
\label{coupling_product}
L_V\,=\,{1\over n!}\,\sigma^{p_1\dots p_n}\,G^{p_1\dots p_n}_{1\dots n}\,
:J^{p_1}_1\dots J^{p_n}_n:\quad.
\end{equation}
Here, $J^p_1$ are quantum field superoperators in Liouville space for the reservoirs, 
defined by (where $A$ is an arbitrary reservoir operator)
\begin{equation}
\label{liouville_field_operators}
J_1^p\,A \,=\,
\left\{
\begin{array}{cl}
a_1\,A\, &\mbox{for }p=+ \\
A\,a_1\, &\mbox{for }p=-
\end{array}
\right.\quad.
\end{equation}
$p$ is the Keldysh index indicating whether the field operator is acting on the
upper or the lower part of the Keldysh contour. $G^{p_1\dots p_n}_{1\dots n}$ is
a superoperator acting in Liouville space of the quantum system, and is defined
by ($A$ is an arbitrary operator of the quantum system)
\begin{equation}
\label{G_vertex_liouville}
G^{p_1\dots p_n}_{1\dots n}\,A\,=\,
\delta_{pp_1}\dots\delta_{pp_n}\,
\left\{
\begin{array}{cl}
1\, &\mbox{for }n\mbox{ even} \\
\sigma^p\, &\mbox{for }n\mbox{ odd}
\end{array}
\right\}\,
\left\{
\begin{array}{cl}
g_{1\dots n}\,A\, &\mbox{for }p=+ \\
-A \,g_{1\dots n}\, &\mbox{for }p=-
\end{array}
\right.\quad.
\end{equation}
We implicitly sum over $p=\pm$ on the r.h.s. of this definition, 
and $\sigma^{p_1\dots p_n}$ is
a sign-superoperator acting in Liouville space of the quantum system, accounting for fermionic
sign factors. For fermions, it is defined by its matrix representation
\begin{equation}
\label{liouville_sign_operator}
(\sigma^{p_1\dots p_n})_{ss',\bar{s}\bar{s}'}\,=\,
\delta_{s\bar{s}}\delta_{s'\bar{s}'}\,
\left\{
\begin{array}{cl}
p_2\cdot p_4\cdot\dots\, &\mbox{for }N_s-N_{s'}\mbox{ even} \\
p_1\cdot p_3\cdot\dots\, &\mbox{for }N_s-N_{s'}\mbox{ odd} 
\end{array}
\right.\quad,
\end{equation}
whereas, for bosons, it is defined by the unity operator. For $n=1$, 
(\ref{liouville_sign_operator}) has to be interpreted as
\begin{equation}
\label{liouville_sign_operator_n=1}
(\sigma^{p})_{ss',\bar{s}\bar{s}'}\,=\,
\delta_{s\bar{s}}\delta_{s'\bar{s}'}\,
\left\{
\begin{array}{cl}
1 \quad &\mbox{for }N_s-N_{s'}\mbox{ even} \\
p \quad &\mbox{for }N_s-N_{s'}\mbox{ odd} 
\end{array}
\right.\quad.
\end{equation}
We note some important properties 
of the sign operator, which will be needed later for some proofs
\begin{eqnarray}
\label{sign_operator_property_decomposition}
\sigma^{p_1\dots p_n p_1^\prime\dots p_m^\prime}\,&=&\,
\sigma^{p_1\dots p_n}\,\sigma^{p_1^\prime\dots p_m^\prime} \quad \mbox{for }n\mbox{ even}\quad,\\
\label{sign_operator_property_G}
\sigma^{p_1\dots p_n}\,G^{\tilde{p}_1\dots\tilde{p}_n}_{1\dots n}
\,\sigma^{p_1^\prime\dots p_m^\prime}\,&=&\,
\sigma^{p_1\dots p_n p_1^\prime\dots p_m^\prime}\,
G^{\tilde{p}_1\dots\tilde{p}_n}_{1\dots n} \quad.
\end{eqnarray}
The proof is quite easy and follows directly from the definition (\ref{liouville_sign_operator})
and from the fact that, for $n$ odd, the vertex $G_{1\dots n}$ changes the parity of the 
particle number difference between the states on the upper and lower part of the Keldysh contour,
and leaves it invariant for $n$ even, i.e. for the matrix element 
$(G_{1\dots n})_{ss',\bar{s}\bar{s'}}$ we have
\begin{equation}
(-1)^{N_s-N_{s'}}\,=\,(-1)^n\,(-1)^{N_{\bar{s}}-N_{\bar{s}'}}\quad.
\end{equation}
\begin{figure}
  \centerline{\psfig{figure=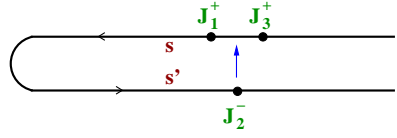,scale=0.43}}
  \caption{Determination of the fermionic sign occuring by moving the
reservoir field operator $J_2$ from the lower to the upper part of the
Keldysh contour such that the sequence $J_1^+ J_2^+ J_3^+ \dots$ is obtained.
All $J_i$ belong to the same coupling vertex. $s$ and $s'$ indicate
the intermediate states for the quantum system on the upper and lower part
of the Keldysh contour just left to the coupling vertex.}
\label{fig:fermionic_sign}
\end{figure}

The proof of (\ref{coupling_product}) is a straightforward exercise of algebra 
and is provided in Appendix B. The definition of the vertex operators 
$G^{p_1\dots p_n}_{1\dots n}$ may at first seem unusual and it deserves some
comments. It relates to the choice of the sign operators, which is in fact
not unique,
especially the distinction between $N_s-N_{s'}$ being even or odd in
(\ref{liouville_sign_operator}) is not necessary (as can be seen by the proof in
Appendix B, where this distinction is not used at all). However, there is a special
reason why the sign operator $\sigma^{p_1\dots p_n}$ has been defined in such a way,
thereby fixing the definition of the coupling vertex $G^{p_1\dots p_n}_{1\dots n}$
in Liouville space. The combination
\begin{equation}
\label{sign_property}
\sigma^{p_1\dots p_n}\,:J^{p_1}_1\,\dots\,J^{p_n}_n:
\end{equation}
has the property that the sign-operator exactly compensates {\it additional} signs due 
to interchanges of fermionic reservoir field operators, which arise due to
the presence of field operators on the lower part of the Keldysh contour,
i.e. for $p_i=-$. The sign from $\sigma^{p_1\dots p_n}$ is precisely the sign
obtained when permuting the later operators to the corresponding position
on the upper part of the Keldysh contour, assuming that all fermionic reservoir 
field operators anticommute.
This means that the determination of the fermionic sign
within Wick's theorem (see below) can be determined as if all fermionic
reservoir field operators lie on the upper part of the contour precisely
in the sequence $J_1^+\dots J_n^+$ and as if the
sign-operator were not present. Although there is a correction sign from the
permutation of field operators belonging to the same contraction to be
considered (see below), this simplifies the determination of 
fermionic signs considerably. To see this, consider the situation depicted 
in Fig.~\ref{fig:fermionic_sign}, where the second field operator $J_2^-$ 
is intended to be moved from the lower to the upper part of the Keldysh
contour along the reversed direction (only virtually to determine the 
corresponding sign from interchanges of fermionic operators). The fermionic sign
is determined by the parity of the number of fermionic reservoir 
field operators it passes. Up to the first field operator $J^+_1$, 
the parity is identical to the parity of $N_s-N_{s'}$, where $s$ and $s'$
are the intermediate states of the quantum system on the upper and lower
part of the Keldysh contour left to the considered coupling vertex. The
reason for this is the fact that the total parity of fermions (reservoirs
plus quantum system) is conserved under the coupling and, consequently, also 
only those matrix elements of the reduced density matrix are unequal to zero 
where the parity of fermions of the quantum system are the same (in other words,
the external operator $R\equiv |s\rangle\langle s'|$ in Fig.~\ref{fig:Keldysh}
does not change the parity of fermion number). Thus, by
moving $J_2$ finally also through $J_1$, we see that the parity of the total
number of interchanges is odd (even) if $N_s-N_{s'}$ is even (odd), which is
precisely the value the sign-operator (\ref{liouville_sign_operator}) produces.
The same is obtained if any $J_k$ with $k$ even is moved virtually to the 
upper part of the contour, whereas for $k$ odd the
total parity is identical to the parity of $N_s-N_{s'}$, again corresponding 
to the definition (\ref{liouville_sign_operator}). The same proof can be
used if several $J_k$ are moved to the upper part, one just has to move all
$J_k$ subsequently to the upper part, starting with the smallest $k$.

From the definitions, the following useful properties follow directly for the
Liouville operators
\begin{eqnarray}
\label{L_S_property}
\mbox{Tr}_S\,L_S\,&=&\,0 \quad,\\
\label{G_property}
\sum_{p_1\dots p_n}\,\mbox{Tr}_S\,G^{p_1\dots p_n}_{1\dots n} &=& 0 \quad,\\
\label{L_res_property}
\mbox{Tr}_{res}\,L_{res}\,&=&\,0 \quad,
\end{eqnarray}
where $\mbox{Tr}_S$ denotes the trace with respect to the states of the
quantum system. The properties (\ref{L_S_property}) and (\ref{G_property}) will turn out
to be crucial for the property of conservation of probability.

Furthermore, from the (anti-)symmetry (\ref{g_symmetry}) of $g_{1\dots n}$
we get
\begin{equation}
\label{G_symmetry}
G^{p_1\dots p_i\dots p_j\dots p_n}_{1\dots i\dots j\dots n}\,=\,
\pm\,G^{p_1\dots p_j\dots p_i\dots p_n}_{1\dots j\dots i\dots n}\quad.
\end{equation}

The hermiticity of the Hamiltonian $H_S=H_S^\dagger$ and the corresponding
condition (\ref{g_hermiticity}) for the coupling vertex imply the relations
\begin{eqnarray}
\label{L_hermiticity}
(L_S)_{ss',\bar{s}\bar{s}'} \,&=&\,
-(L_S)_{s's,\bar{s}'\bar{s}}^* \quad,\\
\label{G_hermiticity}
(G^{p_1\dots p_n}_{1\dots n})_{ss',\bar{s}\bar{s}'} \,&=&\,
-\,\left\{
\begin{array}{cl}
1\, &\mbox{for (}n\mbox{ even) or (}N_s-N_{s'}\mbox{ even)} \\
-1\, &\mbox{for (}n\mbox{ odd) and (}N_s-N_{s'}\mbox{ odd)} 
\end{array}
\right\}\,
(G^{\bar{p}_n\dots \bar{p}_1}_{\bar{n}\dots \bar{1}})_{s's,\bar{s}'\bar{s}}^* \quad,
\end{eqnarray}
where $\bar{p}=-p$. This can be shown by some straightforward algebra. The prefactor in the last 
equality stems from the term $\sigma^p$ in the
definition (\ref{G_vertex_liouville}) of the coupling vertex $G$. Using the
definiton (\ref{liouville_sign_operator_n=1}) of $\sigma^p$, it can
also be written in the form
\begin{equation}
\label{zw_sign_factor}
\left\{
\begin{array}{cl}
1\, &\mbox{for (}n\mbox{ even) or (}N_s-N_{s'}\mbox{ even)} \\
-1\, &\mbox{for (}n\mbox{ odd) and (}N_s-N_{s'}\mbox{ odd)} 
\end{array}
\right\}
\,=\,((\sigma^-)^n)_{ss',ss'}\quad.
\end{equation}
The properties (\ref{L_hermiticity}) and (\ref{G_hermiticity}) can be written 
more elegantly in operator notation by defining the c-transform $A^c$ for 
an arbitrary operator $A$ in Liouville space by 
\begin{equation}
\label{c_transformation}
(A^c)_{ss',\bar{s}\bar{s}'}\,=\,A_{s's,\bar{s}'\bar{s}}^*\quad,
\end{equation}
which, concerning the Keldysh contour, corresponds to interchanging the states
on the upper and lower part of the contour and taking the complex conjugate (note that
this definition has to be distinguished from taking the hermitian conjugate, defined
by $(A^\dagger)_{ss',\bar{s}\bar{s}'}=A_{\bar{s}\bar{s}',ss'}^*$). Using this
formal notation together with (\ref{zw_sign_factor}), (\ref{L_hermiticity}) 
and (\ref{G_hermiticity}) can be written as
\begin{eqnarray}
\label{L_c_transform}
(L_S)^c\,&=&\,-L_S \quad, \\
\label{G_c_transform}
(G^{p_1\dots p_n}_{1\dots n})^c\,&=&\,
-\,(\sigma^-)^n\,G^{\bar{p}_n\dots \bar{p}_1}_{\bar{n}\dots\bar{1}}
\quad.
\end{eqnarray}
Reversing the sequence of all indices, the last property can also be written in the form
\begin{equation}
\label{G_c_2_transform}
(G^{p_1\dots p_n}_{1\dots n})^c
\,=\,-\,\sigma^{--\dots-}\,G^{\bar{p}_1\dots \bar{p}_n}_{\bar{1}\dots\bar{n}}
\quad,
\end{equation}
where $n$ minus signs occur in the superscript of the sign operator. For the
proof we used that, for $n=2r$ ($n$ even) or $n=2r+1$ ($n$ odd), 
\begin{equation}
(\sigma^-)^n\,=\,(\pm)^r\,\sigma^{--\dots-}
\end{equation}
and (see (\ref{G_symmetry}))
\begin{equation}
G_{n\dots 1}^{p_n\dots p_1}\,=\,(\pm)^r\,G_{1\dots n}^{p_1\dots p_n}\quad.
\end{equation}
Furthermore, for later purpose, we note the useful relations 
\begin{eqnarray}
\label{c_property}
(A\,B)^c\,&=&\,A^c\,B^c \quad,\\
\label{c_dagger}
(A\,a)^\dagger\,&=&\,A^c\,a^\dagger \quad,
\end{eqnarray}
where $A,B$ are superoperators and $a$ is an operator.

We turn back to the task to perform the trace over the reservoir degrees of 
freedom in each term (\ref{perturbative_series}) of perturbation theory in $L_V$.
We insert the form (\ref{coupling_product}) for all $L_V$ and 
decompose the whole expression into a product of a part for the quantum system and the 
reservoirs by successively moving all reservoir field operators $J_i^{p_i}$
through the resolvents to the right, starting from the last $L_V$. Thereby, we 
use the property
\begin{equation}
\label{J_L_commute}
J_1^p\,L_{res} \,=\, (L_{res}-x_1)\,J_1^p \quad,
\end{equation}
where we have used the short-hand notation
\begin{equation}
\label{x_notation}
x_i \,=\, \eta_i\,(\omega_i\,+\,\mu_{\alpha_i}) \quad,
\end{equation}
which will be used frequently in the following (all frequencies occur only
in this combination).
From (\ref{J_L_commute}) we get 
\begin{equation}
\label{resolvent_commute}
:J^{p_1}_1\dots J^{p_n}_n:\,{1\over E-L_{res}-L_S}\,=\,
{1\over E+x_1+\dots+x_n-L_{res}-L_S}\,:J^{p_1}_1\dots J^{p_n}_n: \quad.
\end{equation}
In this way, all reservoir field operators can be moved to the right.
When moving the trace $\mbox{Tr}_{res}$ in (\ref{perturbative_series}) to the right, we
use the property (\ref{L_res_property}) and can set $L_{res}\rightarrow 0$ in all
resolvents. As a consequence, the term in order $r+1$ of (\ref{perturbative_series})
can be written symbolically in the product form
\begin{eqnarray}
\nonumber
&&i\,{1\over E-L_S}\,({1\over n!}\sigma\,G)_1\,{1\over E+X_1-L_S}
\,({1\over n!}\sigma\,G)_2\dots
{1\over E+X_{r}-L_S}\,({1\over n!}\sigma\,G)_{r+1}\,{1\over E-L_S}\,\rho_S(t_0)\\
&&\nonumber\\
\label{product_form}
&&\hspace{2cm}
\cdot\,\langle (:J\dots J:)_1\,(:J\dots J:)_2\,\dots\,(:J\dots J:)_{r+1} \rangle_{\rho_{res}}
\quad,
\end{eqnarray}
where $({1\over n!}\sigma G)_i$ indicates the sign and vertex operator at the $i$-th position
and $(:J\dots J:)_i$ is the corresponding sequence of reservoir field operators. The energies
$X_i$ are given by the sum of all $x_k$-variables from the field operators $J_k^{p_k}$
which occured left to the $i$-th resolvent.

\begin{figure}
  \centerline{\psfig{figure=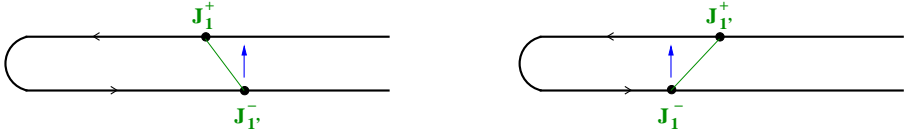,scale=0.43}}
  \caption{Illustration why an additional fermionic sign is obtained when
two field operators are contracted with the right field operator being on the
lower part of the Keldysh contour (left figure). Moving $J_{1'}$ from the
lower to the upper part along the reversed direction of the contour, it
passes through $J_1$, giving rise to an additional fermionic sign compared
to the sign from Wick's theorem. In contrast, if the 
right vertex lies on the upper part (right figure), the two field operators
of the contraction do not pass through each other when moving $J_1$ to
the upper part.}
\label{fig:additional_sign}
\end{figure}

\begin{figure}
  \centerline{\psfig{figure=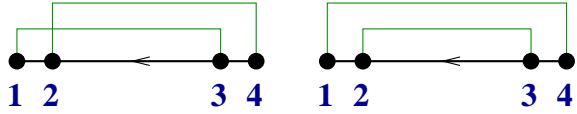,scale=0.43}}
  \caption{Diagrammatic representations of the two contributions of
Eq.~(\ref{Wick}). A dot with index $i$ represents a field operator $J_i^{p_i}$.
Dots standing close to each other belong to the same vertex $G$.
For clarity they are separated to
indicate their sequence in the original expression.
This is important for the determination of the correct fermionic sign.
Contractions are indicated by green lines connecting the dots. The horizontal black 
lines between the vertices represent the propagation of the local quantum system in
Liouville space.}
\label{fig:Wick}
\end{figure}

The reservoir part of (\ref{product_form}) can easily be decomposed into product of 
pair contractions by using Wick's theorem. We define the following contraction for the
Liouville field operators, which can easily be calculated from (\ref{contraction})
\begin{equation}
\label{liouville_contraction}
\gamma_{11'}^{pp'}\,=\,{J_1^p\,J_{1'}^{p'}
  \begin{picture}(-20,11) 
    \put(-22,8){\line(0,1){3}} 
    \put(-22,11){\line(1,0){12}} 
    \put(-10,8){\line(0,1){3}}
  \end{picture}
  \begin{picture}(20,11) 
  \end{picture}
}
\,\equiv\,
\left\{
\begin{array}{cl}
1 \\ p'
\end{array}
\right\}\,
\langle J_1^p J_{1'}^{p'}\rangle_{\rho_{res}}
\,\equiv\,
\left\{
\begin{array}{cl}
1 \\ p'
\end{array}
\right\}\,
\mbox{Tr}_{res}\,J_1^p J_{1'}^{p'}\,\rho_{res}
\,=\,\delta_{1\bar{1}'}\,p'
\left\{
\begin{array}{cl}
\eta \\ 1
\end{array}
\right\}\,
\,f_\alpha(\eta p' \omega) \quad,
\end{equation}
or
\begin{equation}
\label{alternative_liouville_contraction}
\gamma_{11'}^{pp'}
\,=\,\delta_{1\bar{1}'}\,p'
\left\{
\begin{array}{cl}
\eta \\ 1
\end{array}
\right\}\,
\rho_\nu(\omega)\,f_\alpha(\eta p' \omega) \quad,
\end{equation}
if the density of states $\rho_\nu(\omega)$ is taken into the contraction
according to the choice (\ref{alternative_field_operators}) and 
(\ref{alternative_contraction}).
The upper (lower) value corresponds as usual to bosons (fermions). 
Note that there is a prefactor $p'$ for fermions in the definition of the contraction
which arises as follows. As explained above the fermionic sign from the sign operators
is compensated by moving all $J_i^{p_i}$ with $p_i=-$ to the upper part of the Keldysh contour.
This means that we can use the rule that each interchange of two $J_i^{p_i}$ gives a minus sign for fermions,
independent of the value of the Keldysh index $p_i$. However, in doing so, we do {\it not}
obtain the correct sign from Wick's theorem where it is not allowed to permute two
field operators belonging to the same contraction. If we consider a contraction with $p'=-$,
we see that we permute the two field operators belonging to this contraction when moving 
$J_{1'}^-$ to the upper part of the contour, see Fig.~\ref{fig:additional_sign} for an illustration.
Therefore, in order to get the correct sign from Wick's theorem, we have to permute these two
field operators back, leading to an additional sign for fermions. A minus sign is only obtained 
for fermions and $p'=-$, leading to the prefactor $\left\{\begin{array}{cl}1 \\ p'\end{array}\right\}\,$
in (\ref{liouville_contraction}).
As a consequence, we use the following rules for the Wick decomposition
\begin{enumerate}
\item
Contract all $J$-operators in (\ref{product_form}) such that no contractions occur within
the normal-ordered parts.
\item
Disentangle the contractions into a product of pair contractions by leaving the sequence of 
$J$-operators within one contraction invariant. For each interchange of $J$-operators, give
a minus sign for fermions.
\item
Translate the contractions with (\ref{liouville_contraction}) and sum over all
possibilites to contract the field operators.
\end{enumerate}
As an example, we obtain for
\begin{eqnarray}
  \label{Wick}
  \langle :J_1^{p_1} J_2^{p_2}: \, :J_3^{p_3} J_4^{p_4}: \rangle_{\rho_{res}} &=&
  J_1^{p_1} J_2^{p_2} J_3^{p_3} J_4^{p_4}
  \begin{picture}(-32,11)
    \put(-55,10){\line(0,1){3}}
    \put(-55,13){\line(1,0){31}}
    \put(-24,10){\line(0,1){3}}
    \put(-40,10){\line(0,1){6}}
    \put(-40,16){\line(1,0){31}}
    \put(-9,10){\line(0,1){6}}
  \end{picture}
  \begin{picture}(32,11)
  \end{picture}
+ J_1^{p_1} J_2^{p_2} J_3^{p_3} J_4^{p_4}
  \begin{picture}(-32,11)
    \put(-55,10){\line(0,1){6}}
    \put(-55,16){\line(1,0){46}}
    \put(-9,10){\line(0,1){6}}
    \put(-40,10){\line(0,1){3}}
    \put(-40,13){\line(1,0){16}}
    \put(-24,10){\line(0,1){3}}
  \end{picture}
  \begin{picture}(32,11)
  \end{picture}
\nonumber \\
&&\nonumber\\
&=& \pm\,\gamma^{p_1 p_3}_{13}\,\gamma^{p_2 p_4}_{24}\,+\,
\gamma^{p_1 p_4}_{14}\,\gamma^{p_2 p_3}_{23}\quad.
\end{eqnarray}
The corresponding diagrammatic representation is shown in Fig.~\ref{fig:Wick}.

\begin{figure}
  \centerline{\psfig{figure=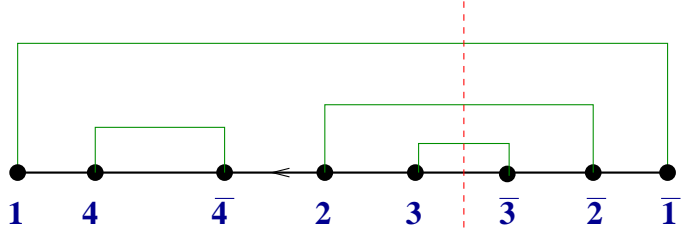,scale=0.43}}
  \caption{Rule how to calculate the energy variables of the resolvents.
The energy $X$ corresponding to the resolvent standing between index $3$ and 
$\bar{3}$ is given by $X=x_1+x_2+x_3$, where $x_i$, $i=1,2,3$ are
the variables corresponding to the contractions which cross the auxiliary
vertical red line. Thereby, the $x$-variable of the field operator standing left
to the resolvent has to be chosen. Note that the variable $x_4$ belonging
to the contraction between index $4$ and $\bar{4}$ does not contribute
because it does not cross the vertical cut.}
\label{fig:energy_rule}
\end{figure}

The energies $X_i$ in (\ref{product_form}) can also be determined by a 
simple rule. Since the contraction (\ref{liouville_contraction}) is only
nonzero for $\eta=-\eta'$, $\omega=\omega'$, and $\alpha=\alpha'$, we
get $x+x'=0$ according to the notation (\ref{x_notation}). Thus, all 
contractions which either stand completely to the left or to the right of the $i$-th
resolvent, do not contribute to the energy $X_i$. Only the $x$-variables from
contractions, which cross over the resolvent, contribute to $X_i$. Thereby, 
the $x$-variable has to be chosen from the 
$J$-operator standing left to the resolvent. This can be visualized by a
simple diagrammatic rule, shown in Fig.~\ref{fig:energy_rule}. At the
position of the resolvent under consideration, draw an auxiliary 
vertical line and consider all $x$-variables of contractions which cross
this line (always taking the $x$-variable from the vertex standing left
to the resolvent). The sum of all these $x$-variables is the energy $X_i$.

Finally, we consider the determination of the prefactor, arising from the
combinatorical factors ${1\over n_i!}$ in (\ref{product_form}). These cancel
almost completely with another factor arising from the number of identical diagrams
when the $J$-operators within each normal-ordered set are permuted. The
value of all these diagrams is exactly the same because the two fermionic signs arising 
from interchanging two reservoir field operators from the same vertex and from interchanging 
the two corresponding indices of the coupling 
vertex $G$ cancel each other, see (\ref{G_symmetry}). However, if $m$
contractions are present which connect the same normal-ordered blocks, 
there are $m!$ ways of permuting 
the $J$-operators of both groups in the same way without giving a new diagram. 
Therefore, in this case, the combinatorical factors can not be omitted completely 
but a symmetry factor $1/m!$ remains.

\begin{figure}
  \centerline{\psfig{figure=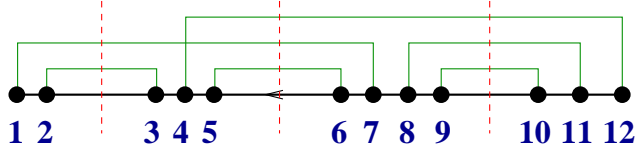,scale=0.43}}
  \caption{Example of a diagram with four vertices. The vertical cuts are
auxiliary lines between the vertices to determine the indices of the 
resolvents $\Pi_{\dots}$
in (\ref{diagram_example}). Precisely those indices occur which belong to the
left field operator of each contraction crossing the vertical cut.}
\label{fig:diagram_example}
\end{figure}

{\bf Summary of diagrammatic rules.} We summarize the diagrammatic rules derived in this section
to calculate the reduced density matrix of the quantum system
in Laplace space. The value of a diagram is symbolically given by
\begin{eqnarray}
\nonumber
\tilde{\rho}_S(E) &\,\rightarrow\,& {i\over S} \, (\pm)^{N_p} \, \left(\prod\gamma\right)
\,\\
\label{value_diagram}
&& \cdot\,{1\over E-L_S}\,G\,{1\over E+X_1-L_S}\,G\,\dots \,G\,
{1\over E+X_r-L_S}\,G\,{1\over E-L_S}\,\rho_S(t_0)
\quad,
\end{eqnarray}
where we use the following rules to calculate the various parts
\begin{enumerate}

\item
$S=\prod_{i<j} m_{ij}!$ is a symmetry factor, where $m_{ij}$ is the number of contractions
between vertex $i$ and $j$. Two diagrams are considered to be different if they
can not be mapped on each other by permuting only the field operators of each vertex (the
field operators are indicated by dots in a diagram, where dots standing close to each
other belong to the same vertex).

\item
$(\pm)^{N_p}$ is a fermionic sign factor, where $N_p$ is the number of interchanges
of fermionic field operators $J^p_i$ in Liouville space which are needed to write the
contractions in product form.

\item
$\prod\gamma$ stands for the product of all contractions. If  $J_1^p$ and $J_{1'}^{p'}$
are contracted, and $J_1^p$ stands left to $J_{1'}^{p'}$, the contraction is given
by $\gamma^{pp'}_{11'}$, with (see (\ref{liouville_contraction}))
\begin{equation}
\gamma^{pp'}_{11'}\,=\,\delta_{1\bar{1}'}\,p'
\left\{
\begin{array}{cl}
\eta \\ 1
\end{array}
\right\}\,
\,f_\alpha(\eta p' \omega) \quad,
\end{equation}
where $f(\omega)=1/(e^{\omega/T_\alpha}\mp 1)$ is the Bose (Fermi) function, and
$\delta_{1\bar{1}'}=\delta_{\eta,-\eta'}\delta_{\nu\nu'}\delta(\omega-\omega')$.

\item
To determine the energy argument $X_i$ of resolvent $i$, we draw an auxiliary vertical
cut at the position of that resolvent. $X_i$ is the sum of all $x$-variables of the
contractions which cross the vertical cut.
The $x$-variable of a contraction $\gamma^{pp'}_{11'}$ is defined as 
$x=\eta(\omega+\mu_\alpha)$, i.e. refers to the {\it left} $J_1^p$-operator of the contraction.

\item
$G\equiv G^{p_1\dots p_n}_{1\dots n}$ are the coupling vertices acting on the quantum system,
defined by (\ref{G_vertex_liouville}). $L_S=[H_S,\cdot]_-$ is the Liouville 
operator of the quantum system.

\item
The formal indices $1\equiv \eta\nu\omega$ and $\nu\equiv\alpha\sigma\dots$ contain
the index $\eta=\pm$ for creation/annihilation operators, the energy $\omega$ of the
reservoir state (relative to the chemical potential $\mu_\alpha$), the reservoir 
index $\alpha$, the spin index $\sigma$, and possible other quantum numbers 
characterizing the reservoir state. We sum (integrate) over all these indices
implicitly.  

\end{enumerate}
To write the resolvents in a compact way, we use the short-hand notation
\begin{equation}
\label{pi_resolvent}
\Pi_{12\dots n}\,\equiv\,{1\over E+x_1+x_2+\dots x_n-L_S}\quad.
\end{equation}
With this convention, the diagram example of Fig.~\ref{fig:diagram_example}
is given by
\begin{eqnarray}
\nonumber
{i\over 2}\,(\pm)\,\gamma_{17}\,\gamma_{23}\,\gamma_{4,12}\,\gamma_{56}\,
\gamma_{8,11}\,\gamma_{9,10}&&\\
\label{diagram_example}
&&\hspace{-3.5cm}\cdot\,{1\over E-L_S}\,G_{12}\,\Pi_{12}\,G_{345}\,\Pi_{145}\,
G_{6789}\,\Pi_{489}\,G_{10,11,12}\,{1\over E-L_S}\,\rho_S(t_0)\quad,
\end{eqnarray}
where we have omitted for simplicity the obvious Keldysh indices at the
couping vertices $G$ and the contractions $\gamma$.
The factor ${1\over 2}$ arises from the fact that the third and fourth
vertex are connected by two contractions, and the sign factor $\pm$ stems from
taking the contraction $\gamma_{4,12}$ out of the rest.

\begin{figure}
  \centerline{\psfig{figure=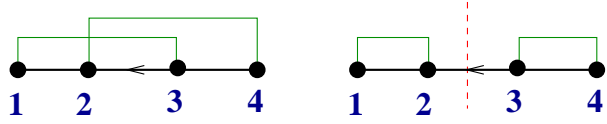,scale=0.43}}
  \caption{Irreducible (left) and reducible (right) diagrams. In the left figure
any vertical cut between the vertices hits at least one reservoir contraction.
In the right figure, the vertical cut between the second and third vertex does
not hit any contraction and corresponds to a resolvent of the form ${1\over E-L_S}$.}
\label{fig:irreducible_reducible}
\end{figure}

{\bf Kinetic equation.} The diagrammatic expansion can be formally resummed and written
in the form of a kinetic equation by distinguishing between irreducible and
reducible diagrams. Irreducible diagrams are those diagrams where any vertical
cut hits at least one reservoir contraction, i.e. in each resolvent at least one
$x$-variable occurs. In contrast, in reducible diagrams there are vertical cuts
crossing no contraction, corresponding to a resolvent of the form $1/(E-L_S)$
without any $x$-variable, see Fig.~\ref{fig:irreducible_reducible} for illustration.

We denote the sum over all irreducible diagrams by the irreducible kernel $\Sigma(E)$.
Using (\ref{value_diagram}), we obtain the following value for any diagram of the kernel
\begin{equation}
\label{value_sigma}
\Sigma(E) \,\rightarrow\, {1\over S} \, (\pm)^{N_p} \, \left(\prod\gamma\right)_{irr}
\,G\,{1\over E+X_1-L_S}\,G\,\dots \,G\,{1\over E+X_r-L_S}\,G \quad,
\end{equation}
where $(\prod\gamma)_{irr}$ means that we only consider irreducible diagrams. 
Compared to (\ref{value_diagram}), we have omitted in this definition the prefactor $i$,
the first and last resolvent, and the initial density matrix $\rho_S(t_0)$.

Each diagram for $\tilde{\rho}_S(E)$ can be written as a sequence
of irreducible parts with resolvents $1/(E-L_S)$ in between. Similiar to Dyson equations
within Green's function methods, all irreducible diagrams can be formally resummed to
define the kernel $\Sigma(E)$, and the total sum of all diagrams can be written as a geometric series
\begin{equation}
\label{resummation}
\tilde{\rho}_S(E)\,=\,i\,{1\over E-L_S}\,\sum_{n=0}^\infty\,
\left(\Sigma(E){1\over E-L_S}\right)^n\,\rho_S(t_0)\,=\,
{i\over E-L_S-\Sigma(E)}\,\rho_S(t_0)\quad,
\end{equation}
leading to the final result
\begin{equation}
\label{solution_rd}
\tilde{\rho}_S(E)\,=\,{i\over E-L_S^{eff}(E)}\,\rho_S(t_0)\quad,
\end{equation}
where
\begin{equation}
\label{L_eff}
L_S^{eff}(E)\,=\,L_S\,+\,\Sigma(E)
\end{equation}
is an effective Liouville operator of the quantum system, which depends on the Laplace 
variable $E$. From (\ref{G_property}) and the fact that any diagram for $\Sigma(E)$ starts
with a coupling vertex $G$, we get the important property
\begin{equation}
\label{L_eff_property}
\mbox{Tr}_S\,\Sigma(E)\,=\,\mbox{Tr}_S\,L_S^{eff}(E)\,=\,0\quad,
\end{equation}
which is important for conservation of probability (see below). Furthermore, in analogy to
(\ref{L_c_transform}), we have
\begin{equation}
\label{L_eff_c_transform}
(L_S^{eff}(E))^c\,=\,-L_S^{eff}(-E^*) \quad,\quad
(\Sigma(E))^c\,=\,-\Sigma(-E^*) \quad.
\end{equation}
The proof of this relation is provided in Appendix C.

\begin{figure}
  \centerline{\psfig{figure=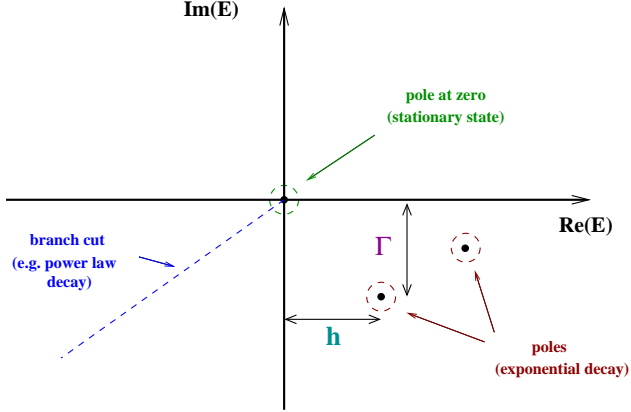,scale=0.43}}
  \caption{Analytic properties of $\tilde{\rho}_S(E)$. Poles  
occur in the lower half plane leading to exponential decay. $\Gamma$ denotes the
relaxation or dephasing rate, $h$ is the oscillation frequency. The pole at
$E=0$ is always present and unique, it corresponds to the stationary state.
Generically, branch cuts will also occur, leading e.g. to power law decay.}
\label{fig:analytic_structure}
\end{figure}

Eq.~(\ref{solution_rd}) is the central result of this section. It shows very clearly
the effect of the coupling to the reservoirs. In the absence of a coupling to the reservoirs, the kernel
$\Sigma(E)$ is zero, and $L_S^{eff}$ is identical to the bare Liouvillian $L_S$, which
is hermitian. As a consequence, the poles of $\tilde{\rho}_S(E)$ lie on the real axis,
corresponding to coherent Rabi oscillations of the quantum system in time space. In
contrast, when the coupling to the reservoirs is nonzero, a dissipative part $\Sigma(E)$ has to be
added to the effective Liouvillian and the analytic structure of the reduced density
matrix changes in Laplace space as illustrated in Fig.~\ref{fig:analytic_structure}.
Generically, a branch cut will occur on the real axis due to the continuous spectrum
of the reservoirs. Analogous to the theory of quantum decay, we turn this branch cut 
into the lower half of the complex plane by analytic
continuation. This leads to poles in the lower half plane, which originally were on the
real axis in the absence of the coupling to the reservoirs. These poles correspond to exponential
decay, the negative imaginary part is the decay rate $\Gamma$ (relaxation or dephasing 
rate, depending on whether the mode corresponds to decay of diagonal or nondiagonal 
matrix elements of the reduced density matrix of the quantum system), and the real
part corresponds to the oscillation frequency $h$ (e.g. an effective magnetic field).
The remaining branch cuts can e.g. lead to power law decay, but usually their prefactor 
is smaller than the one of the exponential decay modes, and they dominate only  
the long-time behaviour. Generically, there will be always a single pole at $E=0$, which 
corresponds to the stationary state (for certain symmetries or in the case of symmetry 
breaking, it may not be unique accidentally). It is determined by the eigenvalue 
equation
\begin{equation}
\label{eigenvalue_zero}
L_S^{eff}(i0^+)\,\rho_S^{st}\,=\,0\quad.
\end{equation}
This pole will play an essential role within the RG formalism presented in Sec.~\ref{sec:4},
since it does not decay. It will be shown that it can be included perturbatively into
the initial condition of effective vertices, whereas the decay rates $\Gamma_i$ and the oscillation
frequencies $h_i$ provide the cutoff of the RG flow. An important advantage of the present
formalism is that the physical decay rates, determining the time evolution of the reduced 
density matrix of the quantum system, follow directly from the poles of 
Eq.~(\ref{solution_rd}). Once the kernel $\Sigma(E)$ has been calculated within perturbation
theory (using the diagrammatic rules (\ref{value_sigma})) or even nonperturbatively using
the RTRG-FS scheme set up in Sec.~\ref{sec:4}, the decay modes can easily be found. 
In contrast, within slave particle formalism, 
the physical decay rates have to be calculated in a complicated way by combining self-energy 
insertions and vertex corrections \cite{paaske_rosch_kroha_woelfle_PRB04}, and within 
flow-equation methods the decay rates follow from the energy scale where certain 1-loop and 
2-loop contributions become of equal order on the r.h.s. of the flow equations.

Conservation of probability follows from (\ref{L_eff_property}).
Acting with the trace $\mbox{Tr}_S$ over the quantum system on Eq.~(\ref{solution_rd}),
we obtain
\begin{equation}
\label{zw_conservation}
\mbox{Tr}_S\,\tilde{\rho}_S(E)\,=\,{i\over E}\,\mbox{Tr}_S\,\rho_S(t_0)\quad,
\end{equation}
which, after transforming back to time space, gives
\begin{equation}
\label{conservation}
\mbox{Tr}_S\,\rho_S(t)\,=\,\mbox{Tr}_S\,\rho_S(t_0)\quad,
\end{equation}
i.e. the normalization of the reduced density matrix is invariant and stays unity
if it is normalized initially $\mbox{Tr}_S\rho(t_0)=1$. Note that this property
holds within any diagrammatic approximation, because any diagram for the kernel
$\Sigma(E)$ starts with a coupling vertex $G$. We will also see in Sec.~\ref{sec:4}
that the RG flow within RTRG-FS preserves conservation of probability in any 
approximation scheme.

Furthermore, by applying the property (\ref{c_dagger}) to (\ref{solution_rd}), we
can show from (\ref{L_eff_c_transform}) and $\rho_S(t_0)^\dagger=\rho_S(t_0)$ that 
$\tilde{\rho}_S(E)$ fulfils the condition
\begin{equation}
\label{rho_laplace_hermiticity}
\tilde{\rho}_S(E)^\dagger\,=\,\tilde{\rho}_S(-E^*)\quad,
\end{equation} 
which is equivalent to the hermiticity of the reduced density matrix in time space
\begin{equation}
\label{rho_time_hermiticity}
\rho_S(t)^\dagger\,=\,\rho_S(t)\quad.
\end{equation}

Eq.~(\ref{solution_rd}) can also be written in time space leading to a kinetic equation.
Multiplying Eq.~(\ref{solution_rd}) with $-i(E-L_S-\Sigma(E))$, we obtain
\begin{equation}
\label{zw_kinetic_equation}
\left[-i\,E\,\tilde{\rho}_S(E)\,-\,\rho_S(t_0)\right]\,+\,i\,L_S\,\tilde{\rho}_S(E)\,=\,
-i\,\Sigma(E)\,\tilde{\rho}_S(E)\quad.
\end{equation}
Using (\ref{laplace}) and the definition
\begin{equation}
\label{sigma_laplace}
\Sigma(E)\,=\,\int_0^\infty dt\,e^{iEt}\,\Sigma(t)
\end{equation}
for the kernel in time space, we see that (\ref{zw_kinetic_equation}) is
equivalent to the following kinetic equation in time space
\begin{equation}
\label{kinetic_equation}
{d\over dt}\,\rho_S(t)\,+\,i\,L_S\,\rho_S(t)\,=\,
-i\,\int_{t_0}^t dt'\,\Sigma(t-t')\,\rho_S(t')\quad.
\end{equation}
The second term on the l.h.s. corresponds to the von Neumann equation of the isolated
quantum system, whereas the r.h.s. describes the non-Markovian dissipative influence
of the coupling to the reservoirs.

Kinetic equations of the form (\ref{kinetic_equation}) are not new in dissipative 
quantum mechanics and can also be derived by other methods, e.g. with projection 
operators \cite{zwanzig} or within slave particle techniques \cite{rosch_paaske_kroha_woelfle_PRL03}.
However, the crucial point is not the form of the kinetic equation (which is trivial and
obvious on physical grounds) but the way the kernel $\Sigma$ is calculated.
Using projection operator techniques a purely formal but quite compact expression of 
the kernel is obtained where certain projectors $Q=1-P$, with $P=\rho_{res}\mbox{Tr}_{res}$,
occur between the Liouville operators $L_V$ projecting on the irreducible part. However, this does not
help at all for the calculation of $\Sigma$ because the reservoir degrees of freedom
are still present in $L_V$. Only after the insertion of $Q=1-P$,
an explicit calculation can be started but an artificial decomposition into reducible
parts is created, induced by the projection operator $P$. All the reducible terms finally cancel
in a complicated way, leaving only those terms of the Wick decomposition which are irreducible.
In contrast, the diagrammatic rule (\ref{value_sigma}) derived in this 
section considers directly the irreducible terms, and the reservoir
degrees of freedom are already integrated out. Using slave-particle techniques and
Keldysh-formalism, the derivation of a kinetic equation (or quantum Boltzmann equation)
is quite complicated (even in lowest order perturbation theory), and the kernel $\Sigma$
is a complicated mixture of self-energy contributions and vertex corrections. Therefore, we
believe that the calculation of the kernel via the diagrammatic rule (\ref{value_sigma})
is very efficient. This has been demonstrated recently within perturbation theory up to
fourth order in the coupling vertex for problems of molecular electronics \cite{leinjse_wegewijs}. 
In Sec.~\ref{sec:4} we will see that it is especially useful for setting up nonequilibrium RG methods
which incorporate the physics of relaxation and dephasing.

\subsection{Observables}
\label{sec:3.2}

{\bf Observables.} The time evolution of an arbitrary observable $R$ can be calculated starting 
from Eqs.~(\ref{observable}) and (\ref{Liouville_observable})
\begin{equation}
\label{zw_observable}
\langle R \rangle(t) \,=\,
\mbox{Tr}_S\,\mbox{Tr}_{res}\,(-iL_R)e^{-iL(t-t_0)}\,\rho_S(t_0)\rho_{res}\quad,\quad 
L_R\,=\,{i\over 2}\,[R,\cdot]_+ \quad.
\end{equation}
The observable $R$ is written in the same generic form (\ref{coupling}) as the 
coupling $V$
\begin{equation}
\label{form_observable}
R\,=\,{1\over n!}\,r_{1\dots n}\,:a_1\dots a_n:\quad,
\end{equation}
with $n=0,1,2,\dots$ (in contrast to $V$, where the $n=0$ case can be incorporated in
$H_S$, this is not possible for $R$). 
Due to (anti-)symmetry and hermiticity, we get similiar to (\ref{g_symmetry})
and (\ref{g_hermiticity}) the properties
\begin{eqnarray}
\label{r_symmetry}
r_{1\dots i\dots j\dots n}\,&=&\,\pm\,r_{1\dots j\dots i\dots n}\quad,\\
\label{r_hermiticity}
(r_{12\dots n})^\dagger \,&=&\, r_{\bar{n} \dots \bar{1}} \quad.
\end{eqnarray}

Analogous to (\ref{coupling_product}), we obtain a corresponding form for the operator $L_R$
in Liouville space
\begin{equation}
\label{liouville_observable}
L_R\,=\,{1\over n!}\,\sigma^{p_1\dots p_n}\,R^{p_1\dots p_n}_{1\dots n}\,
:J^{p_1}_1\dots J^{p_n}_n:\quad,
\end{equation}
with the vertex of the observable given by
\begin{equation}
\label{vertex_observable}
R^{p_1\dots p_n}_{1\dots n}\,A\,=\,{i\over 2}\,
\delta_{pp_1}\dots\delta_{pp_n}\,
\left\{
\begin{array}{cl}
1\, &\mbox{for }n\mbox{ even} \\
\sigma^p\, &\mbox{for }n\mbox{ odd}
\end{array}
\right\}\,
\left\{
\begin{array}{cl}
r_{1\dots n}\,A\, &\mbox{for }p=+ \\
A \,r_{1\dots n}\, &\mbox{for }p=-
\end{array}
\right.\quad.
\end{equation}
The difference to (\ref{G_vertex_liouville}) stems from the form $L_R={i\over 2}[R,\cdot]_+$ 
where the anticommutator and a different prefactor occurs in comparism to $L_V=[V,\cdot]_-$. The
prefactor $i/2$ is taken into the definition of the vertex $R$ to get diagrammatic rules 
similiar to the ones for the reduced density matrix and the kernel $\Sigma$ (see below). We
note that the trace over the states of the quantum system 
always occurs left to $L_R$ in the average (\ref{zw_observable}), i.e. we get 
\begin{equation}
\mbox{Tr}_S\,L_R\dots\,=\,\sum_s\,(L_R)_{ss,\cdot\cdot}\dots \quad.
\end{equation}
Therefore, according to the definition (\ref{liouville_sign_operator_n=1}), 
the sign operator $\sigma^p$ 
can be dropped in (\ref{vertex_observable}) for the calculation of $\langle R\rangle(t)$,
and $\sigma^{p_1\dots p_n}$ can be replaced by
\begin{equation}
\sigma^{p_1\dots p_n}\,\rightarrow\,p_2 p_4 \dots \quad.
\end{equation}
Nevertheless, for the formal identity (\ref{vertex_observable}) the sign operators
have to be used in their general form.

Similiar to (\ref{G_c_transform}), one can easily show from the definition (\ref{vertex_observable})
and the hermiticity condition (\ref{r_hermiticity}) the relation
\begin{equation}
\label{R_c_transform}
(R^{p_1\dots p_n}_{1\dots n})^c\,=\,
-\,(\sigma^-)^n\,R^{\bar{p}_n\dots \bar{p}_1}_{\bar{n}\dots\bar{1}}
\quad.
\end{equation}

In Laplace space we obtain from (\ref{zw_observable})
\begin{equation}
\label{observable_laplace}
\tilde{\langle R\rangle}(E)\,=\,\int_{t_0}^\infty dt\,e^{iE(t-t_0)}\,
\langle R\rangle (t)\,=\,
\mbox{Tr}_S\,\mbox{Tr}_{res}\,L_R\,{1\over E-L_{res}-L_S-L_V}\,\rho_S(t_0)\rho_{res}\quad,
\end{equation}
which, after expanding in $L_V$, leads to terms of the form
\begin{equation}
\label{perturbative_series_observable}
\mbox{Tr}_S\,\mbox{Tr}_{res}\,L_R\,{1\over E-L_{res}-L_S}\,L_V\,{1\over E-L_{res}-L_S}\,L_V\,
\dots \,L_V\,{1\over E-L_{res}-L_S}\,\rho_S(t_0)\rho_{res}\quad.
\end{equation}
The trace over the reservoirs can be evaluated in the same way as described in Sec.~\ref{sec:3.1}
for (\ref{perturbative_series}), and we obtain for a certain diagram of the average of an observable
in analogy to (\ref{value_diagram})
\begin{eqnarray}
\nonumber
\tilde{\langle R\rangle}(E) &\,\rightarrow\,& {1\over S} \, (\pm)^{N_p} \, \left(\prod\gamma\right)
\,\\
\label{value_diagram_observable}
&& \cdot\,\mbox{Tr}_S\,R\,{1\over E+X_1-L_S}\,G\,{1\over E+X_2-L_S}\,\dots \,G\,
{1\over E+X_r-L_S}\,G\,{1\over E-L_S}\,\rho_S(t_0)
\quad
\end{eqnarray}
with the essential difference that the first vertex corresponds to the vertex of the observable $R$ 
and the trace $\mbox{Tr}_S$ over the quantum system has to be performed.

Decomposing (\ref{value_diagram_observable}) into reducible and irreducible parts and
resumming formally all diagrams leads to
\begin{equation}
\label{solution_observable}
\tilde{\langle R\rangle}(E)\,=\,\mbox{Tr}_S\,\Sigma_R(E)\,
{1\over E-L_S^{eff}(E)}\,\rho_S(t_0)\,=\,-i\,\mbox{Tr}_S\,\Sigma_R(E)\,\tilde{\rho}_S(E)\quad,
\end{equation}
where we have used (\ref{solution_rd}) in the last equality. Here, the kernel $\Sigma_R(E)$
is defined analogous to $\Sigma(E)$ with the only difference that the first vertex is replaced by
the observable vertex $R$, i.e. in analogy to (\ref{value_sigma}) the observable kernel 
$\Sigma_R(E)$ is given by the diagrams
\begin{equation}
\label{value_sigma_observable}
\Sigma_R(E) \,\rightarrow\, {1\over S} \, (\pm)^{N_p} \, \left(\prod\gamma\right)_{irr}
\,R\,{1\over E+X_1-L_S}\,G\,\dots \,G\,{1\over E+X_r-L_S}\,G \quad.
\end{equation}

Eqs.~(\ref{solution_observable}) and (\ref{value_sigma_observable}) are the
final result for the average of an observable. Once the kernels $\Sigma_R(E)$ and $\Sigma(E)$ have been 
calculated, they can be inserted into Eq.~(\ref{solution_observable}) and the average of an observable
can be calculated for all values of the Laplace variable $E$, i.e. the full time evolution can
be obtained by transforming to time space. The stationary value of the average of an observable
is given by
\begin{equation}
\label{zw_observable_stationary}
\langle R\rangle^{st}\,=\,\lim_{t\rightarrow\infty}\langle R\rangle(t)
\,=\,-i\lim_{E\rightarrow i0^+} E\,\tilde{\langle R\rangle}(E)\quad,
\end{equation}
which, using (\ref{solution_observable}), gives
\begin{equation}
\label{observable_stationary}
\langle R\rangle^{st}\,=\,-i\,\mbox{Tr}_S\,\Sigma_R(i0^+)\,\rho_S^{st}\quad,
\end{equation}
where $\rho_S^{st}$ is the stationary value of the reduced density matrix of the
quantum system, which can be calculated from $L_S^{eff}(i0^+)\rho_S^{st}=0$, 
see (\ref{eigenvalue_zero}).

Finally, we note that similiar to (\ref{L_eff_c_transform}), one can prove
\begin{equation}
\label{sigma_R_c_transform}
(\Sigma_R(E))^c\,=\,-\Sigma_R(-E^*) \quad, 
\end{equation}
which, together with the hermiticity condition
(\ref{rho_laplace_hermiticity}) for the reduced density matrix, proves that
(\ref{solution_observable}) respects the hermiticity of the observable $R$
\begin{equation}
\label{R_laplace_hermiticity}
\langle \tilde{R} \rangle(E)^*\,=\,\tilde{\langle R \rangle}(-E^*)\quad,
\end{equation}
implying that $\langle R \rangle(t)$ is real in time space.

An important example of an observable is the particle current operator $I^\gamma$ 
flowing from reservoir $\gamma$ to the quantum system (for electrons, the charge current 
is obtained by multiplying with $-e$). It is defined by
\begin{equation}
\label{current_definition}
I^\gamma\,=\,-{d\over dt}N_{res}^\gamma \quad,
\end{equation}
where $N_{res}^\gamma$ is the particle number operator for reservoir $\gamma$
and the derivative is calculated via the Heisenberg picture for the 
Hamiltonian (\ref{H_total})
\begin{eqnarray}
\label{Heisenberg}
\nonumber
I^\gamma\,&=&\,-i\,[H,N^\gamma_{res}]\,=\,-i\,[V,N^\gamma_{res}]\,=\,\\
\nonumber
&=&\,-i\,{1\over n!}\,g_{1\dots n}\,[:a_1\dots a_n:,N^\gamma_{res}]\\
\label{zw_current}
&=&\,{1\over n!}\,i\sum_{i=1}^n\,\eta_i\,\delta_{\alpha_i\gamma}\,
g_{1\dots n}\,:a_1\dots a_n: \quad,
\end{eqnarray}
where the form (\ref{coupling}) for $V$ has been inserted, and we used 
the identity $[a_1,N_{res}^\gamma]=-\eta\delta_{\alpha\gamma}a_1$ in the
last step. Note that we mean by the time derivative 
${d\over dt}N_{res}^\gamma$ of the particle number in reservoir $\gamma$
not the total time derivative, including the one from the coupling of the
reservoir to the bath maintaining the temperature $T_\gamma$ and the
chemical potential $\mu_\gamma$ of the reservoir (leading to the
grandcanonical distribution for reservoir $\gamma$). This total time
derivative would be zero on average since the average particle number in the reservoir is
a constant. Therefore, to get a definition of the local current operator at the position
where the reservoir is coupled to the quantum system, we include in the time
derivative only the term $i[V,N_{res}]$ due to the coupling between reservoir 
and quantum system.

In summary, the current operator can be brought into the form (\ref{form_observable})
\begin{equation}
\label{current_operator}
I^\gamma\,=\,{1\over n!}\,i^\gamma_{1\dots n}\,:a_1\dots a_n:\quad,
\end{equation}
with
\begin{equation}
\label{current_vertex}
i^\gamma_{1\dots n}\,=\,i\,\sum_{i=1}^n\,\eta_i\,\delta_{\alpha_i\gamma}
g_{1\dots n}\quad.
\end{equation}
Inserting this form of $i^\gamma_{1\dots n}$ into (\ref{vertex_observable}) 
for $r_{1\dots n}$, and using (\ref{G_vertex_liouville}), we obtain from
(\ref{liouville_observable}) for the current operator in Liouville space
\begin{equation}
\label{current_liouville}
L_{I^\gamma}\,=\,{1\over n!}\,\sigma^{p_1\dots p_n}\,
(I^\gamma)^{p_1\dots p_n}_{1\dots n}\,:J^{p_1}_1\dots J^{p_n}_n: \quad,
\end{equation}
with
\begin{equation}
\label{current_vertex_liouville}
(I^\gamma)^{p_1\dots p_n}_{1\dots n} \,=\,-{1\over 2}\,
\sum_{i=1}^n\,\eta_i\delta_{\alpha_i\gamma}\,
\delta_{p_1 p}\dots\delta_{p_n p}\,p\,G^{p\dots p}_{1\dots n}\quad.
\end{equation}

\section{Nonequilibrium RG in Liouville space}
\label{sec:4}

In this section we develop a nonequilibrium renormalization group method
based on the diagrammatic rules derived in the previous section. It is a
formally exact RG-approach in the sense that an infinite hierarchy of 
RG equations will be set up, which, if solved completely, would provide
the full solution of the problem. Of course , in practice, approximations
in form of truncation schemes have to be used. Nevertheless, it is quite
useful to know what terms have been neglected in order to be able to improve 
the calculations systematically. Formally exact RG approaches are used in many
fields of physics at the moment and are based on different ideas. The most conventional
one is to integrate out energy (or momentum) scales step by step (starting
from high energies) and leaving the whole set of diagrams invariant by
renormalizing certain parameters, based on the pioneering ideas of Wilson 
\cite{rg_wilson}. This basic idea will be also the guideline of our RG procedure, 
although the way we define the cutoff function is quite different. Furthermore, our 
diagrammatic language is more complicated since the vertices are still operators
on the local quantum system and their time-ordering has to be considered.
As a consequence, we have to use the Laplace transform (instead of the usual
Fourier transform) and renormalization group will generate an
additional energy dependence $L_S(E)$ and $G_{1\dots n}^{p_1\dots p_n}(E)$
of the Liouvillian and the vertices, which can only be described within a
diagrammatic representation (in contrast to approaches based on projection
operators). This energy will turn out to be the equivalent of the Laplace
variable, opening the possibility to address the full time-evolution of the
problem (however, also for the calculation of the stationary state, the additional
energy dependence has to be taken into account in the RG equations).
We will formulate the RG approach such that all vertices stay
in normal-ordered form but it can also be formulated without normal-ordering.
In standard quantum field theory (where all degrees of freedom are integrated
out and the vertices are c-numbers), similiar RG approaches have been developed
based on the same idea, see the normal-ordered version by Salmhofer 
\cite{salmhofer} and the non normal-ordered version by Polchinski
\cite{polchinski} (usually the RG equations are derived within path integral
formalism but they can also be obtained on a pure diagrammatic level using
the idea of invariance \cite{hs_private}). In this sense,
the RG method presented in this section is a generalization to nonequilibrium
and to the case where the vertices are operators. Furthermore, we note that
using the Laplace transform has the consequence that
no energy conservation is associated with the vertices but, as we will see,
it has the advantage that the structure of the Keldysh indices becomes
very simple. We will show that by integrating out the symmetric part of the
reservoir distribution from the very beginning, the final RG equations do 
not contain Keldysh indices at all. 
This simplifies the calculations considerably. Furthermore, we will see that
the physical relaxation and dephasing rates (describing the rates associated
with quantum transport and {\it not} with quantum decay or single-particle
life times as in usual Green's function methods) occur quite natural and
can be incorporated directly into the RG equations. Therefore, our RG schemes
seems to be an appropriate way to combine RG with the physics of relaxation and
dephasing, an important subject in mesoscopics, quantum information
processing and cold atom systems.

We mention that there are other formally exact RG approaches in quantum field
theory which do not
use the conventional idea of integrating out energy degrees of freedom and
leaving the sum of all diagrams invariant by renormalization. One of them is
the RG developed by Wetterich \cite{wetterich} where cutoff dependent 1-particle  
irreducible vertex function are defined, such
that the initial condition at cutoff $\Lambda=\infty$ gives the bare vertices whereas for 
$\Lambda=0$ the full physical vertex functions are obtained. Differential equations
are then set up directly for these vertex functions from which they can be 
calculated systematically using appropriate truncation schemes. This scheme is specifically
adapted to the conventional diagrammatic representation of quantum field theories
and, therefore, can not be overtaken directly to the diagrammatic language
used in this paper. The concept of irreducibility is defined completely different
here and the diagrams generated by our RG procedure are automatically irreducible.
Within the RTRG-FS approach the irreducible diagrams, represented by the
kernel $\Sigma(E)$, are fed back into the RG equations for the vertices via an
effective description of the dynamics of the local quantum system. This is analogous to
the Wetterich RG scheme where the propagators include self-energy insertions. 
For problems where a controlled expansion in the Coulomb interaction is possible, the Wetterich 
RG method is very powerful and, recently, has also been generalized to the nonequilibrium
case \cite{jakobs_diplom,jakobs_meden_hs_PRL07,gezzi_pruschke_meden_PRB07}.
Another RG method is the flow equation method developed by Glazek, Wegner, and Wilson
\cite{flow_eq}, which recently has been generalized to the nonequilibrium and time-dependent
case by Kehrein et al. \cite{kehrein_PRL05,moeckel_kehrein_PRL08}. This scheme is
based on a pure Hamiltonian level and the Hamiltonian is diagonalized by 
unitary transformations step by step. This idea is competely different from all
other formally exact RG methods and can be viewed as a technical alternative to the scheme
presented here.

\subsection{Basic ideas and cutoff function}
\label{sec:4.1}

The central quantities of interest are the effective Liouvillian $L_S^{eff}(E)$
(or equivalently the irreducible kernel $\Sigma(E)$) and the kernel $\Sigma_R(E)$,
given by the diagrammatic representation 
(see (\ref{value_sigma}) and (\ref{value_sigma_observable}))
\begin{equation}
\label{diagram_sigma}
\left\{
\begin{array}{cl}
L_S^{eff}(E)
\\
\Sigma_R(E)
\end{array}
\right\}
\,\rightarrow\, 
\left\{
\begin{array}{cl}
L_S
\\
R_{n=0}
\end{array}
\right\}
+{1\over S}  (\pm)^{N_p}  \left(\prod\gamma\right)_{irr}
\left\{
\begin{array}{cl}
G \\ R
\end{array}
\right\}
{1\over E+X_1-L_S}G\dots G{1\over E+X_r-L_S}G \,.
\end{equation}
The first term represents the $n=0$ case and is written explicitly because $G_{n=0}$ is
by convention incorporated in $L_S$. The second term represents all diagrams with more than
one vertex or, equivalently, with at least one reservoir contraction. There, 
the difference is only the first vertex, which is $G$ for $L_S^{eff}(E)$
and $R$ for $\Sigma_R(E)$. From these quantities the reduced density matrix and 
the average of the observable $R$ can be calculated by using (\ref{solution_rd}) 
and (\ref{solution_observable}) 
\begin{equation}
\label{solution_rd_observable}
\tilde{\rho}_S(E)\,=\,{i\over E-L_S^{eff}(E)}\,\rho_S(t_0)\quad,\quad
\tilde{\langle R\rangle}(E)
\,=\,-i\,\mbox{Tr}_S\,\Sigma_R(E)\,\tilde{\rho}_S(E)\quad,
\end{equation}
and the irreducible kernel $\Sigma(E)$ can be obtained from
\begin{equation}
\label{solution_L_eff}
L_S^{eff}(E)\,=\,L_S\,+\,\Sigma(E)\quad.
\end{equation}

To find a mathematically well-defined formulation of renormalization group (RG),
we note that the physical quantities of interest are a functional of the 
reservoir contraction $\gamma$, the vertices $G$ and $R$, and the Liouvillian $L_S$
\begin{eqnarray}
\label{sigma_functional}
L_S^{eff}(E)\,&=&\,L_S\,+\,{\cal{F}}(\gamma,L_S,G)\quad,\\
\label{sigma_R_functional}
\Sigma_R(E)\,&=&\,R_{n=0}\,+\,{\cal{F}}_R(\gamma,L_S,G,R)\quad.
\end{eqnarray}
The functionals $\cal{F}$ and $\cal{F}_R$ represent the diagrammatic rules of 
the second term of Eq.~(\ref{diagram_sigma}). Note that the $n=0$ case is written 
explicitly in the first term, so that the arguments of the functionals do not include 
the case $n=0$ for the vertices $G_{1\dots n}$ and $R_{1\dots n}$.

The idea of RG is to integrate out the reservoir degrees
of freedom and to account for these by a renormalization of the system parameters without
changing the diagrammatic rules.
This leads to an effective theory for the dynamics of the local quantum system. To
achieve this, we replace the reservoir correlation function $\gamma$ in
the functionals (\ref{sigma_functional}) and (\ref{sigma_R_functional}) by a cutoff 
dependent contraction
$\gamma^\Lambda$, and try to find a corresponding $\Lambda$-dependence of the
Liouvillian $L_S\rightarrow L_S^\Lambda$ and the vertices 
$G,R\rightarrow G^\Lambda,R^\Lambda$, such that, without changing the diagrammatic 
rules (i.e. the functionals), the sum of all diagrams stays invariant, which gives
\begin{eqnarray}
\label{rg_sigma_functional}
L_S^{eff}(E)\,&=&\,L_S^\Lambda(E)\,+\,{\cal{F}}(\gamma^\Lambda,L_S^\Lambda,G^\Lambda)\quad,\\
\label{rg_sigma_R_functional}
\Sigma_R(E)\,&=&\,R_{n=0}^\Lambda(E)\,+\,
{\cal{F}}_R(\gamma^\Lambda,L_S^\Lambda,G^\Lambda,R^\Lambda)\quad.
\end{eqnarray}
This is the important property of {\it invariance } since the r.h.s. gives always the same
independent of $\Lambda$.
In this way an effective theory can be formulated where both the reservoirs
and the quantum system are modified, but the physics of the total system
stays invariant (in principle also the functionals $\cal{F}$ and $\cal{F}_R$ 
can be replaced by a cutoff-dependent functional but we will not consider this case here).
However, as we will see in Sec.~\ref{sec:4.2}, invariance can only be maintained
if the Liouvillian and the vertices get an additional energy dependence 
\begin{eqnarray}
\label{E_L_dependence}
L_S\,&\rightarrow&\,L_S^\Lambda(E)\quad,\quad\\
\label{E_G_dependence}
G^{p_1\dots p_n}_{1\dots n}\,&\rightarrow&\,(G^\Lambda)^{p_1\dots p_n}_{1\dots n}(E)\quad,\quad\\
\label{E_R_dependence}
R^{p_1\dots p_n}_{1\dots n}\,&\rightarrow&\,(R^\Lambda)^{p_1\dots p_n}_{1\dots n}(E)\quad,\quad
\end{eqnarray}
and a new diagrammatic rule has to be set up which states that the 
energy argument $E+X_i$ occuring in some resolvent of (\ref{diagram_sigma}) is the same for
the Liouvillian occuring in that resolvent and for the vertex to the right of that resolvent
(the first vertex gets the energy argument $E$), i.e. (\ref{diagram_sigma}) has to
be replaced by
\begin{eqnarray}
\label{rg_diagram_sigma}
\left\{
\begin{array}{cl}
L_S^{eff}(E)
\\
\Sigma_R(E)
\end{array}
\right\}
\,&\rightarrow&\, 
\left\{
\begin{array}{cl}
L_S^\Lambda(E)
\\
R^\Lambda_{n=0}(E)
\end{array}
\right\}
\,+\,{1\over S} \, (\pm)^{N_p} \, \left(\prod\gamma^\Lambda\right)_{irr}
\left\{
\begin{array}{cl}
G^\Lambda(E) \\ R^\Lambda(E)
\end{array}
\right\}\\
\nonumber
&&\hspace{-1.5cm}
\cdot\,{1\over E+X_1-L_S^\Lambda(E+X_1)}\,G^\Lambda(E+X_1)
\,\dots \,{1\over E+X_r-L_S^\Lambda(E+X_r)}\,
G^\Lambda(E+X_r) \quad.
\end{eqnarray}

The advantage of this formally exact scheme is that one can choose the value for 
the parameter $\Lambda$ and the way one defines the $\Lambda$-dependence 
of the reservoir contraction $\gamma^\Lambda$ in an arbitrary way, opening up many possibilities 
for RG schemes. However, only those schemes are of course of practical use, where the $\Lambda$-dependence of
the Liouvillian and the vertices can be found in a systematic way, and where the final 
evaluation of (\ref{rg_diagram_sigma}) for a certain value of $\Lambda$ has some advantage compared
to the original series (\ref{diagram_sigma}). We discuss some of these schemes in the following.

{\bf Continuous RG scheme.} If the parameter $\Lambda$ is a continuous parameter with initial value
$\Lambda_{in}$ and final value $\Lambda_{fi}$, one defines the boundary conditions of
the contraction such that
\begin{equation}
\label{boundary_contraction}
\gamma^{\Lambda_{in}}\,=\,\gamma \quad,\quad
\gamma^{\Lambda_{fi}}\,=\,0\quad,
\end{equation}
i.e. the initial contraction is the original reservoir correlation function and the final 
one is defined as zero. As a consequence the initial values of the Liouvillian and the
vertices are the original ones
\begin{eqnarray}
\label{initial_L}
L_S^{\Lambda_{in}}(E)\,&=&\,L_S\quad,\quad \\
\label{initial_G}
(G^{\Lambda_{in}})^{p_1 \dots p_n}_{1\dots n}(E)\,&=&\,
G^{p_1 \dots p_n}_{1\dots n}\quad,\quad\\
\label{initial_R}
(R^{\Lambda_{in}})^{p_1 \dots p_n}_{1\dots n}(E)\,&=&\,
R^{p_1 \dots p_n}_{1\dots n}\quad,
\end{eqnarray}
such that the original perturbative expansion (\ref{diagram_sigma}) is reproduced. For the
final value $\Lambda=\Lambda_{fi}$, the reservoir contraction is zero, and only the 
diagrams of (\ref{rg_diagram_sigma}) survive where no contractions are present, i.e. the
$n=0$ vertices, represented by $L_S^\Lambda(E)$ and $R^\Lambda_{n=0}(E)$. Thus, the result 
for the physical quantities is given by
\begin{equation}
\label{final_sigma}
L_S^{eff}(E)\,=\,L^{\Lambda_{fi}}_S(E)\quad,\quad
\Sigma_R(E)\,=\,R^{\Lambda_{fi}}_{n=0}(E)\quad.
\end{equation}
We see that $L_S^\Lambda(E)$ and $R^\Lambda_{n=0}(E)$ flow finally into
the physical quantities and, therefore, we interpret them as effective physical quantities at
scale $\Lambda$. If $\Lambda$ is an energy cutting off the high-energy scales of the 
reservoirs (see (\ref{real_frequency_cutoff}) below), one can interpret the 
$\Lambda$-dependent physical quantities as containing all energy scales between
$\Lambda_{in}$ and $\Lambda$.

Using this picture, one can define a
$\Lambda$-dependent reduced density matrix and a $\Lambda$-dependent average of an
observable $R$ by equations similiar to (\ref{solution_rd_observable}) via
\begin{equation}
\label{lambda_rd_observable}
\tilde{\rho}^\Lambda_S(E)\,=\,{i\over E-L_S^\Lambda(E)}\,\rho_S(t_0)\quad,\quad
\langle \tilde{R}\rangle^\Lambda(E)
\,=\,-i\,\mbox{Tr}_S\,\Sigma_R^\Lambda(E)\,\tilde{\rho}_S^\Lambda(E)\quad,
\end{equation}
with
\begin{equation}
\label{lambda_sigma_R}
\Sigma^\Lambda_R(E)\,=\,R^\Lambda_{n=0}(E)\quad.
\end{equation}
$\tilde{\rho}^\Lambda_S(E)$ and $\langle \tilde{R}\rangle^\Lambda(E)$ start with the 
value for the isolated quantum system at 
$\Lambda=\Lambda_{in}$ and flow into the full solution at $\Lambda=\Lambda_{fi}$. 

In summary, the central task is to find the $\Lambda$-dependence of the Liouvillian 
and the vertices with the initial conditions given by (\ref{initial_L})-(\ref{initial_R}).
To express them via differential equations, the so-called {\it RG equations}, one takes
the derivative of the invariance properties (\ref{rg_sigma_functional}) and 
(\ref{rg_sigma_R_functional}) with respect to $\Lambda$. The l.h.s. gives zero since
the physical kernels do not depend on $\Lambda$. On the r.h.s. one contribution
contains the derivative of the contraction and the other ones the derivative of
the Liouvillian and the vertices. All these contributions have to cancel each other
to fulfil invariance and it is a technical task to find this cancellation on a
diagrammatic level using the representation (\ref{rg_diagram_sigma}). We will discuss
in Sec.~\ref{sec:4.2} how to achieve this in an efficient way. As a result we
obtain RG equations symbolically of the form
\begin{eqnarray}
\label{rg_formal_L}
-{d\over d\Lambda}\,L_S^\Lambda(E)\,&=&\,
{\cal{F}}^{RG}_L(\gamma^\Lambda,L_S^\Lambda,G^\Lambda)\quad,\\
\label{rg_formal_G}
-{d\over d\Lambda}\,G^\Lambda(E)\,&=&\,
{\cal{F}}^{RG}_G(\gamma^\Lambda,L_S^\Lambda,G^\Lambda)\quad,\\
\label{rg_formal_R}
-{d\over d\Lambda}\,R^\Lambda(E)\,&=&\,
{\cal{F}}^{RG}_R(\gamma^\Lambda,L_S^\Lambda,G^\Lambda,R^\Lambda)\quad.
\end{eqnarray}
The minus sign on the l.h.s. indicates that the
sum of all terms has to vanish to fulfil invariance. Thus, the RG functionals
$\cal{F}^{RG}$ contain those terms where one derivative of a reservoir contraction occurs.
To solve the RG equations with the initial conditions 
(\ref{initial_L})-(\ref{initial_R}), one usually has to approximate the RG functionals
in a systematic way by expanding in the coupling parameter, the 
so-called {\it perturbative RG scheme}. However, it
is important to note that one expands not in the original coupling parameter, but in
the coupling parameter of the renormalized vertices at scale $\Lambda$ occuring on the 
r.h.s. of the RG equations. This can improve the result considerably compared to bare 
perturbation theory and, therefore, perturbative RG methods are a very important technical tool
to avoid divergencies occuring often in bare higher-order perturbation theory.

{\bf Choice of cutoff function.} An important question within the continuous scheme described before
is of course the most appropriate choice for the $\Lambda$-dependence of the reservoir
contraction, which, so far, has not been specified at all besides the boundary
conditions (\ref{boundary_contraction}). The criterion for the right choice is to achieve
that the r.h.s. of the RG equations is a well-defined series in the
renormalized coupling constant so that a perturbative RG scheme is possible. 
Alternatively, one can also stop the solution of the RG equations at a certain value
of $\Lambda$ and calculate the kernels from perturbation theory in the couplings at
this value of $\Lambda$ using (\ref{rg_diagram_sigma}). Whether this perturbation
theory is well-defined or not depends on the size of the renormalized coupling 
constants (coming out of the solution of the RG equations and can be smaller or
larger than the original couplings), and the low and high frequency behaviour which
can induce divergencies for the frequency integrals on the r.h.s. of the RG equations.
Divergencies at high frequencies are under control by choosing $\Lambda$ as a 
high-frequency cutoff in the contraction, e.g. the most obvious choice would be
\begin{equation}
\label{real_frequency_cutoff}
\gamma^{pp'\Lambda}_{11'}\,=\,\gamma^{pp'}_{11'}\,\theta(\Lambda-|\omega|)\quad,
\end{equation}
meaning that at cutoff scale $\Lambda$ only reservoir energies below $\Lambda$ are
considered. Within this scheme, the initial cutoff is infinity $\Lambda_{in}=\infty$,
and the final one zero $\Lambda_{fi}=0$ to fulfil the boundary conditions 
(\ref{boundary_contraction}) for the contractions.
This is the conventional bandwidth cutoff of poor man scaling approaches. 
All frequency integrals in (\ref{rg_diagram_sigma}) are cut off by $\Lambda$
for high frequencies. For low frequencies, there can be further divergencies induced
by the sharp step of the Fermi functions or the strong increase of the Bose function.
These divergencies are cut off by some low energy scale $\Lambda_c$ provided by 
the smearing of the distribution functions by temperature or by the other energy scales
appearing in the denominator of the resolvents in (\ref{rg_diagram_sigma}), like the
chemical potentials of the reservoirs, the eigenvalues of $L_S^\Lambda(E+X_i)$, or the
Laplace variable $E$. When $\Lambda$ approaches $\Lambda_c$, the high
and low frequency cutoffs are the same and divergencies are absent.
This is the idea of perturbative RG which works provided that the renormalized 
coupling constants are still small at scale $\Lambda_c$. The latter condition is called the
weak-coupling regime which can be solved systematically by perturbative RG.
When the coupling constants are already of order one at scale $\Lambda_c$, a strong
coupling problem occurs and the perturbative RG scheme becomes uncontrolled.
Truncating the series on the r.h.s. of the RG equations is not justified and it is
not clear which approximation scheme can be trusted. Nevertheless, it has turned out
that even for coupling constants of the order of one, only a few terms of the series
have to be taken into account to get reliable results. Whether this is a generic
feature or just accidental for certain models is one of the most interesting questions
in modern renormalization group theory. We will address this issue in Sec.~\ref{sec:5.3}
for the strong-coupling regime of the Kondo problem.

As we have seen, an important issue is whether a low-energy scale $\Lambda_c$ is 
present which cuts off small frequencies in (\ref{rg_diagram_sigma}) (an equivalent
discussion can be performed for the r.h.s. of the RG equations, see Sec.~\ref{sec:4.3}).
When temperature is small and when the real part of all low-energy scales in the
denominator of the resolvents cancel each other (which happens when the sum of the 
real part of the Laplace variable, the chemical potentials, and the real part of the
eigenvalue of $L_S^\Lambda(E+X_i)$ becomes small), the only low-energy cutoff is 
provided by the imaginary part $\Gamma$ of the eigenvalues of the Liouvillian 
$L_S^\Lambda(E+X_i)$. Therefore, if we denote by $T_K$ the energy scale where the coupling 
constants become of order one (the Kondo temperature for the Kondo model),
we expect that a weak-coupling problem occurs if the minimum of all relaxation
and dephasing rates is much larger than $T_K$
\begin{equation}
\label{weak_coupling_criterion}
\min_i\,\left\{\Gamma_i\right\}\,\gg\,T_K\quad \Rightarrow \quad
\mbox{weak coupling} \quad.
\end{equation}
However, to justify this statement technically, it is important to prove that the
zero eigenvalue of the Liouvillian (which generically occurs and corresponds to the
stationary state, see (\ref{eigenvalue_zero})) does not lead to an accidental 
situation where no low-energy scale occurs. To analyse this problem in more detail,
we discuss first the properties of the eigenvector with eigenvalue zero. Except for
maybe some special values of $\Lambda$ and $E$, the Liouvillian can be diagonalized according
to the eigenvalue problem for the right and left eigenvectors in Liouville space 
\begin{eqnarray}
\label{right_eigenvalue_equation}
L_S^\Lambda(E)\,|x_k^\Lambda(E)\rangle\,&=&\,\lambda_k^\Lambda(E)\,
|x_k^\Lambda(E)\rangle \quad,\\
\label{left_eigenvalue_equation}
\langle \bar{x}_k^\Lambda(E)|\,L_S^\Lambda(E)\,&=&\,\lambda_k^\Lambda(E)\,
\langle \bar{x}_k^\Lambda(E)|\quad.
\end{eqnarray}
The right and left eigenvectors $|x_k^\Lambda(E)\rangle$ and 
$|\bar{x}_k^\Lambda(E)\rangle$ will
in general not coincide since the renormalized Liouvillian contains the dissipative
influence of the reservoirs and, therefore, will be non-hermitian. The eigenvectors
are operators with matrix representation
\begin{equation}
\label{matrix_eigenvectors}
\langle ss'|x_k^\Lambda(E)\rangle\,=\,(x_k^\Lambda(E))_{ss'}\quad,
\end{equation}
where $|ss'\rangle$ is the Dirac notation for the basis operators
$|s\rangle\langle s'|$ in Liouville space. The normalization and completeness relations
of the eigenvectors read
\begin{eqnarray}
\label{normalization}
\langle \bar{x}_k^\Lambda(E)|x_l^\Lambda(E)\rangle\,&=&\,\delta_{kl}\quad,\\
\label{completeness}
\sum_k\,|x_k^\Lambda(E)\rangle\,\langle \bar{x}_k^\Lambda(E)| \,&=&\, 1 \quad.
\end{eqnarray}
The eigenvalues $\lambda_k^\Lambda(E)$ consist of a real and an imaginary part, denoted by
\begin{equation}
\label{rg_eigenvalue}
\lambda_k^\Lambda(E)\,=\,h_k^\Lambda(E)\,-\,i\Gamma_k^\Lambda(E) \quad.
\end{equation}
Using (\ref{completeness}), the reduced density matrix (\ref{lambda_rd_observable}) 
can be written as
\begin{equation}
\label{rd_eigenvectors}
\tilde{\rho}_S^\Lambda(E)\,=\,\sum_k\,{i\over E-\lambda_k^\Lambda(E)}\,|x_k^\Lambda(E)\rangle
\langle \bar{x}_k^\Lambda(E) |\quad,
\end{equation}
and the poles $z_k^\Lambda$ of the reduced density matrix follow from the self-consistent equation
\begin{equation}
\label{poles}
z_k^\Lambda\,=\,\lambda_k^\Lambda(z_k^\Lambda)\,=\,h_k^\Lambda(z_k^\Lambda)\,-\,
i\,\Gamma_k^\Lambda(z_k^\Lambda)\quad,
\end{equation}
provided that the energy-dependence of the eigenvectors does not induce additional
poles. The real and imaginary parts of $z_k^\Lambda$ determine the oscillation frequencies
and the relaxation/dephasing rates at scale $\Lambda$.

As shown later in Sec.~\ref{sec:4.3}, the effective Liouville operator at
scale $\Lambda$ fulfils the same properties as the full effective Liouvillian, 
see (\ref{L_eff_property})and (\ref{L_eff_c_transform})
\begin{equation}
\label{lambda_L_eff_property}
\mbox{Tr}_S\,L_S^\Lambda(E)\,=\,0\quad,\quad
L_S^\Lambda(E)^c\,=\,-L_S^\Lambda(-E^*) \quad.
\end{equation}
Applying the second condition to the eigenvalue equations 
(\ref{right_eigenvalue_equation}) and (\ref{left_eigenvalue_equation}), we obtain
\begin{equation}
\label{eigenvector_hermiticity}
x_k^\Lambda(E)^\dagger\,=\,x_k^\Lambda(-E^*)\quad,\quad
\bar{x}_k^\Lambda(E)^\dagger\,=\,\bar{x}_k^\Lambda(-E^*)\quad,\quad
\lambda_k^\Lambda(E)^*\,=\,-\,\lambda_k^\Lambda(-E^*)\quad.
\end{equation}
If we act with the trace $\mbox{Tr}_S$ over the quantum system on
the eigenvalue equations and use the first property of 
(\ref{lambda_L_eff_property}), we get the relation
\begin{equation}
\label{zw_normalization_property}
\lambda_k^\Lambda(E)\,\sum_s\,x_k^\Lambda(E)_{ss}\,=\,0 \quad,
\end{equation}
implying that either the eigenvalue or the sum over
all diagonal elements of the eigenvector must be zero
\begin{equation}
\label{normalization_property}
\lambda_k^\Lambda(E)\,=\,0 \quad \mbox{or} \quad
\sum_s\,x_k^\Lambda(E)_{ss}\,=\,0 \quad.
\end{equation}
Due to the completeness relation, not all eigenvectors can fulfil
the property that the sum over the diagonal elements is zero. 
Therefore there exists at least one eigenvector with zero eigenvalue,
where the sum over the diagonal elements is unequal to zero. We
denote this eigenvector by $k=0$ and assume that it is unique since
it corresponds to the stationary state at scale $\Lambda$. In this case, 
also the left eigenvector with zero eigenvalue is unique and we get
\begin{equation}
\label{lambda_eigenvalue_zero}
L_S^\Lambda(E)\,|x_0^\Lambda(E)\rangle\,=\,0 \quad,\quad
\langle \bar{x}_0^\Lambda(E)|\,L_S^\Lambda(E)\,=\,0 \quad.
\end{equation}
The right eigenvector $|x_0^\Lambda(E)\rangle$
will change as function of $\Lambda$ and $E$, and will also dependent on the model
under consideration because each system has its own stationary state.
However, the important point is that the left eigenvector $|\bar{x}_0^\Lambda(E)\rangle$ is 
always the same for all $\Lambda$ and $E$, and independent of the model, due to the 
first property of (\ref{lambda_L_eff_property})
\begin{equation}
\mbox{Tr}_S\,L_S^\Lambda(E)\,=\,\sum_s\,L_S^\Lambda(E)_{ss,\cdot\cdot}\,=\,0
\end{equation}
This is equivalent to the eigenvalue equation (\ref{lambda_eigenvalue_zero})
for $\bar{x}_0^\Lambda(E)$ if we choose 
\begin{equation}
\label{left_eigenvector_zero}
\bar{x}_0^\Lambda(E)_{ss'}\,=\,\delta_{ss'} \quad,
\end{equation}
up to a normalization constant, which we have chosen to be unity to fulfil 
the normalization condition (\ref{normalization}) for $k=0$
\begin{equation}
\label{lambda_normalization_probability}
\langle \bar{x}_0^\Lambda(E)|x_0^\Lambda(E)\rangle
\,=\,\sum_s\,x_0^\Lambda(E)_{ss}
\,=\,1\quad.
\end{equation}
Thus, acting with the eigenvector $\bar{x}_0^\Lambda(E)$ from the left is
equivalent to acting with the trace over the quantum system
\begin{equation}
\label{trace_eigenvector}
\langle \bar{x}_0^\Lambda(E)|\,G \,=\, \mbox{Tr}_S\,G \quad,
\end{equation}
where $G$ is an arbitrary superoperator. This relation is one of
the most important properties which is preserved under the RG flow since
(\ref{lambda_L_eff_property}) is fulfilled.

We now turn back to the question how to choose the $\Lambda$-dependence of the
contraction such that the zero eigenvalue of $L_S^\Lambda(E)$ can not lead
to a problem at low energies. Inserting the completeness relation
(\ref{completeness}) between the vertices of the diagrammatic expression
(\ref{rg_diagram_sigma}) we see that the left eigenvector acts on the right
vertex and the contribution from $k=0$ leads precisely
to a term of the form (\ref{trace_eigenvector})
\begin{equation}
\label{tr_vertex}
\langle \bar{x}_0^\Lambda(E)|\,(G^\Lambda)^{p_1\dots p_n}_{1\dots n}(E+X_i) \,=\, 
\mbox{Tr}_S\,(G^\Lambda)^{p_1\dots p_n}_{1\dots n}(E+X_i) \quad,
\end{equation}
where the trace over the quantum system acts on the vertex. We now remind ourselves of
the property (\ref{G_property}) of the vertex which is also preserved under the
RG flow (see Sec.~\ref{sec:4.3} for the proof)
\begin{equation}
\label{lambda_G_property}
\sum_{p_1\dots p_n}\,\mbox{Tr}_S\,(G^\Lambda)^{p_1\dots p_n}_{1\dots n}(E) = 0 \quad.
\end{equation}
It means that if the average over the Keldysh indices were taken in 
(\ref{tr_vertex}), the contribution of the zero eigenvalue would be exactly zero!
Therefore, the task is to choose the $\Lambda$-dependence of the contraction
in such a way that for small values of $\Lambda$ only the average over the Keldysh
indices of each vertex occurs in the perturbation series (\ref{rg_diagram_sigma}).

The Keldysh indices occur in (\ref{rg_diagram_sigma}) via the vertices and via
the reservoir contractions given initially by (\ref{alternative_liouville_contraction})
\begin{equation}
\label{zw_alternative_liouville_contraction}
\gamma_{11'}^{pp'}\,=\,
\delta_{1\bar{1}'}\,p'
\left\{
\begin{array}{cl}
\eta \\ 1
\end{array}
\right\}\,
\,\rho_\nu(\omega)\,f_\alpha(\eta p' \omega) \quad,
\end{equation}
where we have taken the form where the density of states is incorporated.
If we were able to avoid the dependence of the contractions on the Keldysh indices,
only the average of the vertices over the Keldysh indices would occur in the 
perturbation series. The contraction depends only on the second Keldysh index $p'$ 
which appears as a prefactor and in the
argument of the distribution function. Therefore, only the symmetic part of
the distribution function leads to a dependence on the Keldysh indices. We now
see trivially what we have to do to avoid the occurence of the zero eigenvalue. 
We have to replace the distribution function $f_\alpha(\omega)$ by a $\Lambda$-dependent
distribution function $f_\alpha^\Lambda(\omega)$ such that for small $\Lambda$ of the
order of the relaxation and dephasing rates, we get an antisymmetric distribution
function
\begin{equation}
\label{antisymmetry_condition}
f_\alpha^\Lambda(-\omega)\,=\,-f_\alpha^\Lambda(\omega) \quad \mbox{for  }
\Lambda\,\sim\,\min_i\,\left\{\Gamma_i\right\}\quad.
\end{equation}
We conclude that not only the density of states should depend on $\Lambda$ but also
the distribution function, leading to a choice of the form
\begin{equation}
\label{lambda_contraction}
\gamma_{11'}^{pp'\Lambda}\,=\,
\delta_{1\bar{1}'}\,p'
\left\{
\begin{array}{cl}
\eta \\ 1
\end{array}
\right\}\,
\,\rho_\nu^\Lambda(\omega)\,f^\Lambda_\alpha(\eta p' \omega) \quad.
\end{equation}

In summary, we have now two requirements to choose the cutoff function to avoid
problems at high and small frequencies in the renormalized perturbation series
(\ref{rg_diagram_sigma}). First we need that only energy scales smaller than
$\Lambda$ should contribute to the frequency integrals, demanding a cutoff scheme
similiar to the bandwidth cutoff proposed in (\ref{real_frequency_cutoff}).
This corresponds to a $\Lambda$-dependent density of states in 
(\ref{lambda_contraction}) given e.g. by
\begin{equation}
\label{lambda_dos}
\rho_\nu^\Lambda(\omega)\,=\,\rho_\nu(\omega)\,\theta(\Lambda-|\omega|)\quad,
\end{equation}
where $\theta(\omega)$ can be a sharp step function or, if needed,
a certain smearing procedure can be introduced to get simple analytic properties or
to improve numerical stability. Secondly, we demand the distribution function
to be antisymmetric for small values of $\Lambda$ to avoid the occurence of the
zero eigenvalue leading to problems at small frequencies in the perturbative 
series. There are several possibilities to achieve the latter property which we
will discuss in the following.

{\bf Integration by discrete RG step.} The easiest way to get rid of the symmetric part of 
the distribution function is to integrate it out in one step before the continuous
RG starts. This means that we decompose $f$ into a symmetric and an antisymmetric part
\begin{eqnarray}
\nonumber
f_\alpha(\omega)\,&=&\,{1\over 2}\left[f_\alpha(\omega)\,+\,f_\alpha(-\omega)\right]\,+\,
{1\over 2}\left[f_\alpha(\omega)\,-\,f_\alpha(-\omega)\right] \\
\label{decomposition_sym_antisym}
&=&\,\mp\,{1\over 2}\,+\,\left[f_\alpha(\omega)\pm {1\over 2}\right] \quad.
\end{eqnarray}
The symmetric part is a constant given by $\mp{1\over 2}$. The corresponding
decomposition of the contraction (\ref{zw_alternative_liouville_contraction}) reads 
\begin{equation}
\label{contraction_decomposition}
\gamma_{11'}^{pp'}\,=\,(\gamma^s)_{11'}^{pp'}\,+\,\gamma^a_{11'} \quad,
\end{equation}
with
\begin{equation}
\label{contraction_sym_antisym}
(\gamma^s)_{11'}^{pp'}\,=\,
\delta_{1\bar{1}'}\,{1\over 2}\,p'
\left\{
\begin{array}{cl}
-\eta \\ 1
\end{array}
\right\}\,
\,\rho_\nu(\omega) \quad,\quad
\gamma^a_{11'} \,=\,
\delta_{1\bar{1}'}\,
\left\{
\begin{array}{cl}
1 \\ \eta
\end{array}
\right\}\,
\,\rho_\nu(\omega)\,\left[f_\alpha(\omega)\pm {1\over 2}\right] \quad.
\end{equation}
As expected, we see that only the symmetric part depends on the Keldysh indices.

We replace now the contraction $\gamma$ in the functionals (\ref{sigma_functional})
and (\ref{sigma_R_functional}) by the antisymmetric part $\gamma^a$ and renormalize
at the same time the Liouvillian $L_S\rightarrow L_S^a$ and the vertices
$G\rightarrow G^a$, $R\rightarrow R^a$ in such a way that invariance holds, i.e.
we get in analogy to (\ref{rg_diagram_sigma}) the new perturbation series
\begin{eqnarray}
\label{antisym_series}
\left\{
\begin{array}{cl}
L_S^{eff}(E)
\\
\Sigma_R(E)
\end{array}
\right\}
\,&\rightarrow&\, 
\left\{
\begin{array}{cl}
L_S^a(E)
\\
R^a_{n=0}(E)
\end{array}
\right\}
\,+\,
{1\over S} \, (\pm)^{N_p} \, \left(\prod\gamma^a\right)_{irr}
\left\{
\begin{array}{cl}
\bar{G}^a(E) \\ \bar{R}^a(E)
\end{array}
\right\}\\
\nonumber
&&\hspace{-1.5cm}
\cdot\,{1\over E+X_1-L_S^a(E+X_1)}\,\bar{G}^a(E+X_1)
\,\dots \,{1\over E+X_r-L_S^a(E+X_r)}\,
\bar{G}^a(E+X_r) \quad,
\end{eqnarray}
where only the antisymmetric part of the distribution function occurs which is
independent of the Keldysh indices. Therefore, only the average of the vertices
over the Keldysh indices occurs which is denoted by $\bar{G}$ and $\bar{R}$
\begin{equation}
\label{average_Keldysh_indices}
\bar{G}_{1\dots n}(E)\,=\,\sum_{p_1\dots p_n}\,G^{p_1\dots p_n}_{1\dots n}(E)
\quad,\quad
\bar{R}_{1\dots n}(E)\,=\,\sum_{p_1\dots p_n}\,R^{p_1\dots p_n}_{1\dots n}(E)
\quad.
\end{equation}
In this way the symmetric part of the distribution function has been integrated
out and has been shifted into the initial condition of the Liouvillian and
the vertices. As we will describe in the following sections, the quantities $L^a_S$,
$G^a$ and $R^a$ can be calculated from a well-defined perturbation theory
because the symmetric part of the distribution function is a frequency-independent
constant, leading to no divergencies in the perturbation series.

After having integrated out the symmetric part of the distribution function,
the zero eigenvalue of the Liouvillian can no longer occur in the resolvents and
we can subsequently apply a well-defined continuous RG scheme to integrate out the 
energy scales of the antisymmetric part. This can be done e.g. by the bandwidth
cutoff leading to a $\Lambda$-dependent antisymmetric contraction of the form
\begin{equation}
\label{lambda_antisym}
\gamma^\Lambda_{11'} \,=\,
\delta_{1\bar{1}'}\,
\left\{
\begin{array}{cl}
1 \\ \eta
\end{array}
\right\}\,
\,\rho^\Lambda_\nu(\omega)\,\left[f_\alpha(\omega)\pm {1\over 2}\right] \quad,
\end{equation}
with $\rho^\Lambda_\nu(\omega)$ given by (\ref{lambda_dos}). However, as we will
see in the following, also this choice of a real-frequency cutoff is not the most
convenient one. Instead, one should again choose a $\Lambda$-dependent 
antisymmetric distribution function instead of a $\Lambda$-dependent density of
states.

{\bf Cutoff function on the imaginary frequency axis.} The real-frequency bandwidth cutoff scheme to
avoid divergencies for high energies has two important problems. As we will see
in Sec.~\ref{sec:4.2} the r.h.s. of the RG equations (\ref{rg_formal_L})-(\ref{rg_formal_R})
has a similiar structure to the perturbation series (\ref{rg_diagram_sigma}) but
in all resolvents one of the frequencies is replaced by the cutoff $\Lambda$. This
has the consequence that if $\Lambda$ crosses the rest of the real part of the
denominator of the resolvents, the real part of the resolvent rapidly changes sign 
on the scale of the relaxation/dephasing rate $\Gamma$ and becomes quite large 
$\sim{1\over \Gamma}$ in the vicinity of this point. Although these negative and positive 
contributions almost cancel each other (like for a principle value integral)
numerical problems occur, leading to low accuracy and slow algorithms, 
see Refs.~\cite{jakobs_diplom,korb_phd,gezzi_pruschke_meden_PRB07} for more details. 
Another problem is the fact, that in each RG step $\Lambda\rightarrow \Lambda-d\Lambda$,
an infinitesimal energy shell $\Lambda>|\omega|>\Lambda-d\Lambda$ is integrated out in
the reservoirs. For $\Lambda$ large compared to all other physical energy scales, this means
that no energy conserving processes are integrated out. As a consequence, the generation of 
relaxation and dephasing rates happens only for values of $\Lambda$ of the order of
some physical energy scale like temperature or voltage, see 
Ref.~\cite{korb_reininghaus_hs_koenig_PRB07}. Therefore, for large $\Lambda$, 
the imaginary part in the denominators of the resolvents is very small and the resolvent
becomes very large (or can even diverge) when $\Lambda$ cancels the other real parts of the
denominator (note that combinations of real frequencies occur in the denominator which can become
of the same order as $\Lambda$). Of course, this is not a real problem because the different
signs on both sides of the divergence cancel each other but a stable numerical solution
is very problematic.

To avoid these problems another cutoff scheme on the imaginary axis has been proposed 
for a nonequilibrium system \cite{jakobs_meden_hs_PRL07}. Instead of integrating out the 
real frequencies of the contraction (\ref{zw_alternative_liouville_contraction}) step
by step, one analyses the analytic structure of the frequency dependence and tries to
integrate out the poles in the complex plane step by step. For a broad density of states
depending weakly on frequency of the form (\ref{cutoff})
\begin{equation}
\label{dos_broad}
\rho(\omega)\,=\,{D^2\over \omega^2+D^2}\,=\,{D\over 2i}
({1\over \omega-iD}-{1\over \omega+iD})\quad
\end{equation}
the poles of $\rho(\omega)$ occur at $\pm iD$ and have a very large imaginary part.
Therefore, we disregard them in the following, and analyse only the analytic properties
of the distribution function (for more complicated problems also the analytic structure
of the density of states has to be taken into account, see the discussion at the end
of Sec.~\ref{sec:6}).

Using the exact representation of $f_\alpha$ in terms
of the Matsubara frequencies $\omega_n^\alpha=2n \pi T_\alpha$ 
($\omega_n^\alpha=(2n+1)\pi T_\alpha$) for bosons (fermions)
\begin{equation}
\label{distribution_matsubara}
f_\alpha(\omega)\,=\,\pm T\,\sum_n\,{e^{i\,\omega^\alpha_n \,\epsilon}\over \omega-i\omega^\alpha_n}
\quad,
\end{equation}
with $\epsilon=0^+$, we see that the distribution function has poles at the Matsubara 
frequencies with residuum $\pm T$. The convergence factor $e^{i\,\omega^\alpha_n\,\epsilon}$
determines the symmetric part of the distribution function, for $\epsilon=0^+$ we obtain
$f_\alpha(\omega)$, whereas for $\epsilon=0^-$ we get $-f_\alpha(-\omega)=\pm(1\pm f_\alpha(\omega))$.

\begin{figure}
  \centerline{\psfig{figure=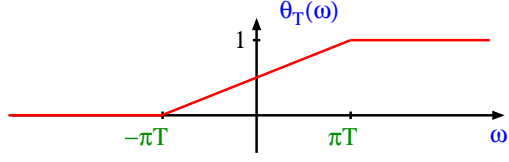,scale=0.5}}
  \caption{Theta function smeared by temperature.}
\label{fig:theta}
\end{figure}

To integrate out the Matsubara poles step by step, we introduce a $\Lambda$-dependent
distribution function of the form
\begin{equation}
\label{lambda_matsubara_distribution}
f^\Lambda_\alpha(\omega)\,=\,
\pm T\,\sum_n\,{e^{i\,\omega^\alpha_n \,\epsilon}\over \omega-i\omega^\alpha_n}
\,\,\theta_{T_\alpha}(\Lambda-|\omega_n^\alpha|)\quad,
\end{equation}
where
\begin{equation}
\label{theta_T}
\theta_T(\omega)\,=\,
\left\{
\begin{array}{cl}
\theta(\omega) &\quad\mbox{for }|\omega|\,>\,\pi T \\
{1\over 2}+{\omega \over 2\pi T} &\quad\mbox{for }|\omega|\,<\,\pi T
\end{array}
\right.
\end{equation}
is a theta function smeared by temperature, sketched in Fig.~\ref{fig:theta}.
Via (\ref{lambda_contraction}), this leads to the following $\Lambda$-dependent contraction
\begin{equation}
\label{lambda_contraction_matsubara}
\gamma_{11'}^{pp'\Lambda}\,=\,
\delta_{1\bar{1}'}\,
\left\{
\begin{array}{cl}
\eta \\ -1
\end{array}
\right\}\,
\,\rho_\nu(\omega)\,T\,\sum_n\,{e^{i\,\omega_n^\alpha\,p'\,\epsilon}
\over x\,-\,\bar{\mu}_\alpha\,-\,i\omega_n^\alpha} 
\,\,\theta_{T_\alpha}(\Lambda-|\omega_n^\alpha|)\quad,
\end{equation}
with $x=\eta(\omega+\mu_\alpha)$ according to (\ref{x_notation}), and 
\begin{equation}
\label{mu_bar}
\bar{\mu}_\alpha\,\equiv\,\eta\,\mu_\alpha \quad.
\end{equation}
If convenient, one can also add a $\Lambda$-dependent density of states,
but we do not consider this case here, assuming that the density of states has
a well-defined analytic structure, as e.g. given by (\ref{dos_broad}).

With the choice (\ref{lambda_matsubara_distribution}) the parameter $\Lambda$ cuts 
off the Matsubara poles of the distribution function.
Therefore, as we will see in Sec.~\ref{sec:4.2}, the complex parameter $i\Lambda$ will
occur in the denominator of all resolvents on the r.h.s of the RG equations and all 
other imaginary parts are also positive. For this reason the resolvents will stay
small and the numerical problems that occur if one uses the real-frequency cutoff can be avoided. 
This can also be seen by rewriting the original perturbation series (\ref{diagram_sigma}) 
directly in Matsubara space before starting the RG procedure. If we assume that all
frequencies are allowed (i.e. the bosonic case with particle number conservation
is not considered here) and the
frequency dependence of the density of states and of the vertices is weak, we can close
all integrations over the $x$-variables in the upper half of the complex plane and
see from (\ref{lambda_contraction_matsubara}) (with $\Lambda=\infty$, i.e. leaving
out the $\theta_{T_\alpha}$ function initially) that only the poles at 
$x=\bar{\mu}_\alpha + i\omega_n^\alpha$ with positive Matsubara frequency
$\omega_n^\alpha>0$ have to be considered. Note that all energies $X_i$ in the resolvents of the 
perturbation series (\ref{diagram_sigma}) are sums of $x$-variables, so that all of them
have positive imaginary parts when the integrations are closed in the upper half of the
complex plane. Furthermore, also the Laplace variable $E$ has positive imaginary part.
Performing all integrations over the $x$-variables in this way, we get a new series of
the same form as (\ref{diagram_sigma}) but with the replacements
\begin{eqnarray}
\label{x_matsubara_replace}
x\,&\rightarrow&\,\bar{\mu}_\alpha\,+\,i\omega_n^\alpha \quad,\quad
\eta\omega\,\rightarrow\,i\omega_n^\alpha \quad,\\
\label{contraction_matsubara_replace}
\gamma^{pp'}_{11'}\,&\rightarrow&\,\delta_{1\bar{1}'}\,
\left\{
\begin{array}{cl}
\eta \\ -1
\end{array}
\right\}\,
\rho_\nu(i\eta\omega_n^\alpha)\,2\pi i\,T\,e^{i\,\omega_n^\alpha\,\epsilon} \quad,
\end{eqnarray}
and we sum only over positive Matsubara frequencies.
Starting from this series, we can now introduce the same ideas of renormalization
group as before and introduce a cutoff function $\theta_{T_\alpha}(\Lambda-|\omega_n^\alpha|)$
into the contraction (\ref{contraction_matsubara_replace}). We obtain a renormalized 
perturbation series of the form (\ref{rg_diagram_sigma}) with the same replacements
(\ref{x_matsubara_replace}) and (\ref{contraction_matsubara_replace}). A similiar
series occurs on the r.h.s. of the RG equations, see Sec.~\ref{sec:4.3}. Assuming
that the imaginary parts of the eigenvalues of $L_S^\Lambda(E)$ are all negative
(corresponding to positive relaxation and dephasing rates, see (\ref{rg_eigenvalue})), 
we see that the imaginary part of all terms in the denominator of the resolvents
are positive so that no cancellations can occur. As a consequence, this form is very
stable for numerical calculations.

We now show that also all other requirements for a suitable $\Lambda$-dependent
contraction are fulfilled by choosing the Matsubara cutoff function 
(\ref{lambda_matsubara_distribution}). This concerns the requirement  
(\ref{antisymmetry_condition}) of antisymmetry for sufficiently small $\Lambda$,
the behaviour for large frequencies,
and the question at what value of $\Lambda$ relaxation and dephasing rates are
generated. To discuss this, we first analyse the analytic form of the 
$\Lambda$-dependent distribution function.

\begin{figure}
  \centerline{\psfig{figure=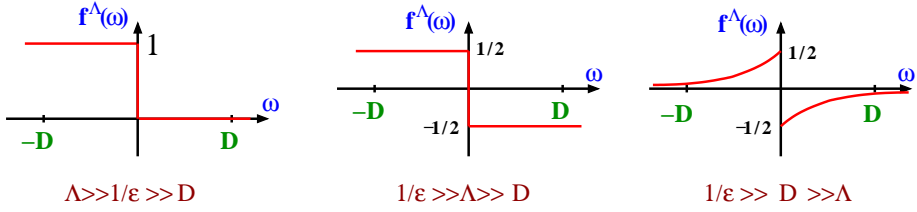,scale=0.5}}
  \caption{Sketch of the $\Lambda$-dependence of the Fermi function for zero
temperature. $D$ is the bandwidth setting the scale which frequencies are important.
Since $\epsilon$ is arbitrarily small, we take ${1\over\epsilon}\gg D$ and three
limiting cases can be considered for $\Lambda$. For $\Lambda$ above ${1\over\epsilon}$,
we have the original Fermi function which is a step function. For $\Lambda$ between
${1\over\epsilon}$ and $D$ the Fermi function has been moved down by ${1\over 2}$,
i.e. only the antisymmetric part of the original Fermi function remains. For
$\Lambda$ below $D$ the high-energy scales get a smaller weight and the Fermi function
is given by $-{1\over\pi}\arctan({\Lambda\over\omega})$.}
\label{fig:fermi}
\end{figure}

For zero temperature, $f^\Lambda(\omega)$ is depicted in Fig.~\ref{fig:fermi}
for the fermionic case (the bosonic case is obtained from changing the sign,
disregarding the divergence occuring for $|\omega|$ smaller than temperature).
Analytically we get
\begin{equation}
\label{lambda_distribution_T_0}
f^\Lambda(\omega)\,=\,\pm\,{1\over 2\pi}\,\int_{-\Lambda}^{\Lambda}d\omega'\,
{e^{i\omega'\epsilon} \over \omega-i\omega'} \quad,
\end{equation}
which can be calculated in two limiting cases
\begin{eqnarray}
\label{large_Lambda} 
|\omega|\ll \Lambda,{1\over\epsilon}\quad &\Rightarrow& \quad
f^\Lambda(\omega)\,=\,\mp\,{1\over\pi}\,\mbox{Si}(\Lambda\epsilon)
\,\pm\,{1\over 2}\,\mbox{sign}(\omega) \quad,\\
\label{small_Lambda} 
|\omega|,\Lambda\ll{1\over\epsilon}\quad &\Rightarrow& \quad
f^\Lambda(\omega)\,=\,\pm\,{1\over \pi}\,\arctan(\Lambda/\omega) \quad,
\end{eqnarray}
with $\mbox{Si}(x)=\int_0^x dy {\sin(y)\over y}$. Using $\mbox{Si}(\infty)={\pi\over 2}$,
we see that for $|\omega|\ll \Lambda,{1\over\epsilon}$, the distribution function is just moved 
up (down) by ${1\over 2}$ for bosons (fermions) when $\Lambda$ crosses 
${1\over \epsilon}$, leaving the antisymmetric part 
\begin{equation}
\label{antisym_T_0}
f^\Lambda(\omega)\,=\,\pm\,{1\over 2\pi}\,\int_{-\Lambda}^{\Lambda}d\omega'\,
{1\over \omega-i\omega'}
\,=\,\pm\,{1\over \pi}\,\arctan(\Lambda/\omega) \quad \mbox{for  }
|\omega|,\Lambda\ll{1\over\epsilon}\quad.
\end{equation}
This is the form for $\epsilon\rightarrow 0^+$ and $\Lambda$ finite. The regime
$\epsilon\rightarrow 0^+$ can also be calculated analytically for finite 
temperature. We obtain 
\begin{eqnarray}
\label{antisym_T_finite}
\lim_{\epsilon\rightarrow 0^+}\,f^\Lambda_\alpha(\omega)\,&=&\,
\pm T\,\sum_n\,{1 \over \omega-i\omega^\alpha_n}
\,\,\theta_{T_\alpha}(\Lambda-|\omega_n^\alpha|) \\
\nonumber
&&\hspace{-1cm}=\,
\left\{
\begin{array}{cl}
-{T_\alpha\over\omega}\,-\,{1\over 2\pi i}\left\{
\psi({\Lambda_{T_\alpha}+i\omega \over 2\pi T_\alpha})-\psi({i\omega\over 2\pi T_\alpha})
+{\Lambda-\Lambda_{T_\alpha}+\pi T_\alpha \over \Lambda_{T_\alpha}+i\omega}
\,-(\omega\rightarrow -\omega)\right\}
&\quad \mbox{for bosons}\\
{1\over 2\pi i}\left\{
\psi({\Lambda_{T_\alpha}+i\omega \over 2\pi T_\alpha})-
\psi({1\over 2}+{i\omega\over 2\pi T_\alpha})
+{\Lambda-\Lambda_{T_\alpha}+\pi T_\alpha \over \Lambda_{T_\alpha}+i\omega}
\,-(\omega\rightarrow -\omega)\right\}
&\quad\mbox{for fermions}
\end{array}
\right.
\quad,
\end{eqnarray}
where $\Lambda_{T_\alpha}$ is the value of the Matsubara frequency which lies
closest to $\Lambda$, and $\psi(z)$ denotes the Digamma function with asymptotic values
\begin{eqnarray}
\label{psi_z_infty}
\psi(z)\,&\rightarrow&\,\ln(z)\,-\,{1\over 2z}\,+\,O({1\over z^2})
\quad \mbox{for  }|z|\rightarrow\infty \quad,\\
\label{psi_z_zero}
\psi(z)\,&\rightarrow&\,-\gamma\,-\,{1\over z}\,+\,O({z})
\quad \mbox{for  }|z|\rightarrow 0 \quad,
\end{eqnarray}
where $\gamma$ is Euler's constant.

Thus, we see that it is important to take $\epsilon$ finite because otherwise
the symmetric part of the distribution function is missing at any finite $\Lambda$, or,
in other words, the symmetric part is integrated out when $\Lambda$ crosses
${1\over \epsilon}$. Since $\epsilon$ can be taken arbitrarily small, this means
that for $\Lambda$ reaching any physical scale, the distribution function is
antisymmetric to any desired accuracy, i.e. the requirement 
(\ref{antisymmetry_condition}) is fulfilled. 

Instead of integrating out the symmetric part smoothly using a finite value of $\epsilon$,
it is more convenient for analytical calculations to integrate it out
in one single step as described above (see (\ref{antisym_series})) and,
subsequently, use the choice (\ref{antisym_T_finite}) for the $\Lambda$-dependence
of the antisymmetric part of the distribution function. In this case, the parameter
$\epsilon$ does not occur at all which simplifies the calculation. Thus, for the 
continuous RG flow, we set $\epsilon=0$ in (\ref{lambda_contraction_matsubara})
and get
\begin{eqnarray}
\label{lambda_contraction_matsubara_antisym}
\gamma_{11'}^\Lambda\,&=&\,
\delta_{1\bar{1}'}\,
\left\{
\begin{array}{cl}
\eta \\ -1
\end{array}
\right\}\,
\,\rho_\nu(\omega)\,T\,\sum_n\,{1\over x\,-\,\bar{\mu}_\alpha\,-\,i\omega_n^\alpha} 
\,\,\theta_{T_\alpha}(\Lambda-|\omega_n^\alpha|)\\
&\begin{array}[b]{c}
\vspace{-1.5mm} T\rightarrow 0 \\ 
\longrightarrow 
\end{array}&
\delta_{1\bar{1}'}\,
\left\{
\begin{array}{cl}
\eta \\ -1
\end{array}
\right\}\,
\,\rho_\nu(\omega)\,{1\over 2\pi}\,\int_{-\Lambda}^\Lambda d\omega'
\,{1\over x\,-\,\bar{\mu}_\alpha\,-\,i\omega'}
\end{eqnarray}
for the $\Lambda$-dependent contraction, and the perturbation series (\ref{antisym_series})
turns into the renormalized series
\begin{eqnarray}
\label{antisym_series_matsubara}
\left\{
\begin{array}{cl}
L_S^{eff}(E)
\\
\Sigma_R(E)
\end{array}
\right\}
\,&\rightarrow&\, 
\left\{
\begin{array}{cl}
L_S^\Lambda(E)
\\
R^\Lambda_{n=0}(E)
\end{array}
\right\}
\,+\,
{1\over S} \, (\pm)^{N_p} \, \left(\prod\gamma^\Lambda\right)_{irr}
\left\{
\begin{array}{cl}
\bar{G}^\Lambda(E) \\ \bar{R}^\Lambda(E)
\end{array}
\right\}\\
\nonumber
&&\hspace{-1.5cm}
\cdot\,{1\over E+X_1-L_S^\Lambda(E+X_1)}\,\bar{G}^\Lambda(E+X_1)
\,\dots \,{1\over E+X_r-L_S^\Lambda(E+X_r)}\,
\bar{G}^\Lambda(E+X_r) \quad,
\end{eqnarray}
where all vertices are now averaged over the Keldysh indices, as defined in 
(\ref{average_Keldysh_indices}), and the zero eigenvalue of the Liouvillian
can no longer occur. This 2-stage procedure of RG, i.e. first integrating out the symmetric
part of the distribution function and then choosing the Matsubara cutoff for the
antisymmetric part, is the procedure which is most suitable and will be used
in the following sections.

Concerning the behaviour at high energies we see from Fig.~\ref{fig:fermi} that the 
$\Lambda$-dependent distribution
function always contains all energy scales and falls off at high frequencies like 
$\Lambda/\omega$. This means that the scale where high frequencies become less 
important is set by $\Lambda$ and, thus, the divergencies of the original 
perturbation theory at high energies are eliminated during the RG flow. Furthermore, in each
infinitesimal step $\Lambda\rightarrow\Lambda-d\Lambda$ of the RG procedure,
reservoir energies on all scales are integrated out, leading to the effect that
relaxation and dephasing rates are generated from the very beginning of the
RG flow (even the discrete RG step at the beginning integrating out the symmetric
part of the distribution function leads to an initial condition for the rates).
This also helps to improve the stability of the flow.

In summary, we conclude that the Matsubara cutoff is a very suitable choice of
a cutoff function for fermionic problems with a flat density of states. 
However, when complicated frequency dependencies of the density
of states appear, one has to consider the analytic structure of the density
of states as well and other cutoff schemes might be more appropriate. 
We summarize the criteria for a suitable $\Lambda$-dependent 
contraction, which we have found in this section

\begin{enumerate}
\item
High frequencies should become less important during the RG flow. This means that
the contraction should suppress frequencies above $\Lambda$.

\item
The distribution function should become antisymmetric during the RG flow to avoid
the appearance of the zero eigenvalue of the Liouvillian corresponding to the
stationary state. Closely connected with this property is the fact that the
effective vertices can be averaged over the Keldysh indices.

\item
The resolvents should not become large during the RG flow leading to numerical
problems. 

\item
Relaxation and dephasing rates should be generated already at the beginning of the
RG flow.

\end{enumerate}

\subsection{Derivation of RG equations}
\label{sec:4.2}

{\bf Single RG step.} In this section we describe in detail of how the renormalized 
Liouvillian and the renormalized vertices have to be defined in order to fulfil the central
properties of invariance, given by (\ref{rg_sigma_functional}) and (\ref{rg_sigma_R_functional}),
together with the renormalized perturbation series as shown in Eq.~(\ref{rg_diagram_sigma}).
Instead of directly aiming at the derivation of the continuous RG equations
(\ref{rg_formal_L})-(\ref{rg_formal_R}), we consider first a single discrete RG step,
where a certain finite part of the reservoir contraction is integrated out in
one step. We consider an arbitrary decompostion of the contraction in two 
parts
\begin{equation}
\label{AB_contraction_decomposition}
\gamma\,=\,\gamma^A\,+\,\gamma^B \quad,
\end{equation}
and want to integrate out the $\gamma^A$-part, i.e. the aim is to replace
$\gamma\rightarrow\gamma^B$ in the diagrams and define a renormalized Liouvillian
$L_S^B(E)$ and renormalized vertices $G^B(E)$ and $R^B(E)$ such that 
(\ref{rg_diagram_sigma}) holds
\begin{eqnarray}
\label{rg_B_diagram_sigma}
\left\{
\begin{array}{cl}
L_S^{eff}(E)
\\
\Sigma_R(E)
\end{array}
\right\}
\,&\rightarrow&\, 
\left\{
\begin{array}{cl}
L_S^B(E)
\\
R^B_{n=0}(E)
\end{array}
\right\}
\,+\,
{1\over S} \, (\pm)^{N_p} \, \left(\prod\gamma^B\right)_{irr}
\left\{
\begin{array}{cl}
G^B(E) \\ R^B(E)
\end{array}
\right\}\\
\nonumber
&&\hspace{-1.5cm}
\cdot\,{1\over E+X_1-L_S^B(E+X_1)}\,G^B(E+X_1)
\,\dots \,{1\over E+X_r-L_S^B(E+X_r)}\,
G^B(E+X_r) \quad.
\end{eqnarray}
An example of the decomposition (\ref{AB_contraction_decomposition}) is the
splitting (\ref{contraction_decomposition}) of the contraction into a symmetric 
and an antisymmetric part, with the task to integrate out the symmetric part in
one single step. Furthermore, we can also view the continuous RG as a sequence
of infinitesimal steps $\Lambda\rightarrow\Lambda-d\Lambda$, where each step 
can be brought into the form (\ref{AB_contraction_decomposition}) by writing
\begin{equation}
\label{AB_infinitesimal_decomposition}
\gamma^\Lambda\,=\,{d\gamma^\Lambda\over d\Lambda}\,d\Lambda \,+\, 
\gamma^{\Lambda-d\Lambda} \quad,
\end{equation}
and interpreting ${d\gamma^\Lambda\over d\Lambda}\,d\Lambda$ as the $\gamma^A$-part
which has to be integrated out. We can then write for the $B$-quantities
\begin{eqnarray}
\label{relation_discrete_continuous_L}
L^B_S(E)\,&\equiv&\,L^{\Lambda-d\Lambda}_S(E)\,=\,L^\Lambda_S(E)\,-\,
{dL_S^\Lambda(E)\over d\Lambda}\,d\Lambda \quad,\\
\label{relation_discrete_continuous_G}
G^B(E)\,&\equiv&\,G^{\Lambda-d\Lambda}(E)\,=\,G^\Lambda(E)\,-\,
{dG^\Lambda(E)\over d\Lambda}\,d\Lambda \quad,\\
\label{relation_discrete_continuous_R}
R^B(E)\,&\equiv&\,R^{\Lambda-d\Lambda}(E)\,=\,R^\Lambda(E)\,-\,
{dR^\Lambda(E)\over d\Lambda}\,d\Lambda \quad,
\end{eqnarray}
from which the RG equations can be read off once a formula for the $B$-quantities
has been derived.

\begin{figure}
  \centerline{\psfig{figure=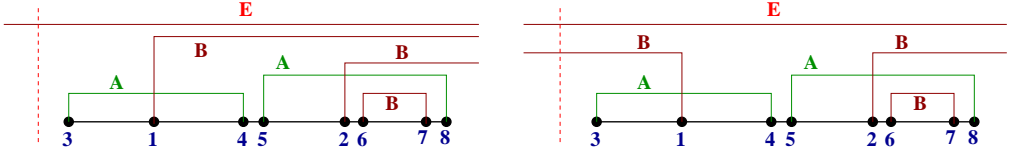,scale=0.27}}
  \caption{Example of an $A$-irreducible block. $A$- and $B$-contractions 
are distinguished by green and red lines. Vertices standing close to each 
other (the pairs $(45)$, $(26)$ and $(78)$) belong to the same $G$. There is 
one internal $B$-contraction connecting the vertices $6$ and $7$ of the block. Two 
$B$-contractions are leaving the block, which are connected to the vertices
$1$ and $2$, running either to the right (vertices $1$ and $2$ for the left 
diagram and vertex $2$ for the right diagram) or to the left (vertex $1$ for the right 
diagram). All the $B$-contractions which cross the block are summarized by the upper
line. $E$ is the sum of the Laplace variable $E$ plus the sum of all
$x$-variables of the $B$-contractions crossing the block. Therefore, for the
left diagram, the resolvent corresponding to the left vertical cut is given by 
${1\over E-L_S(E)}$, and the diagram corresponds to a contribution
for $(G^B)_{12}^{p_1 p_2}(E)$. In contrast, for the right diagram, we get
${1\over E-x_1-L_S(E-x_1)}$ for the resolvent, giving a contribution
for $(G^B)_{12}^{p_1 p_2}(E-x_1)$. Obviously, by shifting $E\rightarrow E+x_1$,
the contribution from the right diagram coincides with the one of the left diagram.}
\label{fig:A_block_lr}
\end{figure}

\begin{figure}
  \centerline{\psfig{figure=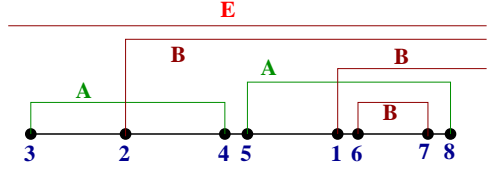,scale=0.3}}
  \caption{The same diagram as shown in Fig.~\ref{fig:A_block_lr} but
the vertices $1$ and $2$ are interchanged. Obviously, since the vertices
$1$ and $2$ do not belong to the same vertex $G$, this is a different diagram
to the effective vertex $(G^B)^{p_1p_2}_{12}(E)$, which has to be counted
separately. An additional minus sign has to be considered for fermions since
the effective vertex assumes the sequence $12$ for the reservoir lines 
leaving the vertex.}
\label{fig:A_block_permute}
\end{figure}

To be general, we start from a perturbation series 
of the form (\ref{rg_B_diagram_sigma}), where the Liouvillian and the vertices have
already a dependence on the energy variable $E$. We denote the
initial values by $L_S(E)$, $G(E)$ and $R(E)$ (corresponding to $L_S^\Lambda(E)$,
$G^\Lambda(E)$ and $R^\Lambda(E)$ if we consider the infinitesimal step
$\Lambda\rightarrow\Lambda-d\Lambda$ of a continuous RG scheme). Replacing each
contraction by the sum of $\gamma^A$ and $\gamma^B$, we obtain
diagrams where the vertices are either connected by $A$- or $B$-contractions. 
We can group each diagram into sequences of $A$-irreducible blocks, where a vertical
cut between two vertices hits at least one $A$-contraction. A single
$A$-irreducible block can have additional $B$-contractions, which either 
connect vertices within this block, or connect this block with another block, or
cross over this block, see Fig.~\ref{fig:A_block_lr} for illustration.
Let us label the indices of the vertices of those $B$-contractions, which connect the
block to other blocks, by $1,2,\dots n$ (in the sequence from left to right) and 
call them external indices. Obviously,
for $n\ge 1$, we interpret the diagram as a contribution to the effective vertex 
$(G^B)^{p_1\dots p_n}_{1\dots n}(E)$ (if the first vertex is $G$) or to 
$(R^B)^{p_1\dots p_n}_{1\dots n}(E)$ (if the first vertex is $R$). 
The energy variable $E$ is chosen such that the
resolvent standing left to the block is of the form ${1\over E-L_S(E)}$, according to
the diagrammatic rule that the energy variable of the effective vertex must be identical to
the one of the resolvent standing left to the vertex.
Thereby we get the same result, independent on whether the $B$-contractions run to blocks to the left 
or to the right of the considered one, see Fig.~\ref{fig:A_block_lr}. Therefore, we determine
the diagram by the convention that all $B$-contractions go to the right. In this case,
the energy variable $E$ is identical to the one corresponding to all $B$-contractions 
crossing the block. Subsequently, we have to sum over all possibilities to commute the 
external indices $1,2,\dots,n$ within the block, see Fig.~\ref{fig:A_block_permute}, except when the 
indices belong to the same vertex $G$ or $R$. To each permutation we associate a corresponding
fermionic sign for two reasons. First, the effective vertex 
$(G^B)^{p_1\dots p_n}_{1\dots n}(E)$ should be antisymmetric for fermions under permutation 
of two indices. Secondly, if the external vertex is used as input for the perturbation
series (\ref{rg_B_diagram_sigma}), it is assumed that the $B$-contractions leave the 
effective vertex in the sequence $1,2,\dots n$ for the determination of the
correct sign when this effective vertex is connected to other effective vertices.
Therefore, if the original sequence differs from the one used in the effective vertex,
as is the case e.g. in Fig.~\ref{fig:A_block_permute}, the corresponding fermionic sign 
has to be corrected by appropriate rules determining the effective vertex. Furthermore,
we attribute to the effective vertex all minus signs which occur when the $A$- and 
$B$-contractions are separated within the $A$-irreducible blocks from each other and from
the external indices (again this is an information 
which is no longer available when the effective vertex is used in the perturbation series
(\ref{rg_B_diagram_sigma})). In this way, we get for the diagrams depicted in 
Figs.~\ref{fig:A_block_lr} and \ref{fig:A_block_permute} the following result
\begin{eqnarray}
\label{effective_vertex_example}
(G^B)_{12}(E)\,&\rightarrow&\,(\pm)\,(\pm)\,\gamma^A_{34}\,\gamma^A_{58}\,\gamma^B_{67}\,
G_3(E)\,\Pi_3\,G_1(E+x_3)\,\Pi_{13}\\
\nonumber
&& \hspace{-1.5cm}\cdot\,G_{45}(E+x_{13})\,\Pi_{15}\,
G_{26}(E+x_{15})\,\Pi_{1256}\,G_{78}(E+x_{1256}) \quad \pm \quad (1\leftrightarrow 2) \quad,
\end{eqnarray}
where we have omitted for simplicity the obvious notation of the Keldysh indices. The 
resolvents and energy variables are defined via the short-hand notations
\begin{equation}
\label{short_notation_Pi_E}
\Pi_{1\dots n}\,=\,{1\over E\,+\,x_{1\dots n}-L_S^B(E\,+\,x_{1\dots n})} \quad,\quad
x_{1\dots n}\,= x_1\,+\dots +\,x_n \quad,
\end{equation}
where we used the full effective Liouvillian in the denominator, see below.
The two sign factors in (\ref{effective_vertex_example}) arise when the two $A$-contractions
are separated. As one can see, the rule to translate the
effective vertex is quite simple. The indices of the energy variables of all vertices
is always identical to the indices of the left preceeding resolvent. The resolvents get the
indices of all the left vertices of contractions crossing the vertical cut at the position
of the resolvent (the indices of all $B$-contractions crossing the diagram and
not belonging to the block are contained in the energy variable $E$).

\begin{figure}
  \centerline{\psfig{figure=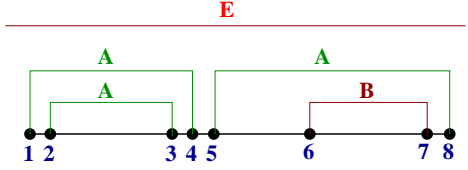,scale=0.3}}
  \caption{An example of an $A$-irreducible diagram contributing to 
the effective Liouvillian $L^B_S(E)$.}
\label{fig:A_block_L}
\end{figure}

If no $B$-contractions leave the $A$-irreducible block, we interpret the diagram as
a contribution to the effective Liouvillian $L_S^B(E)$ (if the first vertex is $G$)
or to $R_{n=0}^B(E)$ (if the first vertex is $R$). An example is shown in
Fig.~\ref{fig:A_block_L}, which translates to
\begin{equation}
\label{effective_L_example}
L^B_S(E)\,\rightarrow\,{1\over 2}\,\gamma^A_{14}\,\gamma^A_{23}\,\gamma^A_{58}\,\gamma^B_{67}\,
G_{12}(E)\,\Pi_{12}\,G_{345}(E+x_{12})\,\Pi_5\,G_6(E+x_5)\,
\Pi_{56}\,G_{78}(E+x_{56})\quad,
\end{equation}
where the factor ${1\over 2}$ is a symmetry factor since two vertices are connected by 
the same contraction within the $A$-irreducible block. As indicated by the definition
of the resolvent (\ref{short_notation_Pi_E}), such diagrams can be resummed on each line
connecting the vertices within the $A$-irreducible blocks, leading to the full 
effective Liouville operator $L_S^B(E)$ occuring in the denominator of the resolvent.
In this way a self-consistent equation is obtained.

Summing up all $A$-irreducible diagrams, we obtain the following perturbation
series for the effective Liouvillian and the effective vertices
\begin{eqnarray}
\label{effective_B_series}
\left\{
\begin{array}{cl}
L_S^B(E)
\\
G^B(E)
\\
R^B(E)
\end{array}
\right\}
\,&\rightarrow&\, 
\left\{
\begin{array}{cl}
L_S(E)
\\
G(E)
\\
R(E)
\end{array}
\right\}
\,+\,
{1\over S} \, (\pm)^{N_p} \, \left(\prod\gamma^A\prod\gamma^B\right)_{A-irr}
\left\{
\begin{array}{cl}
G(E) \\ G(E) \\ R(E)
\end{array}
\right\}\\
\nonumber
&&\hspace{-1.5cm}
\cdot\,{1\over E+X_1-L_S^B(E+X_1)}\,G(E+X_1)
\,\dots \,{1\over E+X_r-L_S^B(E+X_r)}\,
G(E+X_r) \quad,
\end{eqnarray}
where the first term on the r.h.s. represents the original quantities, and
the second term contains all $A$-irreducible contributions, indicated by
the sub-index $A-irr$ at the product of the contractions. As derived above,
we summarize the additional rules

\begin{enumerate}
\item
All free $B$-contractions are directed to the right. Free $A$-contractions are
not allowed.

\item
The external indices of the free $B$-contractions are numerated as $1,2,\dots,n$ from
left to right.

\item
Sum over all permutations of the external indices and assign a minus sign for
fermions according to the parity of the permutation. Omit permutations of 
indices belonging to the same vertex.

\item
Associate a fermionic sign which arises when separating all $A$- and $B$-contractions 
from each other and from the external indices.

\end{enumerate}

Inserting the result (\ref{effective_B_series}) for the $B$-quantities
into the effective perturbation series (\ref{rg_B_diagram_sigma}) for the
physical quantities of interest, we obtain three different terms. The first term
on the r.h.s. of (\ref{rg_B_diagram_sigma}) contains via the two terms on the
r.h.s. of (\ref{effective_B_series}) the original quantities $L_S(E)$, $G(E)$
and $R(E)$, together with all $A$-irreducible diagrams. The second term on
the r.h.s. of (\ref{rg_B_diagram_sigma}) contains all $A$-reducible but
$B$-irreducible diagrams. In this way one can see that the derivation of a 
single RG step is nothing else than an obvious classification into different
topological sectors, combined with a convenient resummation of $A$-irreducible
contributions into effective quantities.

We note that the effective vertices $G^B$ and $R^B$ are again in normal-ordered
form, i.e. in the effective perturbation series (\ref{rg_B_diagram_sigma}) no
$B$-contractions are allowed between field operators belonging to the same
$B$-vertex. The reason is that we have incorporated all such terms 
already into the definition of the effective vertices and the effective 
Liouvillian by allowing for $B$-contractions on the r.h.s. of Eq.~(\ref{effective_B_series}).
In this sense, our RG scheme is quite similiar to the RG scheme
developed by Salmhofer \cite{salmhofer} for quantum field-theoretical
problems described by usual Feynman diagrams. However, besides the fact that
we have to deal with operator vertices here and have to use the Laplace instead
of Fourier transform, an important difference is also that we use the full
effective Liouville operator $L_S^B(E)$ as input for the dynamics of the local
quantum system. In this way a self-consistent equation arises for the determination
of $L_S^B(E)$ via (\ref{effective_B_series}). Transferred to the Salmhofer RG scheme
this would mean in a rough sense that self-energy insertions should be resummed on all
propagators connecting the vertices, similiar to the RG scheme of Wetterich
\cite{wetterich} (disregarding the fact that both schemes are quite different
for many other reasons).

Finally, we mention that one can also set up a non normal-ordered version of the
RG procedure. This can be easily achieved by just forbidding the occurence of
$B$-contractions on the r.h.s. of (\ref{effective_B_series}). This means that no
$B$-contractions are allowed within the $A$-irreducible blocks 
which connect the block with itself, see e.g. the $B$-contraction 
between vertex $6$ and $7$ in Fig.~\ref{fig:A_block_lr}. These diagrams are 
then considered in the non normal ordered effective perturbation series (\ref{rg_B_diagram_sigma})
by allowing for $B$-contractions connecting field operators from the same $B$-vertex. 
Concerning usual quantum field-theoretical problems, this would correspond to
the RG scheme developed by Polchinski \cite{polchinski} together with taking
the full propagators between the vertices including self-energy insertions.
However, as one can see, the normal-ordered version includes more diagrams
into the effective quantities and, therefore, the results are expected to be better.

\begin{figure}
  \centerline{\psfig{figure=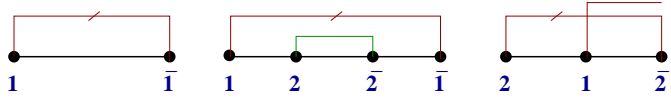,scale=0.3}}
  \caption{RG diagrams for the Liouvillian and the vertex up to 2-loop order for a
model with charge fluctuations. The slash indicates
the derivative of the contraction ${d\gamma\over d\Lambda}$.}
\label{fig:LG_charge}
\end{figure}

\begin{figure}
  \centerline{\psfig{figure=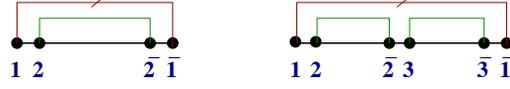,scale=0.3}}
  \caption{RG diagrams for the Liouvillian up to 2-loop order for a
model with spin and/or orbital fluctuations. The slash indicates
the derivative of the contraction ${d\gamma\over d\Lambda}$.}
\label{fig:L_kondo}
\end{figure}

\begin{figure}
  \centerline{\psfig{figure=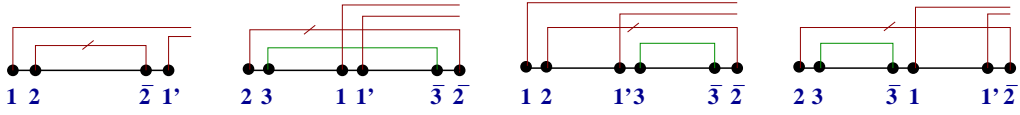,scale=0.3}}
  \caption{RG diagrams for the vertex up to 2-loop order for a
model with spin and/or orbital fluctuations. The slash indicates
the derivative of the contraction ${d\gamma\over d\Lambda}$.}
\label{fig:G_kondo}
\end{figure}

{\bf Continuous RG.} We can now easily derive the rules for setting up the RG equations
(\ref{rg_formal_L})-(\ref{rg_formal_R}) for the continuous RG flow. For a single
infinitesimal step $\Lambda\rightarrow\Lambda-d\Lambda$, we use the decomposition
(\ref{AB_infinitesimal_decomposition}) with
\begin{equation}
\label{gamma_AB_infinitesimal}
\gamma^A\,=\,{d\gamma^\Lambda \over d\Lambda}\,d\Lambda \quad,\quad
\gamma^B\,=\,\gamma^{\Lambda-d\Lambda} \quad.
\end{equation}
Comparing (\ref{relation_discrete_continuous_L})-(\ref{relation_discrete_continuous_R}) 
with (\ref{effective_B_series}) (with $L_S\equiv\L_S^\Lambda$, $G\equiv G^\Lambda$ 
and $R\equiv R^\Lambda$), we obtain immediately
\begin{eqnarray}
\label{rg_equation_continuous}
-\,{d \over d\Lambda}
\left\{
\begin{array}{cl}
L_S^\Lambda(E)
\\
G^\Lambda(E)
\\
R^\Lambda(E)
\end{array}
\right\}
\,&\rightarrow&\, 
{1\over S} \, (\pm)^{N_p} \, 
\left({d\gamma \over d\Lambda}\prod\gamma^\Lambda
\right)_{{d\gamma \over d\Lambda}-irr}
\,\left\{
\begin{array}{cl}
G^\Lambda(E) \\ G^\Lambda(E) \\ R^\Lambda(E)
\end{array}
\right\}\\
\nonumber
&&\hspace{-1.5cm}
\cdot\,{1\over E+X_1-L_S^\Lambda(E+X_1)}\,G^\Lambda(E+X_1)
\,\dots \,{1\over E+X_r-L_S^\Lambda(E+X_r)}\,
G^\Lambda(E+X_r) \quad,
\end{eqnarray}
where, for $d\Lambda\rightarrow 0$, we can omit all terms $\sim (d\Lambda)^k$ 
with $k>1$ on the r.h.s., so that 
only one contraction ${d\gamma \over d\Lambda}$ can occur, and we can replace 
\begin{equation}
L_S^B\equiv L_S^{\Lambda-d\Lambda}\rightarrow L_S^\Lambda \,\,,\,\,
G^B\equiv G^{\Lambda-d\Lambda}\rightarrow G^\Lambda \,\,,\,\,
R^B\equiv R^{\Lambda-d\Lambda}\rightarrow R^\Lambda \,\,,\,\,
\gamma^{\Lambda-d\Lambda}\rightarrow \gamma^\Lambda 
\end{equation}
in all other parts of the diagram.
Since the total expression has to be irreducible with respect to ${d\gamma\over d\Lambda}$
and only one such contraction is allowed to occur, we get the simple rule that the
contraction ${d\gamma\over d\Lambda}$ must contract the first with the last vertex
in the diagram, see Figs.~\ref{fig:LG_charge}, \ref{fig:L_kondo} and \ref{fig:G_kondo} for 
examples. Diagrammatically this contraction is indicated by an additional slash.

To illustrate the rules, we have shown in Fig.~\ref{fig:LG_charge} the RG diagrams
up to 2-loop order for an arbitrary model with charge fluctuations where the vertex
depends only on one index. Omitting the index $\Lambda$ and the trivial Keldysh indices,
and using the notation
\begin{equation}
\label{delta_separation}
\gamma_{11'}\,=\,\delta_{1\bar{1}'}\,\gamma_1
\end{equation}
to exhibit explicitly the $\delta_{1\bar{1}'}$-part of each contraction, 
the RG equations are given by (using again the elegant notation (\ref{short_notation_Pi_E}))
\begin{eqnarray}
\label{rg_L_charge_1}
-\,{dL_S(E)\over d\Lambda}\,&=&\,
{d\gamma_{1}\over d\Lambda}\,G_{1}(E)\,\Pi_{1}\,G_{\bar{1}}(E+x_{1})\\
\label{rg_L_charge_2}
&& + \,{d\gamma_{1}\over d\Lambda}\,\gamma_{2}\,
G_{1}(E)\,\Pi_{1}\,G_2(E+x_1)\,\Pi_{12}\,G_{\bar{2}}(E+x_{12})\,\Pi_{1}\,G_{\bar{1}}(E+x_{1}) 
\quad,\\
\label{rg_G_charge}
-\,{dG_{1}(E)\over d\Lambda}\,&=&\,
\pm\,{d\gamma_{2}\over d\Lambda}\,G_{2}(E)\,\Pi_{2}\,G_1(E+x_2)\,\Pi_{12}\,G_{\bar{2}}(E+x_{12})
\quad.
\end{eqnarray}
Note the sign factor in front of (\ref{rg_G_charge}) due to separating out the
contraction between index $2$ and $\bar{2}$ from the external vertex $1$.

Figs.~\ref{fig:L_kondo} and \ref{fig:G_kondo} show 
the RG diagrams up to 2-loop order for an arbitrary model with spin and orbital 
fluctuations (\ref{spin_orbital}), the Kondo problem (\ref{kondo_vertex}) being
a special example. In this case we get the RG equations
\begin{eqnarray}
\label{rg_L_so_1}
-\,{dL_S(E)\over d\Lambda}\,&=&\,
{d\gamma_{1}\over d\Lambda}\,\gamma_{2}\,G_{12}(E)\,\Pi_{12}\,G_{\bar{2}\bar{1}}(E+x_{12})\\
\label{rg_L_so_2}
&& + \,{d\gamma_{1}\over d\Lambda}\,\gamma_{2}\,\gamma_{3}
\,G_{12}(E)\,\Pi_{12}\,G_{\bar{2}3}(E+x_{12})\,\Pi_{13}\,G_{\bar{3}\bar{1}}(E+x_{13}) 
\end{eqnarray}
for the Liouvillian, and
\begin{eqnarray}
\label{rg_G_so_1}
-\,{dG_{11'}(E)\over d\Lambda}\,&=&\,
\left\{{d\gamma_{2}\over d\Lambda}\,G_{12}(E)\,\Pi_{12}\,G_{\bar{2}1'}(E+x_{12})
\,\, \pm \,\, (1\leftrightarrow 1')\right\}\\
\label{rg_G_so_2}
&& \hspace{-1.5cm} + \, {d\gamma_{2}\over d\Lambda}\,\gamma_{3}
\,G_{23}(E)\,\Pi_{23}\,G_{11'}(E+x_{23})\,\Pi_{11'23}\,G_{\bar{3}\bar{2}}(E+x_{11'23})\\ 
\label{rg_G_so_3}
&& \hspace{-1.5cm} \pm \,\left\{ {d\gamma_{2}\over d\Lambda}\,\gamma_{3}
\,G_{12}(E)\,\Pi_{12}\,G_{1'3}(E+x_{12})\,\Pi_{11'23}\,G_{\bar{3}\bar{2}}(E+x_{11'23})
\,\, \pm \,\,  (1\leftrightarrow 1')\right\} \\ 
\label{rg_G_so_4}
&& \hspace{-1.5cm} + \, \left\{{d\gamma_{2}\over d\Lambda}\,\gamma_{3}
\,G_{23}(E)\,\Pi_{23}\,G_{\bar{3}1}(E+x_{23})\,\Pi_{12}\,G_{1'\bar{2}}(E+x_{12})
\,\, \pm \,\, (1\leftrightarrow 1')\right\} 
\end{eqnarray}
for the vertex. The corresponding equations for $R$ follow from replacing the
first vertex $G\rightarrow R$. Note the sign factors in front of (\ref{rg_G_so_3}) 
due to separating out the contraction between index $2$ and $\bar{2}$, and the missing
permutation of $1$ and $1'$ in (\ref{rg_G_so_2}) since both indices belong to
the same vertex. We again note that the energy arguments of the vertices are
always such that the indices of the $x$-variable coincide with the indices
of the left resolvent. Using these rules, it should be obvious
how to write down RG equations in arbitrary order for generic models 
nearly \lq\lq without thinking''. However, the problem of course is to solve
these RG equations which will be the subject for the rest of the paper.

\subsection{Summary of RG equations and general properties}
\label{sec:4.3}

{\bf Summary.} We first summarize the RG procedure described in the previous sections.
As outlined in Sec.~\ref{sec:4.1}, we use a 2-stage RG procedure. In the
first step we integrate out the symmetric part of the distribution function
by a discrete RG step using the decomposition (\ref{contraction_sym_antisym})
\begin{eqnarray}
\label{contraction_sym}
(\gamma^s)_{11'}^{pp'}\,=\,\delta_{1\bar{1}'}\,p'\,\gamma^s_1
\quad,\quad
\gamma^s_1\,=\,{1\over 2}
\left\{
\begin{array}{cl}
-\eta \\ 1
\end{array}
\right\}\,
\,\rho_\nu(\eta\bar{\omega})\quad,\\ 
\label{contraction_antisym}
\gamma^a_{11'} \,=\,\delta_{1\bar{1}'}\,\gamma^a_1 \quad,\quad
\gamma^a_1\,=\,
\left\{
\begin{array}{cl}
\eta \\ 1
\end{array}
\right\}\,
\,\rho_\nu(\eta\bar{\omega})\,\left[f_\alpha(\bar{\omega})\pm {1\over 2}\right] \quad,
\end{eqnarray}
where we use from now on the conventions
\begin{equation}
\label{conventions}
\bar{\omega}\,\equiv\,\eta\omega\quad,\quad
\bar{\mu}_\alpha\,\equiv\,\eta\mu_\alpha\quad,\quad
x\,\equiv\,\eta(\omega+\mu_\alpha)\,=\,\bar{\omega}\,+\,\bar{\mu}_\alpha\quad,
\end{equation}
and
\begin{equation}
\label{index_convention}
\bar{\omega}_{1\dots n}\,=\,\bar{\omega}_1+\dots+\bar{\omega}_n\quad,\quad
\bar{\mu}_{1\dots n}\,=\,\bar{\mu}_1+\dots+\bar{\mu}_n\quad,\quad
x_{1\dots n}\,=\,x_1+\dots+x_n\quad.
\end{equation}
For fermions and $T=0$ we have
\begin{equation}
\label{fermion_antisym_T=0}
f_\alpha(\bar{\omega})- {1\over 2}\,=\,-\,{1\over 2}\,\mbox{sign}(\bar{\omega}) \quad.
\end{equation}
The effective Liouvillian $L_S^a$ and the effective vertices $G^a$ and $R^a$  
are given by (\ref{effective_B_series})
\begin{eqnarray}
\label{effective_antisym_series}
\left\{
\begin{array}{cl}
L_S^a(E)
\\
G^a(E)
\\
R^a(E)
\end{array}
\right\}
\,&\rightarrow&\, 
\left\{
\begin{array}{cl}
L_S
\\
G
\\
R
\end{array}
\right\}
\,+\,
{1\over S} \, (\pm)^{N_p} \, \left(\prod\gamma^s\prod\gamma^a\right)_{s-irr}
\left\{
\begin{array}{cl}
G \\ G \\ R
\end{array}
\right\}\\
\nonumber
&&\hspace{-1.5cm}
\cdot\,{1\over E+X_1-L_S^a(E+X_1)}\,G
\,\dots \,{1\over E+X_r-L_S^a(E+X_r)}\,
G \quad,
\end{eqnarray}
where $s-irr$ means irreducibility with respect to the symmetric contractions
$\gamma^s$. The vertices on the r.h.s. are the initial vertices having no
dependence on the energy variable $E$. The energies $X_i\equiv x^{(i)}_{1\dots n}$ 
contain all indices of the left vertex of all contractions crossing the $i$-th resolvent. 
We also use the convention
\begin{equation}
\label{E_convention}
E_{1\dots n}\,=\,E\,+\,\bar{\mu}_{1\dots n}
\end{equation}
such that the energies $X_i$ can be written as
\begin{equation}
\label{E_x_relation}
E\,+\,X_i\,\rightarrow\,E\,+\,x_{1\dots n}\,=\,E_{1\dots n}\,+\,\bar{\omega}_{1\dots n}\quad,
\end{equation}
separating more clearly the dependence on the integration variables $\bar{\omega}_i$
and the physical parameters $E_{1\dots n}$ on which the final solution after integration
will depend.

For the next continuous RG procedure we need only the vertices averaged over the Keldysh 
indices. Therefore, we sum over all Keldysh indices on the r.h.s. of
(\ref{effective_antisym_series}). Since the Keldysh indices of the initial vertices
are all the same, see (\ref{G_vertex_liouville}), we obtain only the two possible
combinations
\begin{equation}
\label{G_bar_tilde}
\bar{G}_{1\dots n}\,=\,\sum_p\,G^{pp\dots p}_{1\dots n} \quad,\quad
\tilde{G}_{1\dots n}\,=\,\sum_p\,p\,G^{pp\dots p}_{1\dots n} 
\end{equation}
on the r.h.s. of (\ref{effective_antisym_series}). Due to the form (\ref{contraction_sym})
of the symmetric contraction, $\tilde{G}$ occurs if an odd number of
symmetric and left-going contractions are attached to this vertex.

In the second step we replace the remaining antisymmetric contraction $\gamma^a$
by the $\Lambda$-dependent contraction (\ref{lambda_contraction_matsubara_antisym}) 
with a cutoff defined on the imaginary frequency axis. By convention we write it in
the form
\begin{equation}
\label{form_contraction}
\gamma_{11'}^\Lambda\,=\,\delta_{1\bar{1}'}\,\gamma_1^\Lambda\quad
\end{equation}
with
\begin{eqnarray}
\label{matsubara_antisym}
\gamma_1^\Lambda\,&=&\,
\left\{
\begin{array}{cl}
\eta \\ -1
\end{array}
\right\}\,
\,\rho_\nu(\eta\bar{\omega})\,T\,\sum_n\,{1\over \bar{\omega}\,-\,i\omega_n^\alpha} 
\,\,\theta_{T_\alpha}(\Lambda-|\omega_n^\alpha|)\\
&\begin{array}[b]{c}
\vspace{-1.5mm} T\rightarrow 0 \\ 
\longrightarrow 
\end{array}&
\left\{
\begin{array}{cl}
\eta \\ -1
\end{array}
\right\}\,
\,\rho_\nu(\eta\bar{\omega})\,{1\over 2\pi}\,\int_{-\Lambda}^\Lambda d\omega'
\,{1\over \bar{\omega}\,-\,i\omega'} \quad.
\end{eqnarray}
The corresponding RG equations are given by (\ref{rg_equation_continuous})
\begin{eqnarray}
\label{rg_equations}
-\,{d \over d\Lambda}
\left\{
\begin{array}{cl}
L_S^\Lambda(E)
\\
\bar{G}^\Lambda(E)
\\
\bar{R}^\Lambda(E)
\end{array}
\right\}
\,&\rightarrow&\, 
{1\over S} \, (\pm)^{N_p} \, 
\left({d\gamma \over d\Lambda}\prod\gamma^\Lambda
\right)_{{d\gamma \over d\Lambda}-irr}
\,\left\{
\begin{array}{cl}
\bar{G}^\Lambda(E) \\ \bar{G}^\Lambda(E) \\ \bar{R}^\Lambda(E)
\end{array}
\right\}\\
\nonumber
&&\hspace{-1.5cm}
\cdot\,{1\over E+X_1-L_S^\Lambda(E+X_1)}\,\bar{G}^\Lambda(E+X_1)
\,\dots \,{1\over E+X_r-L_S^\Lambda(E+X_r)}\,
\bar{G}^\Lambda(E+X_r) \quad,
\end{eqnarray}
where only the vertices (\ref{average_Keldysh_indices}) averaged over the 
Keldysh indices occur since the antisymmetric contraction (\ref{matsubara_antisym}) 
does not depend on the Keldysh indices. The initial condition for these
differential equations at $\Lambda=\infty$ is given by 
\begin{eqnarray}
\label{initial_condition_L}
L_S^{\Lambda=\infty}(E)\,&=&\,L_S^a(E) \quad,\\
\label{initial_condition_G}
(\bar{G}^{\Lambda=\infty})_{1\dots n}(E)\,&=&\,
\bar{G}^a_{1\dots n}(E) \quad,\\
\label{initial_condition_R}
(\bar{R}^{\Lambda=\infty})_{1\dots n}(E)\,&=&\,
\bar{R}^a_{1\dots n}(E) \quad,
\end{eqnarray}
where $L_S^a$, $\bar{G}^a$, and $\bar{R}^a$ are given by (\ref{effective_antisym_series})
from the first discrete step.

Solving the RG equations down to $\Lambda=0$ gives the final result for the 
physical quantities via (\ref{final_sigma})
\begin{equation}
\label{final_result}
L_S^{eff}(E)\,=\,L^{\Lambda=0}_S(E)\quad,\quad
\Sigma_R(E)\,=\,R^{\Lambda=0}_{n=0}(E)\quad,
\end{equation}
or, alternatively, one can stop the RG at any arbitrary value of $\Lambda$
and use the effective perturbation series (\ref{antisym_series_matsubara}) to
calculate the physical quantities
\begin{eqnarray}
\label{effective_perturbation_series}
\left\{
\begin{array}{cl}
L_S^{eff}(E)
\\
\Sigma_R(E)
\end{array}
\right\}
\,&\rightarrow&\, 
\left\{
\begin{array}{cl}
L_S^\Lambda(E)
\\
R^\Lambda_{n=0}(E)
\end{array}
\right\}
\,+\,
{1\over S} \, (\pm)^{N_p} \, \left(\prod\gamma^\Lambda\right)_{irr}
\left\{
\begin{array}{cl}
\bar{G}^\Lambda(E) \\ \bar{R}^\Lambda(E)
\end{array}
\right\}\\
\nonumber
&&\hspace{-1.5cm}
\cdot\,{1\over E+X_1-L_S^\Lambda(E+X_1)}\,\bar{G}^\Lambda(E+X_1)
\,\dots \,{1\over E+X_r-L_S^\Lambda(E+X_r)}\,
\bar{G}^\Lambda(E+X_r) \quad.
\end{eqnarray}
Of course this makes only sense if this perturbation theory is well-defined at
the scale $\Lambda$, i.e. if all large terms have already been eliminated by the previous RG flow
(which happens when $\Lambda$ reaches some physical low energy scale and the couplings
are still small, see the discussion in the following sections).

{\bf Symmetry relations.} All the symmetry properties from (anti-)symmetry
(see (\ref{G_symmetry})), conservation of probability (see (\ref{L_eff_property})
and (\ref{G_property})), and those following from the hermiticity of the 
original Hamiltonian (see (\ref{L_eff_c_transform}), (\ref{G_c_transform}), 
(\ref{sigma_R_c_transform}) and (\ref{R_c_transform})), are also fulfilled 
for the effective quantities. In summary we have the following relations
\begin{eqnarray}
\label{rg_G_symmetry}
&& \hspace{1cm} G^{p_1\dots p_i\dots p_j\dots p_n}_{1\dots i\dots j\dots n}(E)\,=\,
\pm\,G^{p_1\dots p_j\dots p_i\dots p_n}_{1\dots j\dots i\dots n}(E)\quad\\
\label{rg_R_symmetry}
&& \hspace{1cm} R^{p_1\dots p_i\dots p_j\dots p_n}_{1\dots i\dots j\dots n}(E)\,=\,
\pm\,R^{p_1\dots p_j\dots p_i\dots p_n}_{1\dots j\dots i\dots n}(E)\quad,\\
\label{rg_LG_property}
&& \hspace{1.2cm} \mbox{Tr}_S\,L_S(E)\,=\,0\quad , \quad
\sum_{p_1\dots p_n}\,\mbox{Tr}_S\,G^{p_1\dots p_n}_{1\dots n}(E) = 0 \quad,\\
\label{rg_L_c_transform}
&& \hspace{2.6cm} L_S(E)^c\,=\,-L_S(-E^*) \quad,\\
\label{rg_G_c_transform}
&& \hspace{-1cm} G^{p_1\dots p_n}_{1\dots n}(E)^c\,=\,
-\,(\sigma^-)^n\,G^{\bar{p}_n\dots \bar{p}_1}_{\bar{n}\dots\bar{1}}(-E^*)
\,=\,-\,\sigma^{--\dots -}\,G^{\bar{p}_1\dots \bar{p}_n}_{\bar{1}\dots\bar{n}}(-E^*)\quad,\\
\label{rg_R_c_transform}
&& \hspace{-1cm} R^{p_1\dots p_n}_{1\dots n}(E)^c\,=\,
-\,(\sigma^-)^n\,R^{\bar{p}_n\dots \bar{p}_1}_{\bar{n}\dots\bar{1}}(-E^*)
\,=\,-\,\sigma^{--\dots -}\,R^{\bar{p}_1\dots \bar{p}_n}_{\bar{1}\dots\bar{n}}(-E^*)\quad,
\end{eqnarray}
where we can either use $L_S^a,G^a,R^a$ or 
$L_S^\Lambda,\bar{G}^\Lambda,\bar{R}^\Lambda$ for $L_S,G,R$ (for $\bar{G}$ and
$\bar{R}$, the Keldysh indices are of course omitted), a convention we will also
use for the following discussions. These identities follow
directly from the initial symmetries and the RG equations 
(\ref{effective_antisym_series}) and (\ref{rg_equations}), see Appendix D for
the proof.

We see that all symmetry properties are preserved under the RG flow.
Moreover, they are even invariant within all truncation schemes, since they
hold for each term on the r.h.s. of the RG equations separately, provided
the complete sum over all indices is taken. 

{\bf Analytic properties.} For a subsequent discussion of the frequency integrations, we
first study the analytic properties of the Liouvillian and
the vertices in the variables $E$ and $x_1\dots x_n$. For this, we need the essential 
ingredient that  
\begin{equation}
\label{resolvent_analytic}
{1\over E-L_S(E)}\quad \mbox{is analytic in $E$ in the upper half plane}\quad,
\end{equation}
which is equivalent to the property that the effective reduced density matrix,
defined by 
\begin{equation}
\label{rd_analytic}
\tilde{\rho}_S(E)\,=\,{i\over E-L_S(E)}\,\rho_S(t_0) \quad,
\end{equation}
has the analytic properties depicted in Fig.~\ref{fig:analytic_structure}, i.e.
contains no exponentially increasing solutions in time space. 
To prove this statement generically is difficult since it is not clear whether
certain approximations for $L_S(E)$ can not lead to poles in the upper half plane.
The question is ultimately related to whether the relaxation and
dephasing rates generated by the RG flow are positive. So far, we have not seen any
mathematical proof to show this property from the structure of the RG equations.
Of course we know that, if all
terms are taken into account on the r.h.s. of the RG equations, we get the exact
solution and at least at the end of the flow, the analytic property 
(\ref{resolvent_analytic}) must be fulfilled due to physical reasons. Whether it
is also fulfilled throughout the RG flow and within certain approximations of the
RG equations is not clear and an important question for future research. We will
assume in the following that it holds.

From (\ref{resolvent_analytic}) and the analytic structure of the RG equations
(\ref{effective_antisym_series}) and (\ref{rg_equations}), we obtain immediately that
\begin{equation}
\label{E_analytic}
L_S(E), G(E), R(E) \quad \mbox{are analytic in $E$ in the upper half plane}\quad.
\end{equation}

\begin{figure}
  \centerline{\psfig{figure=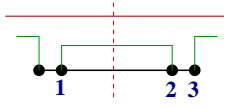,scale=0.3}}
  \caption{Illustration for the structure of the energy arguments of
the vertices. According to the diagrammatic rules, the resolvent and 
the right vertex have the structure ${1\over E+x_1-L_S(E+x_1)}G(E+x_1;x_2x_3)$.
Integrating out the $\delta_{12}\sim \delta(x_1+x_2)$
part of the contraction $\gamma_{12}$, the energy arguments of the vertex
get the form $G(E+x_1;-x_1x_3)$.}
\label{fig:analytic}
\end{figure}

Furthermore, we can also study the analytic properties of the vertices in the
frequency variables $x_1\dots x_n$ with $x_i=\eta_i(\omega_i+\mu_{\alpha_i})$ (a 
better choice than the frequencies $\omega_i$ because always this combination
occurs in the perturbative series). Writing the vertices as
function of the $x$-variables
\begin{equation}
\label{vertex_x}
G_{1\dots n}(E)\,\equiv\, G_{\eta_1\nu_1\dots\eta_n\nu_n}(E;x_1\dots x_n)\quad,
\end{equation}
and assuming that the vertices are analytic in the $x_i$-variables initially, 
we can prove the following analytic property which is preserved under the RG
\begin{eqnarray}
\nonumber
&&G(E\,+\,x_1+\dots+\,x_m;-x_1\dots -x_m\,x_{m+1}\dots x_n) \\
\label{vertex_analyticity}
&&\mbox{is analytic if all variables}\quad E,x_1\dots x_n\quad
\mbox{lie in the upper half plane}\quad,
\end{eqnarray} 
where $1\le m\le n$ can take any value. The same holds if any permutation of the 
indices of the $x$-variables is chosen. It means that the vertices stay
analytic in the upper half plane if the sign of some $x_i$-variable is changed, provided
one replaces $E\rightarrow E+x_i$. The property follows from the
structure of the RG equations since only the frequency combinations of 
(\ref{vertex_analyticity}) occur on the r.h.s. Consider e.g. the first term of
the RG equation (\ref{rg_G_so_1}), which we write in the representation (\ref{vertex_x})
and integrate out the $\delta$-functions of the contractions by using (\ref{form_contraction}).
This gives
\begin{eqnarray}
\label{G_energy_structure_1}
{d\over d\Lambda}\,G_{\eta_1\nu_1,\eta_1'\nu_1'}(E;x_1 x_1')\,&\rightarrow&\,
{d\gamma_2\over d\Lambda}\,\cdot\\
\nonumber
&&\hspace{-3.5cm}
\cdot\,G_{\eta_1\nu_1,\eta_2\nu_2}(E;x_1x_2)\,
{1\over E+x_1+x_2-L_S(E+x_1+x_2)}\,G_{-\eta_2\nu_2,\eta_1'\nu_1'}(E+x_1+x_2;-x_2x_1')
\quad,
\end{eqnarray}
or for $E\rightarrow E+x_1$ and $x_1\rightarrow -x_1$
\begin{eqnarray}
\label{G_energy_structure_2}
{d\over d\Lambda}\,G_{\eta_1\nu_1,\eta_1'\nu_1'}(E+x_1;-x_1x_1')\,&\rightarrow&\,
{d\gamma_2\over d\Lambda}\,\cdot\\
\nonumber
&&\hspace{-3.5cm}
\cdot\,G_{\eta_1\nu_1,\eta_2\nu_2}(E+x_1;-x_1x_2)\,
{1\over E+x_2-L_S(E+x_2)}\,G_{-\eta_2\nu_2,\eta_1'\nu_1'}(E+x_2;-x_2x_1')
\quad.
\end{eqnarray}
In all cases we see that the analyticity in $E$, $x_1$ and $x_1'$ is
preserved. This holds always because the sign of a $x$-variable can
only change due to the $\delta_{12}\sim \delta(x_1+x_2)$ parts of the
contractions. Integrating out the right index of each contraction
$\gamma_{12}$ leads to the argument $-x_1$ for $x_2$. However, the
vertex $G$ containing the index $2$ stands to the right
of the resolvent which the contraction $\gamma_{12}$ crosses, see
Fig.~\ref{fig:analytic} for illustration. Therefore, the energy
arguments of this vertex are $G(E+x_1+\dots;\dots -x_1 \dots)$. This
shows that the structure of the energy arguments is generically
of the form shown in (\ref{vertex_analyticity}).

These considerations prove that the resolvents and vertices are
even analytic with respect to all the internal $x$-variables, e.g. the
$x_2$-variable in the example above. This will turn out to be quite useful
to discuss the frequency integrations over the internal variables.
A generic discussion of the frequency integrations is of course not
possible since the frequency dependence of the density of states
$\rho_\nu(\omega)$ and the initial vertices can differ for each model,
see e.g. the spin boson model discussed in Sec.~\ref{sec:2.2}. We 
discuss here the fermionic case with a flat density of states, i.e. the
frequency integrations are performed from minus to plus infinity and
we cut off the frequencies of all contractions by the Lorentzian function
(\ref{cutoff}) 
\begin{equation}
\label{dos_simple}
\rho(\omega)\,=\,{D^2\over \omega^2+D^2}\,=\,{D\over 2i}
({1\over \omega-iD}-{1\over \omega+iD})\quad.
\end{equation}
Furthermore, the initial vertices are assumed to be analytic functions
of the frequencies. 

{\bf Frequency integrations for the discrete step.} We start with the discussion of the 
frequency integrations for the first discrete RG step where the
symmetric part of the contraction is integrated out. Using the analytic
properties (\ref{resolvent_analytic}) of the resolvents, the integrations
can be performed analytically. When a symmetric contraction crosses over $n$ 
resolvents in the series (\ref{effective_antisym_series}), we get an
integral of the type ($z_i$ are variables with positive imaginary part)
\begin{equation}
\label{integral_symmetric}
\int d\bar{\omega}\,\rho(\bar{\omega})\,\prod_{i=1}^n\,
{1\over \bar{\omega}\,+z_i-L_S^a(\bar{\omega}+z_i)}
\,=\,\pi\,D\,\prod_{i=1}^n\,
{1\over iD\,+z_i-L_S^a(iD+z_i)}\quad,
\end{equation}
i.e. $\bar{\omega}$ is just replaced by $iD$ in all resolvents. This is a very
large imaginary part for each denominator and, therefore, this leads to a 
well-defined perturbation series which can be expanded systematically in the
vertices and in $1/D$. Therefore, it is not necessary to take the
full Liouvillian $L_S^a$ in the denominator and solve the equations self-consistently,
but one can replace it by the initial value $L_S$ and consider the additional
diagrams from s-irreducible insertions perturbatively.

\begin{figure}
  \centerline{\psfig{figure=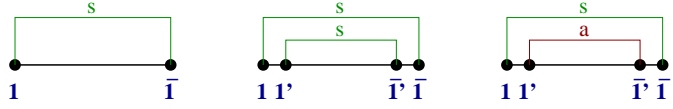,scale=0.3}}
  \caption{Examples for diagrams of the effective Liouvillian when the
symmetric part of the contraction is integrated out in one step. $s$ ($a$) denote
the symmetric (antisymmetric) contraction $\gamma^s$ ($\gamma^a$). Whereas the
first diagram is finite for $D\rightarrow\infty$, the other two diagrams contain
also terms $\sim D$.}
\label{fig:discrete_L}
\end{figure}

Examples of diagrams for the effective Liouvillian are shown in Fig.~\ref{fig:discrete_L}. 
The first diagram correponds to charge fluctuations and reads for fermions, 
using the notations (\ref{conventions}), (\ref{index_convention}) and (\ref{E_convention})
\begin{eqnarray}
\nonumber
L_S^a(E)\,&\rightarrow&\,p'\,\gamma_1^s\,G^p_1\,{1\over E_1\,+\,\bar{\omega}_1\,-\,L_S}
G^{p'}_{\bar{1}} \,=\, {1\over 2}\,\bar{G}_1\,\int d\bar{\omega}\,\rho(\bar{\omega})\,
{1\over \bar{\omega}\,+\,E_1\,-\,L_S}\,\tilde{G}_{\bar{1}} \quad \\
\label{L_initial_charge}
&=&\, {1\over 2}\,\bar{G}_1\,
{\pi D \over iD\,+\,E_1\,-\,L_S}\,\tilde{G}_{\bar{1}} \,=\,
-i\,{\pi \over 2}\,\bar{G}_1\,\tilde{G}_{\bar{1}}\,+\,O({1\over D})\quad.
\end{eqnarray}
Neglecting the contributions $\sim {1\over D}$,
we find a result independent of $D$, indicating that such models of charge
fluctuations (where the vertex $G_1$ has only one index) have a well-defined 
limit for $D\rightarrow\infty$ if all
charge states of the quantum system are taken into account. If the 
symmetric contraction crosses over more than one resolvent, the result
is $\sim {1\over D}$ and can be neglected. Thus, for such models, the
determination of the initial condition is very simple.

In contrast, the second and third diagrams of Fig.~\ref{fig:discrete_L} are associated with
a model where only spin and/or orbital fluctuations are considered, the Kondo
model being a special example. For these models, certain charge excitations 
have already been integrated out, so that a finite bandwidth $D$ has to be
introduced and the limit $D\rightarrow\infty$ is not allowed. In the second
diagram the following integral occurs
\begin{eqnarray}
\nonumber
\int d\bar{\omega}\,\rho(\bar{\omega})\,
\int d\bar{\omega}'\,\rho(\bar{\omega}')\,
{1\over \bar{\omega}\,+\,\bar{\omega}'\,+\,z}\,&=&\,
\int d\bar{\omega}\,\rho(\bar{\omega})\,
{\pi D\over \bar{\omega}\,+\,iD\,+\,z}\,=\,
{(\pi D)^2 \over z\,+\,2iD} \\
\label{expansion_D_1}
&&\hspace{-3cm}\,=\,-i\,{\pi^2 \over 2}\,{D \over 1\,-\,{iz\over 2D}}\,=\,
-i\,{\pi^2\over 2}\,D\,+\,{\pi^2\over 4}\,z\,+\,O({1\over D}) \quad,
\end{eqnarray}
which gives for the diagram the value
\begin{eqnarray}
\nonumber
L_S^a(E)\,&\rightarrow&\,{1\over 2}\,\gamma_1^s\,\gamma_{1'}^s\,G^{pp}_{11'}
\,{1\over E_{11'}\,+\,\bar{\omega}_1\,+\,\bar{\omega}_1^\prime\,-\,L_S}
G^{p'p'}_{\bar{1}'\bar{1}} \\
\nonumber
&=&\, {1\over 8}\,\bar{G}_{11'}\,\int d\bar{\omega}\,\rho(\bar{\omega})\,
\int d\bar{\omega}'\,\rho(\bar{\omega}')\,
{1\over \bar{\omega}\,+\,\bar{\omega}'\,+\,E_{11'}\,-\,L_S}\,
\bar{G}_{\bar{1}'\bar{1}} \quad \\
\label{L_initial_spin_ss}
&=&\,-i\,{\pi^2 \over 16}\,D\,\bar{G}_{11'}\,\bar{G}_{\bar{1}'\bar{1}}
\,+\,{\pi^2 \over 32}\,\bar{G}_{11'}\,(E_{11'}\,-\,L_S)\,
\bar{G}_{\bar{1}'\bar{1}}\,+\,O({1\over D})\quad.
\end{eqnarray}
As we can see also terms $\sim D$ are generated which are unphysical since one
can prove that the model should contain only logarithmic divergencies $\sim (\ln(D))^n$ in
higher-order perturbation theory. Such terms occur because the high temperature 
limit is not well-defined (only temperatures $T<<D$ are allowed since otherwise
charge excitations become important which are not considered in the model). As we will see
later for the Kondo model in Sec.~\ref{sec:5}, the terms $\sim D$ will be 
cancelled by corresponding terms generated in the second continuous RG flow.

Similiarly, also the antisymmetric part of the contraction can occur in the 
diagrams, see e.g. the third diagram of Fig.~\ref{fig:discrete_L}. It contains more
complicated integrals of the form (evaluated here at zero temperature but the same
result comes also out at finite $T\ll D$)
\begin{eqnarray}
\nonumber
\int d\bar{\omega}\,\rho(\bar{\omega})\,
\left[f_\alpha(\bar{\omega})-{1\over 2}\right]\,
\int d\bar{\omega}'\,\rho(\bar{\omega}')\,
{1\over \bar{\omega}\,+\,\bar{\omega}'\,+\,z}\,&=&\,
\int d\bar{\omega}\,\rho(\bar{\omega})\,
\left[f_\alpha(\bar{\omega})-{1\over 2}\right]\,
{\pi D\over \bar{\omega}\,+\,iD\,+\,z}\\
\nonumber
&&\hspace{-5cm} =\,{\pi D^3\over D^2\,+\,(z+iD)^2}\,\ln(1-{iz\over D})
\,=\,{\pi D\over 2}\,{1\over 1\,-\,{iz\over 2D}}\,\,{D\over iz}\,
\ln(1-{iz\over D})\\
\label{expansion_D_2}
&&\hspace{-5cm}=\,-{\pi\over 2}\,D\,-i\,{\pi\over 2}\,z
\,+\,O({1\over D}) \quad.
\end{eqnarray}
This leads to the following value for the third diagram of Fig.~\ref{fig:discrete_L}
\begin{eqnarray}
\nonumber
L_S^a(E)\,&\rightarrow&\,p'\,\gamma_1^s\,\gamma_{1'}^a\,G^{pp}_{11'}
\,{1\over E_{11'}\,+\,\bar{\omega}_1\,+\,\bar{\omega}_1^\prime\,-\,L_S}
G^{p'p'}_{\bar{1}'\bar{1}} \\
\nonumber
&=&\, {1\over 2}\,\bar{G}_{11'}\,\int d\bar{\omega}\,\rho(\bar{\omega})\,
\left[f_\alpha(\bar{\omega})-{1\over 2}\right]\,
\int d\bar{\omega}'\,\rho(\bar{\omega}')\,
{1\over \bar{\omega}\,+\,\bar{\omega}'\,+\,E_{11'}\,-\,L_S}\,
\tilde{G}_{\bar{1}'\bar{1}} \quad \\
\label{L_initial_spin_sa}
&=&\,-\,{\pi \over 4}\,D\,\bar{G}_{11'}\,\tilde{G}_{\bar{1}'\bar{1}}
\,-\,i\,{\pi\over 4}\,\bar{G}_{11'}\,(E_{11'}\,-\,L_S)\,\tilde{G}_{\bar{1}'\bar{1}}
\,+\,O({1\over D})\quad.
\end{eqnarray}
A similiar structure as compared to the other diagram (\ref{L_initial_spin_ss}) occurs. 
This illustrates that
the symmetric part can be integrated out using perturbation theory in the coupling
vertices. No logarithmic divergencies $\sim \ln(D/|E_{1\dots n}-L_S|)$
can occur in any order since the upper limit of all integrals and the cutoff provided in the
denominators of all resolvents is given by the bandwidth $D$. 
\begin{figure}
  \centerline{\psfig{figure=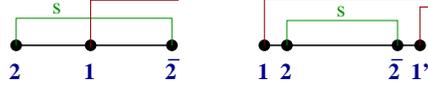,scale=0.3}}
  \caption{Examples for diagrams of the effective vertices when the
symmetric part of the contraction is integrated out in one step. $s$ ($a$) denote
the symmetric (antisymmetric) contraction $\gamma^s$ ($\gamma^a$). Whereas the
left diagram is negligible for $D\rightarrow\infty$, the right diagram is finite.}
\label{fig:discrete_G}
\end{figure}

Examples for diagrams for the determination of the effective vertices are shown in
Fig.~\ref{fig:discrete_G}. The first diagram corresponds to charge fluctuations. It is
of order $O({1\over D})$ and can be neglected
\begin{eqnarray}
\label{G_initial_charge_s}
\bar{G}_{1}(E)\,&\rightarrow&\,
p_2^\prime\,\gamma^s_2\,G^{p_2}_{2}\,
{1\over E_{2}\,+\,\bar{\omega}_2\,-\,L_S}\, G^{p}_{1}\,
{1\over E_{12}\,+\,\bar{\omega}_1\,+\,\bar{\omega}_2\,-\,L_S}\, G^{p_2^\prime}_{\bar{2}}\\
\nonumber
&=&\,{1\over 2}\,\int d\bar{\omega}_2\,\rho(\bar{\omega}_2)\,\bar{G}_{2}\,
{1\over \bar{\omega}_2\,+\,E_{2}\,-\,L_S}\, \bar{G}_{1}\,
{1\over \bar{\omega}_2\,+\,E_{12}\,+\,\bar{\omega}_1\,-\,L_S}\,\tilde{G}_{\bar{2}}
\,\sim\,O({1\over D})\quad.
\end{eqnarray}
The second diagram corresponds to spin/orbital fluctuations and is finite in the 
limit $D\rightarrow\infty$
\begin{eqnarray}
\nonumber
\bar{G}_{11'}(E)\,&\rightarrow&\,p'\,\gamma^s_2\,G^{pp}_{12}\,
{1\over E_{12}\,+\,\bar{\omega}_1\,+\,\bar{\omega}_2\,-\,L_S}\, G^{p'p'}_{\bar{2}1'}
\,-\,(1\leftrightarrow 1')\\
&=&\,{1\over 2}\,\bar{G}_{12}\,
\nonumber
\int d\bar{\omega}_2\,\rho(\bar{\omega}_2)\,
{1\over \bar{\omega}_2\,+\,E_{12}\,+\,\bar{\omega}_1\,-\,L_S}\, \tilde{G}_{\bar{2}1'}
\,-\,(1\leftrightarrow 1')\\
\nonumber
&=&\,{1\over 2}\,\bar{G}_{12}\,
{\pi D\over iD\,+\,E_{12}\,+\,\bar{\omega}_1\,-\,L_S}\, \tilde{G}_{\bar{2}1'}
\,-\,(1\leftrightarrow 1')\\
\label{G_initial_spin_s}
&=&\,-i\,{\pi\over 2}\,\left(\bar{G}_{12}\,\tilde{G}_{\bar{2}1'}\,-\,
\bar{G}_{1'2}\,\tilde{G}_{\bar{2}1}\right)\,+\,O({1\over D})\quad.
\end{eqnarray}

In conclusion, we see that for initial vertices with one index (charge fluctuations), the
effective Liouvillian is finite for $D\rightarrow\infty$ and the effective vertices
are identical to the initial ones. For initial vertices with two indices (spin and/or orbital
fluctuations), the effective Liouvillian contains also terms $\sim D$ and the effective
vertices get a finite correction to the initial ones. 
Initial vertices with three indices correspond to quite artificial models, in that case
also terms $\sim D^2$ can occur for the effective Liouvillian and terms
$\sim D$ for the effective vertices.

Furthermore, we note that the effective Liouvillian can contain terms which are linear in the
energy $E$, see (\ref{L_initial_spin_ss}) and (\ref{L_initial_spin_sa}). This does
not lead to a problem for arbitrarily large frequencies because the effective Liouvillian
occurs always in the denominator of all resolvents of the effective perturbation series 
or the r.h.s. of the RG equations. In contrast, the effective vertices 
are independent of frequency, see (\ref{G_initial_spin_s}), which is important for a 
well-defined large frequency behaviour of all expressions.

{\bf Frequency integrations for the continuous RG equations.} We now turn to the frequency integrations of the 
RG equations (\ref{rg_equations}) of the continuous RG flow. From the form of
the contraction (\ref{matsubara_antisym}), we see that by closing the integrations
over $\bar{\omega}$ in the upper half of the complex plane, only the pole from the
Matsubara frequency $\omega_n^\alpha$ and the pole of the Lorentzian cutoff function
(\ref{dos_simple}) at $iD$ contributes (we again replace $\rho_\nu(\omega)$ by
(\ref{dos_simple}) in all contractions). 
All other resolvents and vertices are analytic in the upper
half plane, according to (\ref{resolvent_analytic}) and (\ref{vertex_analyticity}).
Most importantly, the contraction ${d\gamma\over d\Lambda}$, which is differentiated
with respect to $\Lambda$, crosses over all resolvents (to make the expression
${d\gamma\over d\Lambda}$-irreducible), i.e. the frequency variable of that contraction
occurs in {\it all} resolvents! Moreover, due to the derivative, we get for this
contraction from (\ref{matsubara_antisym})
\begin{eqnarray}
\label{matsubara_antisym_derivative_sum}
{d\gamma_1^\Lambda \over d\Lambda}\,&=&\,
\left\{
\begin{array}{cl}
\eta \\ -1
\end{array}
\right\}\,
\,\rho(\bar{\omega})\,T\,\sum_n\,{1\over \bar{\omega}\,-\,i\omega_n^\alpha} 
\,\,\delta_{T_\alpha}(\Lambda-|\omega_n^\alpha|)\\
\label{matsubara_antisym_derivative}
&=&\,\left\{
\begin{array}{cl}
\eta \\ -1
\end{array}
\right\}\,
\,\rho(\bar{\omega})\,{1\over 2\pi}\,
\,\left({1\over \bar{\omega}\,-\,i\Lambda_{T_\alpha}}\,+\,
{1\over \bar{\omega}\,+\,i\Lambda_{T_\alpha}}\right) \quad, 
\end{eqnarray}
where $\delta_T(\omega)$ is a delta function smeared by temperature, defined
as the derivative of $\theta_T(\omega)$, given by (\ref{theta_T}), i.e.
\begin{equation}
\label{delta_T}
\delta_T(\omega)\,=\,{d\over d\omega}\,\theta_T(\omega)\,=\,
\left\{
\begin{array}{cl}
0 &\quad\mbox{for }|\omega|\,>\,\pi T \\
{1\over 2\pi T} &\quad\mbox{for }|\omega|\,<\,\pi T
\end{array}
\right.\quad.
\end{equation}
As a consequence, only the two Matsubara frequencies $\pm\Lambda_{T_\alpha}$ lying 
closest to $\pm\Lambda$ contribute to the sum in Eq.~(\ref{matsubara_antisym_derivative_sum}),
and we obtain exactly Eq.~(\ref{matsubara_antisym_derivative}). Therefore, we get
the important result that, by closing the frequency integration of
${d\gamma\over d\Lambda}$ in the upper half plane, the contribution from the Matsubara
poles gives rise to the replacement 
\begin{equation}
\label{lambda_T_occurence}
\bar{\omega}\,\rightarrow\,i\Lambda_{T_\alpha} \quad 
\mbox{in {\it all} resolvents of the r.h.s. of the RG equation (\ref{rg_equations}).}
\end{equation}
Therefore, a large positive imaginary part occurs in all resolvents so that the
perturbative series on the r.h.s. of the RG equations is well-defined provided that
the coupling constants remain small.
In contrast, the Matsubara poles from all the other contractions occur only in those
resolvents where the contractions cross over.

The pole at $iD$ from the density of states is only important if one wants to study the
precise behaviour of the RG flow when $\Lambda$ crosses the bandwidth $D$. However,
nothing really interesting is happening in that regime, except for the fact that $D$
sets the scale where the divergencies of the original perturbation theory are cut off
at high energies, i.e. at $\Lambda \sim D$ the RG flow starts to renormalize the 
effective parameters significantly. To see this analytically, we perform the integration
over the frequency variable $\bar{\omega}$ of the ${d\gamma\over d\Lambda}$ contraction analytically.
Using (\ref{matsubara_antisym_derivative}) with (\ref{dos_simple}) we obtain by
closing in the upper half of the complex plane
\begin{eqnarray}
\nonumber
\int d\bar{\omega}\,\rho(\bar{\omega})\,{1\over 2\pi}\,
({1\over \bar{\omega}\,-\,i\Lambda_{T_\alpha}}\,+\,
{1\over \bar{\omega}\,+\,i\Lambda_{T_\alpha}})\,
\left\{\dots\right\}_{\bar{\omega}}\,=\\
\label{integration_deriv_gamma}
&&\hspace{-5cm} =\,-i\,{D^2\over \Lambda_{T_\alpha}^2-D^2}\,
\left[\left\{\dots\right\}_{\bar{\omega}=i\Lambda_{T_\alpha}}\,-\,
\left\{\dots\right\}_{\bar{\omega}=iD}\right]\quad,
\end{eqnarray}
where $\{\dots\}_{\bar{\omega}}$ stands for the rest of the diagram. We see that for
$\Lambda\gg D$, there is small prefactor $\sim ({D\over\Lambda})^2\ll 1$ in front of
the r.h.s. of the RG equations leading to a negligible renormalization. In contrast,
for $\Lambda\ll D$, the prefactor is $i$ and the renormalization becomes important.
In this regime, only the term $\{\dots\}_{\bar{\omega}=i\Lambda_{T_\alpha}}$ has to
be considered on the r.h.s. of (\ref{integration_deriv_gamma}) and the r.h.s. of the
RG equation becomes independent of $D$.

In summary, the RG flow will roughly start at $\Lambda\sim D$ but the precise value where
it starts is not so important since the cutoff $D$ occurs logarithmically in the divergencies
of perturbation theory. Therefore, it is a good approximation to evaluate the r.h.s. of the 
RG equations for $\Lambda\ll D$, and use $D$ just as the starting value of the flow parameter $\Lambda$,
where the initial values of the Liouvillian and the vertices are taken from the first
discrete RG step. Moreover, as we will see in more detail in connection with the application
to the Kondo problem in Sec.~\ref{sec:5}, the precise ratio of the initial value of $\Lambda$ 
and the bandwidth $D$ should be chosen in such a way that the terms linear in $D$ of the initial 
Liouvillian cancel out (as it should be since such terms do not exist in the original 
perturbation theory).

As a consequence we consider from now on only the regime $\Lambda\ll D$ and omit the
cutoff function $\rho(\bar{\omega})$ in all contractions. We then close the integrations
over all variables $\bar{\omega}_i$ in the upper half plane and replace them by the
pole of the Matsubara frequencies in all resolvents and vertices of the diagram. 
As a result we find the same form of the RG equations but with the replacements
\begin{equation}
\label{replace_contraction}
\gamma_1^\Lambda\,,\,{d\gamma_1^\Lambda\over d\Lambda}\rightarrow\,i\,
\left\{
\begin{array}{cl}
\eta \\ -1
\end{array}
\right\}
\end{equation}
for all contractions, together with
\begin{equation}
\label{frequency_replacement_contraction}
\bar{\omega}\,\rightarrow\,i\omega_n^\alpha \quad,\quad
\int d\bar{\omega}\,\rightarrow\,2\pi T_\alpha\sum_{0<\omega_n^\alpha}
\theta_{T_\alpha}(\Lambda-\omega_n^\alpha) 
\end{equation}
for the frequency variables of all contractions $\gamma^\Lambda_1$, and
\begin{equation}
\label{frequency_replacement_contraction_derivative}
\bar{\omega}\,\rightarrow\,i\Lambda_{T_\alpha}
\end{equation}
for the frequency variable of the contraction ${d\gamma^\Lambda_1\over d\Lambda}$.
Note that only positive Matsubara frequencies are allowed in the sum.

{\bf RG on Matsubara axis.} 
When the integrations over all real frequencies are performed,
the energy variables $E+X_i$ of the RG equations (\ref{rg_equations}) get the form
\begin{equation}
\label{E_X_matsubara}
E\,+\,X_i\,\rightarrow\,E\,+\,\bar{\mu}_{1\dots n}\,+\,
i\Lambda_{T_{\alpha_1}}\,+\,i\omega_{2\dots n}
\quad,
\end{equation}
where index $1$ is assumed to be the variable of ${d\gamma^\Lambda_1\over d\Lambda}$, and
\begin{equation}
\label{matsubara_frequency_notation}
\omega_{1\dots k}\,=\,\omega_{n_1}^{\alpha_1}\,+\dots+\,\omega_{n_k}^{\alpha_k}
\end{equation}
is the sum over the Matsubara frequencies in analogy to the notation (\ref{index_convention}).
We see that the value for the energies $E$ appearing in the Liouvillian $L_S(E)$ and the
vertices $\bar{G}_{1\dots n}(E)$ are complex numbers with the imaginary
part being some positive Matsubara frequency. Thus, we can reformulate the whole
RG equations for frequencies on the Matsubara axis by replacing all variables
$\bar{\omega}$ by Matsubara frequencies and using for $E$ a real part plus
some Matsubara frequency. With this, only the following quantities occur in
the RG equations
\begin{eqnarray}
\nonumber
L_S(E;\omega)\,&=&\,L_S(E+i\omega)\quad,\\
\label{LG_matsubara_representation}
\bar{G}_{1\dots n}(E,\omega,\omega_1\dots\omega_n)&\equiv&
\bar{G}_{\eta_1\nu_1\dots\eta_n\nu_n}(E,\omega,\omega_1\dots\omega_n)\,=\,
\bar{G}_{1\dots n}(E+i\omega)|_{\bar{\omega}_i\rightarrow i\omega_i}\quad,
\end{eqnarray}
where $\omega,\omega_1\dots\omega_n$ are positive Matsubara frequencies and $E$ is real.
By convention, if the frequencies are written explicitly in the arguments, the index
$1\equiv\eta\nu$ does no longer contain the frequency. The RG equations in Matsubara
space for $L_S(E,\omega)$ and $\bar{G}_{\dots}(E,\omega,\dots)$  are obtained from
the original RG equations by the replacements
\begin{eqnarray}
\label{replacements_rg_matsubara_contraction}
\gamma_1^\Lambda\,,\,{d\gamma_1^\Lambda\over d\Lambda}&\rightarrow&\,i\,
\left\{
\begin{array}{cl}
\eta \\ -1
\end{array}
\right\}\quad,\\
\label{replacements_rg_matsubara_contraction_pi}
\Pi_{1\dots n}\,&\rightarrow&\,\Pi(E_{1\dots n},\omega+\omega_{1\dots n})\quad,\\
\label{replacements_rg_matsubara_contraction_G}
\bar{G}_{1\dots n}(E+x_{1'\dots n'})\,&\rightarrow&\,
\bar{G}_{1\dots n}(E_{1'\dots n'},\omega+\omega_{1'\dots n'},\omega_1\dots\omega_n)
\quad,\\
\label{replacements_rg_matsubara_lambda}
\omega_1\,&\rightarrow&\,\Lambda_{T_{\alpha_1}}\quad\mbox{for }{d\gamma_1\over d\Lambda}
\quad,\\
\label{replacements_rg_matsubara_frequencies}
\int d\omega\,&\rightarrow&\,2\pi {T_{\alpha}}\,\sum_{0<\omega^\alpha_n}
\theta_{T_\alpha}(\Lambda-\omega_n^\alpha)\,
\quad\mbox{for }\gamma_1
\quad,
\end{eqnarray}
where all frequencies correspond to Matsubara frequencies 
$\omega_i\equiv\omega_{n_i}^{\alpha_i}$, and we have used the convention 
$E_{1\dots n}=E+\bar{\mu}_{1\dots n}$ together with the definition
\begin{equation}
\label{pi_matsubara}
\Pi(E,\omega)\,=\,{1\over E\,+\,i\omega\,-\,L_S(E,\omega)}\quad.
\end{equation}
Furthermore, the explicit frequency arguments of the resolvent and the vertices in Matsubara space
can be avoided by the compact notation
\begin{eqnarray}
\label{G_M}
\bar{G}^M_{1\dots n}(E_{1'\dots n'})\,&\equiv&\,\bar{G}_{1\dots n}
(E_{1'\dots n'},\omega+\omega_{1'\dots n'},\omega_1\dots\omega_n)\quad,\\
\label{pi_M}
\pi^M_{1\dots n}\,&\equiv&\,\Pi(E_{1\dots n},\omega+\omega_{1\dots n})\quad,
\end{eqnarray}
which is especially useful for writting down higher-order terms of the RG equations.
In this notation the Matsubara RG equations have nearly the same form as the original
RG equations besides the replacements (\ref{replacements_rg_matsubara_contraction})
for the contractions. However, one should note that the external frequency variable 
$\omega$ is hidden in this notation and one has to specify explicitly which variable 
is set on the parameter $\Lambda_{T_\alpha}$.

As an example consider the RG equations (\ref{rg_L_charge_1})-(\ref{rg_G_charge}) for
a model with charge fluctuations. Using the above rules in the representation
(\ref{LG_matsubara_representation}), we get immediately
\begin{eqnarray}
\label{rg_L_charge_1_matsubara}
-\,{dL_S(E,\omega)\over d\Lambda}\,&=&\,
-i\,\bar{G}_{1}(E,\omega,\Lambda_{T_{\alpha_1}})\,\Pi(E_1,\omega+\Lambda_{T_{\alpha_1}})\,
\bar{G}_{\bar{1}}(E_1,\omega+\Lambda_{T_{\alpha_1}},-\Lambda_{T_{\alpha_1}})\\
\nonumber
&&\hspace{-2.5cm} - \,\bar{G}_{1}(E,\omega,\Lambda_{T_{\alpha_1}})\,\Pi(E_1,\omega+\Lambda_{T_{\alpha_1}})\,
\bar{G}_2(E_1,\omega+\Lambda_{T_{\alpha_1}},\omega_2)\,
\Pi(E_{12},\omega+\Lambda_{T_{\alpha_1}}+\omega_2)\,\cdot\\
\label{rg_L_charge_2_matsubara}
&&\hspace{-1.5cm}\cdot\,\bar{G}_{\bar{2}}(E_{12},\omega+\Lambda_{T_{\alpha_1}}+\omega_2,-\omega_2)\,
\Pi(E_1,\omega+\Lambda_{T_{\alpha_1}})\,
\bar{G}_{\bar{1}}(E_1,\omega+\Lambda_{T_{\alpha_1}},-\Lambda_{T_{\alpha_1}})
\quad,\\
\nonumber
-\,{d\bar{G}_{1}(E,\omega,\omega_1)\over d\Lambda}\,&=&\,
i\,\bar{G}_{2}(E,\omega,\Lambda_{T_{\alpha_2}})\,\Pi(E_{2},\omega+\Lambda_{T_{\alpha_2}})\,
\bar{G}_1(E_2,\omega+\Lambda_{T_{\alpha_2}},\omega_1)\,\cdot \\
\label{rg_G_charge_matsubara}
&&\cdot\,\Pi(E_{12},\omega+\omega_1+\Lambda_{T_{\alpha_2}})\,
\bar{G}_{\bar{2}}(E_{12},\omega+\omega_1+\Lambda_{T_{\alpha_2}},-\Lambda_{T_{\alpha_2}})
\quad.
\end{eqnarray}

Using the compact notation (\ref{G_M}) and (\ref{pi_M}), the RG equations 
(\ref{rg_L_so_1})-(\ref{rg_G_so_4}) for a model with arbitrary spin and/or orbital 
fluctuations read for fermions
\begin{eqnarray}
\label{rg_L_so_1_matsubara}
-\,{dL_S(E,\omega)\over d\Lambda}\,&=&\,
-\,\bar{G}^M_{12}(E)\,\pi^M_{12}\,\bar{G}^M_{\bar{2}\bar{1}}(E_{12})
\,|_{\omega_1=\Lambda_{T_{\alpha_1}}}\\
\label{rg_L_so_2_matsubara}
&& + \,i
\,\bar{G}^M_{12}(E)\,\pi^M_{12}\,\bar{G}^M_{\bar{2}3}(E_{12})\,
\pi^M_{13}\,\bar{G}^M_{\bar{3}\bar{1}}(E_{13})
\,|_{\omega_1=\Lambda_{T_{\alpha_1}}}\quad, \\
\label{rg_G_so_1_matsubara}
-\,{d\bar{G}_{11'}(E,\omega,\omega_1,\omega_1^\prime)\over d\Lambda}\,&=&\,
-i\,\left\{\bar{G}^M_{12}(E)\,\pi^M_{12}\,\bar{G}^M_{\bar{2}1'}(E_{12})
\,\, - \,\, (1\leftrightarrow 1')\right\}
\,|_{\omega_2=\Lambda_{T_{\alpha_2}}} \\
\label{rg_G_so_2_matsubara}
&& \hspace{-1.5cm} -  
\,\bar{G}^M_{23}(E)\,\pi^M_{23}\,\bar{G}^M_{11'}(E_{23})\,
\pi^M_{11'23}\,\bar{G}^M_{\bar{3}\bar{2}}(E_{11'23})
\,|_{\omega_2=\Lambda_{T_{\alpha_2}}}  \\
\label{rg_G_so_3_matsubara}
&& \hspace{-1.5cm} + \,\left\{ 
\,\bar{G}^M_{12}(E)\,\pi^M_{12}\,\bar{G}^M_{1'3}(E_{12})\,
\pi^M_{11'23}\,\bar{G}^M_{\bar{3}\bar{2}}(E_{11'23})
\,\, - \,\,  (1\leftrightarrow 1')\right\} 
\,|_{\omega_2=\Lambda_{T_{\alpha_2}}}  \\
\label{rg_G_so_4_matsubara}
&& \hspace{-1.5cm} - \, \left\{
\,\bar{G}^M_{23}(E)\,\pi^M_{23}\,\bar{G}^M_{\bar{3}1}(E_{23})\,
\pi^M_{12}\,\bar{G}^M_{1'\bar{2}}(E_{12})
\,\, - \,\, (1\leftrightarrow 1')\right\}
\,|_{\omega_2=\Lambda_{T_{\alpha_2}}}  \quad.
\end{eqnarray}

For the final physical quantities, we need according to (\ref{final_result})
\begin{equation}
\label{final_result_matsubara}
L_S^{eff}(E)\,=\,L^{\Lambda=0}_S(E,\omega=0)\quad,\quad
\Sigma_R(E)\,=\,R^{\Lambda=0}_{n=0}(E,\omega=0)\quad,
\end{equation}
i.e. we can fix $E$ in the RG equations and see that only $E$ shifted by the 
discrete set of values $\bar{\mu}_{1\dots n}$ occurs for the first frequency variable.
This means that the RG equations are local in Laplace space up to a shift by multiples
of the chemical potentials of the reservoirs. This simplifies the calculation of the
time evolution considerably.
E.g. if only two reservoirs with $\mu_L=-\mu_R={V\over 2}$ are present, we get $n{V\over 2}$ with
$n=0,\pm 1,\pm 2, \dots$ for $\bar{\mu}_{1\dots n}$, i.e. only the fixed Laplace variable
shifted by multiples of half the
bias voltage can occur for the first frequency variable of the Liouvillian and
the vertices. 

In summary, we find that the RG can be formulated on the Matsubara axis but in contrast
to equilibrium Matsubara formalism, there is a whole set of Matsubara axis shifted by
the real quantities
\begin{equation}
\label{shift_matsubara}
E\,+\,n_1\mu_1\,+\dots +\,n_Z\,\mu_Z \quad,\quad n_i=0,\pm 1,\pm 2,\dots 
\end{equation}
where $Z$ is the number of reservoirs, $\mu_k$ is the chemical potential of reservoir $k$,
and $E$ is the real part of the Laplace variable. This is the price one has to pay 
compared to equilibrium
to calculate the time evolution and the effect from different chemical potentials. 
If different temperatures of the reservoirs occur (leading to nonequilibrium
heat currents), the Matsubara frequency axis has to be defined for each reservoir 
individually.

Finally, we note that within our formalism no analytic continuation to real frequencies
is necessary to calculate the time evolution of the average
over an arbitrary observable (the same holds for correlation functions not shown in this
paper). This is an advantage compared to the usual linear response formalism \cite{kubo}
or to recently developed nonequilibrium formalism with complex chemical potentials \cite{QMC},
where an analytic continuation is necessary. In this way we can avoid the numerical problems
associated with analytic continuations.

{\bf Cutoff parameters for the RG flow.}
Using the form (\ref{E_X_matsubara}) together with the representation 
(\ref{LG_matsubara_representation}), and
replacing the effective Liouvillian by its eigenvalue via (\ref{rg_eigenvalue}), we find 
the following structure of each resolvent of the RG equation
\begin{equation}
\label{resolvent_structure}
{1\over E\,+\,\bar{\mu}_{1\dots n}\,-\,h_k^\Lambda\,+\,i\Lambda_{T_{\alpha_1}}\,+\,i\omega\,+\,
i\omega_{2\dots n}\,+\,i\Gamma_k^\Lambda} \quad,
\end{equation}
where $\omega$ is the Matsubara frequency associated with the energy variable $E$. 
The energy arguments of the real and imaginary parts $h_k^\Lambda,\Gamma_k^\Lambda$ of the 
eigenvalues of the effective Liouvillian at scale $\Lambda$ are not indicated (which is the
same energy as occuring in the rest of the resolvent). 
As already mentioned in Sec.~\ref{sec:4.1}, this form of the
resolvents is very useful for stable numerical calculations since all imaginary parts in the
denominator are positive. Furthermore, it shows generically what the 
various cutoff parameters of the RG flow are. If the flow parameter $\Lambda$ falls 
roughly below the value ($E$ is assumed to be real here)
\begin{equation}
\label{rg_cutoff}
\Lambda\,<\,\mbox{max}\left\{T_\alpha\,,\,
|E+\bar{\mu}_{1\dots n}-h_k^\Lambda|\,,\,\omega\,,\,\omega_2\,,\,\dots\,,\,\omega_n
\,,\,\Gamma_k^\Lambda\right\} \quad,
\end{equation}
the resolvent (\ref{resolvent_structure}) becomes approximately a constant, cutting off
logarithmic divergencies for low energies. In this way the temperatures $T_\alpha$ of
the reservoirs, the Laplace variable $E$, the chemical potentials $\mu_\alpha$ of the 
reservoirs, the oscillation frequencies $h_k$, the Matsubara frequencies 
$\omega_n^\alpha$, and the relaxation and dephasing
rates $\Gamma_k$ serve as cutoff parameters. The cutoff occurs in a natural way and in an
intuitively expected form in the RG equations. However, in contrast to temperature and
the imaginary parts of the resolvent, one should note that only the
combination $E+\bar{\mu}_{1\dots n}-h_k^\Lambda$ is the cutoff parameter in the
real part of the denominator. Therefore, there are interesting points
when this combination is zero
\begin{equation}
\label{resonances}
E\,+\,\bar{\mu}_{1\dots n}\,-\,h_k\,=\,0\quad,
\end{equation}
where enhanced renormalizations are expected. These points correspond physically to
resonances. E.g. for the Kondo model they lead to an additional logarithmic enhancement 
of the differential conductance when the external magnetic field is identical to the 
bias voltage, see e.g. Ref.~\cite{reininghaus_hs_preprint} for a detailed study of this problem 
using RTRG-FS or 
Refs.~\cite{rosch_paaske_kroha_woelfle_PRL03,paaske_rosch_woelfle_PRB04,glazman_pustilnik_05}
where other methods have been used). A nice feature of the RTRG-FS procedure presented here is
that, without specifying the model at all, we can generically predict that logarithmic
enhancements are expected when the condition (\ref{resonances}) is fulfilled.

In contrast, temperature is always a cutoff scale because when $\Lambda$ falls below 
$T_\alpha$, the last Matsubara frequency has been integrated out and the RG flow stops. 
Compared to equilibrium, this means
for nonequilibrium problems that the voltage (inducing nonequilibrium stationary states)
and the Laplace variable $E$ (determining the time evolution of an initially out of equilibrium
state) induce different cutoff behaviours of the RG equations than temperature. All 
resolvents on the r.h.s. of the RG equations have a different structure concerning the
combination of energy variables $E+\bar{\mu}_{1\dots n}-h_k$. In equilibrium,
only the eigenvalues $h_k$ can occur but they can not be cancelled by other energy
variables. This opens up many new interesting physical phenomena for nonequilibrium 
systems to be studied in the future.

As temperature, also the Matsubara frequencies and the relaxation and dephasing rates
will cut off the RG flow and can not be cancelled in the denominator since they are
all positive. Most importantly, as has already been discussed in all detail in Sec.~\ref{sec:4.1},
the zero eigenvalue of the effective Liouvillian can not occur in the resolvents. This means
that we have found a generic proof that relaxation and dephasing rates will always cut off
the RG flow, irrespective of the truncation scheme used and of the specific model under 
consideration.

\subsection{Weak coupling limit}
\label{sec:4.4}

{\bf Definition of weak coupling.}
We start with the definition
of weak coupling by defining the dimensionless \lq\lq coupling constant'' $J_\Lambda$ at
scale $\Lambda$
\begin{equation}
\label{coupling_dimensionless}
J_\Lambda\,\equiv\,\Lambda^{{n\over 2}-1}\,G^\Lambda_{1\dots n} \quad,
\end{equation}
where, from now on, we use the definition (\ref{continuum}) for the field operators
such that $a_\nu(\omega)$ has dimension ${1\over \sqrt{E}}$.
The connection between the RG equations for $J_\Lambda$ and $G^\Lambda$ is
\begin{equation}
\label{rg_J}
{dJ_\Lambda\over dl}\,=\,(1-{n\over 2})\,J_\Lambda\,-\,\Lambda^{n\over 2}\,{dG^\Lambda\over d\Lambda}
\quad,\quad l\,=\,\ln({D\over\Lambda})\quad,
\end{equation}
where $l$ is a dimensionless flow parameter running from $l=0$ at $\Lambda=D$ to $l=\infty$
at $\Lambda=0$. Here, the first term describes the trivial renormalization of $J_\Lambda$ by the rescaling
factor in (\ref{coupling_dimensionless}), whereas the second one stems from the renormalization of 
$G^\Lambda$. If we rescale all energies in the RG equation for $G^\Lambda$ by $\Lambda$, the
second term on the r.h.s. of (\ref{rg_J}) is a power series in $J_\Lambda$ starting
at $J_\Lambda^2$ for $n>1$ and at $J^3_\Lambda$ for $n=1$, 
see the RG diagrams of Fig.~\ref{fig:LG_charge} for $n=1$ and
Fig.~\ref{fig:G_kondo} for $n=2$. For a term $\sim J^k_\Lambda$, there are $k-1$ resolvents in between, so
we estimate the term in this order as
\begin{equation}
\label{rg_estimate_order_k}
\Lambda^{n\over 2}\,{dG^\Lambda\over d\Lambda}\,\rightarrow\,
J_\Lambda^k\,\left\{
\begin{array}{cl}
({\Lambda\over\Delta})^{k-1}\quad &\mbox{for }\Lambda\ll\Delta \\
O(1)\quad &\mbox{for }\Lambda\gg\Delta
\end{array}
\right.\quad,
\end{equation}
where $\Delta$ stands symbolically for all the physical scales occuring in the denominator
of the resolvents according to (\ref{resolvent_structure}), i.e.
\begin{equation}
\label{cutoff_possibilities}
\Delta\,\sim\,T\,,\,E+\bar{\mu}_{1\dots n}-h\,,\,\Gamma \quad,
\end{equation}
except for the frequencies which are rescaled by $\Lambda$ and will give something of order
$O(1)$ in the denominator of the resolvent (note that all frequency integrations are cut off
by $\Lambda$, they all give a positive contribution to the imaginary part of the denominator,
and $i\Lambda$ occurs in all resolvents).

Weak coupling means that the expansion (\ref{rg_estimate_order_k}) of the expression 
$\Lambda^{n\over 2}\,{dG^\Lambda\over d\Lambda}$ in a power series in $J_\Lambda$ is under control for
all values of the flow parameter, which is the case if the condition
\begin{equation}
\label{weak_coupling_condition}
J_\Lambda \ll 1 \quad \mbox{or} \quad \Lambda\ll\Delta
\end{equation}
is always fulfilled. Only in this case truncation schemes on the r.h.s. of
the RG equations are justified, leading to a well-defined renormalized 
perturbation series in $J$. We discuss this condition now for the possible values of $n$
in (\ref{rg_J}), where $n$ is the number of field operators of the considered vertex 
$G^\Lambda_{1\dots n}$.

For $n>2$, the weak-coupling condition is certainly not a problem
since already the first term of (\ref{rg_J}) leads to a reduction of $J$.
Therefore these terms are called irrelevant in the renormalization group sense.
 
For $n=2$ (e.g. spin and/or orbital fluctuations) we have $J_\Lambda=G^\Lambda$ and the first
term on the r.h.s. of (\ref{rg_J}) is zero. By convention, this is called the marginal case. The
weak-coupling condition (\ref{weak_coupling_condition}) will be fulfilled if the renormalized
coupling $J_\Lambda$ stays small until $\Lambda$ reaches the cutoff scale $\Delta$. Under this
condition, $J_\Lambda$ will also stay small for $\Lambda < \Delta$, since an additional factor 
$\sim ({\Lambda\over\Delta})^{k-1}$ appears in this regime, see (\ref{rg_estimate_order_k}). Since
the smallest cutoff scale is the minimum of all relaxation and dephasing rates of the problem,
we obtain the condition (\ref{weak_coupling_criterion})
\begin{equation}
\label{weak_coupling_criterion_n=2}
\min_i\,\left\{\Gamma_i\right\}\,\gg\,T_K\quad \Rightarrow \quad
\mbox{weak coupling} \quad,
\end{equation}
where $T_K$ is defined as the energy scale where the coupling $J_{\Lambda=T_K}\sim O(1)$ if
all physical cutoff scales $\Delta$ are disregarded.

For $n=1$ (e.g. charge fluctuations) the first term on the r.h.s. of (\ref{rg_J}) leads to an 
increase of $J$, therefore this situation is called the relevant case. Due to the relations
(\ref{coupling_tunneling}) and (\ref{gamma}), we expect $G^\Lambda\sim\sqrt{\Gamma^\Lambda}$
and we obtain
\begin{equation}
\label{J_n=1}
J_\Lambda\,=\,{1\over \sqrt{\Lambda}}\,G^\Lambda\,\sim\,\sqrt{\Gamma^\Lambda/ \Lambda}\quad.
\end{equation}
When $\Lambda$ reaches the physical relaxation or dephasing rate $\Gamma$, we expect
$\Gamma^\Lambda\sim\Gamma$, and thus
\begin{equation}
\label{J_at_scale_Gamma_n=1}
J_{\Lambda=\Gamma} \sim \sqrt{\Gamma/\Gamma} \sim O(1)\quad.
\end{equation}
Therefore, already before $\Lambda$ reaches $\Gamma$, it may happen that $J$ becomes of order
one and a truncation of the series is not strictly justified. For $\Lambda<\Gamma$, 
we expect $\Gamma^\Lambda$ to saturate to $\Gamma$ so that $J_\Lambda\sim\sqrt{\Gamma/ \Lambda}$
with $\Gamma$ being independent of $\Lambda$. Therefore, in this regime, we estimate 
(\ref{rg_estimate_order_k}) to be
\begin{equation}
\label{estimate_n=1}
J_\Lambda^k\,({\Lambda/\Gamma})^{k-1}\,\sim\,
\left(\sqrt{\Gamma/ \Lambda}\right)^k\,({\Lambda/\Gamma})^{k-1}\sim 
\left(\sqrt{\Lambda/\Gamma}\right)^{k-2}
\quad,
\end{equation}
so that, for $k>2$, we expect the terms to become small again. Since, for $n=1$, at least three 
vertices are necessary for a renormalization of the vertex, see Fig.~\ref{fig:LG_charge}, the
condition $k>2$ is fulfilled. As a consequence, there is only a small region around
$\Lambda\sim\Gamma$, where it is not clear whether truncation schemes are justified. Moreover,
$\Gamma$ is not the only possibility for the cutoff scale $\Delta$. 
Each resolvent has its own cutoff scale and, for $n=1$, the eigenvalue $h$ alternates from
spin/orbital-excitation energies to charge excitation energies for successive resolvents.
Therefore, for $\Lambda\sim\Gamma$, we expect further suppression factors
$\sim{\Lambda/\Delta}\sim{\Gamma/\Delta}\ll 1$ in many terms of the RG equations, provided
that
\begin{equation}
\label{weak_coupling_criterion_n=1}
\Gamma\,\ll \, \Delta 
\end{equation}
for some other physical cutoff scale $\Delta$. So also for $n=1$ there will be many
situations where the weak-coupling regime can be defined in a well-controlled way but
a detailed study for all cases remains to be done, see e.g. Ref.~\cite{kurz_hs_wegewijs_preparation}.

{\bf Generic procedure.}
In the weak coupling limit a systematic procedure to solve the RG equations of RTRG-FS has
been developed in Ref.~\cite{reininghaus_hs_preprint} for the case $n=2$, i.e. for 
models with spin or orbital fluctuations (similiar procedures for models with charge
fluctuations, i.e. $n=1$, are in progress \cite{kurz_hs_wegewijs_preparation}). 
For $n=2$, the idea is to define first a reference solution $\bar{G}^{(1)}_{12}$ for the 
vertex $\bar{G}_{12}$ from the lowest order term of the RG equation
when $\Lambda$ is larger than all physical cutoff scales. This means that we 
neglect the frequencies and the Liouvillian on the r.h.s. of the RG equation and truncate 
at the first term. This gives the poor man scaling equation, which reads according to 
(\ref{rg_G_so_1_matsubara})
\begin{equation}
\label{reference_solution_so}
{d\bar{G}_{11'}^{(1)}\over d\Lambda}\,=\,
{1\over\Lambda}\,\left\{\bar{G}^{(1)}_{12}\,\bar{G}^{(1)}_{\bar{2}1'}\,-\,
(1\leftrightarrow 1')\right\}\, \quad.
\end{equation}
We use the Matsubara frequency representation (\ref{LG_matsubara_representation}) but
have set all frequency arguments to zero (we implicitly assume this in the following when
no argument is written). The initial condition for the leading order RG equation is the 
bare vertex. The order of magnitude of the leading order solution is denoted by the
dimensionless parameter $J\sim G^{(1)}_{12}$. If an observable is 
considered the first vertex has to be replaced by $R$
\begin{equation}
\label{reference_solution_observable_so}
{d\bar{R}_{11'}^{(1)}\over d\Lambda}\,=\,
{1\over\Lambda}\,\left\{\bar{R}^{(1)}_{12}\,\bar{G}^{(1)}_{\bar{2}1'}\,-\,
(1\leftrightarrow 1')\right\}\, \quad.
\end{equation}
This trivial replacement holds for all equations in the following, therefore we do not
specify it in all cases. For an observable only the combination 
$\mbox{Tr}_S \bar{R}_{1\dots n}\dots$ occurs for the calculation of the average. 
This has to be kept in mind because some properties of the vertex are only fulfilled if
this combination is taken.

The leading order solution is a good approximation when $\Lambda$ is much larger than the
maximal physical cutoff scale. We denote this scale by $\Lambda_c$, i.e. we define
\begin{equation}
\label{maximal_cutoff}
\Lambda_c\,=\,\mbox{max}\{|E|,|\mu_\alpha|,|h_k|\}\quad.
\end{equation}
Except for the Laplace variable $E$, no other frequencies are included in this definition because 
they do not enter into the final solution of the physical quantities. Of course, the leading
order solution is only good for frequencies below $\Lambda$ but since all frequency integrals
are cut off by $\Lambda$ in the RG equations this is not dangerous. Furthermore, we do not use any
combinations of $E$, $\mu_\alpha$, and $h_k$ in this definition (like they occur in the
resolvents) because each resolvent of the RG equation has its own combination. Temperature is also
not included in the definition of $\Lambda_c$ since temperature is a trivial cutoff scale. The 
effect of some finite temperature $T_\alpha$ of reservoir $\alpha$ is just that the Matsubara
frequencies of this reservoir do no longer occur in the RG equations for $\Lambda<T_\alpha$.
Since the scale of the relaxation and dephasing rates is given by $\Gamma_k\sim\Lambda_c J_c^2$,
with $J_c$ being the scale of the vertex at $\Lambda=\Lambda_c$, we get $\Gamma_k<\Lambda_c$ due
to the weak coupling condition $J_c\ll 1$. Therefore, also the relaxation and dephasing rates
do not enter the definition (\ref{maximal_cutoff}).

The next step is to expand the full solution for the Liouvillian and the vertices systematically 
around the leading-order solution for $\Lambda>\Lambda_c$ using the exact RG equations.
This leads to an expansion of the form
\begin{eqnarray}
\nonumber
L_S(E,\omega)\,&=&\,L_S^{(0)}\,+\,L_S^{(1)}(E,\omega)
\,+\,L_S^{(2)}(E,\omega)\,+\,\dots\quad\\
\nonumber
\bar{G}_{12}(E,\omega,\omega_1,\omega_2)\,&=&\,
\bar{G}_{12}^{(1)}\,+\,
\bar{G}_{12}^{(2)}(E,\omega,\omega_1,\omega_2)\,+\,\dots\quad,\\
\nonumber
\bar{G}_{1234}(E,\omega,\omega_1\dots\omega_4)\,&=&\,
\bar{G}_{1234}^{(2)}(E,\omega,\omega_1\dots\omega_4)\,+\,
\bar{G}_{1234}^{(3)}(E,\omega,\omega_1\dots\omega_4)\,+\,\dots\quad,\\
\label{LG_expansion}
&&\hspace{-0.5cm}\mbox{etc.}
\end{eqnarray}
where $L_S^{(k)},\bar{G}^{(k)}\sim J^k$ and $L_S^{(0)}=[H_S,\cdot]$ is the bare Liouvillian. The
expansion for the vertex $G_{1\dots n}$ starts at $k=n/2$.
As shown in Ref.~\cite{reininghaus_hs_preprint} it turns out that this expansion is not a 
pure power series in $J$ but can also contain terms $\sim J^k (\ln J)^{k-1}$. However, for $J\ll 1$,
the series is also well-defined in this case. 
We stop this procedure at scale $\Lambda=\Lambda_c$ and obtain a series of the
Liouvillian and the vertices in powers of the leading-order solution at the
same scale $\Lambda_c$. The order of
magnitude of the leading-order solution at $\Lambda_c$ is denoted by $J_c\ll 1$ 
which is assumed to be a small quantity for the weak-coupling regime.

If all temperatures $T_\alpha$ are larger than $\Lambda_c$ we can stop the
RG at the minimum of all temperatures. In all other cases we solve the exact 
RG equations for $\Lambda<\Lambda_c$ perturbatively in
$J_c$ using the Liouvillian and the vertices from (\ref{LG_expansion})
at scale $\Lambda_c$ as initial condition. This perturbation theory is well-defined
because, looking at the exact RG equation (\ref{rg_J}) for $n=2$, we see
that only terms $\sim (J_c)^k\,(\ln(\Lambda_c/\Delta))^{k-1}$ can be produced, with 
$\Delta$ being some other physical low-energy cutoff scale of the form (\ref{cutoff_possibilities}).
The lowest possible value for $\Delta$ is the minimum $\Gamma$ of all physical
relaxation and dephasing rates. If we assume that the order of magnitude of 
$\Gamma$ is identical to its value $\Gamma_c$ at scale $\Lambda=\Lambda_c$,
we can estimate the maximal value of the logarithm as
\begin{equation}
\label{estimate_logarithm}
\ln({\Lambda_c\over\Gamma_c})\,\sim\,\ln(J_c) \quad,
\end{equation}
where we have used $\Gamma_c\sim \Lambda_c (J_c)^2$ due to a dimensional analysis
(except for $\Lambda$, there is no other physical energy scale available for $\Lambda>\Lambda_c$).
Therefore, similiar to the series (\ref{LG_expansion}), we get correction terms 
$\sim (J_c)^k\,(\ln J_c)^{k-1}$ which are not dangerous for $J_c\ll 1$. If the
cutoff scale is not $\Gamma$, some deviation from a resonance position occurs for
$\Delta$, leading to interesting logarithmic enhancements at resonance positions.
This has been evaluated in Ref.~\cite{reininghaus_hs_preprint} within a
consistent 2-loop formalism for the anisotropic Kondo model in a finite magnetic field.

{\bf 1-loop treatment.} Here, we summarize the procedure of Ref.~\cite{reininghaus_hs_preprint} in 1-loop,
i.e. we will consider all terms which are proportional to either $J_c$, $J_c^2$ or $J_c^2\ln(J_c)$.
Thereby, it will turn out that the real part of the eigenvalue of the effective 
Liouvillian contains terms of order $J_c$ and $J_c^2\ln(J_c)$, so that corrections of
the order $J_c^2$ can be neglected. For the imaginary part of the eigenvalue and for the current
kernel, the series starts at $J_c^2$. For the rates, terms proportional to $J_c^3\ln(J_c)$ can be 
calculated consistently only within a 2-loop analysis, see Ref.~\cite{reininghaus_hs_preprint}
for more details.

First, we note that the leading-order solutions $\bar{G}^{(1)}_{11'}$ and
$\bar{R}^{(1)}_{11'}$ do not generate
any new structure in Liouville space but leave the form (\ref{G_vertex_liouville})
of the original vertex and the form (\ref{vertex_observable}) of the original observable
invariant (the latter holds only for $\mbox{Tr}_S \bar{R}^{(1)}_{11'}$). 
As shown in Appendix E we get
\begin{eqnarray}
\label{ansatz_G}
\bar{G}^{(1)}_{11'}\,&=&\,\sum_p\,(G^{(1)})^{pp}_{11'}\,=\,{[g_{11'},\cdot]}_-
\quad,\\
\label{ansatz_R}
\bar{R}^{(1)}_{11'}\,&=&\,\sum_p\,(R^{(1)})^{pp}_{11'}\,=\,{i\over 2}\,{[r_{11'},\cdot]}_+
\quad,
\end{eqnarray}
with
\begin{eqnarray}
\label{rg_g}
{dg_{11'}\over d\Lambda}\,&=&\,
{1\over\Lambda}\,\left\{g_{12}\,g_{\bar{2}1'}\,-\,(1\leftrightarrow 1')\right\} \quad,\\
\label{rg_r}
{dr_{11'}\over d\Lambda}\,&=&\,
{1\over\Lambda}\,\left\{r_{12}\,g_{\bar{2}1'}\,+\,g_{12}\,r_{\bar{2}1'}
\,-\,(1\leftrightarrow 1')\right\} \quad.
\end{eqnarray}
These are the so-called poor man scaling equations. As a consequence (see Appendix E), the
adjoint of the leading-order vertices is given by 
\begin{eqnarray}
\label{hermiticity_gG}
(g_{11'})^\dagger\,&=&\,g_{\bar{1}'\bar{1}}\quad,\quad
(G^{(1)}_{11'})^\dagger\,=\,G^{(1)}_{\bar{1}'\bar{1}}\quad,\\
\label{hermiticity_rR}
(r_{11'})^\dagger\,&=&\,r_{\bar{1}'\bar{1}}\quad,\quad
(R^{(1)}_{11'})^\dagger\,=\,-R^{(1)}_{\bar{1}'\bar{1}}\quad.
\end{eqnarray}
Note the difference to the property (\ref{G_c_transform}), where the c-transformation
has been considered.

Furthermore, using (\ref{ansatz_G}), we define for arbitrary $\Lambda$ 
\begin{equation}
\label{tilde_G}
\tilde{G}^{(1)}_{11'}\,=\,\sum_p\,p\,(G^{(1)})^{pp}_{11'}\,=\,{[g_{11'},\cdot]}_+
\quad.
\end{equation}
As is shown in Appendix E the RG equation for $\tilde{G}^{(1)}_{11'}$ reads
\begin{eqnarray}
\nonumber
{d\tilde{G}_{11'}^{(1)}\over d\Lambda}\,&=&\,
{1\over\Lambda}\,\left\{\tilde{G}^{(1)}_{12}\,\bar{G}^{(1)}_{\bar{2}1'}\,-\,
\bar{G}^{(1)}_{1'2}\,\tilde{G}^{(1)}_{\bar{2}1}\right\}\\
\label{G_tilde_leading}
&=&\,
{1\over\Lambda}\,\left\{\bar{G}^{(1)}_{12}\,\tilde{G}^{(1)}_{\bar{2}1'}\,-\,
\tilde{G}^{(1)}_{1'2}\,\bar{G}^{(1)}_{\bar{2}1}\right\}\, \quad,
\end{eqnarray}
and the leading order result for the current operator can be written as
\begin{equation}
\label{current_operator_leading}
\bar{I}^{\gamma(1)}_{11'}\,=\,c^\gamma_{11'}\,\tilde{G}_{11'}^{(1)}\quad,\quad
c^\gamma_{11'}\,=\,-{1\over 2}\,(\eta_1\delta_{\alpha_1\gamma}+\eta_2\delta_{\alpha_2\gamma})
\quad.
\end{equation}

Next, we consider the 1-loop RG equations (\ref{rg_L_so_1_matsubara})
and (\ref{rg_G_so_1_matsubara}) for zero temperature (finite $T$ just leads to a trivial
cutoff of the RG flow)
\begin{eqnarray}
\label{rg_L_so_zw}
-\,{dL_S(E,\omega)\over d\Lambda}\,&=&\,
-\,\bar{G}_{12}\,
\int_0^\Lambda d\omega_2\,{1 \over i\omega_2+i\omega+i\Lambda+E_{12}-L_S(E_{12},\omega_2+\omega+\Lambda)}
\,\bar{G}_{\bar{2}\bar{1}}\quad,\\
\label{rg_G_so_zw}
-\,{d\bar{G}_{11'}\over d\Lambda}\,&=&\,
-i\,\left\{\bar{G}_{12}\,{1\over i\Lambda}\,\bar{G}_{\bar{2}1'}
\,\, - \,\, (1\leftrightarrow 1')\right\}
\quad.
\end{eqnarray}
We have neglected the frequency dependence 
of the vertices and the influence of the Liouvillian on
the RG equation for the vertex. These corrections are of the same order
as the other 2-loop terms (\ref{rg_L_so_2}), (\ref{rg_G_so_2})-(\ref{rg_G_so_4})
on the r.h.s. of the exact RG equations and
have been analysed in detail in Ref.~\cite{reininghaus_hs_preprint}.
It can be shown that they do not influence the final result up to second order in $J_c$
(see a more detailed discussion of this point at the end of this section).
To perform the frequency integration in (\ref{rg_L_so_zw}), we use
the form (coming out of the RG flow described below, see (\ref{L_first_decomposition}))
\begin{equation}
\label{form_L}
L_S(E,\omega)\,=\,L_S^{(0)}\,+\,L_S^{(1)}\,-\,(E+i\omega)\,Z^{(1)}\,+\,L_S^{(2)}(E)
\quad,
\end{equation}
where $L_S^{(1)}$ and $Z^{(1)}$ are terms linear in $J$ which are defined via the differential
equations (\ref{L_first_rg}). The dependence
of $L_S^{(2)}(E,\omega)$ on the frequency $\omega$ has been neglected, which is a term
of higher order. With this representation the frequency integral in (\ref{rg_L_so_zw})
can easily be calculated with the result
\begin{eqnarray}
\label{rg_L_so_approx}
{dL_S(E,\omega)\over d\Lambda}\,&=&\,
-i\,\bar{G}_{12}\,{\cal{L}}_\Lambda(E_{12}+i\omega-\tilde{L}_S(E_{12}))\,{1\over 1+Z^{(1)}}
\,\bar{G}_{\bar{2}\bar{1}}\quad,\\
\label{rg_G_so_approx}
{d\bar{G}_{11'}\over d\Lambda}\,&=&\,
{1\over \Lambda}\,\left\{\bar{G}_{12}\,\bar{G}_{\bar{2}1'}
\,\, - \,\, (1\leftrightarrow 1')\right\}
\quad,
\end{eqnarray}
where we have defined the important function
\begin{equation}
\label{L_function}
{\cal{L}}_\Lambda(z)\,=\,\ln({2\Lambda-iz \over \Lambda-iz})\quad,
\end{equation}
and
\begin{equation}
\label{L_tilde}
\tilde{L}_S(E)\,=\,{1\over 1+Z^{(1)}}\,\left(L_S^{(0)}\,+\,L_S^{(1)}\,+\,L_S^{(2)}(E)\right)\quad.
\end{equation}
The initial conditions at $\Lambda=\Lambda_0\sim D$ from the integration over the symmetric part
of the distribution function have already been derived in (\ref{L_initial_spin_ss}),
(\ref{L_initial_spin_sa}) and (\ref{G_initial_spin_s}) 
\begin{eqnarray}
\nonumber
L_S(E,\omega)|_{\Lambda=\Lambda_0}\,&=&\,L_S^{(0)}\,-\,
i\,{\pi^2 \over 16}\,D\,\bar{G}^{(1)}_{11'}\,\bar{G}^{(1)}_{\bar{1}'\bar{1}}
\,+\,{\pi^2 \over 32}\,\bar{G}^{(1)}_{11'}\,(E_{11'}\,+\,i\omega\,-\,L_S^{(0)})\,
\bar{G}^{(1)}_{\bar{1}'\bar{1}}\\
\label{L_initial_weak_coupling}
&&\hspace{1cm}\,-\,
{\pi \over 4}\,D\,\bar{G}^{(1)}_{11'}\,\tilde{G}^{(1)}_{\bar{1}'\bar{1}}
\,-\,i\,{\pi\over 4}\,\bar{G}^{(1)}_{11'}\,(E_{11'}\,+\,i\omega\,-\,L_S^{(0)})\,\tilde{G}^{(1)}_{\bar{1}'\bar{1}}
\quad,\\
\label{G_initial_weak_coupling}
\bar{G}_{11'}|_{\Lambda=\Lambda_0}\,&=&\,\bar{G}_{11'}^{(1)}\,-\,
i\,{\pi\over 2}\,\left(\bar{G}^{(1)}_{12}\,\tilde{G}^{(1)}_{\bar{2}1'}\,-\,
\bar{G}^{(1)}_{1'2}\,\tilde{G}^{(1)}_{\bar{2}1}\right)\quad.
\end{eqnarray}
Analogous RG equations and
initial conditions hold for $\Sigma_R(E)$ and $\bar{R}_{11'}$ by replacing the first vertex 
on the r.h.s. of all equations by $R$.

The RG equation (\ref{rg_G_so_approx}) for the vertex $\bar{G}_{11'}$ has the same form as 
the RG equation (\ref{reference_solution_so}) for the leading-order vertex $\bar{G}^{(1)}_{11'}$.
However, the two vertices are not identical because the initial condition (\ref{G_initial_weak_coupling})
is different. Whereas $\bar{G}^{(1)}_{11'}$ is the bare vertex initially, the vertex $\bar{G}_{11'}$
has an additional second order correction $\sim J^2$. Therefore, in the spirit
of the series (\ref{LG_expansion}), we write
\begin{equation}
\label{G_decomposition}
\bar{G}_{11'}\,=\,\bar{G}^{(1)}_{11'}\,+\,\bar{G}^{(2)}_{11'}\quad,
\end{equation}
with
\begin{equation}
\label{rg_G_second_order}
{d\bar{G}^{(2)}_{11'}\over d\Lambda}\,=\,
{1\over \Lambda}\,\left\{\bar{G}^{(2)}_{12}\,\bar{G}^{(1)}_{\bar{2}1'}\,+\,
\bar{G}^{(1)}_{12}\,\bar{G}^{(2)}_{\bar{2}1'}
\,\, - \,\, (1\leftrightarrow 1')\right\}
\quad,
\end{equation}
and initial condition
\begin{equation}
\label{solution_G_second_order}
\bar{G}^{(2)}_{11'}\,=\,
-i\,{\pi\over 2}\,\left(\bar{G}^{(1)}_{12}\,\tilde{G}^{(1)}_{\bar{2}1'}\,-\,
\bar{G}^{(1)}_{1'2}\,\tilde{G}^{(1)}_{\bar{2}1}\right)\quad.
\end{equation}
Interestingly, it can be shown that this form holds not only initially but
for all values of $\Lambda$, i.e. (\ref{solution_G_second_order}) is the
solution to the differential equation (\ref{rg_G_second_order}). This holds
also for the case of the observable if $\mbox{Tr}_S$ is applied from the
left. For the proof we refer to Appendix E. As one can see, the correction has
an additional factor $i$, so that it leads to the lowest order result for the
complex part of the vertex.

In contrast to all the other second order corrections, which have already been
neglected in the approximate form (\ref{rg_G_so_approx}) of the RG equation
for the vertex, the correction $\bar{G}_{11'}^{(2)}$ is important to calculate
even the lowest order result in $J$. To see this we insert the expansion
(\ref{G_decomposition}) into the RG equation (\ref{rg_L_so_approx}) for the
Liouvillian, expand ${1\over 1+Z^{(1)}}\approx 1-Z^{(1)}$, and 
neglect all terms $\sim J^4$ on the r.h.s. leading to
\begin{eqnarray}
\nonumber
{dL_S(E,\omega)\over d\Lambda}\,&=&\,\\
\nonumber
&&\hspace{-2cm}=\,-i\,\bar{G}^{(1)}_{12}\,{\cal{L}}_\Lambda(E_{12}+i\omega-\tilde{L}_S(E_{12}))
\,\bar{G}^{(1)}_{\bar{2}\bar{1}}\,
+i\,\bar{G}^{(1)}_{12}\,{\cal{L}}_\Lambda(E_{12}+i\omega-\tilde{L}_S(E_{12}))\,Z^{(1)}
\,\bar{G}^{(1)}_{\bar{2}\bar{1}}
\\
\label{rg_L_so_second_third}
&&\hspace{-2cm}-\,i\,\bar{G}^{(2)}_{12}\,{\cal{L}}_\Lambda(E_{12}+i\omega-\tilde{L}_S(E_{12}))
\,\bar{G}^{(1)}_{\bar{2}\bar{1}}\,-\,
i\,\bar{G}^{(1)}_{12}\,{\cal{L}}_\Lambda(E_{12}+i\omega-\tilde{L}_S(E_{12}))
\,\bar{G}^{(2)}_{\bar{2}\bar{1}} \quad.
\end{eqnarray}
We have to be careful not to count the powers in $J$ incorrectly. As we can
already see from the RG equations (\ref{reference_solution_so}) and
(\ref{rg_G_second_order}) an addtional factor ${1\over\Lambda}$ on the r.h.s.
can lead to one order less in $J$ when compared to the l.h.s. Therefore, in
the regime $\Lambda > \Lambda_c$, we have to check if additional factors
${1\over\Lambda}$ can occur on the r.h.s. of the RG equation for the Liouvillian.
To see this we subtract from the function ${\cal{L}}_\Lambda(z)$ the asymptotic
part $\sim {z\over\Lambda}$ and define
\begin{equation}
\label{L_tilde_function}
{\tilde{\cal{L}}}_\Lambda(z)\,=\,{\cal{L}}_\Lambda(z)\,-\,{iz \over 2\Lambda}\,=\,
\ln({2\Lambda-iz\over \Lambda-iz})\,-\,{iz \over 2\Lambda} \quad,
\end{equation}
so that ${\cal{L}}_\Lambda(z)$ and ${\tilde{\cal{L}}}_\Lambda(z)$ are integrated
by the functions
\begin{eqnarray}
\label{F_function}
{\cal{L}}_\Lambda(z)\,&=&\,{d\over d\Lambda}\,F_\Lambda(z)\quad,\quad
F_\Lambda(z)\,=\,\Lambda\,\ln({2\Lambda-iz\over \Lambda-iz})\,-\,{iz\over 2}
\ln({(2\Lambda-iz)(-iz)\over (\Lambda-iz)^2}) \quad,\\
\label{F_tilde_function}
{\tilde{\cal{L}}}_\Lambda(z)\,&=&\,{d\over d\Lambda}\,\tilde{F}_\Lambda(z)\quad,\quad
\tilde{F}_\Lambda(z)\,=\,F_\Lambda(z)\,-\,{iz\over 2}\,(\ln({\Lambda\over -2iz})\,+\,1)
\quad,
\end{eqnarray}
with the following asymptotic behaviours
\begin{eqnarray}
\label{F_asymptotic}
F_\Lambda(z)\,&\rightarrow&\,
\left\{
\begin{array}{cl}
\Lambda\,(\ln(2)\,+\,O({z\over\Lambda})^2)\,+\,{iz\over 2}\,(\ln({\Lambda\over -2iz})\,+\,1)
\quad &\mbox{for }\Lambda \gg |z| \\
{1\over 2}\,{\Lambda^2\over iz}
\quad &\mbox{for }\Lambda \ll |z|
\end{array}
\right.\quad,\\
\label{F_tilde_asymptotic}
\tilde{F}_\Lambda(z)\,&\rightarrow&\,
\left\{
\begin{array}{cl}
\Lambda\,(\ln(2)\,+\,O({z\over\Lambda})^2)
\quad &\mbox{for }\Lambda \gg |z| \\
{iz\over 2}\,(\ln({\Lambda\over -2iz})\,+\,1\,-\,{\Lambda^2\over z^2})
\quad &\mbox{for }\Lambda \ll |z|
\end{array}
\right.\quad.
\end{eqnarray}
We now insert 
${\cal{L}}_\Lambda(z)={d\over d\Lambda}{\tilde{\cal{F}}}_\Lambda(z)+{iz\over 2\Lambda}$
into the RG equation (\ref{rg_L_so_second_third}) for the Liouvillian and start with the
contribution from the part ${d\over d\Lambda}{\tilde{\cal{F}}}_\Lambda(z)$. For the first
term on the r.h.s. of (\ref{rg_L_so_second_third}) we perform a partial integration and get
\begin{eqnarray}
\nonumber
-i\,\bar{G}^{(1)}_{12}\,\left\{{d\over d\Lambda}{\tilde{\cal{F}}}_\Lambda(z)\right\}
\,\bar{G}^{(1)}_{\bar{2}\bar{1}}\,&=&\,
{d\over d\Lambda}\left\{-i\,\bar{G}^{(1)}_{12}\,{\tilde{\cal{F}}}_\Lambda(z)
\,\bar{G}^{(1)}_{\bar{2}\bar{1}}\right\}\\
\label{partial_integration}
&&\hspace{-3cm}
+\,i\,\left\{{d\over d\Lambda}\bar{G}^{(1)}_{12}\right\}\,{\tilde{\cal{F}}}_\Lambda(z)
\,\bar{G}^{(1)}_{\bar{2}\bar{1}}\,+\,
i\,\bar{G}^{(1)}_{12}\,{\tilde{\cal{F}}}_\Lambda(z)
\,\left\{{d\over d\Lambda}\bar{G}^{(1)}_{\bar{2}\bar{1}}\right\} 
\quad,
\end{eqnarray}
with $z\equiv E_{12}+i\omega-\tilde{L}_S(E_{12})$. Here, we have neglected terms which differentiate with 
respect to $\tilde{L}_S(E_{12})$ occuring in $z$, which are terms of higher order when inserting the
RG equation for the Liouvillian. The first term on the 
r.h.s. of (\ref{partial_integration}) leads to the following contribution for the Liouvillian 
\begin{equation}
\label{F_tilde_contribution}
L^{(2a)}_S(E,\omega)\,=\,-i\,\bar{G}^{(1)}_{12}\,
{\tilde{\cal{F}}}_\Lambda(E_{12}+i\omega-L_S(E_{12}))
\,\bar{G}^{(1)}_{\bar{2}\bar{1}}
\quad.
\end{equation}
Its initial value cancels with the second term on the r.h.s. of the
initial condition for the Liouvillian (\ref{L_initial_weak_coupling}) by choosing approximately
\begin{equation}
\label{value_initial_cutoff}
\Lambda_0\,=\,{\pi^2\over 16 \ln(2)}\,D
\end{equation}
for the initial value of the parameter $\Lambda$ (note that $\tilde{F}_\Lambda(z)$ can
be approximated by $\ln(2)\Lambda$ in this regime, according to (\ref{F_tilde_asymptotic})).
In this way the terms $\sim D$ vanish. We note that there is no term arising from the
RG which cancels the fourth term on the r.h.s. of (\ref{L_initial_weak_coupling}) which is
also $\sim D$. This term is real and leads to a renormalization of the real part of the
eigenvalues of the Liouvillian. If it is nonzero the model is not well defined and the 
frequency dependence of the original vertices is important. As we will see in Sec.~\ref{sec:5}
the term vanishes for the Kondo model.

(\ref{F_tilde_contribution}) is a contribution $\sim |E_{12}+i\omega-\tilde{L}_S(E_{12})|J^2$ to the 
Liouvillian for $\Lambda\rightarrow |E_{12}+i\omega-\tilde{L}_S(E_{12})|$. Therefore, it is in fact a second order 
contribution to $L_S(E,\omega)$. The last two terms on the
r.h.s. of (\ref{partial_integration}) can be estimated by inserting the RG equation
(\ref{reference_solution_so}) for the leading-order vertex, which gives
${d\over d\Lambda}G^{(1)}\sim {1\over\Lambda}J^2$. Using $\tilde{F}_\Lambda(z)\sim \Lambda$
for $\Lambda\gg z$, we get terms $\sim J^3$ which are of third order and are neglected.
If we perform the same analysis for the other terms on the r.h.s. of (\ref{rg_L_so_second_third}),
one of the vertices in (\ref{F_tilde_contribution}) is
replaced by $G^{(2)}$ or we get an additional factor $Z^{(1)}$. Thus, also these terms
are of third order in $J$ and are neglected.

The remaining terms
in the RG equation stem from the part ${iz\over \Lambda}$ of ${\cal{L}}_\Lambda(z)$, 
giving rise to contributions to the Liouvillian in first and second order.
Using (\ref{L_tilde}), we can systematically expand up to third order in $J$ on the
r.h.s. of (\ref{rg_L_so_second_third}), and can group the various terms by the 
following differential equations
\begin{eqnarray}
\label{L_first_order}
{d L^{(1)}_S(E,\omega)\over d\Lambda}\,&=&\,{1\over 2\Lambda}
\,\bar{G}^{(1)}_{12}\,(E_{12}+i\omega-L_S^{(0)})\,\bar{G}^{(1)}_{\bar{2}\bar{1}}\quad,\\
\label{L_b_second_order}
{d L^{(2b)}_S(E,\omega)\over d\Lambda}\,&=&\,{1\over 2\Lambda}\,\left\{
\,\bar{G}^{(2)}_{12}\,(E_{12}+i\omega-L_S^{(0)})\,\bar{G}^{(1)}_{\bar{2}\bar{1}}\,+
\,\bar{G}^{(1)}_{12}\,(E_{12}+i\omega-L_S^{(0)})\,\bar{G}^{(2)}_{\bar{2}\bar{1}}\right\}\quad,\\
\label{L_c_second_order}
{d L^{(2c)}_S(E,\omega)\over d\Lambda}\,&=&\,{1\over 2\Lambda}
\,\bar{G}^{(1)}_{12}\,\left\{-L_S^{(1)}-(E_{12}+i\omega)Z^{(1)}+Z^{(1)} L_S^{(0)}
+L_S^{(0)}Z^{(1)}\right\}\,
\bar{G}^{(1)}_{\bar{2}\bar{1}}\quad,
\end{eqnarray}
where the initial condition for $L_S^{(1)}(E,\omega)$ is zero, and for $L_S^{(2b)}(E,\omega)$ and
$L_S^{(2c)}(E,\omega)$ it is given by the fifth and third term on the r.h.s. of 
(\ref{L_initial_weak_coupling}), respectiviely,
\begin{eqnarray}
\label{L_1_initial}
L_S^{(1)}(E,\omega)|_{\Lambda=\Lambda_0}\,&=&\,0\quad,\\
\label{L_b_initial}
L_S^{(2b)}(E,\omega)|_{\Lambda=\Lambda_0}\,&=&\,
-\,i\,{\pi\over 4}\,\bar{G}^{(1)}_{11'}\,(E_{11'}\,+\,i\omega\,-\,L_S^{(0)})\,
\tilde{G}^{(1)}_{\bar{1}'\bar{1}}\quad.\\
\label{L_c_initial}
L_S^{(2c)}(E,\omega)|_{\Lambda=\Lambda_0}\,&=&\,
{\pi^2 \over 32}\,\bar{G}^{(1)}_{11'}\,(E_{11'}\,+\,i\omega\,-\,L_S^{(0)})\,
\bar{G}^{(1)}_{\bar{1}'\bar{1}}\quad,
\end{eqnarray}
There is an additional factor ${1\over\Lambda}$ on the r.h.s. of these RG equations and, 
therefore, one gets one order less in $J$ when integrating them. 
$L^{(1)}_S(E,\omega)$ is a contribution to the
renormalization of the real part of the eigenvalues of the Liouvillian in first
order in $J$, e.g. for the Kondo model it leads to a renormalization of the
magnetic field. It has a linear frequency dependence and can be decomposed as
\begin{equation}
\label{L_first_decomposition}
L_S^{(1)}(E,\omega)\,=\,L_S^{(1)}\,-\,(E+i\omega)\,Z^{(1)}\,
\end{equation}
with $L_S^{(1)}\equiv L_S^{(1)}(E=0,\omega=0)$ and
\begin{equation}
\label{L_first_rg}
{d L^{(1)}_S \over d\Lambda}\,=\,{1\over 2\Lambda}
\,\bar{G}^{(1)}_{12}\,(\bar{\mu}_{12}-L_S^{(0)})\,\bar{G}^{(1)}_{\bar{2}\bar{1}}\quad,\quad
{d Z^{(1)}\over d\Lambda}\,=\,-\,{1\over 2\Lambda}
\,\bar{G}^{(1)}_{12}\,\bar{G}^{(1)}_{\bar{2}\bar{1}}\quad.
\end{equation}

$L_S^{(2c)}(E,\omega)$ is a contribution to the renormalization of the real part of 
the eigenvalues in second order in $J$ and can be neglected compared
to $L_S^{(1)}(E,\omega)$. It is not calculated consistently here because the higher order
terms of the RG equations and the frequency dependence of the vertices give rise to
terms of the same order, see the complete 2-loop analysis of Ref.~\cite{reininghaus_hs_preprint}.
Nevertheless, we have included it here for completeness.

$L_S^{(2b)}(E,\omega)$ is a contribution to the rate since $G^{(2)}$ is
complex according to the solution (\ref{solution_G_second_order}). As we will
see for the special example of the Kondo problem in Sec.~\ref{sec:5} it is the
only term giving rise to a finite current in lowest order in $J$. 
The important difference between $L_S^{(2a)}$ and $L_S^{(2b)}$ is that the generation
of $L_S^{(2b)}$ is finished at $\Lambda=\Lambda_c$. As we explained above, for
$\Lambda<\Lambda_c$, we solve the RG equations perturbatively in $J_c$ which is
the scale of the leading order vertex at $\Lambda=\Lambda_c$. Therefore, in this
regime, all quantities are systematically expanded in $J_c$ and usual power 
counting applies. Therefore, the terms involving $G^{(2)}$ on the r.h.s. of the 
RG equation (\ref{rg_L_so_second_third}) for the Liouvillian lead to corrections
$\sim J_c^3$ for $\Lambda<\Lambda_c$ and do not influence the result up to $J_c^2$. 

In contrast, the first term on the r.h.s. of (\ref{rg_L_so_second_third}) gives
rise to a contribution $\sim J_c^2$ in the regime $\Lambda<\Lambda_c$. Replacing there
the leading order vertices by their value $G^{(1)c}_{11'}$ at $\Lambda=\Lambda_c$,
and using (\ref{F_function}), we can integrate this term from $\Lambda_c$
to $\Lambda$ with the result
\begin{eqnarray}
\nonumber
L_S(E,\omega)_\Lambda\,&=&\,L_S(E,\omega)_{\Lambda_c}\,-\\
\label{integration_below_c}
&&\hspace{-2cm}-\,i\,\bar{G}^{(1)c}_{12}\,\left\{{\cal{F}}_{\Lambda}
(E_{12}+i\omega-\tilde{L}_S(E_{12})_{\Lambda})-{\cal{F}}_{\Lambda_c}
(E_{12}+i\omega-\tilde{L}_S(E_{12})_{\Lambda_c})\right\}
\,\bar{G}^{(1)c}_{\bar{2}\bar{1}}\quad.
\end{eqnarray}
Thereby we have neglected the derivative of $\tilde{L}_S(E_{12})_\Lambda$ with 
respect to $\Lambda$ (see the discussion below). The term involving 
${\cal{F}}_{\Lambda_c}$ can be
taken together with $L^{(2a)}_S(E,\omega)_{\Lambda_c}$ by using the result 
(\ref{F_tilde_contribution}) at $\Lambda=\Lambda_c$. Together with the relation 
(\ref{F_tilde_function}) we get
\begin{eqnarray}
\label{final_solution_Lambda_finite}
\nonumber
L_S(E,\omega)_\Lambda\,&=&\,
L_S^{(0)}\,+\,L_S^{(1)c}(E,\omega)\,+\,L_S^{(2b)c}(E,\omega)\,+\,L_S^{(2c)c}(E,\omega)\,+\\
&&\hspace{0.5cm}+\,L_S^{(2\tilde{a})c}(E,\omega)
-\,i\,\bar{G}^{(1)c}_{12}\,{\cal{F}}_{\Lambda}
(E_{12}+i\omega-\tilde{L}_S(E_{12})_{\Lambda})\,\bar{G}^{(1)c}_{\bar{2}\bar{1}}\quad,
\end{eqnarray}
where the index $c$ indicates always the value at $\Lambda=\Lambda_c$, and 
$L_S^{(2\tilde{a})c}(E,\omega)$ is defined by 
\begin{eqnarray}
\label{L_2a_below_c}
L_S^{(2\tilde{a})c}(E,\omega)\,&=& \\
\nonumber
&&\hspace{-2cm}=\,-\,{1\over 2}\,\bar{G}^{(1)c}_{12}\,
(E_{12}+i\omega-\tilde{L}_S(E_{12})_{\Lambda_c})\,
\left(\ln{\Lambda_c\over -2i(E_{12}+i\omega-\tilde{L}_S(E_{12})_{\Lambda_c})}\,+\,1\right)
\,\bar{G}^{(1)c}_{\bar{2}\bar{1}}\quad
\end{eqnarray}
in the regime $\Lambda<\Lambda_c$. 
Setting $\Lambda=0$ and $\omega=0$, we get the effective Liouvillian $L_S^{eff}(E)$. 
Using $F_{\Lambda=0}(z)=0$, we get
\begin{equation}
\label{final_solution}
L_S^{eff}(E)\,=\,L_S^{(0)}\,+\,L_S^{(1)c}(E)\,+\,L_S^{(2\tilde{a})c}(E)\,+\,L_S^{(2b)c}(E)
\,+\,L_S^{(2c)c}(E)
\quad.
\end{equation}
The problem is the precise value of $\tilde{L}_S(E_{12})$ in (\ref{L_2a_below_c}). The 
integration can not be performed analytically if the $\Lambda$-dependence
of $\tilde{L}_S(E_{12})_\Lambda$ is taken into account in the regime $\Lambda<\Lambda_c$. 
Since no terms are generated linear in $J_c$ in the regime $\Lambda<\Lambda_c$,  
we get according to (\ref{L_tilde}),
\begin{equation} 
\label{L_tilde_below_c}
\tilde{L}_S(E)_\Lambda\,=\,{1\over 1+Z^{(1)c}}\,\left(L_S^{(0)}\,+\,L_S^{(1)c}
\,+\,L_S^{(2)}(E)_\Lambda\right)\quad,
\end{equation}
with $L_S^{(2)}(E)_\Lambda=L_S^{(2)}(E,\omega=0)_\Lambda$ which are given by the second order terms
of (\ref{final_solution_Lambda_finite}) in first approximation. Using (\ref{F_function})
we see that the last term on the r.h.s. of (\ref{final_solution_Lambda_finite}) has a
complicated logarithmic dependence on $\Lambda$ which can only be integrated numerically.
However, even from a more precise numerical analysis we do not expect that the variation of
$\tilde{L}_S(E)_\Lambda$ will change the result for the effective Liouvillian significantly.
The important part of $\tilde{L}_S(E)_\Lambda$ is the hermitian part which determines the position of
the resonances where the logarithmic function in (\ref{L_2a_below_c}) becomes maximal. According
to (\ref{L_tilde_below_c}), this part can be cut off in good approximation by the terms up to
linear order in $J_c$ which are independent of $\Lambda$. Expanding (\ref{L_tilde_below_c}) up
to this order gives
\begin{equation}
\label{h_tilde}
\tilde{h}_c\,=\,L_S^{(0)}\,+\,L_{Sd}^{(1)c}\,-\,Z^{(1)c}_d\,L_S^{(0)}\quad,
\end{equation}
where the additional index $d$ means that we take only the diagonal part with respect to
the Liouvillian $L_S^{(0)}$ (otherwise the various terms do not commute and we consider 
higher-order corrections which are not calculated consistently). Terms of order $J_c^2$
are neglected in $\tilde{h}_c$. Note that $\tilde{h}_c$ can depend implicitly on $E$ via
the cutoff $\Lambda_c$ when $E$ is the maximal low-energy scale. Only in a small region
of order $\Gamma$ around the resonances the antihermitian part of $\tilde{L}_S^{(2)}$ cuts of
the logarithmic divergencies. The prefactor of the imaginary part occurs then under the
logarithm and gives only a very small perturbative correction. Therefore, in performing
the integral, the error is quite small when neglecting the $\Lambda$-dependence of 
$L_S^{(2)}(E)_\Lambda$. We have taken the value at $\Lambda=\Lambda_c$ in
(\ref{L_2a_below_c}), but, alternatively, one can also take the value at $\Lambda=0$. 
We denote the antihermitian part of $L_S^{(2)}(E)_{\Lambda_c}$
by $-\tilde{\Gamma_c}(E)$, which according to (\ref{integration_below_c}), is given by
\begin{equation}
\label{Gamma_tilde}
\tilde{\Gamma}_c(E)\,=\,i\,L_{Sd}^{(2a)c}(E)\,+\,i\,L_{Sd}^{(2b)c}(E)\quad,
\end{equation}
where we take again only the diagonal part with respect to $L_S^{(0)}$. In summary, 
we replace $\tilde{L}_S(E_{12})$ in (\ref{L_2a_below_c}) by 
$\tilde{h}_c-i\tilde{\Gamma}_c(E_{12})$. Using
\begin{eqnarray}
\nonumber
(E_{12}-\tilde{h}_c+i\tilde{\Gamma}_c(E_{12}))\,
\ln{\Lambda_c\over -2i(E_{12}-\tilde{h}_c+i\tilde{\Gamma}_c(E_{12}))}\,&\approx&\,\\
\label{logarithm_decomposition}
&&\hspace{-5cm}\,\approx\,H_{\tilde{\Gamma}_c(E_{12})}(E_{12}-\tilde{h}_c)
\,+\,i{\pi\over 2}\,|E_{12}-\tilde{h}_c|\quad,
\end{eqnarray}
with
\begin{equation}
\label{H_function}
H_\Gamma(E)\,=\,E\,\left(\ln{\Lambda_c\over 2\sqrt{|E|^2+\Gamma^2}}\,+\,1\right)\quad,
\end{equation}
we can decompose the final solution (\ref{final_solution}) in hermitian and antihermitian parts 
(which holds when $E$ is real according to the property (\ref{hermiticity_gG}))
\begin{eqnarray}
\label{final_solution_decomposition}
L_S^{eff}(E)\,&=&\,h^{eff}(E)\,-\,i\,\Gamma^{eff}(E)\quad,\\
\label{final_solution_real}
h^{eff}(E)\,&=&\,L_S^{(0)}\,+\,L_S^{(1)c}\,-\,E\,Z^{(1)c}
\,+\,L_S^{(2\tilde{a})Re}(E)\,+\,L_S^{(2c)c}(E)\quad,\\
\label{final_solution_imaginary}
\Gamma^{eff}(E)\,&=&\,i\,L_S^{(2\tilde{a})Im}(E)\,+\,i\,L_S^{(2b)c}(E)\quad,
\end{eqnarray}
with
\begin{eqnarray}
\label{L_2a_real}
\,L_S^{(2\tilde{a})Re}(E)\,&=&\,
-\,{1\over 2}\,\bar{G}^{(1)c}_{12}\,H_{\tilde{\Gamma}_c(E_{12})}(E_{12}-\tilde{h}_c)
\bar{G}^{(1)c}_{\bar{2}\bar{1}}\quad,\\
\label{L_2a_imag}
i\,L_S^{(2\tilde{a})Im}(E)\,&=&\,
\,{\pi\over 4}\,
\bar{G}^{(1)c}_{12}\,|E_{12}-\tilde{h}_c|\,\bar{G}^{(1)c}_{\bar{2}\bar{1}}\quad.
\end{eqnarray}
In these formulas $L_S^{(1)c}$, $Z^{(1)c}$, $L_S^{(2b)c}(E)$ and $L_S^{(2c)c}(E)$ are 
determined by solving the RG equations (\ref{L_first_rg}), (\ref{L_b_second_order}) and
(\ref{L_c_second_order}) up to the scale $\Lambda_c$. The quantities $\tilde{h}_c$ 
and $\tilde{\Gamma}_c$ are defined by (\ref{h_tilde}) and (\ref{Gamma_tilde}). 
$\tilde{h}_c$ is identical to $h^{eff}$ up to the terms linear in $J_c$, with 
$E\rightarrow L_S^{(0)}$ and taking the diagonal part
with respect to $L_S^{(0)}$. As discussed above, without significant error one can also
calculate $\tilde{\Gamma}_c(E)$ from $\Gamma^{eff}(E)$ at $\Lambda=0$, i.e.
\begin{equation}
\label{Gamma_tilde_alternative}
\tilde{\Gamma}_c(E)\,\approx\,\left\{\Gamma^{eff}(E)\right\}_d\,=\,
\left\{i\,L_S^{(2\tilde{a})Im}(E)\,+\,i\,L_S^{(2b)c}(E)\right\}_d
\quad.
\end{equation}
We note that a precise justification to take the values at $\Lambda=0$ is given in
Ref.~\cite{reininghaus_hs_preprint}.
Similiar formulas hold for the kernel $\Sigma_R(E)$ of the observable $R$ if the first
vertex is replaced by $R$ in all equations. Note that all formulas contain an additional
implicit dependence on $E$ via the cutoff $\Lambda_c$ defined by (\ref{maximal_cutoff}).

The final results (\ref{final_solution_real}) and (\ref{final_solution_imaginary})
are generic formulas including all terms $\sim J_c\,,\,J_c^2\,,\,J_c^2\ln(J_c)$ for a 
fermionic model with spin and/or orbital fluctuations. Terms $\sim\ln(J_c)$ occur at 
resonance $|E_{12}-\tilde{h}_c)|=0$, where the logarithmic term in $H_{\tilde{\Gamma}_c}$ 
becomes of order $\ln(\Lambda_c/\tilde{\Gamma}_c)\sim \ln(J_c)$ since 
$\tilde{\Gamma}_c\sim J_c^2\Lambda_c$ according to (\ref{Gamma_tilde}). 
Although the prefactor of $H_{\tilde{\Gamma}_c}(E)$ becomes very small for $E\rightarrow 0$,
derivatives of the effective Liouvillian with respect to some energy scale like voltage or 
magnetic field will show the logarithmic increase at resonance. However, we note that the
precise position and the broadening of the resonance is not exactly given by the poles of
the reduced density matrix, but can only be calculated numerically. Nevertheless, the 
difference is not very large and hardly visible. We note that the logarithmic
part $\sim J_c^2\ln(J_c)$ of (\ref{final_solution_real}) appears only in the renormalization
of the hermitian part of the Liouvillian, i.e. it does not contribute to
the rates, e.g. to the stationary transport current. However, for the time evolution
of physical quantities, it will play a crucial role due to its logarithmic dependence on
the Laplace variable $E$, leading to branch cuts in the complex plane. As we have shown,
the logarithmic terms arise from the solution of the RG equations between $\Lambda_c$
and some other cutoff scale $\Delta\equiv|E_{12}-\tilde{h}_c|$. This is generically the case 
because the solution of the RG equations at scale $\Lambda_c$ involves only the maximal
cutoff scale $\Lambda_c$ and no logarithmic terms $\sim\ln(\Lambda_c/\Delta)$ with some
smaller cutoff scale $\Delta$ can occur. Therefore, the solution at scale $\Lambda_c$
is a power series in $J_c$ with possible logarithmic contributions $\sim J_c^k(\ln J_c)^{k-1}$
from higher orders (but independent of $\Delta$).

What we have not analysed here are the 2-loop terms 
(\ref{rg_L_so_2}), (\ref{rg_G_so_2})-(\ref{rg_G_so_4}) of the RG equations and the
influence of the frequency dependence of the vertices. They provide further terms of
order $J^3$ on the r.h.s. of the RG equations and can, as we have seen above, influence
the result for the Liouvillian in order $J_c^2$ at scale $\Lambda_c$. This has been
analysed in detail in Ref.~\cite{reininghaus_hs_preprint} with the result that
all these terms nearly cancel, the only effect is a small correction for the definition
of the function $H$, given by (\ref{H_function}), which has to be replaced by
\begin{equation}
\label{H_function_corr}
\tilde{H}_\Gamma(E)\,=\,E\,\left(\ln{\Lambda_c\over \sqrt{|E|^2+\Gamma^2}}\,+\,1\right)\quad,
\end{equation}
i.e. only the factor $2$ in the argument of the logarithm is absent. This leads 
only to a small correction of order $\sim J_c^2$
for the hermitian part of the Liouvillian. Neither the terms $\sim J_c$ or 
$\sim J_c^2\ln(J_c)$ for
the hermitian part nor the terms $\sim J_c^2$ for the antihermitian part are changed.

However, the 2-loop analysis of Ref.~\cite{reininghaus_hs_preprint} is very 
important if one is interested in corrections to the rates of order
$J_c^3\ln(\Lambda_c/\Delta)$ which become $\sim J_c^3\ln(J_c)$ at resonance. Such terms
lead e.g. to experimentally accessible resonances for the differential conductance 
when the bias voltage matches with certain energy excitations like magnetic fields.
It turns out that especially the frequency dependence of the vertices is important for the
generation of such terms below $\Lambda_c$. Moreover, for $\Lambda>\Lambda_c$, even the
frequency independent part of the vertices is corrected by various terms in second order 
in $J$. For the Kondo model, it can be shown that these corrections lead to a redefinition
of the Kondo temperature. However, for more generic models with orbital fluctuations, 
it might be the case that the vertices in second order become a new structure in Liouville 
space and will also influence the prefactor of some $J_c^3\ln(\Lambda_c/\Delta)$ terms.

\subsection{Strong coupling limit}
\label{sec:4.5}

When the conditions of weak coupling are not fulfilled no systematic truncation scheme can
be applied to solve the RG equations. Nevertheless, since the RG equations are a set of
fully self-consistent equations with the renormalized Liouvillian appearing in the 
denominator of all resolvents, there is some hope that the equations can also lead to
reliable results when the coupling constants become of order one. A divergence of the
coupling constant like in poor man scaling equations is not expected since the 
relaxation and dephasing rates will also increase for increasing coupling and will cut
of any divergence. However, whether quantitatively reliable results can be expected is
not at all clear and certainly an interesting field for future research.

Using RTRG-FS, the problem of solving the full RG equations numerically in any truncation 
scheme lies in the frequency dependence of the vertices and the Liouvillian. Therefore, to
start with a minimal ansatz, one possibility is to retain only the dependence on the
Laplace variable $E$ since this variable takes discrete values shifted by the
chemical potentials of the reservoirs, see (\ref{shift_matsubara}). So e.g. for the
fermionic problem of spin/orbital fluctuations, where the RG equations are given by
(\ref{rg_L_so_1_matsubara})-(\ref{rg_G_so_4_matsubara}), a minimal approach would
consist in omitting the dependence on all other frequencies and take only the first
term on the r.h.s. of the RG equations into account. At zero temperature, this leads 
to the equations
\begin{eqnarray}
\label{rg_L_so_strong_zw}
-{dL_S(E)\over d\Lambda}\,&=&\,
-\bar{G}_{12}(E)\,
\int_0^\Lambda d\omega_2\,{1 \over E_{12}+i\Lambda+i\omega_2-L_S(E_{12})}
\,\bar{G}_{\bar{2}\bar{1}}(E_{12})\quad,\\
\label{rg_G_so_strong_zw}
-{d\bar{G}_{11'}(E)\over d\Lambda}\,&=&\,
-i\bar{G}_{12}(E)\,{1\over E_{12}+i\Lambda-L_S(E_{12})}\,\bar{G}_{\bar{2}1'}(E_{12})
\,\, - \,\, (1\leftrightarrow 1')
\quad.
\end{eqnarray}
Performing the integral leads to 
\begin{eqnarray}
\label{rg_L_so_strong}
{dL_S(E)\over d\Lambda}\,&=&\,
-i\,\bar{G}_{12}(E)\,{\cal{L}}_\Lambda(E_{12}-L_S(E_{12}))
\,\bar{G}_{\bar{2}\bar{1}}(E_{12})\quad,\\
\label{rg_G_so_strong}
{d\bar{G}_{11'}(E)\over d\Lambda}\,&=&\,
\bar{G}_{12}(E)\,{1\over \Lambda-iE_{12}+i L_S(E_{12})}\,\bar{G}_{\bar{2}1'}(E_{12})
\,\, - \,\, (1\leftrightarrow 1')
\quad,
\end{eqnarray}
where ${\cal{L}}_\Lambda(z)=\ln({2\Lambda-iz\over \Lambda-iz})$ has been defined in (\ref{L_function}).
The initial conditions are given by (\ref{L_initial_weak_coupling}) and (\ref{G_initial_weak_coupling}).

At finite temperature a similiar set of equations can be set up, the only difference is that
integrals have to be replaced by sums over Matsubara frequencies, and we have to use the smeared
theta function $\theta_T(\omega)$ defined in (\ref{theta_T}). Instead of
(\ref{rg_L_so_strong_zw}) and (\ref{rg_G_so_strong_zw}), we obtain the equations (for simplicity, 
we assume that the temperatures of all reservoirs are the same)
\begin{eqnarray}
\label{rg_L_so_strong_finite_T_zw}
-{dL_S(E)\over d\Lambda}\,&=&\,
-\bar{G}_{12}(E)\,
\left\{2\pi T \,\sum_{n=0}^\infty \,
\,{\theta_T(\Lambda-|\omega_n|) \over E_{12}+i\Lambda_T+i\omega_n-L_S(E_{12})}\right\}
\,\bar{G}_{\bar{2}\bar{1}}(E_{12})\quad,\\
\label{rg_G_so_strong_finite_T_zw}
-{d\bar{G}_{11'}(E)\over d\Lambda}\,&=&\,
-i\bar{G}_{12}(E)\,{1\over E_{12}+i\Lambda_T-L_S(E_{12})}\,\bar{G}_{\bar{2}1'}(E_{12})
\,\, - \,\, (1\leftrightarrow 1')
\quad,
\end{eqnarray}
where $\omega_n$ are the discrete Matsubara frequencies and $\Lambda_T$ is the Matsubara
frequency lying closest to $\Lambda$. The sum in (\ref{rg_L_so_strong_finite_T_zw}) can
be performed analytically with the result 
\begin{eqnarray}
\label{rg_L_so_strong_finite_T}
{dL_S(E)\over d\Lambda}\,&=&\,
-i\,\bar{G}_{12}(E)\,{\cal{L}}_\Lambda(E_{12}-L_S(E_{12}))
\,\bar{G}_{\bar{2}\bar{1}}(E_{12})\quad,\\
\label{rg_G_so_strong_finite_T}
{d\bar{G}_{11'}(E)\over d\Lambda}\,&=&\,
\bar{G}_{12}(E)\,{1\over \Lambda_T-iE_{12}+i L_S(E_{12})}\,\bar{G}_{\bar{2}1'}(E_{12})
\,\, - \,\, (1\leftrightarrow 1')
\quad,
\end{eqnarray}
i.e. the same result as for $T=0$ but $\Lambda$ is replaced by $\Lambda_T$ in the
RG equation for the vertex, and the function ${\cal{L}}_\Lambda(z)$ has to be
replaced by the more general expression
\begin{equation}
\label{L_function_finite_T}
{\cal{L}}_\Lambda(z)\,=\,
\Psi({2\Lambda_T-iz\over 2\pi T})\,-\,\Psi({\Lambda+\pi T-iz\over 2\pi T})\,+\,
{\Lambda-\Lambda_T+\pi T\over 2\Lambda_T-iz}\quad,
\end{equation}
where $\Psi(z)$ is the Digamma function with asymptotic properties given by 
(\ref{psi_z_infty}) and (\ref{psi_z_zero}). Furthermore, it can be shown that
the initial conditions for the Liouvillian and the vertices are the same
as those for zero temperature.

These sets of RG equations will be analysed in Sec.~\ref{sec:5.3} for the 
isotropic Kondo model in the absence of a magnetic field. As expected the
coupling constants stay finite and a quite promising agreement with results
from NRG calculations in equilibrium is obtained. However, it turns out that
the equations are unstable against exponentially small changes for the
inital condition for the relaxation rate $\Gamma$. Therefore, it has to be
analysed in future how the equations can be improved by higher order terms
in order to stabilize them. One possibility is to consider the
frequency dependence of the vertices in leading order by taking the  
frequency independent vertices
on the r.h.s. of the RG equations, together with the consideration of 
the next order terms of the RG equations. A numerical analysis of the 
full frequency dependence is certainly very time consuming but it has to
be expected that the numerical solution will be very stable since all
the imaginary parts of the denominators are positive. All these investigations
will be the subject of future research.

\section{Application: The nonequilibrium Kondo model}
\label{sec:5}

In this section we will apply the renormalization group formalism developed in
Sec.~\ref{sec:4} to the nonequilibrium isotropic Kondo model in the absence of a magnetic field.
For weak coupling, we evaluate the general equations of Sec.~\ref{sec:4.4} to get
all physical quantities up to second order in the coupling. The results presented are
a special case of the more general treatment in Ref.~\cite{reininghaus_hs_preprint}
where the anisotropic Kondo model in the presence of a magnetic field has been considered,
and all quantities were calculated up to third order in the coupling. In Sec.~\ref{sec:5.3}
we will discuss some preliminary results in the strong coupling regime.

\subsection{Model and algebra of basis operators in Liouville space}
\label{sec:5.1}

{\bf Model.} We discuss here the Kondo model introduced in Sec.~\ref{sec:2.2}. The quantum
system is a spin ${1\over 2}$ which is coupled to the spins of several reservoirs by 
the exchange coupling term 
\begin{equation}
\label{V_kondo}
V\,=\,{1\over 2}\,g_{11'}\,:a_1 a_{1'}:\quad,
\end{equation}
with
\begin{equation}
\label{g_kondo}
g_{11'}\,=\,{1\over 2}\,\left\{
\begin{array}{cl}
J_{\alpha\alpha'}\,\underline{S}\cdot\underline{\sigma}_{\sigma\sigma'}\, 
&\mbox{for }\eta=-\eta'=+ \\
-J_{\alpha'\alpha}\,\underline{S}\cdot\underline{\sigma}_{\sigma'\sigma}\, 
&\mbox{for }\eta=-\eta'=- 
\end{array}
\right.\quad.
\end{equation}
$J_{\alpha\alpha'}=J_{\alpha'\alpha}$ are the exchange coupling constants, $\underline{S}$ is
the spin ${1\over 2}$ operator of the quantum system, and $\underline{\sigma}$
are the Pauli matrices. We have taken here the isotropic case where the 
exchange couplings are the same for all spatial directions. $\alpha$ is the
reservoir index. In principle the formalism presented here is applicable to
an arbitrary number of reservoirs kept at chemical potentials $\mu_{\alpha}$,
but sometimes we refer to the case of two reservoirs ($\alpha=\pm\equiv L,R$)
with chemical potentials given by
\begin{equation}
\mu_{\alpha}\,=\,\alpha\,{V\over 2}\quad,
\end{equation}
where $V$ is the applied voltage and we use units $e=\hbar=1$. 

We assume here the case of zero magnetic field, i.e. the Hamiltonian and
the Liouvillian of the quantum system are initially set to zero 
\begin{equation}
\label{H_L_quantum_system}
H_S\,=\,0\quad,\quad L_S^{(0)}\,=\,0\quad.
\end{equation}

Using (\ref{G_vertex_liouville}) we get for the vertex in Liouville space
\begin{equation}
\label{G_kondo}
G^{pp}_{11'}\,=\,{1\over 2}\,\left\{
\begin{array}{cl}
J_{\alpha\alpha'}\,\underline{L}^p\cdot\underline{\sigma}_{\sigma\sigma'}\, 
&\mbox{for }\eta=-\eta'=+ \\
-J_{\alpha'\alpha}\,\underline{L}^p\cdot\underline{\sigma}_{\sigma'\sigma}\, 
&\mbox{for }\eta=-\eta'=- 
\end{array}
\right.\quad,
\end{equation}
where $\underline{L}^p$ are two fundamental spin vector operators in
Liouville space defined by ($A$ is an arbitrary operator in Hilbert space)
\begin{equation}
\label{spin_operators_liouville}
\underline{L}^+\,A\,=\,\underline{S}\,A\quad,\quad
\underline{L}^-\,A\,=\,-\,A\,\underline{S}\quad.
\end{equation}

For convenience, we will consider in the following always the case
$\eta=-\eta'=+$ when discussing some vertex $G_{11'}$, since the
antisymmetry $G_{11'}=-G_{1'1}$ gives trivially the other case $\eta=-\eta'=-$.

For $\eta=-\eta'=+$ and using (\ref{current_vertex_liouville}), 
the vertex of the current operator in Liouville space can be written as
\begin{equation}
\label{current_kondo}
(I^{\gamma})^{pp}_{11'}\,=\,c^\gamma_{\alpha\alpha'}\,p\,G_{11'}^{pp}\quad,
\end{equation}
with
\begin{equation}
\label{c_gamma_symbol}
c^\gamma_{\alpha\alpha'}\,=\,
-{1\over 2}\,(\delta_{\alpha\gamma}-\delta_{\alpha'\gamma}) 
\,=\,-\,c^\gamma_{\alpha'\alpha} \quad.
\end{equation}
The kernel (\ref{value_sigma_observable}) for the current is denoted by $\Sigma_\gamma(E)$.

Since the Kondo model emerges after integrating out the
charge degrees of freedom, a high-frequency cutoff is needed, which we
describe by the cutoff function (\ref{cutoff})
\begin{equation}
\label{cutoff_kondo}
\rho(\omega)\,=\,{D^2\over \omega^2+D^2}\quad,
\end{equation}
where $D$ is the physical bandwidth (which is fixed here and is not changed
during the RG procedure).

{\bf Basis operators in Liouville space.} Due to spin rotational symmetry not all $16$ matrix
elements of each operator in Liouville space are needed. To get a minimal set
for the isotropic case, we first define two scalar operators by 
\begin{equation}
\label{scalar_operators}
L^a\,=\,{3\over 4}\,+\,\underline{L}^+\cdot\underline{L}^-\quad,\quad
L^b\,=\,{1\over 4}\,-\,\underline{L}^+\cdot\underline{L}^-\quad,
\end{equation}
and three vector operators by
\begin{eqnarray}
\label{vector_operator_1}
\underline{L}^1\,&=&\,{1\over 2}\,(\underline{L}^+\,-\,\underline{L}^-\,-\,
2i\,\underline{L}^+\wedge\underline{L}^-)\quad,\\
\label{vector_operator_2}
\underline{L}^2\,&=&\,-{1\over 2}\,(\underline{L}^+\,+\,\underline{L}^-)\quad,\\
\label{vector_operator_3}
\underline{L}^3\,&=&\,{1\over 2}\,(\underline{L}^+\,-\,\underline{L}^-\,+\,
2i\,\underline{L}^+\wedge\underline{L}^-)\quad.
\end{eqnarray}
Using spin rotational invariance together with
spin conservation (i.e. the parity of $s-s'$ and $\bar{s}-\bar{s}'$
must be the same for any matrix element $(L_S)_{ss',\bar{s}\bar{s}'}$) and the property 
$\mbox{Tr}_S\,L_S(E)=\mbox{Tr}_S\,L^a=0$,
the Liouvillian $L_S(E)$ and the kernel $\Sigma_\gamma(E)$ can
be represented as
\begin{eqnarray}
\label{representation_liouvillian}
L_S(E)\,&=&\,(h(E)\,-\,i\Gamma(E))\,L^a \quad,\quad\\
\label{representation_sigma_gamma}
\mbox{Tr}_S\,\Sigma_\gamma(E)\,&=&\,i\Gamma_\gamma(E)\,\mbox{Tr}_S\,L^b \quad,
\end{eqnarray}
where the trace over the quantum system has been written in the second equation
since only this combination occurs for the observables. Many equations for the
observable are only valid in this sense. Therefore, we will use in the following
the convention that each time the current vertex or the current kernel 
occurs, we act implicitly with the trace over the quantum system from the left.
The kernel for the current is written such that $\Gamma_\gamma(0^+)$ gives the
stationary current. Using $\mbox{Tr}_S\,L^b\,A=\mbox{Tr}_S\,A$ (see 
(\ref{trace_system_basis_operators}) below) and 
$\mbox{Tr}_S\,\tilde{\rho}_S(E)=i/E$
(see (\ref{zw_conservation})), we get from (\ref{solution_observable}) and
(\ref{zw_observable_stationary})
\begin{equation}
\label{current_representation}
\langle I^\gamma \rangle(E)\,=\,{i\over E}\Gamma_\gamma(E)\quad,\quad
{\langle I^\gamma \rangle}^{st}\,=\,\Gamma_\gamma(0^+)\quad.
\end{equation}
The quantities $h(E)$ and $\Gamma(E)$ correspond to the real and negative imaginary parts
of the eigenvalues of $L_S(E)$, respectively.

To express the vertices $G_{11'}\equiv G_{+\alpha\sigma,-\alpha'\sigma'}$ (note
that we use implicitly $\eta=-\eta'=+$) in terms of basis operators, we need 
an appropriate representation for the reservoir spin labels $\sigma$ and $\sigma'$.
We introduce the tensor operators
\begin{eqnarray}
\label{tensor_ab}
(L^\chi)_{\sigma\sigma'}\,&=&\,L^\chi\,\delta_{\sigma\sigma'}
\hspace{0.25cm}\quad\mbox{for }\chi\,=\,a,b \quad,\\
\label{tensor_123}
(L^\chi)_{\sigma\sigma'}\,&=&\,\underline{L}^\chi\cdot
\underline{\sigma}_{\sigma\sigma'}\quad\mbox{for }\chi\,=\,1,2,3\quad.
\end{eqnarray}
The operators $L^\chi$ act simultaneously in Liouville space of the quantum system 
and in the space of the reservoir spin labels. We note that the notation for $L^a$ and
$L^b$ is ambigious, however, it should be always clear from
the context whether $L^a$ and $L^b$ also act in reservoir spin space or not (e.g. in 
(\ref{representation_liouvillian}) and (\ref{representation_sigma_gamma}) this is not
the case). The vertices are then represented as (implicitly for $\eta=-\eta'=+$)
\begin{eqnarray}
\label{G_representation}
\bar{G}_{11'}(E)\,&=&\,\sum_{\chi=a,b,1,2,3}\,\bar{G}^\chi_{\alpha\alpha'}(E)\,
(L^\chi)_{\sigma\sigma'}\quad,\\
\label{I_representation}
\bar{I}^\gamma_{11'}(E)\,&=&\,\sum_{\chi=a,b,1,2,3}\,\bar{I}^{\gamma\,\chi}_{\alpha\alpha'}(E)\,
(L^\chi)_{\sigma\sigma'}\quad.
\end{eqnarray}
In this way we only need to solve RG equations for the c-numbers $\bar{G}^\chi_{\alpha\alpha'}(E)$
and $\bar{I}^{\gamma\,\chi}_{\alpha\alpha'}(E)$ (or for the matrices $\bar{G}^\chi(E)$ and 
$\bar{I}^{\gamma\,\chi}(E)$ in reservoir space). The algebra of these representations is
closed as we will see in the following.
\begin{table}
\begin{center}
\begin{tabular}{c|ccccc}
\hline\noalign{\smallskip}
      & $L^a$ & $L^b$ & $L^1$      & $L^2$                   & $L^3$   \\
\noalign{\smallskip}\hline\noalign{\smallskip}
$L^a$ & $L^a$ & $0$   & $0$        & $L^2$                   & $L^3$  \\
$L^b$ & $0$   & $L^b$ & $L^1$      & $0$                     & $0$    \\
$L^1$ & $L^1$ & $0$   & $0$        & $L^1$                   & $3L^b$ \\
$L^2$ & $L^2$ & $0$   & $0$        & ${1\over 2}(L^a + L^2)$ & $L^3$  \\
$L^3$ & $0$   & $L^3$ & $L^a+2L^2$ & $0$                     & $0$    \\
\noalign{\smallskip}\hline
\end{tabular}
\end{center}
\caption{Algebra of Liouville basis operators. The table shows the
product $L^\chi\,L^{\chi'}$. The same comes out for the combination
$\left((L^\chi)^T\,(L^{\chi'})^T\right)^T$ but the sign of the operators
$L^1$, $L^2$, and $L^3$ has to be changed in the table.}
\label{tab:1}       
\end{table}

We note some important transformations and properties of the basis operators which will 
be frequently needed in the following. We define the transpose by only interchanging the 
reservoir spin indices
\begin{equation}
\label{def_transpose}
\left((L^\chi)^T\right)_{\sigma\sigma'}\,=\,(L^\chi)_{\sigma'\sigma}\quad.
\end{equation}
As a consequence, using $\bar{G}_{11'}=-\bar{G}_{1'1}$, we get for the representation 
of the vertex for all cases of $\eta$ and $\eta'$
\begin{equation}
\label{G_representation_all_eta}
\bar{G}_{11'}(E)\,=\,\sum_{\chi=a,b,1,2,3}\,
\left\{
\begin{array}{cl}
\bar{G}^\chi(E)\,L^\chi\quad
&\mbox{for }\eta=-\eta'=+ \\
-\,\bar{G}^\chi(E)^T\,(L^\chi)^T\quad
&\mbox{for }\eta=-\eta'=-
\end{array}
\right.\quad
\end{equation}
where $\bar{G}^\chi(E)$ is considered as a matrix in the reservoir indices.

In contrast, the $c$-transform $(L^\chi)^c$, defined in (\ref{c_transformation}) 
for operators in Liouville space, is defined only with respect to the degrees of 
freedom of the local quantum system. We get
\begin{equation}
\label{c_transpose_basis_operators}
(L^\chi)^c\,=\,L^\chi \quad\mbox{for }\chi\,=\,a,b,1,3\quad,\quad
(L^2)^c\,=\,-L^2 \quad.
\end{equation}
Applying this transformation to the representation of the Liouvillian, the kernel,
and the vertices in terms of the basis operators, and using the properties
(\ref{rg_L_c_transform})-(\ref{rg_R_c_transform}), we obtain the following
helpful relations (note that $h(E)$ and $\Gamma(E)$ are defined as being real for
arbitrary $E$)
\begin{eqnarray}
\nonumber
&& h(E)\,=\,-h(-E^*)\quad,\quad \Gamma(E)\,=\,\Gamma(-E^*)\quad,\quad 
\Gamma_\gamma(E)^*\,=\,\Gamma_\gamma(-E^*)\quad,\\
\nonumber
&& \bar{G}^\chi_{\alpha\alpha'}(E)^*\,=\,-\bar{G}^\chi_{\alpha'\alpha}(-E^*)\quad,\quad
 \bar{I}^{\gamma\,\chi}_{\alpha\alpha'}(E)^*\,=\,
-\bar{I}^{\gamma\,\chi}_{\alpha'\alpha}(-E^*)\quad
\mbox{for }\chi=a,b,1,3 \quad,\\
\label{symmetry_h_gamma}
&& \bar{G}^2_{\alpha\alpha'}(E)^*\,=\,\bar{G}^2_{\alpha'\alpha}(-E^*)\quad,\quad
\bar{I}^{\gamma\,2}_{\alpha\alpha'}(E)^*\,=\,
\bar{I}^{\gamma\,2}_{\alpha'\alpha}(-E^*)\quad.
\end{eqnarray}
Denoting by $\mbox{Tr}_\sigma$ the trace with respect to reservoir spin
indices, we get
\begin{equation}
\label{trace_spin_basis_operators}
\mbox{Tr}_\sigma\,L^\chi\,=\,2\,L^\chi \quad\mbox{for }\chi\,=\,a,b\quad,\quad
\mbox{Tr}_\sigma\,L^\chi\,=\,0 \quad\mbox{for }\chi\,=\,1,2,3\quad.
\end{equation}
If the trace over the quantum system acts left to the basis operators, we obtain
\begin{equation}
\label{trace_system_basis_operators}
\mbox{Tr}_S\,L^\chi\,=\,0 \quad\mbox{for }\chi\,=\,a,2,3\quad,\quad
\mbox{Tr}_S\,L^b\,A\,=\,\mbox{Tr}_S\,A \quad.
\end{equation}
Table \ref{tab:1} shows the closed algebra $L^\chi L^{\chi'}$ of the basis operators 
$L^\chi$. The same algebra holds for the combination 
\begin{equation}
\label{algebra_transpose}
\left((L^\chi)^T\,(L^{\chi'})^T\right)^T\quad,
\end{equation} 
but a different sign occurs for the operators $L^1$, $L^2$, and $L^3$. Note that one is not
allowed to write $L^{\chi'}L^\chi$ for (\ref{algebra_transpose}) because the transpose
is only acting in reservoir spin space but not in Liouville space of the quantum system.

To evaluate the RG equations in lowest order, we will either encounter terms 
of the form $A_{12}B_{\bar{2}1'}$ (where we sum over the index $2$) or
$K(E_{11'})A_{11'}B_{\bar{1}'\bar{1}}$ (where we sum over the indices $1$ and $1'$).
Here, $A_{11'}$ and $B_{11'}$ are arbitrary vertices represented in the form
(\ref{G_representation}), and $K(E_{11'})$ is an arbitrary function of the variable 
$E_{11'}=E+\bar{\mu}_{11'}=E+\eta\mu_\alpha+\eta'\mu_{\alpha'}$ introduced in
(\ref{E_convention}). Using the antisymmetry $A_{11'}=-A_{1'1}$ and $B_{11'}=-B_{1'1}$,
and inserting the representation (\ref{G_representation_all_eta}), one obtains after 
summing over the two possibilities $\eta=-\eta'=\pm$ the helpful identities
\begin{eqnarray}
\label{evaluation_rg_identity_A}
A_{12}\,B_{\bar{2}1'}\,&=&\,\left\{
\begin{array}{cl}
A^\chi\,B^{\chi'}\,L^\chi\,L^{\chi'} \quad
&\mbox{for }\eta=-\eta'=+ \\
(A^\chi)^T (B^{\chi'})^T\,(L^\chi)^T\,(L^{\chi'})^T \quad
&\mbox{for }\eta=-\eta'=- 
\end{array}
\right.\quad,\\
\label{evaluation_rg_identity_B}
K(E_{11'})\,A_{11'}\,B_{\bar{1}'\bar{1}}\,&=&\,2\,K(E_{\alpha\alpha'})\,
A^\chi_{\alpha\alpha'}\,B^{\chi'}_{\alpha'\alpha}\,\mbox{Tr}_\sigma\,L^\chi\,L^{\chi'}\quad,
\end{eqnarray}
where we use the definition
\begin{equation}
\label{E_alpha}
E_{\alpha\alpha'}\,=\,E+\mu_\alpha-\mu_{\alpha'}\quad.
\end{equation}
$A^\chi$ and $B^{\chi'}$ in (\ref{evaluation_rg_identity_A}) are matrices in 
reservoir space and we sum implicitly over all $\chi,\chi',\alpha,\alpha'$.
Using (\ref{trace_spin_basis_operators}) and the algebra of the basis operators, 
we see that (\ref{evaluation_rg_identity_B}) is only nonzero for the combinations
\begin{equation}
\label{unzero_combinations}
(\chi\chi')\,=\,(aa),(bb),(13),(22),(31)\quad.
\end{equation}
For the evaluation of the RG equations in strong coupling we will also need the
identity
\begin{equation}
\label{evaluation_rg_identity_C}
K(E_{12})\,A_{12}\,B_{\bar{2}1'}\,=\,\left\{
\begin{array}{cl}
K(E_{\alpha\alpha_2})\,A^\chi_{\alpha\alpha_2}\,B^{\chi'}_{\alpha_2\alpha'}\,L^\chi\,L^{\chi'} \quad
&\mbox{for }\eta=-\eta'=+ \\
K(E_{\alpha_2\alpha})\,A^{\chi}_{\alpha_2\alpha}\,B^{\chi'}_{\alpha'\alpha_2}
\,(L^\chi)^T\,(L^{\chi'})^T \quad
&\mbox{for }\eta=-\eta'=- 
\end{array}
\right.\quad.
\end{equation}

\subsection{RG in weak coupling}
\label{sec:5.2}

In this section we apply the general scheme described in Sec.~\ref{sec:4.4} to the Kondo
problem. Since no magnetic field is assumed, the parameter $\Lambda_c$ defined in 
(\ref{maximal_cutoff}) is given by
\begin{equation}
\label{maximal_cutoff_kondo}
\Lambda_c\,=\,\mbox{max}\{|E|,\mu_\alpha\}\quad,
\end{equation}
i.e. $\Lambda_c$ is just the maximum of the Laplace variable $E$ and the voltage $V$.
To stay in the weak coupling regime, we assume that this energy scale is much larger than
the Kondo temperature $T_K$ which is the energy scale where the RG equations for the
leading order vertices diverge (see (\ref{solution_leading_order_kondo}) below)
\begin{equation}
\label{weak_coupling_condition_kondo}
\mbox{max}\{|E|,V\}\,\gg\,T_K\,=\,\Lambda_0\,e^{-{1\over 2J_0}}\quad.
\end{equation}
Here, $\Lambda_0$ is the initial value for $\Lambda$ and $J_0$ is the overall scale of 
the initial exchange couplings. 

{\bf Leading order RG.} We start with the RG for the leading order vertices $\bar{G}_{11'}^{(1)}$,
$\tilde{G}_{11'}^{(1)}$ and $\bar{I}^{\gamma(1)}_{11'}$, defined by the RG equations
(\ref{reference_solution_so}) and (\ref{G_tilde_leading}), and the result 
(\ref{current_operator_leading}) for the current vertex.
Using (\ref{G_kondo}) and (\ref{vector_operator_1})-(\ref{vector_operator_3}), we make the
same ansatz as for the initial form ($\eta=-\eta'=+$ is always assumed)
\begin{equation}
\label{leading_order_vertices}
\bar{G}^{(1)}_{11'}\,=\,-J\,L^2 \quad,\quad
\tilde{G}^{(1)}_{11'}\,=\,{1\over 2}\,J\,(L^1\,+\,L^3)\quad,\quad
\bar{I}^{\gamma(1)}_{11'}\,=\,{1\over 2}\,J^\gamma\,L^1\quad,
\end{equation}
where, in the equation for the current vertex, we assume implicitly that $\mbox{Tr}_S$
is taken from the left which cancels the contribution of $L^3$ according to
(\ref{trace_system_basis_operators}). $J=J^T$ is the symmetric matrix of the
exchange couplings in reservoir space and the antisymmetric matrix 
$J^\gamma=-(J^\gamma)^T$ is defined by
\begin{equation}
\label{J_gamma}
J^\gamma_{\alpha\alpha'}\,=\,c^\gamma_{\alpha\alpha'}\,J_{\alpha\alpha'}\quad.
\end{equation}

To prove (\ref{leading_order_vertices}) we use this ansatz together with (\ref{evaluation_rg_identity_A}) 
and the algebra of the basis operators to evaluate the r.h.s. of the RG equations as
\begin{eqnarray}
\label{rg_leading_bar_G}
{d\over d\Lambda}\bar{G}^{(1)}_{11'}\,&=&\,{1\over\Lambda}J^2\left\{L^2 L^2-((L^2)^T(L^2)^T)^T)\right\}
\,=\,{1\over\Lambda} J^2 L^2\quad,\\
{d\over d\Lambda}\tilde{G}^{(1)}_{11'}\,&=&\,-{1\over 2\Lambda}J^2
\left\{(L^1+L^3)L^2-((L^2)^T(L^1+L^3)^T)^T)\right\}
\,=\,-{1\over\Lambda} J^2 (L^1+L^3)\quad,
\end{eqnarray}
We see that (\ref{leading_order_vertices}) is fulfilled provided that the matrix $J$
is defined by the well known poor man scaling equation of the Kondo problem
\begin{equation}
\label{poor_man_scaling_kondo}
{dJ\over d\Lambda}\,=\,-{1\over\Lambda}\,J^2\quad.
\end{equation}
The current follows directly from (\ref{current_operator_leading}). The poor man scaling
equation can be easily solved with the ansatz (if this is also fulfilled initially)
\begin{equation}
\label{ansatz_leading_order_kondo}
J_{\alpha\alpha'}\,=\,2\,\sqrt{x_\alpha\,x_{\alpha'}}\,\bar{J}\quad,\quad 
\sum_\alpha\,x_\alpha\,=\,1 \quad,
\end{equation}
with $\bar{J}$ being the overall scale of all exchange couplings given by 
\begin{equation}
\label{solution_leading_order_kondo}
{d\bar{J}\over d\Lambda}\,=\,-{2\,\bar{J}^2\over\Lambda}\quad,\quad
\bar{J}\,=\,{1\over 2\,\ln{\Lambda\over T_K}}\quad,
\end{equation}
where $T_K$ is the Kondo temperature given by (\ref{weak_coupling_condition_kondo}). As a
consequence, under the condition (\ref{weak_coupling_condition_kondo}), the leading order 
vertices will stay small in the regime $\Lambda>\Lambda_c$.

{\bf Vertices in second order.} The vertices $\bar{G}^{(2)}_{11'}$ and $\bar{I}^{\gamma(2)}_{11'}$ follow
directly from (\ref{solution_G_second_order}) by inserting (\ref{leading_order_vertices}) and using
(\ref{evaluation_rg_identity_A}) 
\begin{eqnarray}
\label{rg_G_second_order_kondo}
\bar{G}^{(2)}_{11'}\,&=&\,-i{\pi\over 2}(-{1\over 2})J^2\left\{L^2 (L^1+L^3)-((L^2)^T(L^1+L^3)^T)^T)\right\}
\,=\,i{\pi\over 2} J^2 L^3\quad,\\
\nonumber
\bar{I}^{\gamma(2)}_{11'}\,&=&\,-i{\pi\over 2}{1\over 4}
\left\{J^\gamma J L^1(L^1+L^3)-J J^\gamma ((L^1)^T(L^1+L^3)^T)^T)\right\}\\
\label{rg_I_second_order_kondo}
\,&=&\,-i{3\pi\over 8}(J^\gamma J-JJ^\gamma)L^b\quad.
\end{eqnarray}
As we will see below, $\bar{G}^{(2)}_{11'}$ is essential for the leading order result for 
the current, whereas $\bar{I}^{\gamma(2)}_{11'}$ is not important in this order.

{\bf Liouvillian and current kernel.} To evaluate the Liouvillian $L^{eff}_S(E)$ and the current kernel
$\Sigma_\gamma(E)$ from (\ref{final_solution_decomposition})-(\ref{final_solution_imaginary}), we
need the identity (\ref{evaluation_rg_identity_B}) for $B\equiv \bar{G}^{(1)}_{11'}=-J L^2$ or
$B\equiv \bar{G}^{(2)}_{11'}=i{\pi\over 2} J^2 L^3$. Due to (\ref{unzero_combinations}), the
corresponding possibilities for $A$ are $A=\bar{G}^{(1)}_{11'}=-J L^2$ or 
$A=\bar{I}^{\gamma(1)}_{11'}=(1/2)J^\gamma L^1$ with the result
\begin{eqnarray}
\label{G1_G1_combination}
K(E_{11'})\bar{G}^{(1)}_{11'}\bar{G}^{(1)}_{\bar{1}'\bar{1}}\,&=&\,
2K(E_{\alpha\alpha'})(J_{\alpha\alpha'})^2 L^a \quad,\\
\label{I1_G2_combination}
K(E_{11'})\bar{I}^{\gamma(1)}_{11'}\bar{G}^{(2)}_{\bar{1}'\bar{1}}\,&=&\,
3\pi i K(E_{\alpha\alpha'})c^\gamma_{\alpha\alpha'} 
J_{\alpha\alpha'}(J^2)_{\alpha\alpha'}L^b \quad.
\end{eqnarray}

As a consequence, together with $L^{(0)}_S=0$, we find directly from 
(\ref{L_first_order}), (\ref{L_b_second_order}), (\ref{L_2a_real}) and (\ref{L_2a_imag}) that
\begin{equation}
\label{zero_components}
\Sigma_\gamma^{(1)}(E)\,=\,L_S^{(2b)}(E)\,=\,
\Sigma_\gamma^{(2\tilde{a})Re}(E)\,=\,\Sigma_\gamma^{(2\tilde{a})Im}(E)\,=\,0\quad.
\end{equation}

$L^{(1)}_S$ and $Z^{(1)}$ follow from (\ref{L_first_rg}). For $L_S^{(1)}$ we need 
$K\rightarrow \mu_\alpha-\mu_{\alpha'}$ in (\ref{G1_G1_combination}) which gives zero
due to $\alpha\leftrightarrow\alpha'$. As a consequence we get also $\tilde{h}_c=0$ from (\ref{h_tilde}).
For $Z^{(1)}$ we need $K\rightarrow 1$ in (\ref{G1_G1_combination}). This gives
\begin{equation}
\label{rg_Z_kondo}
{d\over d\Lambda}Z^{(1)}\,=\,-(\mbox{Tr}_\alpha J^2)L^a\,=\,
{d\over d\Lambda} (\mbox{Tr}_\alpha J) L^a \quad,
\end{equation}
where $\mbox{Tr}_\alpha$ is the trace with respect to the reservoir indices, and we have
used the poor man scaling equation (\ref{poor_man_scaling_kondo}) in the last step. As
a result we find
\begin{equation}
\label{L_first_order_kondo}
L^{(1)}_S\,=\,0 \quad,\quad Z^{(1)}\,=\,(\mbox{Tr}_\alpha J)\,L^a\quad.
\end{equation}
To be precise one should subtract from $Z^{(1)}$ a contribution where $J$ is replaced by
its initial value $J_0$. However, this contribution is very small and vanishes in the scaling limit
$J_0\rightarrow 0$, $D\rightarrow \infty$, such that the Kondo temperature remains constant.

$L_S^{(2c)}(E)$ and $\Sigma_\gamma^{(2c)}(E)$ follow from (\ref{L_c_second_order}). We can replace 
$Z^{(1)}\rightarrow \mbox{Tr}_\alpha J$ in this equation since $L^a L^2=L^2 L^a=L^2$.
Thus, we get $\Sigma_\gamma^{(2c)}(E)=0$ since no combination applies according to 
(\ref{unzero_combinations}). For $L_S^{(2c)}(E)$, we get the form (\ref{G1_G1_combination}) with 
$K\rightarrow E_{\alpha\alpha'}$.
The part $\mu_\alpha-\mu_{\alpha'}$ does not contribute due to $\alpha\leftrightarrow\alpha'$.
This gives
\begin{equation}
\label{rg_L_2c_kondo}
{d\over d\Lambda}L_S^{(2c)}(E)\,=\,-{1\over\Lambda}E
(\mbox{Tr}_\alpha J)(\mbox{Tr}_\alpha J^2) L^a
\,=\,{1\over 2}E{d\over d\Lambda}(\mbox{Tr}_\alpha J)^2 L^a \quad,
\end{equation}
where we again used the poor man scaling equation (\ref{poor_man_scaling_kondo}) in the
last step. So we get
\begin{equation}
\label{L_2c_kondo}
L^{(2c)}_S(E)\,=\,{1\over 2}\,E\,(\mbox{Tr}_\alpha J)^2\, L^a \quad,\quad 
\Sigma_\gamma^{(2c)}(E)\,=\,0\quad.
\end{equation}
$L^{(2c)}_S(E)$ is a contribution of second order in $J$ with the same form as the linear order
term $-E Z^{(1)}=-E (\mbox{Tr}_\alpha J)L^a$ for $L^{(1)}_S(E)$ from (\ref{L_first_order_kondo}).
Therefore, it is an unimportant contribution. 

The most important terms arise from $\Sigma_\gamma^{(2b)}(E)$, $L_S^{(2\tilde{a})Re}(E)$ and
$L_S^{(2\tilde{a})Im}(E)$, which can be calculated from (\ref{L_b_second_order}), 
(\ref{L_2a_real}) and (\ref{L_2a_imag}). For $\Sigma_\gamma^{(2b)}(E)$, the combination
(\ref{I1_G2_combination}) with $K\rightarrow E_{\alpha\alpha'}$ applies. We see that
only the vertex $\bar{G}^{(2)}_{11'}$ contributes but not $\bar{I}^{\gamma(2)}_{11'}$. The part
from $E$ gives zero due to $\alpha\leftrightarrow\alpha'$. It remains the part
\begin{equation}
\label{Sigma_2b_kondo}
{d\over d\Lambda}\Sigma_\gamma^{(2b)}(E)\,=\,{1\over 2\Lambda}3\pi i(\mu_\alpha-\mu_{\alpha'})
c^\gamma_{\alpha\alpha'}J_{\alpha\alpha'}(J^2)_{\alpha\alpha'}L^b\,=\,
-i{3\pi\over 4}(\mu_\alpha-\mu_{\alpha'})c^\gamma_{\alpha\alpha'}{d\over d\Lambda}
(J_{\alpha\alpha'})^2 L^b\quad.
\end{equation}
Inserting $c^\gamma_{\alpha\alpha'}=-(1/2)(\delta_{\alpha\gamma}-\delta_{\alpha'\gamma})$ and
interchanging $\alpha\leftrightarrow\alpha'$ in the second term, we get
with $\Sigma_\gamma^{(2b)}(E)=i\Gamma_\gamma(E)$ (see (\ref{representation_sigma_gamma})) the
result
\begin{equation}
\label{sigma_2b_kondo}
\Gamma_\gamma(E)\,=\,{3\pi\over 4}\,\delta_{\alpha\gamma}\,
(\mu_\alpha-\mu_{\alpha'})\,(J_{\alpha\alpha'})^2\quad.
\end{equation}
For $\Lambda=\Lambda_0$, this result agrees with the initial condition (\ref{L_b_initial}).
$L_S^{(2\tilde{a})Re}(E)$ and $L_S^{(2\tilde{a})Im}(E)$ follow directly from
the combination (\ref{G1_G1_combination}) with 
$K\rightarrow H_{\tilde{\Gamma}_c}(E_{\alpha\alpha'})$
and $K\rightarrow E_{\alpha\alpha'}$, respectively, with the result
\begin{eqnarray}
\label{L_2a_real_kondo}
L_S^{(2\tilde{a})Re}(E)\,&=&\,
-\tilde{H}_{\tilde{\Gamma}_c(E_{\alpha\alpha'})}(E_{\alpha\alpha'})\,(J^c_{\alpha\alpha'})^2\,L^a\quad,\\
\label{L_2a_imaginary_kondo}
L_S^{(2\tilde{a})Im}(E)\,&=&\,
-i{\pi\over 2}\,|E_{\alpha\alpha'}|\,(J^c_{\alpha\alpha'})^2\,L^a\quad,
\end{eqnarray}
where $\tilde{H}_\Gamma(E)$ is defined in (\ref{H_function_corr}) (differing from 
(\ref{H_function}) by an unimportant factor $2$ inside the logarithm arising from
a full 2-loop analysis, see the discussion at the end of Sec.~\ref{sec:4.4}).
Using (\ref{Gamma_tilde_alternative}), $\tilde{\Gamma}_c$ can be calculated
from $L_S^{(2\tilde{a})Im}(E)=-i\tilde{\Gamma}_c(E) L^a$ (due to $L^a L^a=L^a$ the operator
$L^a$ is not needed for $\tilde{\Gamma}_c(E)$).

{\bf Summary.} Collecting all terms and inserting in 
(\ref{final_solution_decomposition})-(\ref{final_solution_imaginary}) and
(\ref{current_representation}), we obtain the
following result for the effective Liouvillian and the current kernel
\begin{eqnarray}
\label{effective_liouvillian}
L^{eff}_S(E)\,&=&\,(h^{eff}(E)\,-\,i\Gamma^{eff}(E))\,L^a \quad,\quad\\
\label{current_kernel}
\mbox{Tr}_S\,\Sigma_\gamma(E)\,&=&\,i\Gamma^{eff}_\gamma(E)\,\mbox{Tr}_S\,L^b \quad,
\end{eqnarray}
with
\begin{eqnarray}
\label{h_eff}
h^{eff}(E)\,&=&\,-E\,(\mbox{Tr}_\alpha J^c)\,+\,{1\over 2}\,E\,(\mbox{Tr}_\alpha\,J^c)^2 \,-\,
H_{\tilde{\Gamma}_c(E_{\alpha\alpha'})}(E_{\alpha\alpha'})\,(J^c_{\alpha\alpha'})^2\quad,\\
\label{Gamma_eff}
\Gamma^{eff}(E)\,&=&\,{\pi\over 2}\,|E_{\alpha\alpha'}|\,(J^c_{\alpha\alpha'})^2
\,\approx\,\tilde{\Gamma}_c(E)\quad,\\
\label{Gamma_current_eff}
\Gamma_\gamma^{eff}(E)\,&=&\,\langle I^\gamma \rangle(E)\,=\,
{3\pi\over 4}\,\delta_{\alpha\gamma}\,
(\mu_\alpha-\mu_{\alpha'})\,(J^c_{\alpha\alpha'})^2
\quad,
\end{eqnarray}
where $E_{\alpha\alpha'}=E+\mu_\alpha-\mu_{\alpha'}$ and
\begin{equation}
\label{H_function_kondo}
\tilde{H}_\Gamma(E)\,=\,E\,\left(\ln{\mbox{max}\{|E|,V\}\over 
\sqrt{|E|^2+\Gamma^2}}\,+\,1\right)\quad.
\end{equation}
$J^c$ is the solution of the poor man scaling equation (\ref{poor_man_scaling_kondo})
at $\Lambda=\Lambda_c=\mbox{max}\{|E|,V\}$, given by (\ref{ansatz_leading_order_kondo}) and 
(\ref{solution_leading_order_kondo}) as
\begin{equation}
\label{J_c_kondo}
J^c_{\alpha\alpha'}\,=\,2\,\sqrt{x_\alpha\,x_{\alpha'}}\,\bar{J}^c\quad,\quad\sum_\alpha x_\alpha=1 
\quad,\quad
\bar{J}^c\,=\,{1\over 2\,\ln{\mbox{max}\{|E|,V\}\over T_K}}\quad.
\end{equation}
Note that $J^c$ depends implicitly on $E$ for $|E|>V$.

If only two reservoirs with $\mu_\alpha=\alpha V/2$ are present, the result can be 
written as
\begin{eqnarray}
\nonumber
h^{eff}(E)\,&=&\,-E\,(J^c_L+J^c_R)\,+\,{1\over 2}\,E\,(J^c_L+J^c_R)^2 \,-\,\\
\label{h_eff_two_reservoirs}
&&-\,H_{\tilde{\Gamma}_c(E)}(E)\,((J^c_L)^2+(J^c_R)^2)\,-\,
H_{\tilde{\Gamma}_c(E\pm V)}(E\pm V)\,(J^c_{nd})^2\quad,\\
\label{Gamma_eff_two_reservoirs}
\Gamma^{eff}(E)\,&=&\,{\pi\over 2}\,|E|\,((J^c_L)^2+(J^c_R)^2)\,+\,
{\pi\over 2}\,|E\pm V|\,(J^c_{nd})^2
\,\approx\,\tilde{\Gamma}_c(E)\quad,\\
\label{Gamma_current_eff_two_reservoirs}
\Gamma_\gamma^{eff}(E)\,&=&\,\langle I^\gamma \rangle(E)\,=\,
\gamma\,{3\pi\over 4}\,(J^c_{nd})^2\,V\quad,
\end{eqnarray}
where $J_\alpha=J_{\alpha\alpha}$ are the diagonal and $J_{nd}=J_{LR}=J_{RL}$ the
nondiagonal exchange couplings, and we sum implicitly over the two possibilities $\pm$
on the r.h.s. of the equations.

For $E=0$, (\ref{Gamma_eff_two_reservoirs}) and (\ref{Gamma_current_eff_two_reservoirs})
are the well-known golden rule results for the Korringa spin relaxation rate and
the current, with the exchange couplings replaced by the renormalized ones from
the poor man scaling equation cut off at the voltage. However, for $E\ne 0$, 
interesting logarithmic contributions $\,\sim E\ln(V/|E|)$ (for $E\ll V$) or
$\,\sim (E\pm V)\ln(V/|E\pm V|)$ (for $E\sim \mp V$) appear in second order in $J_c$ for $h^{eff}(E)$,
which are present in both regimes $E>V$ and $E<V$. This will generically lead to branch cuts 
in the complex plane for the reduced density matrix. In contrast, up to second order in $J_c$,
the current depends on the Laplace space $E$ only for $E>V$ via $J^c=1/(2\ln(E/T_K))$.
However, as has been discussed in detail in Ref.~\cite{reininghaus_hs_preprint}, this
changes in third order in $J_c$, where also the current gets similiar logarithmic contributions
as $h^{eff}(E)$. An interesting field of future research will be to discuss
in detail the consequences for the time evolution of the spin population and the 
nonequilibrium current.

\subsection{RG in strong coupling}
\label{sec:5.3}

In this section we briefly discuss some preliminary results in the strong coupling regime.
The strong coupling regime is defined by the condition that all low energy scales fall
below the Kondo temperature
\begin{equation}
\label{definition_strong_kondo}
|E|,V,T\,<\,T_K\quad.
\end{equation}
In this regime no systematic truncation scheme can be applied because the vertices become
of order one. Nevertheless, it can be studied what result the RG equations in the minimal approximation
scheme proposed in Sec.~\ref{sec:4.5} give. We will study here the stationary case (i.e. $E=0$) 
for the differential conductance 
\begin{equation}
\label{differential_conductance}
G(T,V)\,=\,{d\over dV}\langle I^\gamma \rangle^{st}
\,=\,{d\over dV}\Gamma_\gamma(E=0)
\end{equation}
for the special case of two reservoirs with $\mu_\alpha=\alpha V/2$. Note that we work in
units $e=\hbar=1$, so that finally the result has to be multiplied by $\pi$ to get the
conductance in units of the universal conductance $G_0=2e^2/h$. From exact solutions
(see e.g. Refs.~\cite{hewson,glazman_pustilnik_05,oguri}) it is known that 
at $T=V=0$, the conductance should be universal and the following Fermi liquid
result should apply for low temperatures and voltages
\begin{equation}
\label{exact_result}
G/G_0\,=\,1\,-\,{\pi^2 T^2\over T_K^2}\,-\,{3 V^2\over 2 T_K^2}\quad,
\end{equation}
where the Kondo temperature $T_K$ in this result is not universal. However, $T_K$ drops
out when considering the universal ratio
of the coefficients of the terms $\sim T^2$ and $\sim V^2$.
Furthermore, for $V=0$, one can compare the linear conductance with results from
numerical renormalization group, where the Kondo temperature is defined such that
the conductance at $T=T_K$ is given by $G=G_0/2$ \cite{costi}. With this numerical
method one can especially study the crossover at $T\sim T_K$ very reliably.

We restrict ourselves to the leading order form of the Liouvillian, the kernel, and the
vertices as they have been used for the weak coupling regime
\begin{eqnarray}
\label{L_representation_strong}
L_S(E)\,&=&\,(h(E)\,-\,i\Gamma(E))\,L^a \quad,\quad\\
\label{Sigma_representation_strong}
\mbox{Tr}_S\,\Sigma_\gamma(E)\,&=&\,i\Gamma_\gamma(E)\,\mbox{Tr}_S\,L^b \quad,\\
\label{G_representation_strong}
\bar{G}_{11'}(E)\,&=&\,\bar{G}^2(E)\,L^2\,+\,\bar{G}^3(E)\,L^3\quad,\\
\label{I_representation_strong}
\mbox{Tr}_S\,\bar{I}^\gamma_{11'}(E)\,&=&\,\bar{I}^{\gamma\,1}(E)\,\mbox{Tr}_S\,L^1\quad.
\end{eqnarray}
However, we include the full dependence on the energy variable $E$ since this is not
very time consuming for a numerical solution. The vertex $\bar{G}^3(E)$ has been included
because, as we have seen within the weak coupling analysis in Sec.~\ref{sec:5.2}, 
the vertex $\bar{G}^{(2)}_{11'}\sim L^3$ is essential to obtain the current in
leading order, whereas $\bar{I}^{\gamma(2)}_{11'}\sim L^b$ does not contribute
in this order, see the discussion before (\ref{Sigma_2b_kondo}).
For convenience we introduce the matrices $J$ and $K$ defined by 
\begin{equation}
\label{J_K_definition}
\bar{G}^2(E)\,=\,-\,J(E)\quad,\quad \bar{G}^3(E)\,=\,i\pi\,K(E)\quad,
\end{equation}
and take only the real part of $J_{\alpha\alpha'}(E)$ and $K_{\alpha\alpha'}(E)$ into
account. For simplicity we calculate only the stationary current so that we can set
$E=0$ for $\bar{I}^{\gamma\,1}$ and $\Gamma_\gamma$. With these approximations we obtain
the following relations from (\ref{symmetry_h_gamma}) for $E$ being real
\begin{eqnarray}
\nonumber
&& h(E)\,=\,-h(-E)\quad,\quad \Gamma(E)\,=\,\Gamma(-E)\quad,\quad 
\Gamma_\gamma^*\,=\,\Gamma_\gamma\quad,\\
\label{symmetries_strong_kondo}
&& J_{\alpha\alpha'}(E)\,=\,J_{\alpha'\alpha}(-E)\quad,\quad
  K_{\alpha\alpha'}(E)\,=\,K_{\alpha'\alpha}(-E)\quad,\quad
 \bar{I}^{\gamma\,1}_{\alpha\alpha'}\,=\,
-\bar{I}^{\gamma\,1}_{\alpha'\alpha}\quad.
\end{eqnarray}
Due to the property 
$ \bar{I}^{\gamma\,1}_{\alpha\alpha'}=-\bar{I}^{\gamma\,1}_{\alpha'\alpha}\,\,$,
we can parametrize the current vertex by
\begin{equation}
\label{J_I}
\bar{I}^{\gamma\,1}_{\alpha\alpha'}\,=\,{1\over 2}\,c^\gamma_{\alpha\alpha'}\,J_I \quad.
\end{equation}
Finally, we use the abbreviations
\begin{eqnarray}
\label{pi_short}
\Pi(E)\,&=&\,\mbox{Re}\left\{{1\over \Lambda_T+\Gamma(E)-iE+ih(E)}\right\}\quad,\\
\label{L_function_short}
{\bar{\cal{L}}}(E)\,&=&\,{\cal{L}}_\Lambda(E-h(E)+i\Gamma(E))\quad,
\end{eqnarray}
where the function ${\cal{L}}_\Lambda(E)$ is defined in (\ref{L_function_finite_T}) for
the general case of finite temperature. $\Lambda_T$ is as usual the Matsubara frequency
lying closest to $\Lambda$. From (\ref{symmetries_strong_kondo}) we obtain
\begin{equation}
\label{pi_L_symmetry}
\Pi(E)\,=\,\Pi(-E) \quad,\quad 
{\bar{\cal{L}}}(E)\,=\,{\bar{\cal{L}}}(-E)^*\quad.
\end{equation}

The vertices are parametrized in such a form that the initial conditions become very
simple. From (\ref{leading_order_vertices}) and (\ref{rg_G_second_order_kondo}) we obtain
\begin{equation}
\label{vertices_initial_condition_kondo}
K_{\alpha\alpha'}\,=\,{1\over 2}(J^2)_{\alpha\alpha'}\quad,\quad
J_I\,=\,J_{nd}\quad,
\end{equation}
and $J_{\alpha\alpha'}$ are identical to the initial exchange couplings.
From the general expression for the initial condition (\ref{L_initial_weak_coupling}) for
the Liouvillian and the current kernel, and using the helpful identity (\ref{evaluation_rg_identity_B}),
we get with the original form (\ref{leading_order_vertices}) for the vertices the following result
for the initial conditions of $h(E)$, $\Gamma(E)$, and $\Gamma_\gamma(E)$
\begin{equation}
\label{L_initial_condition_kondo}
h(E)\,=\,E{\pi^2\over 16}(\mbox{Tr}_\alpha J^2)\quad,\quad
\Gamma(E)\,=\,D{\pi^2\over 8}(\mbox{Tr}_\alpha J^2)\quad,\quad
\Gamma_\gamma(E)\,=\,\gamma V{3\pi\over 4}(J_{nd})^2\quad.
\end{equation}
The initial conditions for $h(E)$ and $\Gamma_\gamma(E)$ are unimportant since they
vanish in the scaling limit $J\rightarrow 0$ and $D\rightarrow \infty$ such that the
Kondo temperature remains constant. In contrast, as already discussed previously, the
terms $\sim D$ for $\Gamma(E)$ are artificial and the 
initial value $\Lambda_0$ for the parameter $\Lambda$ has to be
chosen such that these terms vanish. In Sec.~\ref{sec:4.4} we found
for the weak coupling case that $\Lambda_0$ is approximately given by
\begin{equation}
\label{value_initial_cutoff_strong}
\Lambda_0\,=\,{\pi^2\over 16 \ln(2)}\,D\quad,
\end{equation}
leading to
\begin{equation}
\label{Gamma_initial}
\Gamma(E)\,=\,2\ln(2)\Lambda_0 \,(\mbox{Tr}_\alpha J^2)\quad.
\end{equation}
Whether this is also the correct value for the strong coupling case where the RG equations
are different is not at all clear. In fact we will see below that the strong coupling
solution depends crucially on the inital value of $\Gamma(E)$.

We study now the RG equations (\ref{rg_L_so_strong_finite_T}) and (\ref{rg_G_so_strong_finite_T})
proposed in Sec.~\ref{sec:4.5} as a minimal set to go beyond a weak coupling analysis. 
Using the relations (\ref{evaluation_rg_identity_B}) and (\ref{evaluation_rg_identity_C})
together with the algebra of the basis operators in Liouville space, we obtain after some
straightforward manipulations (similiar to those already explained in all detail in Sec.~\ref{sec:5.2})
the following RG equations
\begin{eqnarray}
\label{rg_h_strong}
{d\over d\Lambda}h(E)\,&=&\,2\,\mbox{Im}\left\{\bar{{\cal{L}}}(E_{\alpha\alpha'})\right\}\,
J_{\alpha\alpha'}(E)\,J_{\alpha'\alpha}(E_{\alpha\alpha'})\quad,\\
\label{rg_Gamma_strong}
{d\over d\Lambda}\Gamma(E)\,&=&\,2\,\mbox{Re}\left\{\bar{{\cal{L}}}(E_{\alpha\alpha'})\right\}\,
J_{\alpha\alpha'}(E)\,J_{\alpha'\alpha}(E_{\alpha\alpha'})\quad,\\
\nonumber
{d\over d\Lambda}J_{\alpha\alpha'}(E)\,&=&\,-{1\over 2}\,
\left\{\Pi(E_{\alpha\alpha_2})\,J_{\alpha\alpha_2}(E)\,J_{\alpha_2\alpha'}(E_{\alpha\alpha_2})
\,+\right.\quad\\
\label{rg_J_strong}
&&\hspace{0.5cm}\left.+\,\Pi(E_{\alpha_2\alpha'})\,J_{\alpha_2\alpha'}(E)\,J_{\alpha\alpha_2}(E_{\alpha_2\alpha'})
\right\}\quad
\end{eqnarray}
for the determination of the values for $h(E)$ and $\Gamma(E)$, and the RG equations
\begin{eqnarray}
\label{rg_Gamma_gamma_strong}
{d\over d\Lambda}\Gamma_\gamma\,&=&\,-12\pi i\,\bar{{\cal{L}}}(\mu_{\alpha}-\mu_{\alpha'})\,
\bar{I}^{\gamma\,1}_{\alpha\alpha'}\,K_{\alpha'\alpha}(\mu_{\alpha}-\mu_{\alpha'})\quad,\\
\nonumber
{d\over d\Lambda}\bar{I}^{\gamma\,1}_{\alpha\alpha'}\,&=&\,-
\left\{\Pi(\mu_{\alpha}-\mu_{\alpha_2})\,\bar{I}^{\gamma\,1}_{\alpha\alpha_2}
\,J_{\alpha_2\alpha'}(\mu_{\alpha}-\mu_{\alpha_2})
\,+\right.\quad\\
\label{rg_I_strong}
&&\hspace{0.5cm}\left.+\,\Pi(\mu_{\alpha_2}-\mu_{\alpha'})\,
\bar{I}^{\gamma\,1}_{\alpha_2\alpha'}\,J_{\alpha\alpha_2}(\mu_{\alpha_2}-\mu_{\alpha'})
\right\}\quad\\
\nonumber
{d\over d\Lambda}K_{\alpha\alpha'}(E)\,&=&\,-
\left\{\Pi(E_{\alpha\alpha_2})\,J_{\alpha\alpha_2}(E)\,K_{\alpha_2\alpha'}(E_{\alpha\alpha_2})
\,+\right.\quad\\
\label{rg_K_strong}
&&\hspace{0.5cm}\left.+\,\Pi(E_{\alpha_2\alpha'})\,J_{\alpha_2\alpha'}(E)\,K_{\alpha\alpha_2}(E_{\alpha_2\alpha'})
\right\}\quad
\end{eqnarray}
to obtain the stationary current rate $\Gamma_\gamma$. Note that $h(E)$ and $\Gamma(E)$ enter
into the resolvent $\Pi(E)$ and the function $\bar{{\cal{L}}}(E)$ via the definitions
(\ref{pi_short}) and (\ref{L_function_short}). In this way, especially $\Gamma(E)$ provides 
an important cutoff parameter for the RG flow in the strong coupling limit so that the
coupling constants do not diverge and a finite conductance comes out.

From (\ref{rg_J_strong}) one can prove the relation
\begin{equation}
\label{J_relation}
J_{\alpha\alpha'}(E)\,=\,J_{\alpha'\alpha}(E_{\alpha\alpha'})\quad.
\end{equation}

We now turn to the case of two reservoirs with $\mu_\alpha=\alpha V/2$.  
Inserting the parametrization (\ref{J_I}) and using the symmetry 
relations (\ref{symmetries_strong_kondo}),
one obtains after some algebra the simplified RG equations
\begin{eqnarray}
\label{rg_Gamma_gamma_two_reservoirs_strong}
{d\over d\Lambda}\Gamma_\gamma\,&=&\,-6\pi\gamma \,\mbox{Im}\bar{{\cal{L}}}(V)\,J_I\,K_{RL}(V)
\quad,\\
\label{rg_J_I_two_reservoirs_strong}
{d\over d\Lambda}J_I\,&=&\,-\Pi(V)\,J_I\,(J_L+J_R)(V)\quad.
\end{eqnarray}
Using the symmetry relation (\ref{J_relation}), the RG equations for $h(E)$, 
$\Gamma(E)$, $J_\alpha(E)=J_{\alpha\alpha}(E)$, and
$J_{\alpha\bar{\alpha}}(E)$, with $\bar{\alpha}=-\alpha$, can be simplified to
\begin{eqnarray}
\label{rg_h_two_reservoirs_strong}
{d\over d\Lambda}h(E)\,&=&\,2\,\mbox{Im}\bar{{\cal{L}}}(E)\,J_{\alpha}(E)^2
\,+\,2\,\mbox{Im}\bar{{\cal{L}}}(E+\alpha V)\,J_{\alpha\bar{\alpha}}(E)^2
\quad,\\
\label{rg_Gamma_two_reservoirs_strong}
{d\over d\Lambda}\Gamma(E)\,&=&\,2\,\mbox{Re}\bar{{\cal{L}}}(E)\,J_{\alpha}(E)^2
\,+\,2\,\mbox{Re}\bar{{\cal{L}}}(E+\alpha V)\,J_{\alpha\bar{\alpha}}(E)^2
\quad,\\
\label{rg_J_two_reservoirs_diagonal_strong}
{d\over d\Lambda}J_{\alpha}(E)\,&=&\,-\Pi(E)\,J_{\alpha}(E)^2\,-\,\Pi(E+\alpha V)\,
J_{\alpha\bar{\alpha}}(E)^2
\quad,\\
\nonumber
{d\over d\Lambda}J_{\alpha\bar{\alpha}}(E)\,&=&\,
-\,\Pi(E)\,{1\over 2}(J_L+J_R)(E)\,J_{\alpha\bar{\alpha}}(E)\,-\\
\label{rg_J_two_reservoirs_nondiagonal_strong}
&&\hspace{0.5cm}-\,\Pi(E+\alpha V)\,{1\over 2}(J_L+J_R)(E+\alpha V)\,J_{\alpha\bar{\alpha}}(E)
\quad.
\end{eqnarray}
The RG equations for $K_{\alpha}(E)=K_{\alpha\alpha}(E)$ and $K_{\alpha\bar{\alpha}}(E)$ read
for two reservoirs
\begin{eqnarray}
\label{rg_K_two_reservoirs_diagonal_strong}
{d\over d\Lambda}K_{\alpha}(E)\,&=&\,-\Pi(E)\,J_{\alpha}(E)\,K_\alpha(E)\,-\,\Pi(E+\alpha V)\,
J_{\alpha\bar{\alpha}}(E)\,K_{\bar{\alpha}\alpha}(E+\alpha V)
\quad,\\
\nonumber
{d\over d\Lambda}K_{\alpha\bar{\alpha}}(E)\,&=&\,
-\,\Pi(E)\,{1\over 2}(J_L+J_R)(E)\,K_{\alpha\bar{\alpha}}(E)\,-\\
\label{rg_K_two_reservoirs_nondiagonal_strong}
&&\hspace{0.5cm}-\,\Pi(E+\alpha V)\,J_{\alpha\bar{\alpha}}(E)\,{1\over 2}(K_L+K_R)(E+\alpha V)
\quad.
\end{eqnarray}

These RG equations can not be solved analytically but some general features can be studied.
First, in the weak coupling regime $V>T_K$, the RG equations 
(\ref{rg_J_two_reservoirs_diagonal_strong}) and
(\ref{rg_J_two_reservoirs_nondiagonal_strong}) reveal how the exchange couplings are cut off
by the voltage. Independent on whether the diagonal or the nondiagonal exchange coupling is
considered, there are terms $\sim \Pi(E)$ which are independent of the voltage, and other terms
$\sim \Pi(E\pm V)$ which contain the voltage. This means that for $E=0$, the diagonal as well as
the nondiagonal coupling are finally cut off by the spin relaxation rate $\Gamma$ and not by
the voltage. However, this does not lead to a problem or an increased conductance because the
current rate $\Gamma_\gamma$ as well as the current vertex $J_I$ are cut off by the voltage, 
as can be seen from (\ref{rg_Gamma_gamma_two_reservoirs_strong}) and (\ref{rg_J_I_two_reservoirs_strong}).
Therefore, the logarithmic increase of the exchange couplings between $V$ and $\Gamma$ does not
influence the conductance considerably. However, it is important to notice that this does not
mean that one can omit the spin relaxation rate $\Gamma$ from the RG equation for the couplings.
In this case, the couplings would diverge at the Kondo temperature and, as a consequence, also
the current vertex and the conductance since the cutoff functions are smooth functions. Therefore,
although the rate $\Gamma$ does not appear in the final result for the conductance for
$V>T_K$, it is still important to have it in the RG equations for the couplings.

\begin{figure}
  \centerline{\psfig{figure=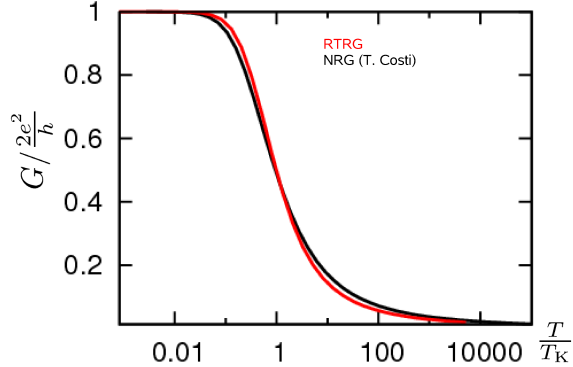,scale=0.3}}
  \caption{Comparism of the linear conductance as function of temperature from RTRG-FS
(red curve) with NRG results from Costi \cite{costi}. The Kondo temperature $T_K$ is
defined such that $G(T=T_K)={1\over 2}G_0$.}
\label{fig:kondo_strong_coupling}
\end{figure}

In the strong coupling regime $|E|,T,V<T_K$, the couplings do not diverge because the rate $\Gamma$
cuts off the RG flow roughly at the Kondo temperature. Since no other energy scale remains, 
$\Gamma$ is expected to approach $T_K\,J^2$ when $\Lambda$ is below the Kondo temperature.
To analyse this question let us consider the simplest case $E=V=T=0$ and all exchange couplings
to be the same $J_{\alpha\alpha'}=J$. Since $h(0)=0$, we get the set of differential equations
\begin{equation}
\label{simplest_case_strong}
{d\Gamma\over d\Lambda}\,=\,8\,\ln\left({2\Lambda+\Gamma\over \Lambda+\Gamma}\right)\,J^2\quad,\quad
{dJ\over d\Lambda}\,=\,-{2J^2\over \Lambda+\Gamma}\quad.
\end{equation}
Analysing this nonlinar set of differential equations numerically, one finds that the solution
is unstable against exponentially small changes in the initial condition for $\Gamma$. Thereby, the
crucial problem is not the elimination of the linear terms in $\Lambda_0$ of the initial condition  
(\ref{Gamma_initial}) for $\Gamma$, but to determine the initial condition such that
$\Gamma$ saturates at the Kondo temperature. Furthermore, if one studies the influence on the
differential conductance, one finds that the conductance at $T=V=0$ can be tuned to any value depending
on the initial condition for $\Gamma$. Due to this ambiguity, we have fitted the initial condition
for $\Gamma$ at $T=V=E=0$ such that the conductance becomes universal $G=G_0=2e^2/h$. Using this
single parameter, we have then solved the RG equations for arbitrary values of voltage and 
temperature. The result for the linear conductance $G(T)/G_0$ is shown in 
Fig.~\ref{fig:kondo_strong_coupling} together with a comparism to numerical renormalization group
results performed by Costi \cite{costi}. The coincidence is quite impressive considering the 
fact that we have made many approximations by neglecting higher order terms and the frequency
dependence of the vertices and the Liouvillian. Especially the fact that the broadening of the
crossover regime comes out correctly shows that the coincidence is nontrivial and is not
done by hand with the fitting parameter. It means that the approximate RG equations 
contain the right correlation between the broadening of the crossover behaviour and the
final height of the conductance at $T=0$. It will be an interesting question for future research
whether the inclusion of higher order terms will avoid the instability and provide the saturation
of $\Gamma$ at the right scale. The conductance can also be calculated as function of voltage but
the result falls nearly on top of the curve shown in Fig.~\ref{fig:kondo_strong_coupling} by using
a simple rescaling. One can also determine the coefficients of the Fermi liquid relation
(\ref{exact_result}) but it turns out that the ratio of the coefficients comes out incorrectly by
roughly a factor of $2$.
\section{Summary and outlook}
\label{sec:6}

In this paper we have presented a formally exact renormalization group scheme to analyse
the time evolution and the stationary state for a generic problem of dissipative
quantum mechanics: An arbitrary local quantum system with a small number of states
coupled to several grandcanonical reservoirs at different temperatures and
chemical potentials. The RG scheme is set up by an expansion in the system-reservoir
coupling using a compact diagrammatic representation which is based on a 
quantum field-theoretical formulation in Liouville space.
It allows for a direct determination of the irreducible kernel 
which is the dissipative part of the kinetic equation for the reduced density matrix
of the local system. This kernel determines at the same time the effective
Liouville operator of the local system which is fed back into the RG equations
to determine the effective propagator between the vertices. The RG scheme is set
up in Laplace space allowing for an analysis of the time evolution in the
presence of inhomogeneous boundary conditions leading to a nonequilibrium
stationary state. Generically, we have shown how to avoid the occurence of the
zero eigenvalue of the effective Liouville operator in the propagator.
This provides a generic proof that the RG flow is ultimatively cut off by relaxation
and dephasing rates. We have shown how to calculate the average of arbitrary
observables. Correlation functions can also be calculated with similiar schemes
which have not been shown here.

The RG equations represent an infinite hierarchy of coupled differential equations
for effective vertices with an arbitrary number of reservoir field operators. The 
important input of this scheme is a cutoff-dependent correlation function for
the reservoirs which can be chosen arbitrarily. However, to be able to truncate the
infinite hierarchy in a systematic way, we discussed several criteria how an
appropriate cutoff function has to be defined. To avoid the problem of the occurence
of the zero eigenvalue in the effective propagators we developed a generic two stage
procedure. First one integrates out the symmetric part of the reservoir distribution
function by using perturbation theory. The result is used as an input for the
following continuous RG flow with a cutoff dependent antisymmetric
reservoir distribution function. The antisymmetry guarantees that the zero eigenvalue
can not occur in the effective propagators between the vertices. Connected with this
property is the fact that only the vertices averaged over the Keldysh indices occur
in the RG scheme which simplifies the calculation. For the choice of
the cutoff dependence of the reservoir correlation function within the continuous
RG scheme, we proposed to integrate out the poles and branch cuts in the complex
plane by defining the cutoff on the imaginary frequency axis. This has three 
advantages. First, high energy scales are integrated out systematically providing
a high-energy cutoff for the frequency integrations in the effective perturbation
theory. Secondly, with respect to the real frequencies, all energy scales are 
considered in each step of the RG procedure so that relaxation and dephasing rates
are generated from the very beginning. Third, the numerical stability of the RG
flow is improved considerably because the resolvents occuring on the r.h.s. of the
RG equation contain large imaginary parts in the denominator which are of the order of 
the cutoff parameter.

The choice of the cutoff function is particularly simple for the special case of
fermionic reservoirs with a flat density of states. In this case, the cutoff is
defined by integrating out the Matsubara poles of the Fermi functions step by step.
All integrations over the real frequencies can then be calculated analytically
and the RG scheme can be set up on the Matsubara axis similiar to exact RG schemes
for equilibrium problems. However, no analytic continuation to real frequencies is
necessary to calculate transport properties. Furthermore, one has to consider a whole 
set of Matsubara axis shifted by the real part of the Laplace variable (to account for the time
evolution) and by multiples of the chemical potentials of the reservoirs (to account
for finite bias voltages). For different temperatures, each reservoir 
defines its own Matsubara axis. We have shown that the stability of the RG flow
is expected to be very good because all imaginary parts in the denominators of the
resolvents are strictly positive so that no cancellations can occur. From the real
part of the denominators one can generically see that enhanced renormalizations are
expected if the resonance condition $E-\bar{\mu}_{1\dots n}-h=0$ is fulfilled, where
$E$ is the real part of the Laplace variable, $\bar{\mu}_{1\dots n}$ consists of the
sum of arbitrary multiples of the chemical potentials of the reservoirs, and $h$
are the renormalized oscillation frequencies of the time evolution of the reduced
density matrix of the local quantum system (e.g. magnetic field, level spacing, 
charging energy, etc.).

For the case of fermionic reservoirs with a flat density of states we have shown how
to solve the RG equations analytically in the weak coupling regime. Following 
Ref.~\cite{reininghaus_hs_preprint}, we considered a system with generic spin 
and/or orbital fluctuations and derived the final result for the effective Liouvillian
and the average of an arbitray observable up to order $J_c^2$. Here, $J_c$ is the
order of magnitude of the effective vertex in leading order at scale 
$\Lambda_c=\mbox{Max}\{|E|,|\bar{\mu}_{1\dots n}|,|h|\}$. For the effective 
oscillation frequencies we found a logarithmic term
$\sim\ln(\Lambda_c/|E-\bar{\mu}_{1\dots n}-h+i\Gamma|)$, where $\Gamma$ is the
relaxation or dephasing rate corresponding to the oscillation frequency $h$.
At resonance $E-\bar{\mu}_{1\dots n}-h=0$, the logarithmic term is cut off by 
$\Gamma$ and an enhancement $\sim J_c^2\ln(J_c)$ is obtained. In arbitrary order one
expects enhancements $\sim J_c^k(\ln{J_c})^{k-1}$ with $k\ge 2$. As shown in 
Ref.~\cite{reininghaus_hs_preprint}, similiar features 
are generically expected for the relaxation and dephasing rates and for averages 
of observables like the current, but the order of magnitude is one power less in $J_c$,
i.e. one expects enhancements $\sim J_c^k(\ln{J_c})^{k-2}$ with $k\ge 3$ at resonance. For
the special case of the nonequilibrium Kondo model at finite voltage $V$, this leads to 
the well-known logarithmic enhancements $\sim J_c^3\ln(V/|V-h+i\Gamma|)$ for the differential 
conductance at $V\sim h$ 
\cite{glazman_pustilnik_05,rosch_kroha_woelfle_PRL01,rosch_paaske_kroha_woelfle_PRL03}.

For strong coupling no systematic truncation scheme can be developed but we have shown
that the presence of effective relaxation and dephasing rates cuts off the RG flow such
that the coupling vertices will not diverge and stay of order one when the cutoff reaches
the scale of the Kondo temperature (where usual poor man scaling equations diverge). 
We proposed a minimal set of RG equations for the strong coupling regime by truncating the
RG equations and using a leading-order parametrization of the vertices. Preliminary results 
have been presented for the isotropic Kondo model, where a comparism to numerical renormalization 
group methods showed a promising coincidence for the linear conductance as function of temperature.
However, within the lowest order approximation scheme, the RG equations showed an instability
against exponentially small changes in the initial condition for the relaxation rate $\Gamma$.
Therefore, a fitting procedure was necessary to fix the conductance at $T=V=0$ to the universal
value. It remains an interesting subject for future research to include higher-order terms
and study the stability of the RG equations against the choice of the initial conditions in the
strong coupling regime in more detail.

Besides the strong coupling regime interesting open questions for the future concern especially the
study of charge fluctuations, the time evolution, frequency-dependent density of states
and bosonic systems. Especially for systems where the density of states has branch cuts in the
upper half of the complex plane which do not coincide with the Matsubara axis, the RG equations
get a different structure compared to fermionic systems with a flat density of states. Although
the cutoff can be always defined by cutting off the imaginary part of all branch cuts, the 
presence of several branch cuts changes the result of the integration over the real frequencies.
As a consequence, not all resolvents of the RG equations will contain the cutoff and the numerical
stability of the equations has to be studied in more detail. However, the general two stage procedure
and the systematic solution in the weak coupling regime can be used as well.
Interesting systems in this
connection are bosonic systems with particle number conservation (e.g. cold atom gases), where the 
condition $\omega>0>\mu_\alpha$ has to be fulfilled for the reservoir energies, or superconducting 
reservoirs where the density of states has a gap. Systems, where the frequency dependence of the
density of states is analytic (besides possible poles lying at very high energies) or when the
branch cuts can be turned into the lower half of the complex plane, can be described by RG equations
with a similiar structure as the one for a flat density of states. The only difference is that
the first step of the RG flow, where the symmetric part of the distribution function is 
integrated out, has to be considered again because the frequency integrals are different.

\section{Acknowledgments}
\label{sec:7}
For the development and application of nonequilibrium RG methods 
I owe special thanks to S. Jakobs, M. Keil, T. Korb, J. K\"onig, V. Meden, 
M. Pletyukhov, and F. Reininghaus. Useful discussions are acknowledged with 
T. Costi, L. Glazman, S. Kehrein, H. Kroha,
M. Kurz, J. Paaske, A. Rosch, D. Schuricht, M. Wegewijs, and P. Woelfle. Finally,
I would like to acknowledge the support from the
DFG-Forschergruppe 723 on \lq\lq Functional Renormalization Group
in Correlated Fermion Systems'' and thank all members of this group
for many fruitful discussions on general aspects of renormalization
group, especially to H. Gies, W. Metzner, J. Pawlowski, M. Salmhofer, 
and K. Sch\"onhammer.

\begin{appendix}
\section{Appendix: Commutation of dot and reservoir operators}
\label{appendix_A}
In this appendix we prove that all signs occuring from interchanges of
fermionic dot and reservoir operators cancel exactly the signs emerging
from the prefactor $\eta_1\dots\eta_n$ and the reversed sequence $:a_n\dots a_1:$
(compared to $:a_1\dots a_n:$) in the form (\ref{coupling_fermions}) of
the coupling
\begin{equation}
\label{zw}
V\,=\, {1\over n!}\,\eta_1\dots\eta_n
\,:a_n a_{n-1} \dots a_1:\,\,g_{12\dots n} \quad.
\end{equation}
This means that after having defined the coupling vertex $g_{1\dots n}$ via
this equation, one can use the simpler form (\ref{coupling}) and
consider dot and reservoir operators as commuting objects. This property
can be proven for any average of a sequence of $V$-operators over the reservoir 
distribution $\rho_{res}$
\begin{equation}
\label{average_reservoir}
\langle V(t_1)V(t_2)\dots V(t_r)\rangle_{\rho_{res}} \quad,
\end{equation}
where $V(t)$ is the interaction picture with respect to $H_{res}+H_S$. These
are the expressions which occur for the dynamics of the reduced density matrix
of the dot. For observables one of the $V$-operators is replaced by the 
observable but the proof 
remains the same since the observable is of the same form as $V$. 

We start the proof for $n=1$. The coupling is of the form $V=\eta_1 a_1 g_1$ 
and (\ref{average_reservoir}) becomes (time-dependence not indicated)
\begin{equation}
\label{zw_1}
\eta_1\eta_2\dots\eta_r\langle a_1 g_1 \,a_2 g_2 \dots a_r g_r\rangle_{\rho_{res}}
\quad.
\end{equation}
This average is only nonzero if $r$ is even and if $\sum_i\eta_i=0$, so that
an equal number of annihilation and creation operators is present. Thus, taking
all reservoir field operators to the right (starting with $a_r$), we get a total
sign $(-1)^{r/2}$ from commuting them through the dot operators, and (\ref{zw_1}) reads
\begin{equation}
\eta_1\eta_2\dots\eta_r\,(-1)^{r/2}\,g_1 g_2\dots g_r\,
\langle a_1 a_2\dots a_r\rangle_{\rho_{res}}\quad.
\end{equation}
Using $\sum_i\eta_i=0$ and $r$ even, we get 
$\eta_1\dots\eta_r\,(-1)^{r/2}=(-1)^{r/2}(-1)^{r/2}=(-1)^r=1$ for the prefactor, and all 
sign factors have cancelled leaving an independent product of dot and reservoir
operators.

For $n>1$, we formally imagine $g_{1\dots n}$ to consist of a product $g_1 g_2\dots g_n$
of fermionic dot operators $g_i$ (which is allowed to determine the fermionic sign factors).
From the proof for $n=1$, we know that if $V$ is of the form
\begin{equation}
\label{zw_2}
V\,=\, {1\over n!}\,\eta_1\eta_2\dots\eta_n\,:(a_1 g_1)(a_2 g_2)\dots (a_n g_n): \quad,
\end{equation}
all sign factors cancel in the end. Moving all reservoir field operators in (\ref{zw_2})
to the left (starting with $a_n$), each time involving an even number of permutations of
fermionic operators, we obtain
\begin{equation}
V\,=\, {1\over n!}\,\eta_1\eta_2\dots\eta_n\,:a_n a_{n-1}\dots a_1:\,g_1 g_2\dots g_n \quad,
\end{equation}
which is precisely of the form (\ref{zw}) after replacing $g_1 \dots g_n$
by $g_{1\dots n}$.

\section{Appendix: Coupling operator in Liouville space}
\label{appendix_B}
In this appendix we prove Eq.~(\ref{coupling_product}) for the coupling vertex, originally
defined by (\ref{L_decomposed}) with $V$ given by (\ref{coupling}), i.e. we have to prove
\begin{equation}
\label{app_coupling_product}
L_V\,=\,{1\over n!}\,[g_{1\dots n}:a_1\dots a_n:\,,\cdot]_-\,=\,
{1\over n!}\,\sigma^{p_1\dots p_n}\,G^{p_1\dots p_n}_{1\dots n}\,
:J^{p_1}_1\dots J^{p_n}_n:\quad.
\end{equation}
Since, due to the definition (\ref{G_vertex_liouville}) of the coupling vertex, 
all $p_i$ in this expression are identical to a common index $p$, we distinguish the
two cases $p=\pm$ and add them up finally. Thereby, $p=+$ produces the first term of
the commutator in (\ref{app_coupling_product}), and $p=-$ the second one. 

For $p=+$, we get by acting on an arbitrary operator $A$
\begin{equation}
\sigma^{+\dots +}\,G^{+\dots +}_{1\dots n}\,:J^+_1\dots J^+_n:\,A \,=\,
g_{1\dots n}\,:a_1\dots a_n:\,A \quad,
\end{equation}
where we have used the definitions (\ref{liouville_sign_operator}), 
(\ref{G_vertex_liouville}), and (\ref{liouville_field_operators}) of the sign-operator,
the coupling vertex and the reservoir field operators in Liouville space. As a consequence,
the first term of the commutator in (\ref{app_coupling_product}) is produced for $p_i=p=+$.

For $p=-$, we first consider bosons, where $\sigma^{-\dots -}=1$. We obtain
\begin{eqnarray}
\sigma^{-\dots -}\,G^{-\dots -}_{1\dots n}\,:J^-_1\dots J^-_n:\,A 
\,&=&\, -A\,:a_n\dots a_1: \,g_{1\dots n}\nonumber \\
\,&=&\, -A\,g_{1\dots n}\,:a_1\dots a_n: 
\quad,
\end{eqnarray}
coinciding with the second part of the commutator of (\ref{app_coupling_product}).

For $p=-$ and fermions, we distinguish between $n$ even and $n$ odd. For $n$ even, we
have $\sigma^{-\dots -}=(-1)^{n/2}$ and the same result is obtained for
\begin{eqnarray}
\sigma^{-\dots -}\,G^{-\dots -}_{1\dots n}\,:J^-_1\dots J^-_n:\,A 
\,&=&\,-(-1)^{n/2}\,A\,:a_n\dots a_1: \,g_{1\dots n}\nonumber\\
\,&=&\,-A\,g_{1\dots n}\,(-1)^{n/2}\,:a_n\dots a_1: \nonumber
\,=\,-A\,g_{1\dots n}\,:a_1\dots a_n: 
\quad.
\end{eqnarray}
Here we have used in the second equality the fact that operators from the reservoirs and
the quantum system are considered to commute (a property which holds finally if all expressions
are averaged with respect to the reservoir degrees of freedom, as explained in detail in Sec.~\ref{sec:2.1}
and appendix A). For $n$ odd, we use $\sigma^{-\dots -}=(-1)^{(n-1)/2}\sigma^-$, and, since
$\sigma^-$ cancels with another $\sigma^-$ occuring in the definition (\ref{G_vertex_liouville})
of the coupling vertex, we obtain again
\begin{eqnarray}
\sigma^{-\dots -}\,G^{-\dots -}_{1\dots n}\,:J^-_1\dots J^-_n:\,A 
\,&=&\,-(-1)^{(n-1)/2}\,A\,:a_n\dots a_1: \,g_{1\dots n}\nonumber\\
\,&=&\,-A\,g_{1\dots n}\,(-1)^{(n-1)/2}\,:a_n\dots a_1: \nonumber
\,=\,-A\,g_{1\dots n}\,:a_1\dots a_n: 
\quad,
\end{eqnarray}
and the proof of Eq.~(\ref{app_coupling_product}) is complete.

\section{Appendix: Kernel properties}
\label{appendix_C}
Here we prove the property (\ref{L_eff_c_transform}) of the kernel $\Sigma(E)$
\begin{equation}
\label{sigma_c_transform}
(\Sigma(E))^c\,=\,-\Sigma(-E^*) \quad.
\end{equation}
We use the diagrammatic representation (\ref{value_sigma}) for the kernel 
\begin{equation}
\label{app_C_value_sigma}
\Sigma(E) \,\rightarrow\, {1\over S} \, (\pm)^{N_p} \, \left(\prod\gamma\right)_{irr}
\,G\,{1\over E+X_1-L_S}\,G\,\dots \,G\,{1\over E+X_r-L_S}\,G \quad,
\end{equation}
and calculate the transformation $(\Sigma(E))^c$ using the relation $(AB)^c=A^c B^c$ 
(see (\ref{c_property})). With the help of the properties 
(\ref{L_c_transform}) and (\ref{G_c_2_transform}), and renaming all summation variables
by $p\rightarrow -p$ and $\eta\rightarrow -\eta$, we obtain the following replacements
in (\ref{app_C_value_sigma})
\begin{eqnarray}
\label{gamma_replace}
\gamma^{pp'}_{11'}\,&\rightarrow&\,\gamma^{\bar{p},\bar{p}'}_{\bar{1}\bar{1}'}
\,=\,\pm\,\gamma^{pp'}_{11'} \quad, \\
\label{G_replace}
G^{p_1\dots p_n}_{1\dots n}\,&\rightarrow&\, -\,\sigma^{--\dots-}\,G^{p_1\dots p_n}_{1\dots n}
\quad, \\
\label{resolvent_replace}
{1\over E-X_i-L_S}\,&\rightarrow&\,{1\over E^*+X_i+L_S}
\,=\,-{1\over -E^*-X_i-L_S} \quad,
\end{eqnarray} 
where we have used the form $(\ref{liouville_contraction})$ for the contraction and the
upper (lower) sign corresponds to bosons (fermions). Note 
the fact that $\eta\rightarrow -\eta$ implies also $x=\eta(\omega+\mu_\alpha)\rightarrow -x$ 
and, consequently, $X_i\rightarrow -X_i$. Besides a factor $\pm$ for each contraction and
a sign operator $\sigma^{--\dots-}$ for each vertex (which can give rise to a minus sign only 
for fermions), we obtain the replacement $E\rightarrow -E^*$
and a total minus sign since the number of vertices is by one larger than the number of
resolvents in (\ref{app_C_value_sigma}). Thus, to complete the proof of (\ref{sigma_c_transform}), we
have to show for fermions that the occurence of the sign operators cancels the minus sign for each
contraction. To show this, we shift all sign operators through the vertices $G$ to the left by
using  the property (\ref{sign_operator_property_G}). As a consequence, all sign operators can
be taken together to a single sign operator of the form $\sigma^{--\dots -}$, where the number
of minus signs in this expression corresponds to the total number $Z$ of reservoir field operators
of the diagram. Since $Z=2r$ must be an even number, the total sign operator is identical to 
$\sigma^{--\dots -}=(-1)^r$ according to the definition (\ref{liouville_sign_operator}). Since
$r$ is precisely the number of contractions, this sign factor cancels exacly one sign for each
contraction.

\section{Appendix: Invariance of symmetry relations}
\label{appendix_D}

Here we show that the symmetry properties (\ref{rg_G_symmetry})-(\ref{rg_R_c_transform})
are invariant under the RG flow. Obviously, they are fulfilled initially. So we assume
that we can use them on the r.h.s. of the RG equations 
(\ref{effective_antisym_series}) and (\ref{rg_equations}), and prove them for the l.h.s.

The (anti-)symmetry relations (\ref{rg_G_symmetry}) and (\ref{rg_R_symmetry}) follow
from the diagrammatic rules stated after (\ref{effective_B_series}). Since we sum in the
RG equations over all permutations of the external indices and assign a corresponding
fermionic sign, the effective vertices become automatically (anti-) symmetric.

The property (\ref{rg_LG_property}) of conservation of probability is also trivially
fulfilled, since by acting with $\mbox{Tr}_S$ on the RG equation and summing over the
external Keldysh indices, we get zero since the first vertex already fulfils this property.
Note that by summing over all Keldysh indices, the first vertex is just averaged over
the Keldysh indices since the contractions do not depend on the left Keldysh index.

The properties (\ref{rg_L_c_transform})-(\ref{rg_R_c_transform}), originally related to
the hermiticity of the Hamiltonian, follow immediately by taking the $c$-transform 
(defined in Eq.~(\ref{c_transformation})) of the 
RG equations and applying a proof in analogy to the one presented in Appendix C. The only
difference is that, for the vertices, the total sign operator $\sigma^{--\dots -}$, 
obtained from shifting all sign operators to the left using (\ref{sign_operator_property_G}),
contains also the minus signs from the external indices. However, by using
(\ref{sign_operator_property_decomposition}), this sign operator can be split into
two sign operators, one containing the minus signs from the external vertices (leading
to the sign operator on the r.h.s. of the properties 
(\ref{rg_L_c_transform})-(\ref{rg_R_c_transform})), the other containing the minus
signs from the internal vertices which cancels the signs from all the contractions, see
Appendix C.

\section{Appendix: Leading order RG equations}
\label{appendix_E}

Here we prove various properties for the vertices $\bar{G}^{(1)}_{11'}$, $\tilde{G}^{(1)}_{11'}$,
and $\bar{G}^{(2)}_{11'}$, together with corresponding properties for the vertices 
$\bar{R}^{(1)}_{11'}$ and $\bar{R}^{(2)}_{11'}$, which are needed
in Sec.~\ref{sec:4.4}. For notational simplicity, we omit in the following the index $(1)$ for
all vertices.

First, we prove that the RG equations (\ref{reference_solution_so}) and (\ref{G_tilde_leading})
for the vertices $\bar{G}^{(1)}_{11'}$ and $\tilde{G}^{(1)}_{11'}$ are consistent with
the RG equation (\ref{rg_g}) for the vertex $g_{11'}$ using the relations
(\ref{ansatz_G}) and (\ref{tilde_G}) between $G$ and $g$. From (\ref{ansatz_G}) and
(\ref{rg_g}) we get
\begin{equation}
\label{eq_1}
{d\over d\Lambda}G^{pp}_{11'}\,=\,{1\over\Lambda}\,
(G^{pp}_{12}G^{pp}_{\bar{2}1'}-G^{pp}_{1'2}G^{pp}_{\bar{2}1})\quad,
\end{equation}
by simply acting on an operator and using $G_{11'}^{pp}=- G_{1'1}^{pp}$. Summing this
equation over $p$ gives the correct RG equation for $\bar{G}_{11'}$
\begin{eqnarray}
\label{eq_2}
{d\over d\Lambda}\bar{G}_{11'}\,
&=&\,{1\over\Lambda}\,\sum_p\,(G^{pp}_{12}G^{pp}_{\bar{2}1'}-G^{pp}_{1'2}G^{pp}_{\bar{2}1})\\
\label{eq_3}
&=&\,{1\over\Lambda}\,\sum_p\,(G^{pp}_{12}G^{pp}_{\bar{2}1'}-G^{pp}_{1'2}G^{pp}_{\bar{2}1}
+G^{pp}_{12}G^{\bar{p}\bar{p}}_{\bar{2}1'}-G^{\bar{p}\bar{p}}_{1'2}G^{pp}_{\bar{2}1})\\
\label{eq_4}
&=&\,{1\over\Lambda}\,\sum_p\,(G^{pp}_{12}\bar{G}_{\bar{2}1'}-\bar{G}_{1'2}G^{pp}_{\bar{2}1})\\
\label{eq_5}
&=&\,{1\over\Lambda}\,(\bar{G}_{12}\bar{G}_{\bar{2}1'}-\bar{G}_{1'2}\bar{G}_{\bar{2}1})\quad,
\end{eqnarray}
where $\bar{p}=-p$, and we have used 
$G^{pp}_{12}G^{\bar{p}\bar{p}}_{\bar{2}1'}=G^{\bar{p}\bar{p}}_{\bar{2}1'}G^{pp}_{12}
=G^{\bar{p}\bar{p}}_{1'2}G^{pp}_{\bar{2}1}$ to arrive at (\ref{eq_3}) (note that $2$ is
a summation index which can be changed to $\bar{2}$).

Multiplying (\ref{eq_1}) with $p$ and then summing over $p$ gives the RG equation
for $\tilde{G}_{11'}$
\begin{eqnarray}
\label{eq_6}
{d\over d\Lambda}\tilde{G}_{11'}\,
&=&\,{1\over\Lambda}\,\sum_p\,(pG^{pp}_{12}G^{pp}_{\bar{2}1'}-pG^{pp}_{1'2}G^{pp}_{\bar{2}1}
+pG^{pp}_{12}G^{\bar{p}\bar{p}}_{\bar{2}1'}-pG^{\bar{p}\bar{p}}_{1'2}G^{pp}_{\bar{2}1})\\
\label{eq_7}
&=&\,{1\over\Lambda}\,\sum_p\,(pG^{pp}_{12}\bar{G}_{\bar{2}1'}-p\bar{G}_{1'2}G^{pp}_{\bar{2}1})\\
\label{eq_8}
&=&\,{1\over\Lambda}\,(\tilde{G}_{12}\bar{G}_{\bar{2}1'}-\bar{G}_{1'2}\tilde{G}_{\bar{2}1})\quad,
\end{eqnarray}
and with a similiar proof we can show the same with $\bar{G}\leftrightarrow\tilde{G}$ on the r.h.s.

Next we show that the RG equation (\ref{reference_solution_observable_so})
for $\bar{R}^{(1)}_{11'}$ is consistent with
the RG equation (\ref{rg_r}) for the vertex $r_{11'}$ using the relation
(\ref{ansatz_R}) between $R$ and $r$. As usual this holds only if one acts with
the trace over the quantum system from the left. Doing this and acting with the
r.h.s. of (\ref{reference_solution_observable_so}) on an arbitrary operator $A$, we
get the correct RG equation for $\bar{R}_{11'}$
\begin{eqnarray}
\label{eq_9}
&&{1\over\Lambda}\mbox{Tr}_S(\bar{R}_{12}\bar{G}_{\bar{2}1'}-\bar{R}_{1'2}\bar{G}_{\bar{2}1})A\\
\label{eq_10}
&&=\,{i\over\Lambda}\mbox{Tr}_S(r_{12}g_{\bar{2}1'}-r_{1'2}g_{\bar{2}1}-
g_{\bar{2}1'}r_{12}+g_{\bar {2}1}r_{1'2})A\\
\label{eq_11}
&&=\,{i\over\Lambda}\mbox{Tr}_S(r_{12}g_{\bar{2}1'}-r_{1'2}g_{\bar{2}1}+
-g_{12}r_{\bar{2}1'}-g_{1'2}r_{\bar{2}1})A\\
\label{eq_12}
&&=\,i{d\over d\Lambda}\mbox{Tr}_S\,r_{11'}A\,=\,{d\over d\Lambda}\mbox{Tr}_S\,\bar{R}_{11'}A\quad.
\end{eqnarray}

For the special case of the current vertex, we prove that the ansatz (\ref{current_operator_leading})
for the current vertex fulfils the correct RG equation. For this we need the property
\begin{equation}
\label{tr_G_tilde_G_bar}
\mbox{Tr}_S\,\tilde{G}_{11'}\bar{G}_{22'}\,=\,
-\,\mbox{Tr}_S\,\tilde{G}_{22'}\bar{G}_{11'}\quad,
\end{equation}
which is based on applying several times the property $\mbox{Tr}_S\bar{G}_{11'}=0$ 
(see (\ref{G_property})) or
\begin{equation}
\label{tr_G_bar}
\mbox{Tr}_S\,G^{pp}_{11'}\,=\,
-\mbox{Tr}_S\,G^{\bar{p}\bar{p}}_{11'}\quad,
\end{equation}
with $\bar{p}=-p$. As a consequence we also get
\begin{equation}
\label{eq_12a}
\mbox{Tr}_S\,G^{pp}_{11'}G^{pp}_{22'}\,=\,
-\mbox{Tr}_S\,G^{\bar{p}\bar{p}}_{11'}G^{pp}_{22'}\,=\,
-\mbox{Tr}_S\,G^{pp}_{22'}G^{\bar{p}\bar{p}}_{11'}\,=\,
\mbox{Tr}_S\,G^{\bar{p}\bar{p}}_{22'}G^{\bar{p}\bar{p}}_{11'}\quad,
\end{equation}
and (\ref{tr_G_tilde_G_bar}) can be proven in the following way
\begin{eqnarray}
\label{eq_12b}
\mbox{Tr}_S\,\tilde{G}_{11'}\bar{G}_{22'}\,&=&\,
\mbox{Tr}_S\,\sum_{pp'}pG^{pp}_{11'}G^{p'p'}_{22'}\\
\label{eq_12c}
&&\hspace{-2cm}=\,
\mbox{Tr}_S\,\sum_{p}pG^{pp}_{11'}G^{pp}_{22'}\,+\,
\mbox{Tr}_S\,\sum_{p}pG^{pp}_{11'}G^{\bar{p}\bar{p}}_{22'}\\
\label{eq_12d}
&&\hspace{-2cm}=\,
\mbox{Tr}_S\,\sum_{p}pG^{\bar{p}\bar{p}}_{22'}G^{\bar{p}\bar{p}}_{11'}\,+\,
\mbox{Tr}_S\,\sum_{p}pG^{\bar{p}\bar{p}}_{22'}G^{pp}_{11'}\\
\label{eq_12e}
&&\hspace{-2cm}=\,
\mbox{Tr}_S\,\sum_{p}pG^{\bar{p}\bar{p}}_{22'}\bar{G}_{11'}\,=\,
-\mbox{Tr}_S\,\sum_{p}pG^{pp}_{22'}\bar{G}_{11'}\,=\,
-\mbox{Tr}_S\,\tilde{G}_{22'}\bar{G}_{11'}\quad,
\end{eqnarray}
where we have applied (\ref{eq_12a}) in (\ref{eq_12c}) and (\ref{eq_12e}).

With the help of (\ref{tr_G_tilde_G_bar}), we can now easily proof the correct RG equation for
the current vertex using the ansatz (\ref{current_operator_leading})
\begin{eqnarray}
\label{eq_12f}
{d\over d\Lambda}\mbox{Tr}_S\,\bar{I}^\gamma_{11'}\,&=&\,
c^\gamma_{11'}\mbox{Tr}_S\,{d\over d\Lambda}\tilde{G}_{11'}\,=\,
{1\over \Lambda}
c^\gamma_{11'}\mbox{Tr}_S\,(\tilde{G}_{12}\bar{G}_{\bar{2}1'}-\bar{G}_{1'2}\tilde{G}_{\bar{2}1})\\
\label{eq_12g}
&=&\,{1\over\Lambda}c^\gamma_{11'}\mbox{Tr}_S\,\tilde{G}_{12}\bar{G}_{\bar{2}1'}\,=\,
{1\over\Lambda}\mbox{Tr}_S\,\left\{c^\gamma_{12}\tilde{G}_{12}\bar{G}_{\bar{2}1'}
+c^\gamma_{\bar{2}1'}\tilde{G}_{12}\bar{G}_{\bar{2}1'}\right\}\\
\label{eq_12h}
&=&\,{1\over\Lambda}\mbox{Tr}_S\,\left\{c^\gamma_{12}\tilde{G}_{12}\bar{G}_{\bar{2}1'}
-c^\gamma_{\bar{2}1'}\tilde{G}_{\bar{2}1'}\bar{G}_{12}\right\}\\
\label{eq_12i}
&=&\,{1\over\Lambda}\mbox{Tr}_S\,\left\{c^\gamma_{12}\tilde{G}_{12}\bar{G}_{\bar{2}1'}
-c^\gamma_{1'2}\tilde{G}_{1'2}\bar{G}_{\bar{2}1}\right\}\\
\label{eq_12j}
&=&\,{1\over\Lambda}\mbox{Tr}_S\,(\bar{I}^\gamma_{12}\bar{G}_{\bar{2}1'}-
\tilde{I}^\gamma_{1'2}\bar{G}_{\bar{2}1})\quad,
\end{eqnarray}
where we have used $c^\gamma_{11'}=c^\gamma_{12}+c^\gamma_{\bar{2}1'}$ in (\ref{eq_12g})
and applied (\ref{tr_G_tilde_G_bar}) to arrive at (\ref{eq_12h}).

We study now the RG equation for the vertices $\bar{G}^{(2)}_{11'}$ and $\bar{R}^{(2)}_{11'}$ 
which are defined by (\ref{solution_G_second_order}). For both cases we arrive with the same
proof at the correct RG equation (\ref{rg_G_second_order}) in the following way (we have 
chosen $R$ for the proof)
\begin{eqnarray}
\label{eq_13}
\mbox{Tr}_S\,{d\over d\Lambda}\,R_{11'}^{(2)}\,&=&\,
-i{\pi\over 2}\mbox{Tr}_S\,\left\{
{d\bar{R}_{12}\over d\Lambda}\tilde{G}_{\bar{2}1'}+
\bar{R}_{12}{d\tilde{G}_{\bar{2}1'}\over d\Lambda}-(1\leftrightarrow 1')\right\}\\
\nonumber
&=&\,-i{\pi\over 2\Lambda}\mbox{Tr}_S\,\left\{
(\bar{R}_{13}\bar{G}_{\bar{3}2}-
\bar{R}_{23}\bar{G}_{\bar{3}1})\tilde{G}_{\bar{2}1'}+\right.\\
\label{eq_14}
&&\hspace{2cm}\left.
+\bar{R}_{12}(\tilde{G}_{\bar{2}3}\bar{G}_{\bar{3}1'}-
\bar{G}_{1'3}\tilde{G}_{\bar{3}\bar{2}})
-(1\leftrightarrow 1')\right\}\\
\nonumber
&=&\,-i{\pi\over 2\Lambda}\mbox{Tr}_S\,\left\{
\bar{R}_{12}(\bar{G}_{\bar{2}3}\tilde{G}_{\bar{3}1'}-
\bar{G}_{1'3}\tilde{G}_{\bar{3}\bar{2}})+\right.\\
\label{eq_15}
&&\hspace{2cm}\left.
+\bar{R}_{13}\tilde{G}_{\bar{3}2}\bar{G}_{\bar{2}1'}
-\bar{R}_{23}\bar{G}_{\bar{3}1}\tilde{G}_{\bar{2}1'}
-(1\leftrightarrow 1')\right\}\\
\nonumber
&=&{1\over \Lambda}\mbox{Tr}_S\,\left\{\bar{R}_{12}\bar{G}^{(2)}_{\bar{2}1'}+
\bar{R}^{(2)}_{12}\bar{G}_{\bar{2}1'}-(1\leftrightarrow 1')\right\}\,-\\
\label{eq_16}
&&\hspace{0.1cm}-i{\pi\over 2\Lambda}\mbox{Tr}_S\,
\bar{R}_{23}\left\{\tilde{G}_{\bar{3}1}\bar{G}_{\bar{2}1'}-
\bar{G}_{\bar{3}1}\tilde{G}_{\bar{2}1'}-(1\leftrightarrow 1')\right\}\quad,
\end{eqnarray}
where we have interchanged $2\leftrightarrow 3$ in the first and third term of the
r.h.s. of (\ref{eq_14}) and have added and subtracted the term 
$\bar{R}_{23}\tilde{G}_{\bar{3}1}\bar{G}_{\bar{2}1'}$ to (\ref{eq_15}). The first term of
(\ref{eq_16}) gives the correct RG equation for $R_{11'}^{(2)}$. The second term is
zero as can be seen from
\begin{eqnarray}
\label{eq_17}
&&\bar{R}_{23}\left\{\tilde{G}_{\bar{3}1}\bar{G}_{\bar{2}1'}-
\bar{G}_{\bar{3}1}\tilde{G}_{\bar{2}1'}-
\tilde{G}_{\bar{3}1'}\bar{G}_{\bar{2}1}+
\bar{G}_{\bar{3}1'}\tilde{G}_{\bar{2}1}\right\}\\
\label{eq_18}
&=&\,\bar{R}_{23}\left\{\tilde{G}_{\bar{3}1}\bar{G}_{\bar{2}1'}-
\bar{G}_{\bar{3}1}\tilde{G}_{\bar{2}1'}+
\tilde{G}_{\bar{2}1'}\bar{G}_{\bar{3}1}-
\bar{G}_{\bar{2}1'}\tilde{G}_{\bar{3}1}\right\}\\
\label{eq_19}
&=&\,\bar{R}_{23}\sum_{pp'}\left\{
p G^{pp}_{\bar{3}1}G^{p'p'}_{\bar{2}1'}-
p'G^{pp}_{\bar{3}1}G^{p'p'}_{\bar{2}1'}+
p G^{pp}_{\bar{2}1'}G^{p'p'}_{\bar{3}1}-
p'G^{pp}_{\bar{2}1'}G^{p'p'}_{\bar{3}1}\right\}\\
\label{eq_20}
&=&\,2\,\bar{R}_{23}\sum_{p}\left\{
pG^{pp}_{\bar{3}1}G^{\bar{p}\bar{p}}_{\bar{2}1'}+
\bar{p}G^{\bar{p}\bar{p}}_{\bar{2}1'}G^{pp}_{\bar{3}1}\right\}\\
\label{eq_21}
&=&\,2\,\bar{R}_{23}\sum_{p}p\left\{
G^{pp}_{\bar{3}1}G^{\bar{p}\bar{p}}_{\bar{2}1'}-
G^{\bar{p}\bar{p}}_{\bar{2}1'}G^{pp}_{\bar{3}1}\right\}\,=\,0\quad,
\end{eqnarray}
where we have interchanged $2\leftrightarrow 3$ 
in the third and fourth term of (\ref{eq_17}) and used $\bar{R}_{23}=-\bar{R}_{32}$.

Finally, we prove the properties (\ref{hermiticity_gG}) and
(\ref{hermiticity_rR}). $(g_{11'})^\dagger=g_{\bar{1}'\bar{1}}$ and 
$(r_{11'})^\dagger=r_{\bar{1}'\bar{1}}$ follow trivially from the RG equations 
(\ref{rg_g}) and (\ref{rg_r}). For the vertices $\bar{G}_{11'}$ and $\bar{R}_{11'}$ 
we use the matrix representation
\begin{eqnarray}
\label{matrix_G}
(\bar{G}_{11'})_{s_1s_1^\prime,s_2s_2^\prime}\,&=&\,
(g_{11'})_{s_1s_2}\delta_{s_1^\prime s_2^\prime}-
\delta_{s_1s_2}(g_{11'})_{s_2^\prime s_1^\prime}\quad,\\
\label{matrix_R}
(\bar{R}_{11'})_{s_1s_1^\prime,s_2s_2^\prime}\,&=&\,{i\over 2}\left\{
(r_{11'})_{s_1s_2}\delta_{s_1^\prime s_2^\prime}+
\delta_{s_1s_2}(r_{11'})_{s_2^\prime s_1^\prime}\right\}\quad,
\end{eqnarray}
and get for $\bar{G}_{11'}^\dagger$ 
\begin{eqnarray}
\nonumber
(\bar{G}^\dagger_{11'})_{s_1s_1^\prime,s_2s_2^\prime}\,&=&\,
(\bar{G}_{11'})^*_{s_2s_2^\prime,s_1s_1^\prime}\,=\,
(g_{11'})^*_{s_2s_1}\delta_{s_1^\prime s_2^\prime}-
\delta_{s_1s_2}(g_{11'})^*_{s_1^\prime s_2^\prime}\\
\nonumber
&=&\,(g^\dagger_{11'})_{s_1s_2}\delta_{s_1^\prime s_2^\prime}-
\delta_{s_1s_2}(g^\dagger_{11'})_{s_2^\prime s_1^\prime}\\
\nonumber
&=&\,(g_{\bar{1}'\bar{1}})_{s_1s_2}\delta_{s_1^\prime s_2^\prime}-
\delta_{s_1s_2}(g_{\bar{1}'\bar{1}})_{s_2^\prime s_1^\prime}\\
\label{eq_22}
&=&\,(\bar{G}_{\bar{1}'\bar{1}})_{s_1s_1^\prime,s_2s_2^\prime}\quad,
\end{eqnarray}
which proves (\ref{hermiticity_gG}). In the same way one proves
(\ref{hermiticity_gG}) for the vertex $\bar{R}_{11'}$.

\end{appendix}

\end{document}